\documentstyle[12pt,epsf]{book}

\advance \topmargin by -\headheight
\advance \topmargin by -\headsep

\evensidemargin \oddsidemargin
 \advance\hoffset by -3mm  
 \advance\voffset by  8mm  

\def\ie{{\em i.e.}}

\def\ie{\hbox{\it i.e.}}

\def\CC{{\mathchoice
{\rm C\mkern-8mu\vrule height1.45ex depth-.05ex 
width.05em\mkern9mu\kern-.05em}
{\rm C\mkern-8mu\vrule height1.45ex depth-.05ex 
width.05em\mkern9mu\kern-.05em}
{\rm C\mkern-8mu\vrule height1ex depth-.07ex 
width.035em\mkern9mu\kern-.035em}
{\rm C\mkern-8mu\vrule height.65ex depth-.1ex 
width.025em\mkern8mu\kern-.025em}}}

\def\RR{{\rm I\kern-1.6pt {\rm R}}}

\def\ZZ{{\rm Z}\kern-3.8pt {\rm Z} \kern2pt}

\def\np{Nucl. Phys.}
\def\pl{Phys. Lett.}
\def\prl{Phys. Rev. Lett.}
\def\pr{Phys. Rev.}

\def\jmp{J. Math. Phys.}

\def\mpl{Mod. Phys. Lett.}

\def\phyrep{Phys. Rep.}

\def\sjnp{Sov. J. Nucl. Phys.}

\def\atmp{Adv. Theor. Math. Phys. }
\def\jhep{J. High Energy Phys.}
\def\ptp{Prog. Theor. Phys.}
\def\jgp{J. Geom. Phys.}
\def\atmp{Adv. Theor. Math. Phys.}
\def\jmp{J. Math. Phys.}

\newcommand{\beq}{\begin{equation}}
\newcommand{\eeq}{\end{equation}}
\newcommand{\rc}{\nonumber\\}
\newcommand{\bear}{\begin{eqnarray}}
\newcommand{\eear}{\end{eqnarray}}

\newcommand\ls{\ell_s}
\newcommand\lp{\ell_p}
\newcommand\hi{{\rm i}}

\def\to{\rightarrow}

\def\dal{\dot\alpha_}

\def\to{\rightarrow}

\def\Pislash{\rlap/\Pi}
\def\sqr#1#2{{\vcenter{\vbox{\hrule height.#2pt
         \hbox{\vrule width.#2pt height#1pt \kern#1pt
            \vrule width.#2pt}
         \hrule height.#2pt}}}}
\def\dal{\mathop{\mathchoice\sqr75\sqr75\sqr{3.75}4\sqr34}\nolimits}

\newfont{\namefont}{cmr10}
\newfont{\addfont}{cmti7 scaled 1440}
\newfont{\boldmathfont}{cmbx10}
\newfont{\headfontb}{cmbx10 scaled 1728}
\renewcommand{\theequation}{{\rm\thesection.\arabic{equation}}}

\def\nonu{\nonumber}
\def\br{\begin{eqnarray}}
\def\er{\end{eqnarray}}
\def\brs{\begin{eqnarray*}}
\def\ers{\end{eqnarray*}}
\def\be{\begin{equation}}
\def\ee{\end{equation}}
\def\bd{\begin{description}}
\def\ed{\end{description}}

\newtheorem{prop}{Proposici\'on}[section]
\newtheorem{defi}{Definici\'on}[section]
\newtheorem{theorem}{Teorema}[section]
\newtheorem{lema}{Lema}[section]
\newtheorem{corolario}{Corolario}[section]

\def\bedef{\begin{defi}}
\def\endef{\end{defi}}

\def\bethe{\begin{theorem} }
\def\enthe{ \end{theorem} }

\def\beprop{\begin{prop} }
\def\enprop{ \end{prop} }

\def\belem{\begin{lema}}
\def\enlem{\end{lema}}

\def\becor{\begin{corolario}}
\def\encor{\end{corolario}}

\def\pa{\partial}

\def\a{\alpha}
\def\b{\beta}
\def\d{\delta}
\def\D{\Delta}

\def\e{\epsilon}
\def\g{\gamma}
\def\G{\Gamma}

\def\l{\lambda}
\def\L{\Lambda}
\def\k{\kappa}
\def\m{\mu}
\def\n{\nu}
\def\o{\omega}
\def\O{\Omega}
\def\ol{\overline}
\def\p{\phi}
\def\P{\Phi}
\def\pa{\partial}
\def\r{\rho}

\def\r{\rho}
\def\s{\sigma}
\def\S{\Sigma}
\def\t{\tau}
\def\th{\theta}

\begin{document}
%


\begin{titlepage}

\begin{center} 
\large \sf  UNIVERSIDADE DE SANTIAGO DE COMPOSTELA

\vspace{.3cm}

\large Departamento de F\'\i sica de Part\'\i culas 
\end{center}

\vspace{5cm}

\begin{center} 
\LARGE  \bf WORLDVOLUME DYNAMICS 

OF BRANES
\end{center}

\vspace{3cm}

\begin{center} 
Tese presentada para optar \'o gao de Doutor en F\'\i sica.
\end{center}

\vspace{4cm}

\begin{center} 
{\sf\bf \large Jos\'e Manuel Camino Mart\'\i nez}

\sf Santiago de Compostela, outubro 2002.
\end{center} 

\end{titlepage}

\pagestyle{empty}

\mbox{}

\newpage

\setcounter{page}{1}

{\Large \sf  Agradecimientos}

\vspace{1cm}

\sf 

En primer lugar quiero agradecer a Alfonso V\'azquez Ramallo la oportunidad 
que me dio de realizar la tesis bajo su direcci\'on. Nunca olvidar\'e el 
apoyo que me brind\'o en todo momento as\'\i\ como su dedicaci\'on, inter\'es y 
paciencia. Quiero profesar mi admiraci\'on personal hacia Alfonso, tanto  por 
su capacidad profesional como por su calidad humana. 
 
Con Jos\'e S\'anchez de Santos y \'Angel Paredes colabor\'e durante la realizaci\'on de 
esta tesis y quiero expresar mi especial aprecio por ambos. Con Josi\~no tuve la 
oportunidad de trabajar codo con codo y guardo muy grato recuerdo de su cordialidad. 
Trabajar con \'Angel fue un verdadero est\'\i mulo. 

Mis compa\~neros de despacho C\'esar, Marta, Max, David, Ver\'onica, \'Angel, Teresa 
y Diego se han convertido en verdaderos amigos para m\'\i . Especialmente C\'esar, Marta, 
David y Max con quienes compart\'\i\ muchas de mis inquietudes y a quienes les debo sabios 
consejos y apoyo en momentos cruciales. Nunca olvidar\'e los piques al futbol\'\i n y 
nuestros combates de ``boxeo'' en los que siempre, como no, ganaba Marta (aunque no 
tuviera la raz\'on). 

Quiero mencionar tambi\'en a todos los dem\'as amigos con los que he podido 
contar durante todos estos a\~nos: Jose, todo un ejemplo que nunca dejar\'a de 
sorprenderme; Luis, que convierte todo instante en una alegr\'\i a; Julio y Manuel, 
que siempre est\'an ah\'\i ; Ana, Elena y Fuco con quien pas\'e tan buenos momentos; 
Jorge, con su optimismo y buen humor; Eduardo, con nuestras interminables conversaciones; 
Patricia y Ba\~na, mis queridos coleguillas de piso que voy a echar de menos; las 
chicas de Qu\'\i mica, Ana y Bel\'en, Antonina y F\'elix  y todos mis amigos de 
F\'\i sica y de Fonseca. Tampoco me olvido de todos mis primos y t\'\i os a los que 
alg\'un d\'\i a les har\'e ver la utilidad de lo que hago.

A mis padres y a mis hermanos, que siempre me han apoyado y a quienes se lo debo todo, 
nunca podr\'e agradec\'erselo lo suficiente. Mis hermanos siempre ser\'an mis mejores 
amigos y todos estos a\~nos en Santiago con Luc\'\i a, Ali y Mar\'\i a Luisa fueron 
incre\'\i bles. 

Dolores le da sentido a todo lo que hago y es en definitiva mi musa. Quiero agradecerle 
el haber estado siempre junto a m\'\i.  

Quiero dedicar especialmente este trabajo a mis abuelos Luisa y Germ\'an, que me han 
mostrado hasta donde puede llegar la bondad humana.

\newpage

\mbox{}
\newpage


\rm

\setcounter{footnote}{0}

\tableofcontents

\pagestyle{headings}

\setcounter{equation}{0}


\chapter*{Introduction}
\addcontentsline{toc}{chapter}{Introduction}

\medskip

\thispagestyle{myheadings}
\markboth{INTRODUCTION}{}

String theory is a candidate for a fundamental theory that unifies 
all the forces present in Nature. Three of them, electromagnetic, weak 
and strong interactions, are well described by a consistent quantum 
theory, the Standard Model, which agrees with experiments. However, the 
fourth interaction, gravity, is not included in the Standard Model. Gravity 
is described as the dynamics of spacetime by the theory of General Relativity, 
which is a classical theory. This interaction can be neglected at the energy 
scale of the interactions of the Standard Model but at higher energies both 
quantum and gravitational effects are relevant, making thus necessary 
a quantum theory of gravity. The energy at which quantum effects are not 
negligible in the description of spacetime defines the Planck scale, which 
is of the order of $10^{19} GeV$. However, when one tries to quantize gravity 
one finds a non-renormalisable quantum field theory, and a new physics is 
expected to appear at this energy scale.

In string theory particles appear as vibration modes 
of the string, and among these modes there is a massless one with spin 2 
which can be identified with the graviton. These strings propagate in spacetime 
describing surfaces, called world-sheets, which generalize the world-line 
described by point particles. The string interactions are constructed by 
joining or splitting these world-sheets, and then the interaction is smeared 
in spacetime. This is an intrinsically perturbative picture which generalizes 
the one of point particles. 

The formulation of a quantum theory of relativistic strings generalizes also the 
one of point particles. The action describing the propagation of a string in 
spacetime has as dynamical fields a set of real functions which describe 
the embedding of the string in spacetime. This theory is required to have 
the symmetries of general relativity and the Standard Model, that is, 
diffeomorphism invariance and gauge symmetry. However, more symmetries appear 
in this formulation such as conformal symmetry and supersymmetry, the latter being 
introduced to include fermions in the world-sheet. Superstring theory is only 
consistent in ten spacetime dimensions and one finds that there are five 
consistent superstring theories, all different at the perturbative level. 
Moreover, in order to get an effective four dimensional physics one has to 
compactify the six extra dimensions, which can be performed in many 
inequivalent ways without a preferred one. Thus, all this is against the 
hope that string theory is a fundamental theory of Nature. 

The problem of the number of string theories is solved by duality. String 
dualities relate different string theories and show that they are unified in 
a single theory. They also provide information on the strong coupling regimes 
of the different string theories, since some of these dualities map the weak 
coupling regime of a superstring theory to the strong coupling regime of 
another one. The different string theories have been conjectured to be different 
limits of an underlying eleven-dimensional theory called M-theory.

One can obtain effective actions describing the dynamics of the string modes. If 
one computes the effective action for massless modes with small energies 
compared to the string tension, one gets the different ten dimensional 
supergravities. These actions have classical solutions which can be identified 
as solitons of the full string theory. Some of these string solitons can be 
interpreted as extended p-dimensional objects which are called p-branes. These 
objects are non-perturbative, since their mass goes as the inverse of the string 
coupling constant, and are important tools to test dualities among different 
regimes of string theories. Indeed, p-branes are useful because they are 
supersymmetric objects and some of their properties do not vary when the 
string coupling constant changes. A particular case of p-branes are the 
Dp-branes, which are extended objects carrying charge under certain fields 
appearing in the supergravity actions.

On the other hand, D-branes can be identified as hyperplanes on which open 
strings with certain boundary conditions can end. Indeed, when one computes 
the masses and charges of these objects they match with the ones of 
solitonic extended Dp-brane solutions. D-branes become then true dynamical objects, 
their fluctuations 
in shape and form being determined by the dynamics of the strings ending on them. 
One can use this to compute an effective action for the fluctuations of the 
branes. In this thesis we will use the low energy effective actions for the 
branes of string and M-theory to study some relevant configurations of brane 
probes in different supergravity backgrounds, and some aspects of the dynamics 
of branes will be discussed.

This Ph.D. thesis is based on papers \cite{Baryon, Flux1, Flux2, NCgrav} and 
\cite{MTgrav}. It is structured as follows. In chapter 1 we give an introduction 
to string theory, discussing briefly the dualities among the different string 
theories and the appearance of M-theory. The low-energy effective actions for 
the two possible closed string theories will be reviewed and some of their BPS 
solutions will be listed. We are interested in extremal solutions corresponding 
to stacks of coincident branes and in some of their non-threshold bound states, 
since these solutions will be used as supergravity backgrounds in which our 
brane probes propagate. To end the chapter some aspects of Dp-branes are 
discussed and the Maldacena conjecture is introduced.

In chapter 2, the supersymmetric effective action for bosonic Dp-branes will be 
constructed. This action posesses a local fermionic symmetry, called kappa 
symmetry, which guarantees that the number of bosonic and fermionic degrees 
of freedom coincide. It will also be presented a proposal for the non-abelian 
generalization of the above action which describes the dynamics of multiple 
coincident D-branes. The dynamics which results from this action implies that 
Dp-branes can be polarized by external fields. This is the so called Myers 
polarization effect. To end the chapter, it will be given a kappa-symmetric 
Born-Infeld-like worldvolume action for the bosonic M5-brane of M-theory.

In chapter 3, we study the embedding of D(8-p)-branes in the background 
geometry of parallel Dp-branes for $p\le 6$. The embeddings we obtain 
represent branes joined by tubes. By analyzing the energy of these tubes we 
conclude that they can be regarded as bundles of fundamental strings. Our 
solution provides an explicit realization of the Hanany-Witten effect.

In chapter 4, we study some wrapped configurations of branes in the near-horizon
geometry of a stack of other branes. The common feature of all  the cases 
analyzed is a quantization rule and the  appearance of a finite number of
static configurations in which the branes  are partially wrapped on
spheres. The energy of these configurations can be given in closed form
and the analysis of their small oscillations shows that they are stable.
The brane configurations found admit the interpretation of bound states 
of strings (or M2-branes in M-theory) which extend along the unwrapped 
directions. We check this fact directly in a particular case by using the 
Myers polarization mechanism.

In chapter 5, we study giant graviton configurations of brane probes in the 
background created by a stack of non-threshold bound states.  We show that
for a particular value of the worldvolume gauge field there exist configurations
of the brane probe which behave as  massless particles and can be interpreted as
gravitons blown up into a fuzzy sphere and extended over a noncommutative plane. 
We check this behaviour by studying the motion and energy of the brane and by 
determining the way in which supersymmetry is broken by the probe as it moves 
under the action of the background.


\setcounter{chapter}{0}
\chapter{Superstrings and M-theory}
\medskip

The aim of this chapter is to make a very short introduction to 
string theory and M-theory. Special attention is paid to p-branes.
Section \ref{secpert} is devoted to the presentation of the five 
different superstring theories and their respective low energy 
effective actions for the massless fields. The standard references to this 
material are \cite{GSW, Polchinski}. In section \ref{secdual} a short 
description of the dualities among the different string theories is 
presented and this leads to the appearance of M-theory in section 
\ref{secMT}. In section \ref{secstrsol} some solitonic extended solutions 
of the superstring effective actions and 11 dimensional supergravity are 
listed. These solutions can be interpreted as extended objects (p-branes) 
and are BPS states preserving some fraction of the target space supersymmetries. 
Some bound states of p-branes are also written, since they will be of interest 
in the following chapters. In section \ref{secbranes} some facts about D-branes 
are discussed and finally, in section \ref{secmaldacena}, we comment about 
the Maldacena conjecture.

\medskip
\section{Perturbative string theory} \label{secpert}
\medskip                                             
\setcounter{equation}{0}

We want a classical and quantum description of a string moving in 
a flat spacetime  of D dimensions. A one-dimensional object sweeps out 
a two-dimensional worldsheet, which can be described in terms of two 
parameters $X^{\m}(\t,\s)$. The simplest reparametrization invariant 
action one can construct, the Nambu-Goto action, is proportional to 
the area of the world-sheet,
\be
S\,=\,{-1\over{2\pi \a^{\prime}}}\,\int_{\S}\,d\t d\s\,
\sqrt{-det(h_{ab})}\,,
\label{strng}
\ee
where $\S$ denotes the world-sheet and $h_{ab}=\partial_aX^{\m}\,
\partial_bX^{\n}\,\eta_{\m\n}$ 
is the induced metric on $\S$. The tension of the string is 
$T_f=1/2\pi\a^{\prime}$. One can simplify the Nambu-Goto action by 
introducing an independent world-sheet metric $\g_{ab}(\t,\s)$. In 
fact, the action
\be
S\,=\,{-1\over{4\pi \a^{\prime}}}\,\int_{\S}\,d\t d\s\,
\sqrt{-\g}\,\g^{ab} \,\partial_aX^{\m}\,\partial_bX^{\n}\,\eta_{\m\n}\,,
\label{strpol}
\ee
is equivalent to (\ref{strng}) when the equation of motion 
of $\g_{ab}$ is used in (\ref{strpol}). The variation $\d_{\g}S=0$ implies 
that $\g_{ab}(\t,\s)$ is proportional to the induced metric:
\be
\g_{ab}(-\g)^{-1/2}=h_{ab}(-h)^{-1/2}\,.
\label{strind}
\ee
The action (\ref{strpol}) is called Polyakov action.
Apart from the reparametrization invariance this action has D-dimensional 
Poincar\'e invariance and two-dimensional Weyl invariance
\br
{X^{\prime}}^{\m}(\t,\s)&=&X^{\m}(\t,\s)\,, \nonu\\
\g^{\prime}_{ab}(\t,\s)&=&
e^{2\omega(\t,\s)}\g_{ab}(\t,\s)\,, 
\er
for arbitrary $\omega(\t,\s)$. The Weyl invariance has no analog in (\ref{strng}) 
and this is because eq. (\ref{strind}) determines $\g_{ab}$ up to a local 
rescaling, so Weyl-equivalent metrics correspond to the same embedding in 
spacetime. One can use two reparametrizations and one Weyl rescaling to fix the 
three independent parameters of $\g_{ab}$ such that $\g_{ab}=\eta_{ab}$, 
where $\eta_{ab}$ is the two dimensional Minkowski metric. The action 
(\ref{strpol}) then becomes:
\be
S\,=\,-{T_f\over 2}\,\int_{\S}\,d\t d\s\,
\,\eta^{ab} \,\partial_a X^{\m}\,\partial_b X^{\n}\,\eta_{\m\n}\,,
\label{strgfpol}
\ee
Varying $X^{\m}$ in (\ref{strgfpol}) one gets
\be
\d S=\frac{1}{2\pi\a^{\prime}}\int d\t d\s\,
\d X_{\m}\,{\dal}\,X^{\m}\,-\,
\frac{1}{2\pi\a^{\prime}} \int d\t \,
\d X^{\m}\partial_{\s}X_{\m}\vert_{\s=0}^{\s=\pi}\,,
\ee
where we have taken the coordinate region corresponding to the world-sheet 
to be $-\infty\le \t\le\infty$, $0\le \s \le\pi$.
The volume term gives the equation of motion
\be
{\dal}\,X^{\m}=0\,.
\ee
The second term is a surface term. It vanishes automatically for 
a closed string, since in this case the fields are periodic:
\br
X^{\m}(\t,\pi)&=&X^{\m}(\t,0) \nonu\\
\partial_{\s}X^{\m}(\t,\pi)&=&\partial_{\s}X^{\m}(\t,0) \nonu\\
\g_{ab}(\t,\pi)&=&\g_{ab}(\t,0)
\er
In this way the endpoints are joined to form a closed loop and there is no 
boundary. For 
the open string one can have two sets of boundary conditions:
\br
\partial_{\s}X^{\m}(\t,0)=\partial_{\s}X^{\m}(\t,\pi)=0\,,\ \ \ \ &&{\rm 
Neumann}\,,\nonu\\
\nonu\\
\d X^{\m}(\s=0)=\d X^{\m}(\s=\pi)=0\,,\ \ \ \ &&{\rm Dirichlet}\,. \nonu\\
\label{strbc}
\er
The Neumann boundary conditions for the open string mean that there is no 
momentum flow through the endpoints of the string. Dirichlet boundary 
conditions break Poincar\'e invariance and require the endpoints of the 
string to be fixed to a hyperplane. 

Different boundary conditions correspond to different spectra. The massless 
spectrum of the closed string contains a particle of spin two which can be 
identified with the graviton, an scalar field $\p$ (the dilaton) and an 
antisymmetric two-form $B_{\m\n}$, the NSNS two-form. The spectrum of an open 
string moving in flat spacetime with Neumann boundary conditions in 
all directions contains a vector field $A_{\m}$. A theory of open strings 
contains necessarily closed strings since one can join the endpoints of 
two open strings to form a closed string.

One can include fermions in the worldsheet by making the action 
supersymmetric. The world-sheet action for the superstring, with 
the world-volume metric gauge fixed to flat Minkowski metric, is
\be
S=-{1\over 4\pi \a'}\int d\t d\s\,
\left\{ \eta^{ab} \partial_a X^\m \partial_b X^\n
+i \bar{\psi}^\m \g^a \partial_a \psi^\n \right\}\,\eta_{\m\n}\,,
\label{strssa}
\ee 
where $\psi^{\m}=\psi^{\m}(\s)$ are the world-sheet spinors carrying 
a vectorial index of the SO(1,9) spacetime Lorentz group. This action 
is invariant under the supersymmetry transformations
\be
\d_{\epsilon}X^{\m}=i\bar\epsilon\psi^{\m}\,,\ \ \ \ \ \ \ \ 
\d_{\epsilon}\psi^{\m}=\g^a\partial_a X^{\m}\epsilon\,.
\ee
This symmetry closes on-shell to the momentum operator. 
In two dimensions the spinors $\psi$ can be Majorana-Weyl, and if 
we denote by $\g_2$ the chirality operator in two dimensions, then the 
spinor $\psi$ splits in two chiral components, $\psi=\psi_L+\psi_R$ or, 
in two component notation
\be
\psi=\left(\begin{array}{c} \psi_R \\ 
                            \psi_L 
           \end{array}
     \right)\,.
\ee
The fermionic term of (\ref{strssa}) can be rewritten as:
\be
-{i\over 2\pi \a'}\int d\t d\s\,\left\{
\psi^\m_R \partial_+ \psi^\n_R+ \psi^\m_L \partial_- \psi^\n_L \right\}
\,\eta_{\m\n}\,,
\label{strfer}
\ee
in terms of the light-cone coordinates $\s^{\pm}=\t\pm\s$. The equations
of motion $\partial_+\psi_R^{\m}=0$, $\partial_-\psi_L^{\m}=0$ 
imply that the fermions with positive and negative chirality
are respectively right and left moving. The boundary condition 
for the case of the open superstring is
\be
0=(\psi_L^{\m}\d{\psi_L}_{\m}-\psi_R^{\m}\d{\psi_R}_{\m})|_{\s=0}^{\s=\pi}\,\,.
\ee
One can impose two different sets of boundary conditions for the fermionic 
fields: $\psi_R=\pm \psi_L$. If at both ends of the open string we have 
the condition with the same sign, the fermionic fields $\psi^\mu$ will
be integer moded. This is called the Ramond (R) sector. If, on the contrary,
they have opposite sign, the $\psi^\mu$ will be half-integer moded. This is 
called the Neveu-Schwarz (NS) sector.

For a closed string we have two possible choices of the periodicity condition 
for the fermions: $\psi^\mu_R(\sigma=2\pi)=\pm \psi^\mu_R(\sigma=0)$, and 
similarly for $\psi^\mu_L$. The $+$ and $-$ sign correspond respectively to the 
R or NS sector respectively. However, we have now two independent choices to 
make. There will be thus four different sectors, namely NS-NS, R-R, R-NS and 
NS-R.

In the R sector of the open string the ground state has to be a 
representation of the algebra of the zero modes of the fermionic 
fields $\psi^\mu_0$, which is actually the Clifford
algebra for the space-time Lorentz group $SO(1,9)$. 
If we classify the states in
terms of the representations of the little group $SO(8)$, we see
that the ground state of the R sector can be chosen to be
either in the ${\bf 8}_+$ or in the ${\bf 8}_-$ representation 
(which differ by their chirality). On the other hand, the ground 
state of the NS sector has a negative vacuum energy of $-{1\over 2}$ 
(when the minimal gap for the bosonic fields is 1), and it is a 
singlet of $SO(8)$. This state is hence a tachyon. When acting on 
it with a $\psi_{-{1\over 2}}$ mode, a massless
state is created with the indices of the ${\bf 8}_v$ vectorial 
representation of $SO(8)$.

Spacetime supersymmetry requires the same number of on-shell bosonic
and fermionic degrees of freedom at each mass level. This is achieved 
when the GSO projection is performed, which basically corresponds to set 
to zero all the states in the NS sector which are created by acting with 
an even number of fermionic creation operators on the vacuum, and fixes 
the chirality of the R sector of the ground state. The tachyon is then 
eliminated from the spectrum. The massless sector of the open superstring 
consists of the states ${\bf 8}_v+{\bf 8}_+$ and realize  
${\cal N}=1$ $D=10$ chiral supersymmetry. They correspond to a 
massless vector boson and its fermionic superpartner. The fermionic 
superpartner of the gauge boson is a ten dimensional spinor with 16 
independent components, which reduce to 8 when the equations of motion 
are used, matching the 8 physical components of the gauge field.

In the closed superstring theories the right moving
and the left moving sectors decouple, and each one is quantized in
exactly the same way as for the open string.
The massless sector of a closed string theory is then a tensor product
of two copies of the open string massless sector. There are however two
different tensor products that one can make, depending on whether one
takes the Ramond ground states of the opposite chirality on the two sides or
of the same chirality. Since the effective theory of the open superstring has 
${\cal N}=1$ spacetime supersymmetry, we will have a theory with ${\cal N}=2$ 
supersymmetry. For the closed string we will have then 32 independent 
spinor components. The GSO projection acts independently on each sector 
of the two copies of the open superstring, and then one can choose the 
R sectors in the left and right movers to have the same or opposite chirality.

\subsubsection{Type IIA}
\medskip  

This theory corresponds to choosing the R states with opposite chirality and 
then the spacetime theory is non-chiral. The
massless particles are classified along the following representations
of $SO(8)$:
\br
({\bf 8}_v+{\bf 8}_+) \otimes ({\bf 8}_v + {\bf 8}_-)&=&
({\bf 1}+ {\bf 28} +{\bf 35}_v)_{NSNS} + ({\bf 8}_v + {\bf 56}_v)_{RR}
\nonu \\
& & +({\bf 8}_+ +{\bf 56}_-)_{NSR} + ({\bf 8}_- + {\bf 56}_+)_{RNS}.
\label{striiaspc}
\er
The bosonic particles of the NSNS sector are respectively a scalar field 
related to the dilaton $\p$, a rank-two antisymmetric tensor field 
$B^{(2)}_{\mu\nu}$ and a rank-two symmetric traceless tensor related to 
the metric $g_{\mu\nu}$, the graviton. 
In the RR sector we find a vector and a 3-index antisymmetric tensor, 
which correspond
to the fields $C_\mu^{(1)}$ and $C_{\m_1\m_2\m_3}^{(3)}$. The space-time 
fermions are in the RNS and NSR sectors, the ${\bf 56}_\pm$ representations 
corresponding to the two gravitini. Thus the theory at the massless 
level has ${\cal N}=2$ spacetime supersymmetry. Moreover, since the 
gravitini have opposite chiralities, the theory is non-chiral. The action 
for a ten dimensional non-chiral theory with ${\cal N}=2$ and gravity is 
completely fixed and is called type IIA supergravity. This action reproduces 
the interactions between the massless fields computed from string theory 
amplitudes. The effective action for the massless modes can be also obtained 
from the string world-sheet action in a curved background
and requiring that conformal invariance is not broken quantum mechanically,
\ie , the $\b$-function must vanish. The action for the bosonic sector is:
\br
I_{IIA}={1\over 16\pi G_N}\left[\int d^{10}x\ \sqrt{-g}\,\left\{ e^{-2\phi}
\left({\cal R}+4(\partial \phi)^2 -{1\over 12} H_{(3)}^2\right) -{1\over 4} 
F_{(2)}^2-{1\over 48} F_{(4)}^{2}\right\}\right.+\rc\rc
\left.+\,{1\over 2} \int B_{(2)} \wedge dC^{(3)} 
\wedge dC^{(3)} \right]\,.\ \ \ \ \ \ \ \ \ \ \ \rc
\label{striiaact}
\er
We have defined the field strengths in the following way: 
\br
&&H_{(3)}=dB_{(2)}\,, \nonu\\
&&F_{(2)}=dC^{(1)}\,, \nonu\\
&&F_{(4)}=dC^{(3)}\,-\,H_{(3)}\wedge C^{(1)} \,. 
\er
The metric in the action above is the string metric, which appears
in the string action in a curved background. The fields coming from 
the NSNS sector have a $e^{-2\phi}$ factor in front of their kinetic term, 
while the RR fields not.
The Newton constant appearing in (\ref{striiaact}) depends on the only 
arbitrary dimensionful parameter of the string action, $\a'$, through 
$G_N \sim (\a')^4$. The string coupling constant is actually the 
vacuum expectation value of the dilaton, $g_s=e^{\langle \phi \rangle}$. 
It is thus determined dynamically. The action (\ref{striiaact})
is the leading term in both the $\a'$ and the $g_s$ expansion.

\subsubsection{Type IIB}
\medskip  

In this theory both left and right moving sectors have the same chirality.
The massless particles are classified in representations of $SO(8)$:
\br
({\bf 8}_v+{\bf 8}_+)\otimes ({\bf 8}_v+{\bf 8}_+)&=&
({\bf 1}+{\bf 28}+{\bf 35}_v)_{NSNS}+({\bf 1}+{\bf 28}+{\bf 35}_+)_{RR}
\nonu\\
& & +({\bf 8}_- + {\bf 56}_+)_{NSR} + ({\bf 8}_- + {\bf 56}_+)_{RNS}.
\label{striibspc}
\er
The NSNS sector is exactly the same as the one of the type IIA superstring.
The RR sector is different. It contains a
scalar (or 0-form) $\chi$, a 2-form $C_{\m_1\m_2}^{(2)}$ and a 4-form 
potential $C_{\m_1\dots\m_4}^{(4)}$. The field strength of the 4-form potential 
has to be self-dual in order that $C^{(4)}$ fits into the ${\bf 35}_+$ 
representation. The space-time fermions of the NS-R and R-NS sectors are two 
gravitini, now of the same chirality. We have thus a chiral ten-dimensional 
${\cal N}=2$ theory with gravity. This is type IIB supergravity \cite{SUSYIIB}. 
One can write 
an action that reproduces the field equations of motion but the condition of 
self-duality must be imposed by hand. The bosonic truncation of the low-energy 
effective action of the type IIB superstring is:
\br
I_{IIB}={1\over 16 \pi G_N}\left[\int d^{10} x\ \sqrt{-g} \left\{ e^{-2\phi}
\left( {\cal R}+4(\partial \phi)^2 -{1\over 12 } H_{(3)}^2\right)-{1\over 2}
(\partial \chi)^2\,- \qquad \qquad \right.\right.\rc\rc
\left. \left. -\,{1\over 12}F_{(3)}^{2}-{1\over 240}F_{(5)}^{2} \right\} +
\int C^{(4)} \wedge dC^{(2)} \wedge dC^{(2)} \right]\,,\ \ \ \ \ \ \ \ 
\ \ \ \ \ \ \ \ \ \ \ \rc
\label{striibact}
\er
where the RR field strengths are defined by 
\br
&&F_{(3)}=dC^{(2)}-\chi H_{(3)}\,, \nonu\\
&&F_{(5)}=dC^{(4)}-\,H_{(3)}\wedge C^{(2)}\,,\ \ \ \ \ F_{(5)}={}^*F_{(5)}\,.
\er

\subsubsection{Type I}
\medskip

This is a theory of open and closed non-oriented strings. The massless 
spectrum of the type I theory contains the spectrum of open non-oriented 
strings and closed non-oriented strings. The type II closed strings 
cannot be coupled consistently to the open strings, since they have ${\cal N}=2$
supersymmetry whereas the massless modes of the open superstring realize 
${\cal N}=1$ supersymmetry. It is necessary to divide by two the supersymmetry 
of the closed theories. This can be done if we require the closed strings to be 
non-oriented. To construct a theory of unoriented strings we define the 
world-sheet parity operator $\O$, which changes the orientation of the string 
($\s\rightarrow 2\pi-\s$ for the closed string and $\s\rightarrow\pi-\s$ for 
the open string). This 
operation interchanges the left and right moving sectors of the world-sheet of 
closed strings and interchanges the two ends of the open string. The parity 
$\O$ is a symmetry of the type IIB superstring, since left and right movers have 
the same chirality. Then if we project the spectrum by $\O$ we are left with 
invariant states only, corresponding to a theory of unoriented strings. Only 
half of the massless states survive:
\be
{({\bf 8}_v+{\bf 8}_+)\otimes ({\bf 8}_v + {\bf 8}_+)\over \O} =
({\bf 1}+{\bf 35}_v)_{NSNS}+({\bf 28})_{RR}+({\bf 8}_- +{\bf 56}_+)_{RNS-
NSR},
\label{strispc}
\ee
We are left with a graviton, a scalar and an antisymmetric tensor in the 
bosonic sector and a Majorana-Weyl gravitino and a Majorana-Weyl fermion 
completing the (chiral) ${\cal N}=1$ supergravity multiplet in ten dimensions.

One must add  open strings to achieve consistency at the quantum level. They 
originate a gauge sector in the theory since they can carry charges at their 
endpoints in way consistent with spacetime Poincar\'e invariance and world-sheet 
conformal invariance. This is performed by adding $n$ non-dynamical degrees of 
freedom at the ends of the string labeled with an index $i=1\dots n$. These are 
called Chan-Paton factors. One can define an action with a $U(n)$ symmetry in 
these factors. If the factors are taken to be in the fundamental representation
of the $U(n)$ group, an open string state becomes an $n\times n$ matrix 
transforming under $U(n)$ in the adjoint representation. The massless vector 
of the string spectrum becomes a $U(n)$ Yang-Mills field in the low energy 
action. If the string is unoriented, the group becomes either $SO(n)$ or 
$Sp(n)$ but the only anomaly-free theory is the one with the group $SO(32)$. 
Thus, the low-energy effective action of the type I superstrings is  
${\cal N}=1$ supergravity in 10 dimensions coupled to a super Yang-Mills 
theory \cite{Sgrav+YM} with gauge group $SO(32)$. It has only one free 
parameter, $\a'$ and the bosonic part of the action has the same structure 
as the type II ones, with a $e^{-2\phi}$ factor in front of the NSNS fields, 
and nothing in front of the RR one.

\subsubsection{Heterotic Strings}
\medskip

This is a theory of closed oriented strings. This construction exploits the 
fact that in closed string theories the quantization of the right and 
left moving sectors can be carried out independently. The heterotic string 
is an hybrid of a bosonic string in one sector and a superstring in the other.
For both sectors, the quantization is consistent if the central charge of the 
corresponding Conformal Field Theory vanishes. 
This means that the central charge for bosons and fermions must cancel the one 
coming from the ghosts, which is -26 for the bosonic string and -15 for the 
superstring. Each boson and each fermion contributes respectively 1 and 
$1\over 2$ to the central charge and then, there must be 26 bosons for the 
bosonic 
string, and 10 bosons for the superstring. One can consider that supersymmetry 
is present only in, say, the right moving sector. As the number of bosons is 
the same as the number of dimensions of the spacetime we have that the 
superstring is only consistent in ten dimensions. However the left sector is 
only consistent in 26 dimensions. We must add something to make this sector 
consistent in ten dimensions. We use the freedom to choose the fields
which will build up the $c=26$ central charge in ten dimensions. Ten of these 
fields are the left-moving part of the bosonic embedding coordinates.
In order to complete $c=26$ we add 32 chiral fermions carrying an $SO(8)$ 
scalar representation. In this new fermionic 
sector one can impose periodic or anti-periodic boundary conditions. If we 
choose the same for all the fermions then the system is invariant under 
$SO(32)$ rotations. This is the $SO(32)$ heterotic string. If we 
had imposed periodic boundary conditions in half of the fermions and 
anti-periodic in the other half we would have the $E_8\times E_8$ heterotic 
string. Performing the tensorial product of the right moving 
sector with the left moving space-time vector we get:
\be
({\bf 8}_v+{\bf 8}_+)_R\otimes ({\bf 8}_v)_L =
({\bf 1}+{\bf 28}+{\bf 35}_v)_B+({\bf 8}_- +{\bf 56}_+)_F\,,
\label{strhetspc}
\ee
where all the particles in the r.h.s. are singlets of the $SO(32)$ or 
$E_8\times E_8$. The remaining left moving massless states are in the
adjoint of one of the two groups above, and are $SO(8)$ scalars.
We obtain the ${\cal N}=1$ supergravity multiplet plus ${\cal N}=1$ vector 
multiplet of an $SO(32)$ or $E_8\times E_8$ gauge theory. The massless state 
content of the heterotic string is the same
as the one of the type I superstring. The low-energy effective action for 
the heterotic strings is then ${\cal N}=1$ supergravity coupled to super Yang-Mills 
with the Yang-Mills fields in the adjoint representation. There is an important 
difference with the other string theories: there are no RR fields in the 
spectrum.

We have briefly presented the five consistent supersymmetric string theories. All 
them seem to be unrelated. However we restricted ourselves to the perturbative 
regime of the theories and we do not have a description at strong coupling. In 
section \ref{secdual} we will see that all the string theories are related to each 
other by dualities. Some of these dualities relate the perturbative description 
of a theory with the non-perturbative regime of the other, thus giving information 
of this regime without an explicit formulation. The picture left by the dualities 
among the different string theories leads to the conjecture that the five 
string theories can be interpreted as different limits of an underlying theory, 
which is M-theory.

\medskip
\section{String dualities} \label{secdual}
\medskip                                             
\setcounter{equation}{0}

Two theories are related by a duality if there is a one-to-one map between 
their physical spectra, and if their dynamics are also equivalent. They 
are thus two different descriptions of the same physics. 

A theory is described by a set of parameters called moduli. For the case of 
string theory these are the string coupling constant, the parameters of the 
background metric and the value of the remaining background fields. Dualities 
are specified by some sets of rules relating the two moduli spaces of the 
theories such that some region in the moduli space of a string theory is 
mapped to another region of the moduli space of the same or another theory. 
In the former case the duality is a symmetry of the theory. Some dualities 
involve a map between a weak and a strongly coupled theory, and thus only 
the knowledge of non-perturbative regime could establish the existence of 
such a duality. 

A particular example of duality is the electric-magnetic duality of the 
Maxwell theory, in which two different regimes of the same theory are related. 
Maxwell's equations of motion are invariant under the change of the electric 
and magnetic fields and the simultaneous change of the sources (if we are not 
in the vacuum). The sources for magnetic flux are the magnetic monopoles. They 
are solutions of the classical equations of motion of Yang-Mills theories. These 
solutions satisfy a BPS bound in their masses which relates the mass 
of the monopole to its charge through a proportionality relation. Otherwise, 
the coexistence of electric and magnetic charges is quantum mechanically 
consistent if the Dirac quatization condition $eg=2\pi n, n\in\ZZ$ is satisfied. 
This means that the monopole is a solitonic particle with its mass going as 
$1/g_{YM}$, where $g_{YM}$ is the coupling constant of the theory. Thus in the 
weakly coupled theory the monopoles are very massive particles. If one goes to 
strong coupling the monopoles become very light. 

The electric-magnetic duality maps particles of the perturbative spectrum such 
as the bosons of the broken gauge symmetry to magnetic monopoles (one has to 
consider ${\cal N}=4$ supersymmetric Yang-Mills theory for this map to be 
consistent). This map takes the coupling constant $g_{YM}$ to its inverse 
$g_{YM}\rightarrow 1/g_{YM}$. Thus there is a duality symmetry of the theory 
which interchanges strong with weak coupling and solitons with perturbative 
states.

The proof of the duality discussed above would involve the knowledge of the 
strong coupling dynamics of the theory. However, supersymmetry gives some 
information about this regime. Extended supersymmetry allows the 
presence of central charges in the superalgebra and this implies the 
existence of massive BPS states which preserve some of the supersymmetries and 
fit into short massive supermultiplets. The key point is that the dimension of 
a supermultiplet cannot depend on a continuous parameter such as, for example, 
the coupling constant, and then these states remain 
at any value of the coupling. If two theories are related by a duality, then 
this duality must relate the BPS states of both theories. Indeed, the relation 
between the mass and the charge of a given state does not change when the 
coupling constant varies.

\medskip
\subsubsection{T-duality}
\medskip

Let us consider a closed string moving in a spacetime in which one of the 
spatial directions is compact. We can take this direction to be $x^9$ and 
if we compactify it on a circle of radius R then we make the following 
identification:
\be
x^9\equiv x^9+2\pi R\,.
\ee
We have seen that the equations of motion for the bosonic fields of the 
string action imply that they split into a right and a left moving part
\be
X^\m=X_L^\m(\s^+)+X_R^\m(\s^-), \qquad\qquad \sigma^\pm=\tau\pm\sigma\,.
\label{sprxrl}
\ee
For the closed string one has to impose periodicity conditions in 
the non-compact directions.
\be
X^\m(\t, \s+2\pi)= X^\m(\t, \s)\,,
\ee
The mode expansions for the closed string consistent with the periodicity 
conditions are
\br
X_L^\m &=&x_L^\m+\a' p^\m_L \sigma^+ +i\sqrt{\frac{\a'}{2}} \sum_{n\neq 0}
{1\over n} \a^\mu_n e^{-2in \s^+}\,, \nonumber \\
X_R^\m&=&x_R^\m+\a' p^\m_R \s^- +i\sqrt{\frac{\a'}{2}} \sum_{n\neq 0}
{1\over n} \tilde{\a}^\m_n e^{-2in \s^-}\,, 
\label{strmod}
\er
where $x_{L,R}^\mu$ and $p_{L,R}^\mu$ are real and $(\alpha^\mu_n)^\dagger
=\alpha^\mu_{-n}$, and similarly for $\tilde{\alpha}^\mu_n$.
The total momentum carried by the string is $p^\mu=p^\mu_L+p^\mu_R$.

On the other hand, the string has the possibility to wrap around the 
compact direction a certain number of times and the periodicity condition 
for this direction changes into
\be
X^9(\t, \s+2\pi)= X^9 (\t, \s)+ 2\pi w R.
\label{strwin}
\ee
The number $w$ is integer valued and is the number of times the string 
winds around the compact direction. The momentum in the compact direction 
is quantized in order for the wave function $e^{ip^9X^9}$ to be single valued. 
We then have $p^9=n/R$ with $n$ an integer. Therefore, the solution of the 
equations of motion allowing for this boundary condition is
\be
X^9=x^9+\a' \frac{n}{R} \t + w R\s + i\sqrt{\frac{\a'}{2}} \sum_{l\neq 0}
{1\over l} \left(\a^9_l e^{-2il \s^+}+{\tilde\a}^9_l e^{-2il \s^-}\right)\,,
\ee
The separation in left and right modes is then
\br
X_L^9 (\s^+)&=&{1\over 2}x^9+\frac{\a'}{2} 
\left(\frac{n}{R}+\frac{Rw}{\a^{\prime}}\right)
 \sigma^+ +i\sqrt{\frac{\a'}{2}} \sum_{l\neq 0}
{1\over l} \a^9_l e^{-2il \s^+}\,, \nonu \\
X_R^9 (\s^-)&=&{1\over 2}x^9+\frac{\a'}{2} \left(\frac{n}{R}-
\frac{Rw}{\a^{\prime}}\right)
 \sigma^-                                                                                                                                                              
+i\sqrt{\frac{\a'}{2}} \sum_{l\neq 0}
{1\over l} {\tilde\a}^9_l e^{-2il \s^-}\,.
\label{strmod2}
\er
There are two quantum numbers, $w$ and $n$, which describe the momentum and 
winding 
modes respectively. The Hamiltonian for the closed string is
\be
H=\a' \frac{P^2}{2} + \frac{\a'}{2}\left[\frac{n^2}{R^2}+\frac{R^2 w^2}{{\a'}^2}+
\frac{2}{\a'}\,\sum_{l\neq 0}(\a_{-n}\a_n+{\tilde\a}_{-n}{\tilde\a}_n)\right]\,,
\label{stren}
\ee
where $P^{\m}$ is the  9-dimensional non-compact spacetime momentum. Using that 
$P^2=-M^2$ and that the Hamiltonian must vanish on-shell, 
one can obtain the mass of a closed string wound $w$ times in the compact 
direction:
\be
M^2=\frac{n^2}{R^2}+\frac{R^2w^2}{{\a^{\prime}}^2}\,.
\label{strtduamass}
\ee
The spectrum of eq. (\ref{strtduamass}) is invariant if we perform the 
simultaneous change:
\br
R & \leftrightarrow & {\a' \over R},  \rc\rc
n & \leftrightarrow & w\,.
\label{strtdual}
\er
Momentum and winding numbers are exchanged. The fixed point of this transformation 
$\ls=\sqrt{\a'}$ is the minimum length in string theory since any string probing 
a smaller distance can be reformulated in a length scale larger than $\ls$.

The left and right moving momenta can be read from eqs. (\ref{strmod}) and 
(\ref{strmod2}):
\br
p^9_L\,=\,\frac{1}{2}\left(\frac{n}{R}\,+\,\frac{Rw}{\a'}\right)\,,\rc\rc
p^9_R\,=\,\frac{1}{2}\left(\frac{n}{R}\,-\,\frac{Rw}{\a'}\right)\,.
\er
They transform in the following way under (\ref{strtdual}):
\be
p^9_L \leftrightarrow p^9_L, \qquad \qquad p^9_R \leftrightarrow -p^9_R.
\ee
Then the spectrum is also invariant under this T-duality transformation 
provided that
\be
\a^9_n \leftrightarrow \a^9_n\,,\ \ \ \ \ \ 
\tilde{\a}^9_n \leftrightarrow -\tilde{\a}^9_n\,.
\ee
Thus T-duality can be seen as a spacetime parity transformation acting on the 
right moving degrees of freedom:
\be
X^9=X_L^9 +X_R^9 \qquad \rightarrow \qquad X^{'9}=X_L^9 - X_R^9\,.
\ee
We can now see that because of the world-sheet supersymmetry,
the fermionic superpartner $\psi^9$ also has to
transform under T-duality, as $\psi^9_L \leftrightarrow \psi^9_L$
and $\psi^9_R \leftrightarrow -\psi^9_R$. The parity transformation has 
also the effect of changing the chirality of the right-moving RR ground 
state, thus mapping type IIA superstring theory to type IIB and viceversa.

We have seen that T-duality is a symmetry of closed string theory since the spectrum 
remains invariant. T-duality is perturbative since it maps the perturbative 
spectrum of type IIA closed string theory compactified on a circle of radius $R$ 
to the perturbative spectrum of type IIB theory compactified on a circle of 
radius $1/R$.  Since there is only one supergravity in nine dimensions, 
the T-duality symmetry is realised at the level of low-energy 
effective actions as a mapping between the fields of type IIA and IIB 
supergravities.  

One can see how T-duality relates the string coupling constants of IIA and 
IIB theories if one requires the Newton constant in nine dimensions to be 
invariant. The form of the supergravity actions (\ref{striiaact}) and 
(\ref{striibact}) determines the relation of the Newton constants in nine and 
ten dimensions:
\be
G_{10}\sim \frac{G_9 R_{10}}{g_s^2}\,.
\ee
The nine-dimensional Newton constant derived from the IIA and IIB theories 
must be the same:
\be
\frac{g_A^2{\a^{\prime}}^4}{R_A}=\frac{g_B^2{\a^{\prime}}^4}{R_B}\,.
\ee
The radius of compactification are related by (\ref{strtdual}) $R_B=
\a^{\prime}/R_A$. Since the string scale $\a^{\prime}$ does not change, we get
\be
g_B=g_A\frac{\sqrt{\a^{\prime}}}{R_A}\,.
\ee
The same relation follows if we exchange $A\rightarrow B$.

\medskip
\subsubsection{T-duality in Type I theory}
\medskip

The open strings have no winding modes and then it is clear that an open 
string cannot be T-dual to itself. However, one can see how T-duality acts on the 
modes of the open string. As we have seen in (\ref{strbc}) one can  impose 
two sets of boundary conditions for the open string:
\br
&&\partial_{\s}X^{\m}|_{\s=0,\pi}=0\,,\ \ \ \ {\rm Neumann}\,,\nonu\\
\nonu\\
&&\partial_{\t}X^{\m}|_{\s=0,\pi}=0\,,\ \ \ \ {\rm Dirichlet}\,.
\er
The Neumann boundary conditions impose 
$\a^{\m}_n={\tilde\a}^{\m}_n$ and $p^\m_L=p^\m_R$ causing 
the left- and right-moving components to combine into standing waves
\be
X^\m=x^\m +2\a' p^\m \tau +i \sqrt{2\a'} \sum_{n\neq 0}{1\over n}
\a^\mu e^{-in\tau} \cos n \s\,.
\label{strnsm}
\ee
On the other hand the Dirichlet boundary conditions at both ends of the string impose 
that $\a^{\m}_n=-{\tilde\a}^{\m}_n$ and the most general expansion  has the 
form
\be
X^\m=x^\m + \frac{1}{\pi}(y^{\m}-x^{\m})\s +i \sqrt{2\a'} 
\sum_{n\neq 0}{1\over n}\a^\m_n e^{-in\tau} \sin n \s\,.
\label{strrrm}
\ee
In terms of the left- and right-movers the boundary conditions are given by:
\br
\partial_+ X^\m_L - \partial_- X^\m_R |_{\s=0,\pi}=0\,, \ \ \ \ &&{\rm Neumann}\,, 
\nonu\\
\nonu\\
\partial_+ X^\m_L + \partial_- X^\m_R |_{\s=0,\pi}=0\,, \ \ \ \ &&{\rm Dirichlet}\,, 
\er
with $\s^{\pm}=\t\pm\s$.

Let us consider type I theory, that is, open strings with Neumann b.c. in all directions. 
The most general open string expansion for left- and right-movers is
\br
X_L^\m &=&\frac{1}{2}(x^\m+y^{\m})+\a' p^\m_L \s^+ +i\sqrt{\frac{\a'}{2}} 
\sum_{n\neq 0}{1\over n} \a^\mu_n e^{-in \s^+}\,, \nonumber \\
X_R^\m&=&\frac{1}{2}(x^\m-y^{\m})+\a' p^\m_R \s^- +i\sqrt{\frac{\a'}{2}} 
\sum_{n\neq 0}
{1\over n} \tilde{\a}^\m_n e^{-in \s^-}\,, 
\er
where we have introduced the constant vector $y^{\m}$ which does not appear 
in $X^{\m}=X^{\m}_L+X^{\m}_R$. Let us compactify the 9th 
direction in a circle of radius R. The momentum along this direction will
be again $p^9=m/R$. As for the closed strings T-duality acts as a world-sheet 
parity transformation
\be
X^9=X^9_L+X^9_R\rightarrow {X^{\prime}}^9=X^9_L-X^9_R\,.
\ee
After T-duality in the 9th direction and taking $y^{\m}=(0,\dots,0,y^9)$ we get
\be
{X^{\prime}}^9=y^9 + 2\a^{\prime}\frac{m}{R}\s +i \sqrt{2\a'} 
\sum_{n\neq 0}{1\over n}\a^9_n e^{-in\tau} \sin n \s\,.
\ee
This component carries winding but no momentum and is of the form (\ref{strrrm}) 
and then it satisfies Dirichlet boundary conditions. Indeed the endpoints of the 
strings are fixed at points
\be
X^9(\s=0)=y^9\,,\ \ \ \ \ X^9(\s=\pi)=y^9+2\pi m\frac{\a^{\prime}}{R}=
y^9+2\pi mR^{\prime}\,,
\ee
where $R^{\prime}$ is the dual radius. 

It is remarkable that T-duality for open strings exchanges Neumann and 
Dirichlet boundary conditions. We have seen that open strings with Dirichlet 
b.c. in some directions have their endpoints constrained to live in hyperplanes. 
D-branes are precisely the hyperplanes on which the open strings can end and, 
as we will see later, they are true dynamical objects. Thus D-branes appear 
naturally when T-duality acts on open strings. If one starts from a theory of 
open superstrings and T-dualizes on $9-p$ directions one gets the string 
endpoints constrained to lie on a $p$-dimensional hyperplane which 
corresponds to a Dp-brane. If there is only one brane the two 
endpoints of the string are constrained to lie on the same hyperplane 
($X(\s=0)=y$ and $X(\s=\pi)=y+2\pi wR$) with $w$ an integer. However only the 
mode $w=1$ is topologically conserved, since for higher $w$ there is the 
possibility for the string to break when it crosses the D-brane. 
\begin{figure}
\centerline{\hskip -.8in \epsffile{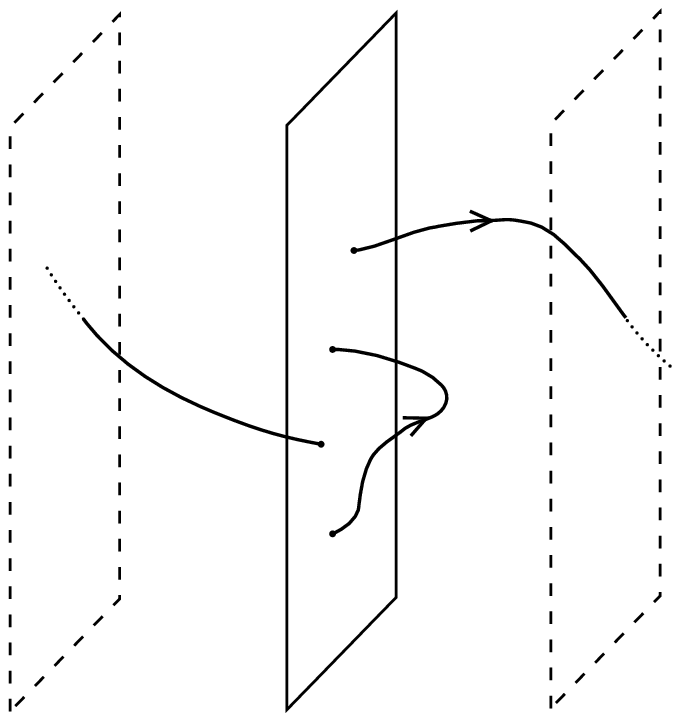}}
\caption{Open strings with endpoints attached to a hyperplane. The dashed 
lines are periodically identified. The strings shown have winding numbers 
zero and one.}
\label{strpol2}
\end{figure}

From the analysis above  we see that when T-duality is performed on a direction 
transverse to the Dp-brane, then this transforms it into a D(p+1)-brane, while 
if the T-duality is performed along a direction longitudinal to the worldvolume 
of the Dp-brane then it becomes a D(p-1)-brane. 

When $R\rightarrow 0$ 
the states of the open string with non-zero momentum along the compact 
direction become infinitely massive. This leaves a theory in $D-1$ dimensions 
whereas the closed strings live in D dimensions. What happens is that the open 
strings have their endpoints constrained to move in a hyperplane and the scalar 
field describing the embedding of the worldsheet into the T-dualized direction 
is fixed by the Dirichlet boundary condition. Thus the string still vibrates 
in D dimensions but with their endpoints constrained to lie on the hyperplane.

\medskip
\subsubsection{S-duality}
\medskip

This is a duality specific of type IIB theory. In the type IIB spectrum 
(\ref{striibspc}) we have two scalars (the dilaton $\p$ and the RR 0-form 
potential $\chi$), and two 2-forms (the RR potential $C_{(2)}$ and the NSNS 
potential $B_{(2)}$). The equations of motion of type IIB theory are invariant 
under a $SL(2,\RR)$ symmetry. If we arrange the RR scalar $\chi$ and the 
dilaton in a complex scalar $\l=\chi + ie^{-\p}$, then the
$SL(2,\RR)$ symmetry acts like:
\be
\l \rightarrow {a\l +b \over c\l +d}\,,
\ee
where the parameters $a, b, c$ and $d$ are real and $ad-bc=1$. The 2-form 
potentials transform as 
\be
\left(\begin{array}{c} B_{(2)}\\ 
                       C_{(2)}
           \end{array}
     \right)
\rightarrow 
\left(\begin{array}{cc} d & -c\\ 
                       -b & a
           \end{array}
     \right)
\left(\begin{array}{c} B_{(2)}\\ 
                       C_{(2)}
           \end{array}
     \right)\,.
\ee
The NSNS 2-form potential $B_{(2)}$ couples to the fundamental string and 
the corresponding charge is the winding number, which is quantized. The 
$SL(2,\RR)$ group is then broken to $SL(2,\ZZ)$.

For the particular case in which the RR scalar is taken to zero and taking 
into account that $g_s=e^{\p_{\infty}}$ the $SL(2,\ZZ)$ transformation 
becomes:
\be
g_s  \rightarrow {1\over g_s}, \qquad B_{(2)}\rightarrow C_{(2)}, \qquad
C_{(2)}  \rightarrow  -B_{(2)}\,.
\ee
Thus, we see that this  duality is non-perturbative, since it exchanges weak 
and strong coupling. The change $B_{(2)}\rightarrow C_{(2)}$ transforms an 
object charged under a NSNS field to another one charged under a RR field. 
There are not such states in the perturbative spectrum and then there should 
exist non-perturbative objects carrying RR charge. The string scale of the 
dual theory is obtained if one requires the 10 dimensional Newton constant 
$G_{10}\sim g_s^2{\a^{\prime}}^4$ to be invariant. We then see that the string 
scale $\a^{\prime}$ transforms as:
\be
\a^{\prime}\rightarrow g_s\a^{\prime}\,.
\ee
Therefore, the fundamental string is then mapped by S-duality to a solitonic 
$(0,1)$-string with its tension going as $1/g_s$.

\medskip
\subsubsection{U-duality}
\medskip

It was conjectured by Hull and Townsend in \cite{HT} that S-duality and 
T-duality fit into a larger group of duality symmetries, called U-duality. 
In particular, they conjectured that the duality group of type II string 
theories compactified on a d-dimensional torus $T^d$ is
\be
G_d(\ZZ) \supset SL(2,\ZZ) \times O(d,d;\ZZ)\,,
\ee
where $G_d(\ZZ)$ is a discrete subgroup of the Cremmer-Julia group of
the $10-d$ dimensional maximal supergravity. All the maximal supergravities 
(except the ten dimensional type IIB) can be derived from 11 dimensional 
supergravity and the Cremer-Julia groups arise as the symmetries of 11 
dimensional supergravity compactified on $T^{d+1}$.

We have seen that type IIA  and type IIB theories compactified on a circle 
are equivalent at the perturbative level, and are related by T-duality. If 
the two theories were compactified in a 
d-dimensional torus $T^d$, the T-duality group extends to $O(d,d,\ZZ)$. This 
can be seen 
in the following way. If there is only one compact direction, the components 
of the left- and right-moving momenta, given in (\ref{strmod}), satisfy the 
relation
\be
(p_L^9)^2\,-\,(p_R^9)^2\,=\,\frac{n\omega}{\a'}\,.
\ee
If there are $d$ compact directions the equation above is generalized to
\be
{\vec p_L}^{\ 2}\,-\,{\vec p_R}^{\ 2}\,=\,\frac{n_i\omega_i}{\a'}\,,\ \ \ \ i=1,\dots d\,,
\ee
which displays $O(d,d,\RR)$ invariance. On the other hand, the S-duality of 
type IIB theory in ten dimensions is inherited by the compactified theory 
and this extends to the $SL(2,\ZZ)$ duality group. The larger group denotes 
the fact that now one has the freedom to change some of the moduli of the 
torus $T^d$, still keeping the $10-d$ dimensional physics invariant.

U-duality maps the fundamental string states to non-perturbative objects. 
In particular, some higher dimensional solitonic objects, called p-branes, 
which saturate a BPS bound in their masses belong to the same orbit as the 
fundamental string, and any of these objects, can then be considered as 
fundamental as the fundamental string itself. Strong and weakly coupled 
regimes of different theories are also mapped by U-duality. This is because 
U-duality mixes all the moduli of the compactified string theory, including 
the coupling constant.

\medskip
\section{11 dimensions and M-theory}  \label{secMT}
\medskip                                             
\setcounter{equation}{0}

It is known that all the maximal supergravities (except the 10 dimensional
type IIB) can be derived from the 11 dimensional supergravity.  The 
Cremmer-Julia groups arise from this point of view as the symmetries of 
11 dimensional supergravity compactified on $T^{d+1}$. The strong coupling 
regimes of any of the string theories can be mapped by means of U-duality to 
the weak coupling limit of another string theory. Indeed, S-duality maps the 
strong and weak coupling regimes of type I and heterotic $SO(32)$ and the S 
self-duality of type IIB theory maps each one of both regimes of the same 
theory to the other. However, the strong coupling of type IIA and 
$E_8\times E_8$ theories is described by an eleven dimensional theory whose 
low energy action is 11 dimensional supergravity. This eleven dimensional 
theory is called M-theory.

The bosonic part of the $D=11$ supergravity has only two fields, the metric and 
the 4-form field strength $F_{(4)}=dC_{(3)}$. The action is \cite{Nahm,Cremmer}
\be
I_{11}={1\over 16\pi G_{11}}\left(\int d^{11}x\ \sqrt{-g} 
\left\{ {\cal R}-\frac{1}{48}{F_{(4)}}^2\right\}\,+\,
{1\over 6} \int C_{(3)} \wedge F^{(4)} \wedge F^{(4)} \right)\,.
\label{str11act}
\ee
The Newton constant $G_{11}$ defines the Plank length ($G_{11}\sim \lp^9$). 
Dimensional reduction of 11 dimensional supergravity yields IIA supergravity. 
Indeed, the number of physical degrees of freedom in 11 dimensional 
supergravity is the same as in IIA supergravity. However, the bosonic sector 
of the IIA theory contains a scalar, the dilaton, whereas in 11 dimensions 
there are no scalars. To explain this let us consider an 11 dimensional spacetime with an 
isometry along 
the $y$ direction and split the spacetime coordinates as $z^{\m}=(x^{\m},y)$. 
By means of a Kaluza-Klein reduction one can obtain an action in 10 dimensions. 
This procedure assumes that the fields do not depend on the $y$ coordinate, 
which is compactified on a circle of radius $R_{11}$. The scalar arises in 
the KK reduction as the metric component along the $y$ direction:
\be
g_{yy}=e^{\frac{4}{3}\p}\,.
\label{strkkr}
\ee
The string coupling constant $g_s=\langle e^{\p}\rangle$ is related to the 
compactification radius in Plank units since from (\ref{strkkr}) one has that 
$g_s{}^{4/3}=\langle g_{yy}\rangle\sim\left(R_{11}/\lp\right)^2$ 
and then
\be
R_{11}\sim g_s^{2/3}\lp\,.
\label{strr11}
\ee
One can relate the 11 dimensional units $\lp$ and $R_{11}$ to string 
units $g_s$ and $\ls=\sqrt{\a^{\prime}}$ since the 10 dimensional effective 
Newton constants have to coincide. Thus
\be
\lp^9\sim G_{11}\sim G_{10}R_{11}\sim g_s^2 \ls^8 R_{11}\,, 
\ee
which together with (\ref{strr11}) implies:
\be
\lp \sim g_s^{1\over 3} \ls, \qquad \qquad R_{11}\sim g_s \ls\,.
\ee
Thus we see that at weak string coupling, $g_s\ll 1$, both the 11 dimensional
Planck scale and the radius of the $11^{\rm th}$ direction are small compared
to the string scale. This means that at perturbative level the theory is 
ten dimensional. When $g_s\gg 1$ the radius $R_{11}$ becomes large and the 
$11^{\rm th}$ dimension effectively decompactifies. In this limit the dynamics 
of type IIA theory is effectively 11 dimensional. Thus the different string 
theories can be seen as different limits of an eleven dimensional theory, 
which is denoted as M-theory.

\medskip
\section{Branes as string solitons} \label{secstrsol}
\medskip                                             
\setcounter{equation}{0}

In this section we will list some classical soliton solutions of the 
supergravity actions which can be interpreted as extended objects 
\cite{supergravity, Argurio}. These solutions are the classical p-branes 
and are solutions of the bosonic truncation of type IIA, IIB 
and 11 dimensional supergravities. The equations of motion 
have an electric-magnetic duality symmetry in terms of the Hodge dual 
field strength ${}^*F$, which is related to $F$ by
\be
{}^*F_{\m_1\dots\m_{D-n}}\,=\,\frac{e^{a\p}}{n!\,\sqrt{|g|}}\,
\epsilon_{\m_1\dots\m_{D-n}\n_1\dots\n_n}\,F^{\n_1\dots\n_n}
\,,\ \ \ \ \ \ \ \ \ \epsilon^{0\,1\dots D-1}=1\,,
\ee
defined in general $D$ dimensions. There are two kinds of ansatze for the field 
strengths compatible with the Killing vectors associated to extended p-branes, 
either electric or magnetic. If the brane is extended along directions $i_1\dots i_p$ 
it will be electrically charged under a $(p+2)$-form $F^{(p+2)}$ if it is of the form
\be
F^{(p+2)}_{ti_1\dots i_pr}=\epsilon_{i_1\dots i_p}\partial_r\,g(r)\,,
\label{elfielsstr}
\ee 
where $r$ is the radial coordinate prametrizing the distance transverse to the 
brane. The electric-magnetic duality of the equations of motion implies that 
if there is a certain solution of the supergravity equations of motion then 
the same solution with $F^{(p+2)}$ replaced  by its Hodge dual field strength will 
also be a solution. The Hodge dual of the $(p+2)$-form (\ref{elfielsstr}) is 
a $(8-p)$-form which has components in 
the sphere transverse to the brane and is proportional to the volume element 
of the unit sphere
\be
{}^*F^{(p+2)}_{\th_1\dots\th_{(8-p)}}=f(r)(\sin \th_1)^{(7-p)}\dots\sin \th_{(8-p)}\,,
\ee
where $\th^1\dots\th^{8-p}$ are spherical coordinates parametrizing the sphere.
This form of $F$ corresponds to a magnetically charged brane. 

The supergravity actions (\ref{striiaact}) and (\ref{striibact}) with the 
dilaton factor multiplying the curvature scalar are said to be in the string 
frame. One can rewrite this actions in the canonical form (or Einstein frame) 
by Weyl rescaling the metric 
\be
g_{\m\n}^S\,=\,e^{\p /2}\,g_{\m\n}^E\,\,,
\label{strSE}
\ee
and then one can give the metric corresponding to some solutions in 
either of both frames and one passes from one description to the other 
with (\ref{strSE}).

We are interested particularlly in solutions which preserve some part of 
the supersymmetries. All the fermion fields 
were set to zero and the requirement of supersymmetry to be preserved is that 
the supersymmetry variation of the fermionic fields vanishes. The fermionic 
fields of type IIA and IIB supergravities are the dilatino $\l$ and the 
gravitino $\psi$. Their supersymmetry variations in the Einstein frame for 
type IIA supergravity are given by \cite{strsolDuff}:
\br
\d\l&=&{1\over 4}~\sqrt{2}~D_{\m}\phi \G^{\m} \G_{11}\,\epsilon + {3\over 16}~
{1\over \sqrt{2}}~e^{3\phi/4} \G^{\m_1\m_2}\,\epsilon\,F^{(2)}_{\m_1\m_2}\,+\rc
&&+\,{1\over 24}~{i\over \sqrt{2}}~e^{-\phi/2}\G^{\m_1\m_2\m_3}\,\epsilon 
\,H^{(3)}_{\m_1\m_2\m_3}\,- \rc
&&-\,{1\over 192}~{i\over \sqrt{2}}~e^{\phi/4}
\G^{\m_1\m_2\m_3\m_4}\,\epsilon\,F^{(4)}_{\m_1\m_2\m_3\m_4}\,, 
\rc\rc
\d\psi_{\m}&=&D_{\m}\,\epsilon\,+ {1\over 64}~e^{3\phi/4}(\G_{\m}\,^{\m_1\m_2} - 
14 \d_{\m}\,^{\m_1} \G^{\m_2})\G_{11}\,\epsilon\,F^{(2)}_{\m_1\m_2}\,+ \rc
&&+\,{1\over 96}~e^{-\phi/2} (\Gamma_{\m}\,^{\m_1\m_2\m_3} - 
9\d_{\m}\,^{\m_1}
\G^{\m_2\m_3}) \G_{11}\,\epsilon\,H^{(3)}_{\m_1\m_2\m_3}\,+ \rc
&&+\,{i\over 256}~e^{\phi/4} (\G_{\m}\,^{\m_1\m_2\m_3\m_4} -
{20\over3}~\d_{\m}\,^{\m_1} \G^{\m_2\m_3\m_4})\,\epsilon\,
F^{(4)}_{\m_1\m_2\m_3\m_4}\,, \rc
\label{strsusyiia}
\er
where $\G^{\m}$ are the $D = 10$ Dirac matrices, which appear in antisymmetrized 
products of the form
\be
\G^{\m_1\m_2\dots\m_n}\,=\,\G^{[\m_1}\G^{\m_2}\dots\G^{\m_n]}\,.
\ee
The ten-dimensional Dirac matrices can be 
written in terms of constant Dirac matrices in flat space $\G_{\underline\m}$ 
as follows:
\be
\G_{\m}\,=\,e_{\m}^{\underline\n}\,\G_{\underline\n}\,,
\label{gammaflat}
\ee
where $e_{\m}^{\underline\n}$ are the viel-beins, which form a local basis of 
orthonormal tangent vectors such that $g_{\m\n}=e_{\m}^{\underline\s}\,
e_{\n}^{\underline\r}\,\eta_{\underline{\s\r}}$. In what follows underlined 
indices will denote flat space indices. The covariant derivative is given by
\be
D_{\m} = \partial_{\m} + {1\over 4}\,\omega_{\m}^{\underline{\n\r}} 
\G_{\underline{\n\r}}\,,
\label{strcode}
\ee
where $\omega_{\m}^{\underline{\n\r}}$ is the Lorentz spin connection, whose 
expression in terms of the viel-beins is 
\br
\omega_{\m}^{\underline{\n\r}}&=&
\frac{1}{2}e^{\s\underline\n}\,(\partial_{\m}e_{\s}^{\underline\r}\,-\,
\partial_{\s}e_{\m}^{\underline\r})\,-\,
\frac{1}{2}e^{\s\underline\r}\,(\partial_{\m}e_{\s}^{\underline\n}\,-\,
\partial_{\s}e_{\m}^{\underline\n})\,-\rc\rc
&-&\frac{1}{2}e^{\s\underline\n}\,e^{\d\underline\r}\,
(\partial_{\s}e_{\d\underline\l}\,-\,\partial_{\d}e_{\s\underline\l})\,
e_{\m}^{\underline\l}\,,
\er
where $e_{\m\underline\n}\,=\,e_{\m}^{\underline\s}\,\eta_{\underline{\s\n}}$ 
and $e^{\m\underline\n}\,=\,g^{\m\s}\,e_{\s}^{\underline\n}$. The chirality 
operator $\G_{11}$ in type IIA supergravity is defined as 
$\G_{11}=\G_{\underline 0}\G_{\underline 1}\dots\G_{\underline 9}$. 

In the type IIB theory the spinor $\epsilon$ is actually composed
by two Majorana-Weyl spinors $\e_L$ and $\e_R$ of well defined
ten-dimensional chirality, which can be arranged as a two-component vector in
the form:
\be
\e\,=\,\pmatrix{\e_L\cr\e_R}\,\,.
\label{striibsp}
\ee
We can use complex spinors instead of working the real 
two-component spinor written in eq.
(\ref{striibsp}).  If 
$\epsilon_R$ and $\epsilon_L$ are the two components of the real spinor written
in eq. (\ref{striibsp}), the complex spinor is simply:
\be
\e\,=\,\e_L\,+\,i\e_R\,\,.
\ee
We have the following rules to pass 
from one notation to the other:
\be
\e^*\,\leftrightarrow\,\s_3\,\e\,\,,
\,\,\,\,\,\,\,\,\,\,\,\,\,\,\,\,\,\,\,
i\e^*\,\leftrightarrow\,\s_1\,\e\,\,,
\,\,\,\,\,\,\,\,\,\,\,\,\,\,\,\,\,\,\,
i\e\,\leftrightarrow\,-i\s_2\,\e\,\,,
\label{striibcn}
\ee
Using complex spinors  
the supersymmetry transformations of the dilatino $\lambda$ and gravitino 
$\psi$ in type IIB supergravity are (Einstein frame) \cite{SUSYIIB, strsolDuff}:
\br
\d\l&=&i\G^{\m}\epsilon^{*}\,P_{\m}\,-\,{i\over 24}\,
\G^{\m_1\m_2\m_3}\,\epsilon\,F_{\m_1\m_2\m_3}\,\,,\rc\rc
\d\psi_{\m}&=&D_{\m}\epsilon\,-\,{i\over 1920}\,\G^{\m_1\dots\m_5}
\G_{\m}\,\epsilon\,F^{(5)}_{\m_1\dots\m_5}\,+\rc\rc
&+&{1\over 96}\,\Big(\,\G_{\m}^{\,\,\,\m_1\m_2\m_3}\,-\,9\,
\d_{\m}^{\m_1}\,\G^{\m_2\m_3}\,\Big)\,\epsilon^*\,F_{\m_1\m_2\m_3}\,\,.
\label{strsusyiib}
\er
In eq. (\ref{strsusyiib}) the $\G^{\m}$'s are again the ten-dimensional Dirac
matrices with curved indices, $F^{(5)}$ is the RR five-form, $D_{\m}$ is the 
covariant derivative given in (\ref{strcode}) and 
$P_{\m}$ and $F_{\m_1\m_2\m_3}$ are given by:
\br
P_{\m}&=&{1\over 2}\,\big[\,\partial_M\phi_D\,+\,ie^{\phi_D}\,
\partial_{\m}\chi\,\big]\,\,,\rc\rc
F_{\m_1\m_2\m_3}&=&e^{-{\phi_D\over 2}}\,H^{(3)}_{\m_1\m_2\m_3}\,+\,
ie^{{\phi_D\over 2}}\,F^{(3)}_{\m_1\m_2\m_3}\,\,,
\label{ncapaccinco}
\er
where $\chi$ is the RR scalar and $H^{(3)}$ and $F^{(3)}$ are, respectively, the NSNS
and RR three-form field strengths.

In 11 dimensional supergravity there is only one fermionic field, the gravitino 
$\psi$, whose supersymmetry variation is given by
\be
\d\psi_{\m}\,=\,D_{\m}\,\epsilon\,+\,{1\over 288}\,
\Bigg(\,\G_{\m}^{\,\,\m_1\dots \m_4}\,-\,8\d_{\m}^{\m_1}\,
\G^{\,\,\m_2\dots \m_4}\,\Bigg)\,\epsilon\,\,
F^{(4)}_{\m_1\dots \m_4}\,\,.
\label{mtapbuno}
\ee
In eq. (\ref{mtapbuno}) the $\Gamma^{M}$'s are  eleven-dimensional Dirac
matrices with curved indices, and $F^{(4)}$ is the four-form field-strength.

\subsubsection{BPS states}
\medskip   

We will be interested in solutions of the supergravity equations of motion 
which preserve some fraction of the supersymmetry. This kind of states 
satisfy a bound in their masses, which are related to their charges, and 
then these states are completely characterized by its charge. This 
kind of solutions are called Bogomoln'yi-Prasad-Sommerfeld (BPS) states 
and the bound in their masses is the Bogomoln'yi bound. However, not all 
the states saturating the BPS bound preserve supersymmetry. BPS 
conditions coincide with the preservation of some supersymmetries only if 
the theory is sufficiently supersymmetric.                        

The relation between the preservation of supersymmetry and the BPS bound 
can be seen if one considers the most general anticommutation rules of the 
supersymmetry charges \cite{Mtfisa}. For the the particular case 
of ${\cal N}=1$ supersymmetry algebra in eleven dimensions they take the 
following form:
\be
\{ Q_\alpha,Q_\beta\} = (\G^{\m}C)_{\a\b} P_{\m} + {1\over2}
(\G_{\m\n}C)_{\a\b} Z^{\m\n} +
{1\over 5!} (\G_{\m_1\dots\m_5}C)_{\a\b} Y^{\m_1\dots\m_5} 
\label{strssach}
\ee
where $P_{\m}$ is the momentum operator, $Q_{\a}$ are the 32 component 
Majorana supercharges, $\G_{\m_1\dots\m_p}$ are 
antisymmetrized products of gamma matrices satisfying the Clifford 
algebra, $C$ is the charge conjugation matrix and, $Z$ and $Y$ are the 
central charges, which commute with the supersymmetry 
generators, and are grouped into representations of the Lorentz group. 
The solitonic solutions will have some charge and then the commutation 
algebra of the supersymmetry generators should contain a central extension 
\cite{Azcarraga}.

Let us take, as an example, the M2-brane of M-theory lying in the 12 plane. 
The central charge associated to it is given by the integral of a current 
density
\be
Z^{\m\n}\,=\,q_{M2}\,\int\,dX^{\m}\wedge dX^{\n}\,.
\ee
Our M2-brane is then associated with a non-zero $Z_{12}$. Let us choose the 
Majorana representation of the Dirac matrices, in which $C=\G^0$. Then, 
for a static membrane the algebra becomes
\be
\{Q,Q\}=P^0+\G^{012}Z_{12}\,.
\label{strqqalg}
\ee
In the Majorana representation $Q$ is real, so the left hand side is 
manifestly positive. Since the sign of $Z_{12}$ can be flipped by replacing 
the membrane by an anti-membrane we must have $P^0\ge 0$ and $P^0\ge |Z_{12}|$ 
which is equivalent to
\be
T_{M2}\ge |q_{M2}|
\label{M2BPSbound}
\ee
In the case that the membrane saturates the bound the anticommutator becomes
\be
\{Q,Q\}=P^0\left[1\pm\G^{012}\right]\,.
\ee 
Spinors $\epsilon$ satisfying
\be
\G^{012}\epsilon=\pm\epsilon\,.
\label{M2pc}
\ee
are eigenspinors of $\{Q,Q\}$ with zero eigenvalue. Since $\G^{012}$ squares 
to the identity and is traceless the condition (\ref{M2pc}) is a projection 
condition and thus, a membrane saturating the bound (\ref{M2BPSbound}) 
preserves $1/2$ of the supersymmetries of the vacuum.

In general, a  BPS state $|\psi\rangle$ will be, by definition, anihilated 
by some linear combination of supercharges, 
\be
Q|\psi\rangle = 0\,.
\ee 
If we take the expectation value of the corresponding commutator of 
supecharges in this state
\be
\langle\psi|\{Q,Q\}|\psi\rangle=0
\ee
and, for a static brane ($P^0=M$), this is equivalent to solve the eigenvalue 
problem
\be
M|\psi\rangle=\G|\psi\rangle\,,
\label{eigbps}
\ee 
where $\G$ involves a certain combination of Dirac matrices and central 
charges which depends on the explicit form of the algebra. The 
possitivity of the commutator $\{Q,Q\}$ implies the bound in the rest mass
\be
M\ge |Z|\,, 
\label{BPSbound}
\ee
where $Z$ is the corresponding central charge, which is saturated by 
the BPS states. Taking $M=|Z|$ in eq. (\ref{eigbps}) we get the  
supersymmetries preserved by the state.

\subsubsection{Extremal branes} \label{sugrasol}
\medskip

A p-brane is said to be extremal if it saturates a BPS bound in its mass. 
Extremal branes are, in theories with enough supersymmetries, BPS states 
and they are conjectured to correspond to fundamental objects such as, 
for example, the fundamental string. In the context of ten and eleven 
dimensional supergravities extremality coincides with supersymmetry for 
the elementary branes \cite{Argurio}. These supersymmetric properties 
of p-branes are crucial in relating the classical solutions to the 
quantum objects. There is a no-force theorem for extremal branes which 
states that two parallel branes can remain static at an arbitrary distance 
from each other without feeling an interaction. The static force between 
parallel branes cancels precisely because of a balance between the atraction 
due to gravity and the electrostatic-type repulsion due to the RR charges of 
the branes. This balance is possible only when $T_p=|q_p|$, where $T_p$ is the 
tension of the brane and $q_p$ is its charge density.

There will be one magnetically and one electrically charged extremal solution 
for each one of the field strengths appearing in the supergravity actions. 
The electrically charged solution under the NSNS 3-form field strength 
$H_{(3)}$ is called the F1-string. It can be identified with the fundamental 
string, since is couples to the NSNS 2-form potential $B_{(2)}$. The 
corresponding magnetically charged solution is a five-brane called NS5-brane. 
The objects charged under the RR forms are the Dp-branes, where $p$ odd 
corresponds to the type IIB theory and $p$ even to the type IIA. The metric 
and dilaton of the F1, NS5 and Dp-brane solutions are given, in the Einstein 
frame by
\br
&&ds_{F1}^2\,=\,f^{-\frac{3}{4}}(-dx_0^2+dx_1^2)+f^{\frac{1}{4}}(dx_2^2+\dots 
+dx_9^2)\,,
\ \ \ \ e^{\p}\,=\,f^{-\frac{1}{2}}\,, \rc\rc
&&ds_{NS5}^2\,=\,f^{-\frac{1}{4}}(-dx_0^2+dx_1^2+\dots +dx_5^2)+f^{\frac{3}{4}}
(dx_6^2+\dots +dx_9^2)\,,
\ \ \ \ e^{\p}\,=\,f^{\frac{1}{2}}\,, \rc\rc
&&ds_{Dp}^2\,=\,f_p^{-\frac{7-p}{8}}(-dx_0^2+dx_1^2+\dots 
+dx_p^2)+f_p^{\frac{p+1}{8}}
(dx_{p+1}^2+\dots +dx_9^2)\,,
\ \ \ \ e^{\p}\,=\,f_p^{\frac{3-p}{4}}\,, \rc\rc
\label{strdpbe}
\er
where $f$ and $f_p$ are harmonic functions, which are given by 
\br
f&=&1\,+\,\frac{R^6}{r^6}\ \ \ \ \ (F1)\,, \rc\rc
f&=&1\,+\,\frac{R^2}{r^2}\ \ \ \ \ (NS5)\,, \rc\rc
f_p&=&1\,+\,\frac{R^{7-p}}{r^{7-p}}\ \ \ \ \ (Dp)\,. \rc
\label{strharmf}
\er
The F1 is extended in the $x_1$ direction, the NS5 in the $x_1\dots x_5$ 
and the Dp-branes in the $x_1\dots x_p$.

In eq. (\ref{strdpbe}) the Dp-branes up to $p=2$ are charged electrically under the 
respective RR 
field strengths, the D3-brane is charged under the $F^{(5)}$, which is 
self-dual and the Dp-branes with $p>3$ are magnetically charged. The 
respective field-strengths are given by
\be
F^{(p+2)}_{x^0x^1\dots x^pr}\,=\,\partial_r(f_p^{-1})\,,
\label{strecDpb}
\ee
for the electrically charged branes and
\be
F^{(8-p)}_{\th^1\dots\th^{8-p}}\,=\,(7-p)\,R^{7-p}\sqrt{\hat g^{(8-p)}}\,,
\label{strecDpbm}
\ee
for the magnetically charged ones. In (\ref{strecDpbm}) we have taken 
$\hat g^{(8-p)}$ to be the metric of the unit $(8-p)$-sphere transverse to the 
Dp-brane and the coordinates $\th^1,\dots,\th^{8-p}$ are spherical coordinates 
parametrizing the sphere. 

The F1-string and the NS5-brane have the following non-vanishing components 
of the NSNS 3-form field strength $H^{(3)}$
\br
&&H^{(3)}_{x^0x^1r}\,=\,\partial_r\, f^{-1}\ \ \ \ \ (F1)\rc\rc
&&H^{(3)}_{\th^1\th^2\th^3}\,=\,2\,R^2\,\sqrt{\hat g^{(3)}}\ \ \ \ \ (NS5)\,,\rc
\er
with the respective radius and with $\hat g^{(3)}$ the metric of the unit 
three-sphere transverse to the NS5-brane.

The solutions (\ref{strdpbe}) can be written in the string frame and have 
the following form
\br
&&ds_{F1}^2\,=\,f^{-1}(-dx_0^2+dx_1^2)+dx_2^2+\dots +dx_9^2\,,\rc\rc
&&ds_{NS5}^2\,=\,-dx_0^2+dx_1^2+\dots +dx_5^2+f(dx_6^2+\dots +dx_9^2)\,,\rc\rc
&&ds_{Dp}^2\,=\,f_p^{-\frac{1}{2}}(-dx_0^2+dx_1^2+\dots +dx_p^2)\,+\,
f_p^{\frac{1}{2}}(dx_{p+1}^2+\dots +dx_9^2)\,. \rc
\label{strdpbs}
\er

All these solutions are BPS states saturating the BPS bound (\ref{BPSbound}). 
The central charge is related to the corresponding p-brane charge $q$ through
the expression
\be
Z\,=\,q\,\int\,dx_1\wedge\dots\wedge dx_p\,,
\ee
for a p-brane extended in the directions $x_1\dots x_p$. The charge of a 
p-brane with an electric field strength $F^{(p+2)}$ is 
\be
q_{e}\,=\,\frac{1}{16\pi G_{10}}\,\int_{S^{8-p}}\,{}^*F^{(p+2)}\,,
\label{strpelch}
\ee
where the integratation is over the unit sphere transverse to the p-brane, 
$G_{10}$ is the ten-dimensional Newton constant and $1/16\pi G_{10}$ is the 
factor appearing in the different supergravity actions. For a p-brane with 
$F^{(8-p)}$ magnetic the charge is
\be
q_{m}\,=\,\frac{1}{16\pi G_{10}}\,\int_{S^{8-p}}\,F^{(8-p)}\,.
\ee

The mass of a p-brane is related to the tension or mass density of the brane 
by 
\be
M\,=\,T_p\,\int\,dx_1\wedge\dots\wedge dx_p\,.
\ee
The saturation of the BPS bound implies the following relation between the 
tension and the charge
\be
T_p=|q|\,.
\ee
This equation determines the value of the radius $R$ for the different p-branes. 
For an electrically charged Dp-brane the RR charge is given by (\ref{strpelch})
with $F^{(p+2)}$ given in eq. (\ref{strecDpb}) and $f_p$ given in eq. 
(\ref{strharmf}). The Hodge-dual field-strength is:
\be
{}^*F^{(p+2)}_{\th^1,\dots,\th^{8-p}}=(-1)^{p+1}\,(7-p)\,R^{7-p}\,
\sqrt{\hat g^{(8-p)}}\,,
\label{strdbF8pf}
\ee
where $\hat g^{(8-p)}$ is the metric of the unit sphere transverse to the 
Dp-brane. The integral of (\ref{strdbF8pf}) over the $(8-p)$-sphere gives
\be
\int_{S^{8-p}}\,{}^*F^{(p+2)}\,=\,(-1)^{p+1}\,(7-p)\,R^{7-p}\,\O_{8-p}\,,
\ee
where $\O_{8-p}$ is the volume of the unit $(8-p)$-sphere, given by
\be
\O_{8-p}\,=\,\frac{2\,\pi^{\frac{9-p}{2}}}{\G(\frac{9-p}{2})}\,.
\ee
The Newton constant $G_{10}$ is related to the string parameters by
\be
G_{10}\,=\,8\pi^6g_s^2(\a^{\prime})^4\,,
\ee
and finally, the result for the charge is
\be
|q_{Dp}|\,=\,\frac{4}{(2\pi)^7(\a^{\prime})^4g_s^2}\,R^{7-p}\,
\frac{\pi^{\frac{9-p}{2}}}{\G(\frac{9-p}{2})}\,.
\ee
For a stack of N coincident Dp-branes the mass density is $N\,T_{Dp}$ and 
the tension is given by
\be
T_{Dp}\,=\,\frac{1}{(2\pi)^p\,g_s\,(\a^{\prime})^{\frac{p+1}{2}}}\,.
\ee
The BPS condition for the stack is then
\be
|q_{Dp}|\,=\,N\,T_{Dp}\,,
\ee
and from this condition we get the following expression for the radius of 
a stack of D-branes:
\be
R^{7-p}\,=\,N\,g_s\,2^{5-p}\,\pi^{\frac{5-p}{2}}\,(\a^{\prime})^{\frac{7-p}{2}}\,
\G(\frac{7-p}{2})\,.
\label{strdbrad}
\ee

One can also check that the Dp-brane solutions do indeed preserve one half of 
the supersymmetries. This is because the condition $\d\l=0$ is a projection 
condition  over the supersymmetry parameter. The explicit form of this 
projection for the Dp-brane solutions of type IIA supergravity is
\be 
\G_{\underline{x_0x_1\dots x_p}}(\G_{11})^{\frac{p-2}{2}}\epsilon=\epsilon\,,
\label{striiapr}
\ee
in terms of the ten-dimensional constant gamma matrices. On the other hand, 
the supersymmetry projection for the Dp-brane solutions of type IIB 
supergravity is given by:
\be 
\G_{\underline {x_0x_1\dots x_p}}(i\s_2)\s_3^{\frac{p-3}{2}}\epsilon=\epsilon\,.
\label{striibpro}
\ee
The condition $\d\psi_{\m}=0$ determines the explicit form of the Killing spinors. 
If one parameterizes the space  transverse to the Dp-brane in terms of polar 
coordinates this equation yields \cite{Rahmfeld}
\be
\epsilon\,=\,g(r)\,e^{\frac{\th}{2}\G_{{\underline {r\th}}}}\prod_{i=1}^{7-p}
e^{-\frac{\th_i}{2}\G_{\underline {\th^i\th^{i+1}}}}\,\epsilon_0\,,
\label{strkisp}
\ee
where $g(r)$ is a radial function which depends on the frame and $\epsilon_0$ 
is a constant spinor satisfying eq. (\ref{striiapr}) or (\ref{striibpro}) 
respectively for type IIA or IIB supergravity.

The fundamental string F1 is electrically charged under the NSNS three-form 
$H_{(3)}$, with $H^{(3)}_{x^0x^1r}=\partial_r\,f^{-1}$ and $f=1+R^6/r^6$. 
This form appears in the type II supergravity actions (string frame) in a term 
with form $\frac{1}{2}e^{-2\p}H_{(3)}^2$ and the charge is, in this case, defined by
\be
q_{F1}\,=\,\frac{1}{16\pi G_{10}}\,\int_{S^7}\,e^{-2\p}\,\,{}^*H_{(3)}\,.
\ee
The BPS condition for a single fundamental string is 
$|q_{F1}|=T_f=1/2\pi\a^{\prime}$ and we get the following value for 
the radius
\be
R^6\,=\,8\,(2\pi)^2\,(\a^{\prime})^3\,g_s^2\,.
\ee

The NS5-brane is magnetically charged under the NSNS three-form $H_{(3)}$  
\be
H_{(3)}\,=\,2\,R^2\epsilon_{(3)}\,.
\label{HNS5}
\ee
The NS5 charge is given by
\be
q_{NS5}\,=\,\frac{1}{16\pi G_{10}}\,\int_{S^3}\,H_{(3)}\,.
\ee
and the tension is 
\be
T_{NS5}\,=\,\frac{1}{(2\pi)^5\,(\a^{\prime})^3\,g_s^2}\,.
\ee
The BPS condition for a stack of NS5-branes, $|q_{NS5}|=NT_{NS5}$, 
determines the radius to be
\be
R^2\,=\,N\,\a^{\prime}\,.
\label{ns5dprad}
\ee

In 11 dimensional supergravity there is only a 4-form field-strength. This 
4-form gives rise to two extremal solutions, the electrically charged M2 and 
the magnetic M5. They are given by
\br
ds^2_{M2}&=&f^{-\frac{2}{3}}\,(-dx_0^2\,+\,dx_1^2\,+\,dx_2^2)\,+\,
f^{\frac{1}{3}}\,(dx_3^2\,+\dots +\,dx_{10}^2)\,,\rc\rc
F^{(4)}_{x^0x^1x^2r}&=&\partial_r(f^{-1})\,,\rc
\er
for the M2-brane solution, with the electric components of $F^{(4)}$ excited, 
and 
\br
ds^2_{M5}&=&f^{-\frac{1}{3}}\,(-dx_0^2\,+\,dx_1^2\,+\dots +dx_5^2)\,+\,
f^{\frac{2}{3}}\,(dx_6^2\,+\dots +\,dx_{10}^2)\,,\rc\rc  
F^{(4)}_{\th^1\dots\th^4}&=&3\,R^3\sqrt{\hat g^{(4)}}\,,\rc
\label{strmtexbr}
\er
for the M5-brane, with magnetic components of $F^{(4)}$. The $\th^1,\dots,\th^4$ 
are spherical coordiantes parametrizing the sphere transverse to the M5. The 
harmonic function $f$ is given by
\be
f\,=\,1\,+\,\frac{R^6}{r^{6}}\ \ \ \ ({\rm M2})\,,\ \ \ \ \ 
f\,=\,1\,+\,\frac{R^3}{r^{3}}\ \ \ \ ({\rm M5})\,.
\ee
In 11 dimensions there is no dilaton and the Einstein metric coincides 
with the string metric. Both solutions are extremal and we can check 
that they preserve one half of the supersymmetries. The requeriment of 
supersymmetry to be preserved is that a supersymmetry variation of the 
gravitino vanishes. The projection conditions are given by
\br
&&\G_{\underline{x^0\,x^1\,x^2}}\,\epsilon\,=\,\epsilon \,,\ \ ({\rm M2})\,,
\rc\rc
&&\G_{\underline{x^0\,\dots x^5}}\,\epsilon\,=\,\epsilon \,,\ \ ({\rm M5})\,.
\er

One can compute the charges and equate them to the respective tensions. The M2 
is electrically charged under $F^{(4)}$ and the corresponding charge is given by:
\be
q_{M2}\,=\,\frac{1}{16\pi G_{11}}\,\int_{S^7}\,{}^*F^{(4)}\,.
\ee
If we parameterize the space transverse to the M2 in spherical coordinates 
the Hodge-dual of the electric $F^{(4)}$ is:
\be
{}^*F^{(4)}\,=\,-6\,R^6\,\epsilon_{(7)}\,,
\ee
where $\hat g^{(7)}$ is the determinat of the metric of the unit 7-sphere 
transverse to the M2. The charge is then given by
\be
q_{M2}\,=\,-\frac{6\,R^6\,\O_7}{(2\pi)^8\,\lp^9}\,,
\label{strqm2}
\ee
where $\O_7$ is the volume of the unit 7-sphere and $\lp$ is the Planck 
length in 11 dimensions. 
The BPS condition is $|q_{M2}|=N\,T_{M2}$, where $T_{M2}$ is the tension of 
a single M2 and is given by
\be
T_{M2}\,=\,\frac{1}{(2\pi)^2\,\lp^3}\,.
\label{strtm2}
\ee
Equating equations (\ref{strqm2}) and (\ref{strtm2}) one obtains the following 
expression for the radius of the M2
\be
R^6\,=\,2^5\,\pi^2\,N\,\lp^6\,.
\ee

The M5 is magnetically charged under $F^{(4)}$. The charge is given by
\be
q_{M5}\,=\,\frac{1}{16\pi G_{11}}\,\int_{S^4}\,F^{(4)}\,.
\ee
For the M5, $F^{(4)}\,=\,3\,R^{3}\epsilon_{(4)}$, where $\epsilon_{(4)}$ 
is the volume form of the unit four-sphere transverse to the M5, and then
\be
q_{M5}\,=\,\frac{3\,R^{3}\,\O_4}{(2\pi)^8\,\lp^9}\,.
\ee
The tension of the M5 is 
\be
T_{M5}\,=\,\frac{1}{(2\pi)^5\,\lp^3}\,,
\ee
and the BPS condition $|q_{M5}|=N\,T_{M5}$ gives the following expression for 
the radius:
\be
R^3\,=\,\pi\,N\,\lp^3\,.
\label{strm5rad}
\ee

\subsection{Intersections of branes and bound states} \label{strsecbs}

There are solutions of the supergravity equations of motion which are charged 
under several antisymmetric tensor fields. These solutions represent elementary 
p-branes which intersect each other with some rules. The intersections can be 
orthogonal, in the sense that each 
brane lies along some definite directions, but there are also configurations 
in which the branes intersect at angles. A set of branes can also 
form a bound state when the different 
branes are not well separated. Some of these bound states are configurations 
of branes which feel an attractive force between them, and then there is 
a non-vanishing binding energy. They are called non-marginal bound states and 
their massess are lower than the sum of the masses of the constituent branes. Indeed, 
if $M_1$ and $M_2$ are the masses of two branes which form a non-marginal bound 
state, then the mass $M$ of the bound state goes like $M\sim\sqrt{M_1^2+M_2^2}$.
In general, the bound state configurations preserve a number of supersymmetries 
determined by the projection conditions of their elementary building blocks. 
However, there are some configurations which are still extremal. Each of the 
`basic', 1/2 supersymmetric, objects of M-theory or type II
superstring theory, with a given orientation, is associated with a constraint 
ofthe form $\G\epsilon=\epsilon$ for some traceless product $\Gamma$ of Dirac
matrices with the property that $\G^2=1$. Given two such objects we have
two matrices with these properties. Let us call them $\G$ and
$\G'$. Let $\zeta$ and $\zeta'$ be the charge/tension ratios of the objects
associated, respectively, with $\G$ and $\G'$. Then the $\{Q,Q\}$
anticommutator takes the form
\be
\{Q,Q\} = P^0\big[ 1 + \zeta \G + \zeta' \G'\big] \,.
\ee
Positivity imposes a bound on $\zeta$ and $\zeta'$, but the form of this bound
depends on whether the two matrices $\G$ and $\G'$ commute or
anticommute. 

If $\{\G,\G'\}=0$ then 
\be
(\zeta \G + \zeta' \G')^2 = \zeta^2 + \zeta'{}^2\, ,
\ee
so the bound is $\zeta^2 + \zeta'{}^2 \le 1$, which is equivalent to a bound of
the type
\be
T \ge \sqrt{Z^2 + Z'{}^2} 
\ee
where $Z$ and $Z'$ are the charges of the two branes. Since the right hand side
is strictly smaller than $|Z|+|Z'|$, unless either $Z$ or $Z'$ vanishes, a
configuration saturating this bound must be a `bound state' with strictly
positive binding energy. It is a non-threshold bound state and is associated 
with a constraint of the form $\G''\epsilon=\epsilon$ where
\be
\G '' = \cos\vartheta\, \G + \sin\vartheta\, \G' 
\ee
for some angle $\vartheta$. Since $\G''$ is traceless and squares to the
identity, the `bound state' is another configuration preserving 1/2
supersymmetry. Such configurations are 
for example the bound states $(Dp,D(p+2))$, $(F1,Dp)$ for $(p\ge 1)$ and 
$(NS5,Dp)$ for $(p\le 5)$, the first bound state denoting a set of Dp-branes 
embedded in another set of D(p+2)-branes and so on. These configurations can 
be interpreted from the point of view of the bigger brane as a flux of a 
field strength on its worldvolume. In this section we will show the explicit 
solutions corresponding to the (D(p-2),Dp), (NS5,Dp) 
and (F, Dp) bound states since they will be used in the following chapters as 
supergravity backgrounds to test brane probes. 

If $[\G,\G']=0$ then 
\be
(\zeta \G + \zeta' \G')^2 = (\zeta + \zeta')^2\, ,
\ee
and then, a configuration saturating the bound 
\be
T\,=\,Z\,+\,Z^{\prime}\,,
\ee
has a vanishing binding energy. These are called threshold bound states and 
correspond to intersections of branes. The supersymmetry preserved by the 
state is determined by the constraints imposed by the separate branes 
$\G\epsilon=\epsilon$ and $\G'\epsilon=\epsilon$. Thus, these states 
preserve $1/4$ of the supersymmetries.

\subsubsection{(D(p-2),Dp) bound state} \label{strsecbsnc}

This supergravity solution corresponds to a stack of N non-threshold bound states of 
Dp and D(p-2) branes for $2\le p\le 6$. The metric and dilaton (in the string 
frame) for such a background are \cite{NCback}
\br
ds^2&=&f_p^{-1/2}\,\Big[\,-(\,dx^0\,)^2\,+\,\cdots\,+\,(\,dx^{p-2}\,)^2\,+\,
\,h_p\,\Big((\,dx^{p-1}\,)^2\,+\,(\,dx^{p}\,)^2\Big)\,\Big]\,+\rc\rc
&&+\,f_p^{1/2}\, \Big[\,dr^2\,+\,r^2\,d\Omega_{8-p}^2\,\Big]\,\,,\rc\rc
e^{\tilde\phi_D}&=&f_p^{{3-p\over 4}}\,\,h_p^{1/2}\,\,,
\label{strncbs1}
\er
where $d\O_{8-p}^2$ is the line element of a unit $(8-p)$ sphere, 
$r$ is a radial coordinate parametrizing the distance to the brane bound state
and $\tilde\phi_D=\phi_D\,-\,\phi_D(r\rightarrow\infty)$. The Dp-brane of the
background extends along the directions $x^0\cdots x^p$, whereas the 
D(p-2)-brane component lies along $x^0\cdots x^{p-2}$. The functions $f_p$ and
$h_p$ appearing in (\ref{strncbs1}) are:
\br
f_p&=&1\,+\,{R^{7-p}\over r^{7-p}}\,\,,\rc\rc
h_p^{-1}&=&\sin^2\varphi\,f_p^{-1}\,+\,\cos^2\varphi\,\,,
\label{strncbs2}
\er
with $\varphi$ being an angle which characterizes the degree of mixing of the 
Dp and D(p-2) branes in the bound state. 

This solution is also endowed with a rank two  NSNS field $B$  directed
along the $x^{p-1}x^{p}$ (noncommutative) plane:
\be
B=\tan\varphi\,f_p^{-1}\,h_p\,\,dx^{p-1}\wedge dx^{p}\,\,,
\label{strncbs4}
\ee
and is charged under RR field strengths,  $F^{(p)}$ and 
$F^{(p+2)}$. The components along the directions parallel to the bound state
are
\br
&&F^{(p)}_{x^0,x^1,\cdots, x^{p-2},r}
\,=\,\sin\varphi\,\partial_r\,f_p^{-1}\,\,,\rc\rc
&&F^{(p+2)}_{x^0,x^1,\cdots, x^{p},r}\,=\,\cos\varphi\,
h_p\partial_r\,f_p^{-1}\,.
\label{strncbs5}
\er
It is understood that the  $F^{(p)}$'s for $p\ge 5$ are the
Hodge duals of those with $p\le 5$, \ie\ $F^{(p)}={}^*F^{(10-p)}$ for $p\le 5$.
In particular this implies that $F^{(5)}$ is self-dual, as is well-known for
the type IIB theory. It is clear from the above equations that for $\varphi=0$
the (D(p-2),Dp) solution reduces to the Dp-brane geometry whereas for 
$\varphi=\pi/2$ it becomes a D(p-2)-brane smeared along the $x^{p-1}x^p$
directions. The Hodge-duals ${}^*F^{(p)}$ and ${}^*F^{(p+2)}$ have the 
following components:
\br
&&{}^*F^{(p)}_{x^{p-1},x^p,\th^1,\dots,\th^{8-p}}\,=\,(-1)^{p+1}(7-p)R^{7-p}
\sin\varphi h_pf_p^{-1}\sqrt{\hat g^{(8-p)}}\,,\rc\rc
&&{}^*F^{(p+2)}_{\th^1,\dots,\th^{8-p}}\,=\,(-1)^{p+1}(7-p)R^{7-p}
\cos\varphi \sqrt{\hat g^{(8-p)}}\,,
\label{strncbsdfs}
\er
where $\th^1,\dots,\th^{8-p}$ are spherical coordinates of the $S^{8-p}$ and 
$\hat g^{(8-p)}$ is the determinant of the metric of the unit $(8-p)$-sphere.

The parameter $R$ is determined by the RR charge and the BPS equation 
$|q_{Dp}|=NT_{Dp}$\,. The charge, computed from the $F^{(p+2)}$ given in 
(\ref{strncbs5}) and (\ref{strpelch}), is given by
\be
|q_{Dp}|\,=\,\frac{4}{(2\pi)^7(\a^{\prime})^4g_s^2}\,R^{7-p}\,\cos\varphi\,
\frac{\pi^{\frac{9-p}{2}}}{\G(\frac{9-p}{2})}\,,
\label{strncbsdpch}
\ee
and the BPS condition determines the radius to be
\be
R^{7-p}\,\cos\varphi\,=\,N\,g_s\,2^{5-p}\,\pi^{{5-p\over 2}}\,
(\,\alpha\,'\,)^{{7-p\over 2}}\,\,
\Gamma\Bigl(\,{7-p\over 2}\Bigr)\,\,,
\label{strncbs3}
\ee
where $N$ is the number of branes of the stack, $g_s$ is the string coupling
constant ($g_s=e^{\phi_D(r\rightarrow\infty)}$) and $\alpha\,'$ is the Regge
slope.

Eq. (\ref{strpelch}) cannot be applied to calculate the charge associated with 
the D(p-2)-branes in the (D(p-2),Dp). The reason is that there is an infinite 
number of parallel D(p-2)-branes in the bound state. In 1 + 3 dimensional 
electrostatics, in order to use the Gauss law to obtain the charge per unit 
length for a uniform line distribution of charge, one has to choose a cylinder 
as the Gauss surface rather than a two-sphere as for a point charge. For the 
present case, let us choose the integration in (\ref{strpelch}) over 
${\cal V}_2\times S^{8-p}$ rather than over an asymptotic $S^{10-p}$, where 
${\cal V}_2=(2\pi\ls)^2$ is a two dimensional area in the $x^{p-1}x^p$ plane. 
Then, we have the following charge, measured asymptotically in the string 
frame metric,
\be
q\,=\,\frac{1}{16\pi G_{10}}\,\int_{{\cal V}_2\times S^{8-p}}\,{}^*F^{(p)}\,
=\,(-1)^{p+1}\,N\,T_{D(p-2)}\,\tan\varphi\,.
\label{ncbscd}
\ee
Since the (D(p-2),Dp) bound state is a BPS state, the total charge associated 
to the D(p-2)-branes  should  be equal to the number of D(p-2) times the 
the tension of a single D(p-2). We therefore conclude there are 
$N\,\tan\varphi$ D(p-2)-branes per ${\cal V}=(2\pi\ls)^2$ of two dimensional 
area in the bound state.

\subsubsection{(NS5,Dp) bound state} \label{strsecNS5Dp}

This is a bound state \cite{NS5Dp} generated by a stack of $N$ 
NS5-branes and Dp-branes for $1\le p\le 5$. The bound state is
characterized by two coprime integers $l$ and $m$ which, respectively, 
determine the number of NS5-branes and Dp-branes which form the bound state. We
shall combine $l$ and
$m$ to form the quantity  $\m_{(l,m)}\,=\,l^2\,+\,m^2\,g_s^2$. Moreover, for a
stack of $N$ (NS5, Dp) bound states we define 
$R^2_{(l,m)}\,=\,N\,\Bigl[\,\mu_{(l,m)}\,\Bigr]^{{1\over 2}}
\,\a'$, in terms of which the harmonic function $H_{(l,m)}(r)$
is defined as
\be
H_{(l,m)}(r)\,=\,1\,+\,{R^2_{(l,m)}\over r^2}\,\,.
\label{strns5dpbs1}
\ee
The solution reduces when $m=0$ to a stack of NS5-branes and the radius is 
then the one given in (\ref{ns5dprad}).
The metric of this background in the string frame is
\br
&&ds^2=\Big[\,h_{(l,m)}(r)\,\Big]^{-{1\over 2}}\,\,
\Bigg[\,\,
\Bigl[\,H_{(l,m)}(r)\,\Bigr]^{-{1\over 4}}\,\,
\Big(\,-dt^2\,+\,(dx^{1})^2\,+\,\dots\,+\,(dx^{p})^2\,\Big)\,+\,\rc\rc
&&+\,h_{(l,m)}(r)\Bigl[\,H_{(l,m)}(r)\,\Bigr]^{{1\over 4}}\,\,
\Big(\,(dx^{p+1})^2\,+\dots\,+\,(dx^{5})^2\,
\Big)\,+\,
\Bigl[\,H_{(l,m)}(r)\,\Bigr]^{{3\over 4}}\,\,
(\,dr^2\,+\,r^2\,d\O_{3}^2\,)\,\Bigg]\,,\rc
\label{strns5dpbs2}
\er
where the function $h_{(l,m)}(r)$ is given by
\be
h_{(l,m)}(r)\,=\,{\mu_{(l,m)}\over
l^2\,\,\Bigl[\,H_{(l,m)}(r)\,\Bigr]^{{1\over 2}}
\,+\,m^2\,g_s^2\,\,
\Bigl[\,H_{(l,m)}(r)\,\Bigr]^{-{1\over 2}}}\,\,.
\label{strns5dpbs3}
\ee

The NS5-branes of this background extend along the $x^1\dots x^5$
coordinates, whereas the Dp-branes lie along $x^1\dots x^p$ and are smeared
in the $x^{p+1}\dots x^5$ coordinates. The integers $l$ and $m$ represent,
respectively, the number of NS5-branes in the bound state and the number of 
Dp-branes in a $(5-p)$-dimensional volume 
${\cal V}_p\,=\,(2\pi \sqrt{\a'})^{5-p}$ in the  $x^{p+1}\dots x^5$
directions. We shall choose spherical coordinates, and
we will represent the $S^3$ line element as 
$d\O_{3}^2\,=\,d\th^2\,+\,(\,{\rm sin}\,\th)^{2}\,\,
d\O_{2}^2$. Other fields in this supergravity solution include the 
dilaton:
\be
e^{-\p}\,=\,g_s^{-1}\,\Bigl[\,H_{(l,m)}(r)\,\Bigr]^{{p-5\over 8}}\,
\Bigl[\,h_{(l,m)}(r)\,\Bigr]^{{p-1\over 4}}\,\,,
\label{strns5dpbs4}
\ee
the NSNS potential $B_{(2)}$:
\be
B_{(2)}\,=\,-l\,N\a'\,C_5(\th)\,\epsilon_{(2)}\,\,,
\label{strns5dpbs5}
\ee
and two RR field strengths $F^{(8-p)}$ and $F^{(6-p)}$, with components
\br 
&&F^{(p+2)}_{t,x^{1},\dots,x^p,r}\,=\,(-1)^{p}\,
\frac{m}{\m_{(l,m)}^{\frac{1}{2}}}\,\Bigl[H_{(l,m)}(r)\Bigr]^{-1}\,,\nonu\\
\nonu\\
&&F^{(8-p)}_{x^{p+1},\dots, x^5,\th,\th^1,\th^2}\,=\,
(-1)^{p+1}\,N\,m\,\a^{\prime}\,\frac{h_{(l,m)}(r)}
{\Bigl[\,H_{(l,m)}(r)\,\Bigr]^{1\over 2}}\,(\sin\th)^2
\,\sqrt{\hat g^{(2)}}\,,\nonu\\
\nonu\\
&&F^{(6-p)}_{x^{p+1},\dots,x^5,r}\,=\,
(-1)^{p}\frac{m}{l}\ \partial_r\,\frac{h_{(l,m)}(r)}
{\Bigl[\,H_{(l,m)}(r)\,\Bigr]^{1\over 2}}\,,
\label{strns5dpbs6}
\er
where $\hat g$  is the determinant of the metric of the unit two-sphere.

The charge of the NS5-branes is esily computed from eq. (\ref{strns5dpbs5}). 
The corresponding value of the NSNS three-form field-strength is 
$H_{(3)}=2lN\a'\epsilon_{(3)}$, and then, looking at eqs. (\ref{HNS5}) and 
(\ref{ns5dprad}) we conclude that 
\be
q_{NS5}=lNT_{NS5}\,.
\ee
In this bound state there is an infinite number of parallel Dp-branes. 
Furthermore, there are $m\,N$ Dp-branes per ${\cal V}_p=(2\pi\ls)^{5-p}$ 
of (5-p)-dimensional volume in the $x^{p+1}\dots x^5$ directions. To see 
this, let us choose the integration in (\ref{strpelch}) over 
${\cal V}_p\times S^3$. The charge, measured asymptotically in the string 
frame metric, is then
\be
q\,=\,\frac{1}{16\pi G_{10}}\,\int_{{\cal V}_p\times S^3}\,{}^*F^{(p)}\,=\,
\,(-1)^{p+1}\,N\,m\,T_{Dp}.
\label{ns5dpcd}
\ee

\subsubsection{(F,Dp) bound state}  \label{strsecFDp}

This bound state is generated by a stack of $N$ Dp-branes and F-strings.
The string frame metric and dilaton for this bound state are \cite{FDp}
\br
ds^2&=&f_p^{-1/2}\,h_p^{-1/2}
\Big[\,-(\,dx^0\,)^2\,+\,(\,dx^{1}\,)^2\,+\,
\,h_p\,\Big((\,dx^{2}\,)^2\,+\,\dots\,+\,\,(\,dx^{p}\,)^2\Big)\,\Big]\,+\rc\rc
&&+\,f_p^{1/2}\, h_p^{-1/2}\,
\Big[\,dr^2\,+\,r^2\,d\O_{8-p}^2\,\Big]\,\,,\rc\rc
e^{\tilde\phi}&=&f_p^{{3-p\over 4}}\,\,h_p^{{p-5\over 4}}\,\,.
\label{strcinueve}
\er
The Dp-brane of the background extends along the directions $x^0\cdots x^p$, 
whereas the F1 component lies along $x^0\,x^1$. The functions $f_p$ and
$h_p$ appearing in are the same as the ones given in (\ref{strncbs2}):
\br
f_p&=&1\,+\,{R^{7-p}\over r^{7-p}}\,\,,\rc\rc
h_p^{-1}&=&\sin^2\varphi\,f_p^{-1}\,+\,\cos^2\varphi\,\,,
\er
with $\varphi$ being an angle which characterizes the degree of mixing of the 
Dp-branes and the F1-strings in the bound state and the radius $R$ is given by 
(\ref{strncbs3}). The NSNS three-form $H^{(3)}$ is $H^{(3)}=-\sin\varphi\,
\partial_r f_p^{-1}\,dx^{0}\wedge dx^{1}\wedge dr$ and the RR field-strengths 
have the following components
\br
&&F^{(p)}_{x^2,\dots,x^p,r}\,=\,(-1)^{p+1}\,\tan\varphi\,\partial_r (h_pf_p^{-1})\,,
\rc\rc
&&F^{(p+2)}_{x^0,x^1,\dots,x^p,r}\,=\,-\,\cos\varphi\,h_p\,\partial_r f_p^{-1}\,,
\rc\rc
&&F^{(8-p)}_{\th,\th^1,\dots,\th^{7-p}}\,=\,(-1)^{7-p}\,(7-p)\,\cos\varphi\,R^{7-p}\,
(\sin\th)^{7-p}\,\sqrt{\hat g^{7-p}}
\er
The value of $F^{(8-p)}_{\th,\th^1,\dots,\th^{7-p}}$ is formally the same as 
the one given in (\ref{strncbsdfs}) for the (D(p-2),Dp) bound state. One can 
compute the RR charge form (\ref{strpelch}) and the result is the same as the 
one in (\ref{strncbsdpch}), thus fixing the radius to be the one in 
(\ref{strncbs3}).

\subsubsection{(M2,M5) bound state} \label{strsecmtbs}

There are also solutions of eleven-dimensional supergravity carrying more 
than one charge as the ones of ten-dimensional supergravities corresponding 
to bound states of branes. 
In particular, we are interested in non-threshold bound state configurations
which preserve one half of the supersymmetries. Such a bound state is the 
one generated by a stack of M2 and M5-branes, which is called the (M2,M5)-
bound state. This solution will be used to study giant graviton configurations 
in the last chapter. This configuration can also be interpreted from the 
point of view of the M5-brane with four-form flux on its worldvolume. 

The corresponding solution of the D=11 supergravity equations was given
in ref. \cite{ILPT} and used as supergravity dual of a non-commutative 
gauge theory in \cite{MR}. The metric generated by a stack of parallel
non-threshold (M2,M5) bound states is
\br
ds^2&=&f^{-1/3}\,h^{-1/3}\,\Big[\,-(\,dx^0\,)^2\,+\,(\,dx^{1}\,)^2\,
+\,(\,dx^{2}\,)^2\,+\,h\, \Big((\,dx^{3}\,)^2\,
+\,(\,dx^{4}\,)^2\,+\,(\,dx^{5}\,)^2\Big) \,\Big]\,+\rc\rc
&&+\,f^{2/3}\,h^{-1/3}\, \Big[\,dr^2\,+\,r^2\,d\O_{4}^2\,\Big]\,\,,
\label{strmtuno}
\er
where $d\O_{4}^2$ is the line element of a unit 4-sphere and the functions
$f$ and $h$ are given by
\br
f&=&1\,+\,{R^{3}\over r^{3}}\,\,,\rc\rc
h^{-1}&=&\sin^2\varphi\,f^{-1}\,+\,\cos^2\varphi\,\,.
\label{strmtdos}
\er
The M5-brane component of this bound state
is extended along the directions $x^0,\dots, x^5$, whereas the M2-brane lies
along $x^0,x^1,x^2$. The angle $\varphi$ in eq. (\ref{strmtdos})
determines the mixing of the M2- and M5-branes in the bound state and the
``radius" $R$ is given by $R^{3}\cos\varphi\,=\,\pi\,N\,{\lp}^3$, where $\lp$ 
is the Planck length in eleven dimensions and $N$ is the number  of bound 
states of the stack. The radius $R$ reduces when $\varphi\rightarrow 0$ to 
the one of eq. (\ref{strm5rad}). The solution of D=11 supergravity is also 
characterized \cite{ILPT} by a non-vanishing value of the four-form field 
strength $F^{(4)}$, namely:
\br
F^{(4)}\,&=&\,\sin\varphi\,\partial_r(f^{-1})\,dx^0\wedge dx^1\wedge
dx^2\wedge dr\,-\,3 R^3\cos\varphi \,\epsilon_{(4)}\,-\cr\cr
&&-\tan\varphi \,\partial_r(hf^{-1})\,dx^3\wedge dx^4\wedge
dx^5\wedge dr\,\,,
\label{strmttres}
\er
where $\epsilon_{(4)}$ represents the volume form of the unit $S^4$.

The parameter $R$ is determined by the charge and the BPS equation. 
Comparing the magnetic component of the $F^{(4)}$ given in (\ref{strmttres}) 
with the one corresponding to the M5, one easily concludes that $R^4$ must 
be the one given above. In the bound state there are $N\,\tan\varphi$ 
M2-branes per ${\cal V}=\,(2\pi\lp)^3$ of three-dimensional volume 
in the $x^3x^4x^5$ directions. Indeed, the Hodge-dual of $F^{(4)}$ has the 
following component:
\be
{}^*F^{(4)}_{x^3x^4x^5\th^1\dots\th^4}\,=\,3\,R^3\,\sin\varphi\,hf^{-1}\,
\sqrt{\hat g^{(4)}}\,,
\label{ncbscd2}
\ee
and then, we have the following charge measured asymptotically,
\be
q\,=\,\frac{1}{16\pi G_{11}}\,\int_{{\cal V}\times S^{4}}\,{}^*F^{(4)}
\,=\,N\,T_{M2}\,\tan\varphi\,.
\ee

\medskip
\section{D-branes} \label{secbranes}
\medskip                                             
\setcounter{equation}{0}

A theory of open strings must also contain closed strings and then D-branes 
should couple to closed strings. It turns out that open strings with Dirichlet 
boundary conditions couple consistently to closed type II strings. Indeed, it 
was shown in \cite{Polchi} that D-branes carry RR charge and that their charge 
is equal to their tension, \ie, they are BPS states. The fact that two parallel 
D-branes can exchange RR bosons which cancel the exchange of NSNS particles 
causes the force between the two static branes to vanish. The Dp-brane tension 
computed in \cite{Polchi} is given by
\be
T_{Dp}={1\over (2\pi)^p g_s \ls^{p+1}}\,.
\label{strten}
\ee
This expression for the tension could also be obtained by applying a chain of 
dualities to the RR charged (0,1)-string (the D1-brane), which was 
obtained in the IIB theory as the S-dual of the fundamental string. The 
Dp-brane tensions satisfy the relation 
\be
T_pT_{6-p}=\frac{2\pi}{16\pi G_{10}}\,,
\label{strttg}
\ee
which agrees with the Dirac-like quantization condition of the Dp-brane 
charges.
The Newton constant in ten dimensions can be determined, in terms of string 
parameters, by using eqs. (\ref{strten}) and (\ref{strttg}). It follows that
\be
G_{10}=8\pi^6g_s^2{\a^{\prime}}^4\,.
\ee

It seems to be inconsistent that D-branes couple to closed string modes because 
open strings with Dirichlet boundary conditions can only have one half of 
the two supersymmetries of the closed strings. What happens is that far away 
from the D-brane we have two spacetime supersymmetries but their generators are 
related to each other on the world-sheet of the open strings. The relation 
between the two supersymmetry generators can be derived by considering the action 
of T-duality on D-branes. Neumann boundary conditions for type I open strings 
impose that the two generators of spacetime supersymmetry $Q_L$ and $Q_R$ must 
be equal, $Q_L=Q_R$. This is consistent if both generators have the same 
spacetime chirality. 

T-duality acts on closed strings as a parity reversal acting on the right-moving 
sector and, on open strings, changes Neumannn and Dirichlet boundary conditions. 
Thus, the effect of introducing Dirichlet boundary conditions on the generators 
(in, say, the $X^9$ direction) is:
\be
Q_L^{\prime}=Q_L\,,\ \ \ \ \ Q_R^{\prime}=P_9Q_R\,,
\ee
where $P_9$ is an operator which anticommutes with $\G_9$ and commutes with all 
the others. It can be taken to be $P_9=\G_9\G_{11}$, where $\G_{11}$ is the 
chirality operator. Acting with T-duality on the directions $X^{p+1}, \dots, 
X^9$ yields a D-brane lying along $X^1, \dots, X^p$ and the following relation 
is imposed on the supersymmetry generators:
\be
Q_L=\pm P_{p+1}\dots P_9Q_R\,.
\ee 
Using that $Q_R$ is chiral or anti-chiral the above relation can be rewritten 
as:
\be
Q_L=\pm\G_0\dots\G_pQ_R\,.
\ee

The massless sector of the open strings contains a vector $A_{\m}$, giving 
rise to the ten dimensional vector multiplet. The effective low-energy 
theory of open strings with Neumann boundary conditions is then ${\cal N}=1$ super 
Yang-Mills (SYM) in ten dimensions. The quantization of open strings with 
Dirichlet boundary conditions is similar. The main difference is that the zero 
modes of open strings with Dirichlet boundary conditions in some directions 
are not dynamical and the low-energy fields corresponding to the massless 
modes of the open superstring are arranged into representations of $SO(1,p)$. 
The vector components along directions with Dirichlet boundary conditions 
become scalars and we are left with a vector $A_{\m}$ with components along 
the directions which have Neumann boundary conditions. These fields are the 
bosonic components  of a vector supermultiplet in $(p+1)$ dimensions. The low 
energy effective theory is then a supersymmetric theory in $(p+1)$ dimensions 
and the fluctuations of D-branes is described by the the scalar 
fields, which are transverse to the D-brane. This action can be computed if 
one requires that the non-linear sigma model describing the propagation of an 
open string with Dirichlet boundary conditions in a general supergravity 
background is conformally invariant. This gives a set of constraints in the 
fields which are the same as the equations of motion derived from the 
following effective action:
\be
S_{Dp}=-T_{p}\int d^{p+1}\s e^{-\p}\sqrt{-det(g+{\cal F})}\,.
\label{strDBIact}
\ee
This is the Dirac-Born-Infled (DBI) action, where $g$ is the induced metric on 
the world-volume and ${\cal F}=dA-B$, with $B$ the pull-back of the NSNS 
two-form and $A$ the U(1) gauge field. $T_{p}$ is the tension of a Dp-brane 
and in what follows we shall use $T_p$ for Dp-branes. Apart from the massless 
modes of the open string this theory contains also their coupling to the
 massless closed string modes.

The DBI action is an abelian $U(1)$ gauge theory which reduces, to leading 
order in $\a^{\prime}$, to YM in $p+1$ dimensions with $9-p$ scalars fields, 
when the target space is flat. To see this let us take the NSNS 2-form $B$ in 
(\ref{strDBIact}) to be zero. Then, the low energy worldvolume action becomes
\be
S_{Dp}=-T_p\int d^{p+1}\s\sqrt{-det g}
\left(1+\frac{1}{4g_{YM}^2}F_{\m\n}F^{\m\n}+
\frac{2}{(2\pi{\a^{\prime}})^2}\partial_{\m}X^i\partial^{\m}X_i+
{\cal O}(F^4)\right)\,.
\ee
The Yang-Mills coupling constant $g_{YM}$ is related to the string parameters 
by
\be
g_{YM}^2=(2\pi)^{p-2}{\a^{\prime}}^{\frac{p-3}{2}}g_s\,.
\ee
If one includes fermions, then the low energy action for the D-branes becomes 
${\cal N}=1$ supersymmetric Yang-Mills theory in $p+1$ dimensions. For $p=9$ we 
have 
a D9-brane filling all the space and the open strings have their endpoints free 
to move in all the spacetime. In this case there are no scalars transverse to 
the D9 and the DBI action becomes $U(1)$ Yang-Mills in 10 dimensions. The DBI 
action for a general Dp-brane can be obtained by dimensional reduction to 
$p+1$ dimensions of this $U(1)$ SYM theory in ten dimensions. In this procedure 
the components of the gauge potential $A_{\m}$ in the directions along which we 
reduce become scalars.

A generalization of this is to consider a stack of N D-branes of the same 
kind. This is possible since the D-branes are BPS objects and then they  
can remain statically at any distance from each other. This introduces a new 
quantum number which distinguishes the branes on which the open strings end. 
This number is precisely the Chan-Paton factor. If one has N parallel D-branes 
labeled by an index $i$ ($i=1,\dots, N$), the massless fields ($A_{\m}, X^i$) 
will carry a pair of indices $i,j$ which refer to the branes to which the 
strings are attached. The strings are oriented and therefore there are $N^2-N$ 
components for each field. If the N D-branes are coincident, the effective 
theory becomes $U(N)$ SYM in $p+1$ dimensions. The  scalars become massless 
in this case and fit into the vector supermultiplet of $U(N)$ gauge theory. 
The scalar fields are then $U(N)$ matrices. The diagonal values of these 
scalars represent the locations of the branes in transverse space. If the 
branes are separated at some distance then the gauge group is broken to 
$U(1)^N$. This is because the scalars acquire an expectation value. Since the 
scalars are in the adjoint representation, the gauge 
group can be broken maximally to $U(1)^N$.

\medskip
\section{The Maldacena Conjecture} \label{secmaldacena}
\medskip                                             
\setcounter{equation}{0}

D-branes are massive objects and then they give rise to a curved geometry. The 
dynamics of the fields in this background is described by the classical supergravity 
equations of motion. On the other hand, the low energy dymamics of D-branes is 
described by the DBI action, which reduces to SYM theory when $\a^{\prime}
\rightarrow 0$. The Maldacena conjecture identifies both descriptions by taking 
into account their respective validity in the string theory moduli space. 

The simplest example is type IIB theory compactified on $AdS_5\times S^5$. 
This corresponds to the near horizon geometry of a stack of parallel D3-branes. 
The boundary of $AdS_5$ is 4 dimensional Minkowski spacetime. Then, the low 
energy dynamics of the massless closed string modes is described by ten 
dimensional supergravity in $AdS_5\times S^5$ and that of the open string 
massless modes by the ${\cal N}=4$ SYM action. The closed string modes propagate in the 
bulk and the open string massless modes are confined to live on the 
four-dimensional worldvolume of the brane. The interaction between the  brane and 
bulk modes is obtained through the explicit dependence of the DBI action on 
the graviton through the pull-back of the background metric. It becomes, in 
the boundary, flat Minkoski metric plus fluctuations associated to 
the gravitons. That is $G=\eta+\kappa_{10}\,h$, where $h$ is the graviton.

One can see that in the limit $\a^{\prime}=\ls^2\rightarrow 0$, with the 
parameters $g_s$ and $N$ fixed, both theories decouple. This is because in 
the limit given above $\kappa_{10}\sim\a^{\prime}\rightarrow 0$. The limit 
$\a^{\prime}\rightarrow 0$ corresponds 
in the supergravity side to small curvature radius in string units. The radius 
is given, for a stack of D3-branes, by $R=4\pi g_sN\ls^4$. In the gauge theory 
side the DBI action reduces to ${\cal N}=4$ U(N) SYM since all the higher 
derivative corrections have possitive powers of $\a^{\prime}$. Thus, we are 
left with two decoupled theories, supergravity on $AdS_5\times S^5$ and 
${\cal N}=4$ 
SYM on the boundary of $AdS$. When we take the limit $\a^{\prime}\rightarrow 0$ 
we are looking at a small portion of spacetime very close to the branes which 
is blown up, since the metric has to be rescaled to get a finite result.

Type IIB string theory on $AdS_5\times S^5$ is a quantum theory with an 
isometry group $SO(2,4)\times SO(6)$. On the other hand ${\cal N}=4$ SYM 
on 3+1 dimensions is a conformal field theory with conformal group $SO(2,4)$. 
Furthermore, the $SO(6)$ of the sphere can be identified with the R-symmetry 
group of the gauge theory. Thus, both theories have the same spacetime 
symmetries. The Maldacena conjecture \cite{Maldacena, MaldaRev} states that type IIB 
superstring theory 
on $AdS_5\times S^5$ and ${\cal N}=4$ U(N) super Yang-Mills (defined on the 
boundary 
of $AdS$) are exactly mathematically equivalent. However both descriptions are 
valid in different regimes of the string moduli space. In terms of the 't Hooft 
coupling $\l\equiv g_{YM}^2N$, which is the relevant parameter for large 
N Yang-Mills theory, the radius $R$ of $AdS$ and of the sphere is given by
\be
R^4\sim 4\pi g_sN\ls^4\sim g_{YM}^2N\ls^4=\ls^4\l\,.
\ee
In the limit $N\rightarrow\infty$ with $\l$ fixed, then $g_s\rightarrow 0$, and 
one can perform calculations in string perturbation theory. We would obtain the quantum 
non-perturbative description of ${\cal N}=4$ SYM 
in terms of the classical limit of string theory. If we furthermore consider 
the strong coupling limit of SYM ($\l\gg 1$) with $R$ fixed, then 
\be
\frac{R^4}{\ls^4}\sim g_sN\sim g_{YM}^2N=\l\gg 1\,,
\ee
and we are therefore dealing with the $\a^{\prime}=\ls^2\rightarrow 0$ limit 
of string theory and, in this limit, classical string theory becomes 
classical supergravity.


\chapter{Effective Brane Actions}
\medskip

The low energy dynamics of the small fluctuations around a classical 
supergravity brane solution is given by a $(p+1)$-dimensional supersymmetric 
field theory. The action describing the dynamics of these fluctuations 
must fulfill certain requirements. It must be invariant under 
world-volume diffeomorphisms and it should be classically equivalent 
when the brane propagates in two target space configurations related by a 
target space gauge symmetry. 

As was mentioned in section \ref{secbranes} the D-brane effective action can be 
obtained by requiring conformal invariance, that is, requiring the 
vanishing of the $\b$-functions for the string ending on a D-brane. 
The relations among the fields implied by this conditions are the same 
as the classical equations of motion derived from the action
\be
S\,=\,-\,\int_{\S_{p+1}}\,d^{p+1}\s\,e^{-\p}\,
\sqrt{-det(g +\cal F)}\ ,
\label{DBIaction}
\ee
which is called the Dirac-Born-Infeld action. Here $g_{\m\n}=\partial_{\m}x^m
\partial_{\n}x^ng_{mn}$ is the pull-back of the spacetime metric 
on the world-volume, where greek indices $\m, \n\dots$ are worlvolume indices 
and indices $m, n\dots$ are target space indices. We will keep this notation for 
the rest of the chapter. The action (\ref{DBIaction}) 
must be multiplied by a constant $T_p$, the D-brane tension, with dimensions 
$mass\times (length)^{-p}$, making the action dimensionless. We see that the 
one-form gauge potential $A$ enters the action through the modified field 
strength ${\cal F}=F-B=dA-B$, where $B$ is the pull-back to the worldvolume of 
the target NSNS two-form potential 
$B_{\m\n}=\partial_{\m}x^m\partial_{\n}x^nB_{mn}$. The form 
${\cal F}=F-B$ is invariant under the combined gauge transformations
\be
\d B\,=\,d\L\,,\ \ \ \ \d A\,=\,\L\,.
\ee
Dp-branes are charged under RR potentials and their action should contain 
a term coupling the brane to these fields. However, the RR fields are subject to 
gauge transformations and the coupling should be gauge invariant \cite{GrHuTo}. 
In order to find a candidate for this coupling let us define $C$ to be the 
formal sum of the background RR fields
\be
C=\sum_{r=0}^{8}C^{(r)}\,,
\ee
where $C^{(r)}$ is a differential form of degree $r$. 
The fields $C^{(r)}$ are the RR gauge potentials of either 
IIA ($r$ odd) or IIB ($r$ even) supergravity. Their gauge transformation 
is
\be
\d_{RR}C=d\L-H\wedge\L+d\L\wedge e^{B}\,,
\ee
where $H=dB$ is the NSNS 3-form field strength and $\L$ is a formal sum of 
arbitrary forms $\L^{(r)}$ with degree $r$ 
\be
\L=\sum_{r=0}^{7}\L^{(r)}.
\ee
The field strengths of the RR fields, given by
\be
R(C)=dC-H\wedge C=\sum_{r=0}^{10}R^{(r)}\,,
\label{RRfs}
\ee
are invariant  under the above 
transformation. The form expansion of $R(C)$ yields 
all the modified field strengths of the IIA and IIB 
potentials and their duals. To relate the potentials 
to their duals we impose
\be
R^{(p)}= ^*R^{(10-p)}\,,\ \ \ p\le 5\,,
\ee
where * is the Hodge dual in ten dimensions. This also 
implies that $R^{(5)}$ is self-dual. The RR field strengths 
given by (\ref{RRfs}) satisfy the Bianchi identity
\be
dR=H\wedge R.
\ee

The requirement of target gauge invariance and the fact that D-branes 
possess electric charge determines the coupling to be of the form
\be
\int_{\S_{p+1}}\,C\wedge e^{\cal F}\,,
\label{WZact}
\ee
where $C$ denotes the pull-back to the worldvolume of the 
corresponding target space potentials.
The integral (\ref{WZact}) over the worlvolume can be written 
as an integral over a $(p+2)$-dimensional manifold 
$M_{p+2}$ whose boundary is the $(p+1)$-dimensional 
worldvolume $\S_{p+1}$. Indeed it can be proved that 
\be
R(C)e^{\cal F}=d\,\left[C\,e^{\cal F}\right]\,,
\ee
and then it follows that (\ref{WZact}) is equivalent to
\be
S_{WZ}=\int_{M_{p+2}}\,R(C)\,e^{\cal F}\,,
\ee
which is manifestly gauge invariant. Moreover, despite of the 
fact that $C\,e^{\cal F}$ is not invariant under the RR 
gauge transformations, its variation is a total derivative, 
thus making (\ref{WZact}) gauge invariant. Actually, 
\be
\d_{RR}[C\wedge e^{\cal F}]=d[\L\,e^{\cal F}]\,.
\ee

The term (\ref{WZact}) is a topological term of the Wess-Zumino 
type and it will be denoted by $S_{WZ}$. If the Dp-brane has RR 
charge $q_{Dp}$ the total action will be the sum of a DBI and a 
WZ term in the form
\be
S\,=\,-\,T_p\,\int_{\S_{p+1}}\,d^{p+1}\s\,e^{-\p}\,
\sqrt{-det(g+\cal F)}\,+\,
q_{Dp}\,\int_{\S_{p+1}}\,C\wedge e^{\cal F}\,.
\ee
Extremal branes satisfy the BPS bound $T_p=|q_{Dp}|$ and their action 
will be given by
\be
S\,=\,-\,T_p\,\int_{\S_{p+1}}\,d^{p+1}\s\,e^{-\p}\,
\sqrt{-det(g+\cal F)}\,\pm\,T_p\,
\int_{\S_{p+1}}\,C\wedge e^{\cal F}\,,
\label{fulldpbact}
\ee
where the $+$ sign is for branes an the $-$ for anti-branes.

\medskip
\section{D-brane actions with kappa symmetry}
\medskip                                             
\setcounter{equation}{0}

We have discussed the coupling of the bosonic degrees of freedom 
of the D-brane to the bosonic background fields, and now we are concerned 
with its supersymmetric extension, appropriate to describe the 
low energy dynamics of the fluctuations around the classical brane 
solution. One can adopt two types of formalisms: 
the Neveu-Schwarz-Ramond formulation, with manifest world-sheet 
supersymmetry and the Green-Schwarz 
(GS) formulation with manifest target space supersymmetry. Both lead to the 
same theory but the second one makes target space supersymmetry manifest, and 
upon quantization, bosonic and fermionic degrees of freedom are unified in a 
single Fock space. The crucial ingredient in the GS formulation is a local 
fermionic symmetry of the world volume theory called ``kappa symmetry''. 
It was first identified by Siegel \cite{siegel} for the superparticle 
\cite{brink}, and subsequently applied to the superstring \cite{green1} 
and the super 2-brane in eleven dimensions \cite{bergshoeff2}, followed 
by all super $p$-branes (without world-volume gauge fields) 
\cite{achucarro1,duff2}.

We want to formulate a generalization for D-branes of the previous description. In the 
case 
of super $p$-branes, whose only degrees of freedom are $(X^m,\theta)$, $(p + 1)$-
dimensional actions in a $d$-dimensional spacetime have been formulated that have 
super-Poincar\'e symmetry 
in $d$ dimensions realized as a global symmetry. In addition they have world-
volume general coordinate invariance, and a local fermionic symmetry called 
``kappa symmetry''. Manifest target space supersymmetry requires the 
introduction of more physical fields than the ones needed to fill in the 
supermultiplet. We can use the local symmetries of the theory to eliminate 
the extra degrees of freedom. Indeed, one can choose a physical gauge in which
all the remaining fields represent dynamical degrees of freedom
of the world-volume theory. Specifically, one can use the $(p+1)$-domensional 
diffeomorphism invariance of the world volume theory to identify $p+1$ of the 
space-time coordinates with the world-volume coordinates. This is called the 
``static gauge''. In such a gauge the remaining spatial coordinates can be
interpreted as $9-p$ scalar fields representing transverse excitations of 
the brane. In fact, they are the Goldstone bosons associated to spontaneously
broken translational symmetries. The main distinction between D-branes and the 
previously studied super $p$-branes is that the field content of the 
world-volume theory includes an abelian vector gauge field $A_\mu$. 
The gauge field $A_\mu$ gives $p - 1$ physical degrees of freedom, so altogether there is 
a total of 
8 bosonic modes. 
 
The role played by $\k$-symmetry is to eliminate the extra fermionic 
degrees of freedom. Indeed, the 32 $\theta$ coordinates are cut in half 
by the equation of motion, and in half again by gauge fixing local 
$\k$-symmetry (gauge fixing $\k$-symmetry involves, as we will see, a projection 
condition), giving rise to 8 physical fermionic degrees of freedom.
The crucial step is to make the right choice of which half of the
$\theta$ coordinates to set to zero in the static gauge. If one doesn't make a
convenient choice, the formulas can become unwieldy. This symmetry reflects the 
fact that the presence of the brane breaks half of the supersymmetry in 
$d$ dimensions, so that half of it is realized linearly and half of it 
nonlinearly in the world-volume theory.  The physical fermions of the 
world-volume theory are the Goldstone fermions associated to the broken 
supersymmetries. 

Fortunately, at least for D-branes in type II theories, there is a gauge choice
for the $\theta$ coordinates that leads to simple tractable formulas. Specifically,
we set one of the two Majorana--Weyl spinors ($\theta_1$ or $\theta_2$) equal
to zero. Then the fields of the remaining gauge-fixed theory are just the
gauge field $A_{\mu}$, the remaining $\theta$,
and the $9-p$ scalar fields mentioned above.

\medskip
\subsubsection{Flat superspace}
\medskip

We will obtain kappa symmetric actions for
D-branes embedded in a flat non-compactified 10 dimensional 
spacetime before generalizing them to the case of curved backgrounds. 
The DBI part of the action is not invariant under kappa symmetry 
and this determines the WZ part of the action such that the full 
action is invariant. 

As before, the $X^m$ are the coordiantes of the ten dimensional spacetime. 
The fermionic coordinates of $N=2$ Type IIA superspace, $\th_1$ and 
$\th_2$, are Majorana-Weyl spinors of opposite chirality and 
we can group them in a unique fermion $\th=\th_1+\th_2$, 
where now, $\th$ is Majorana but not Weyl. In the IIB case there are 
two Majorana-Weyl spinors $\theta_\alpha$ $ (\alpha = 1,2)$ 
of the same chirality. The group that naturally acts on it is 
SL(2,R), whose generators we denote by
Pauli matrices  $\s_1,\s_2, \s_3$.

Superspace forms may be expanded on the coordinate basis of 1-forms $dZ^M$, or 
on the inertial frame basis $E^{\underline M}= dZ^ME_M{}^{\underline M}$, where 
$E_M{}^{\underline M}$ is the supervielbein and $\underline M$ are 
flat indices. The basis $E^{\underline M}$ 
decomposes under the action of the Lorentz group into a Lorentz vector 
$E^{\underline m}$ and a Lorentz spinor $E^{\underline\a}$. The latter 
is a 32-component Majorana spinor for IIA 
superspace and a pair of chiral Majorana spinors for IIB superspace. Thus
\be
E^{\underline M} =\cases{(E^{\underline m}, E^{\underline\a}) & (IIA)\,,\cr
                         (E^{\underline m}, E^{{\underline\a}\ I})\, I=1,2 & (IIB)\,.}
\ee
The pull-back to the worldvolume of the super-vielbeins is given by
\be
E_{\m}{}^{\underline M} = \partial_{\m} Z^M E_M{}^{\underline M}\,.
\label{actpull2}
\ee
The superfields in the NSNS sector of type IIA supergravity are the 
superspace forms 
\be
E^m = \Pi^m  = dX^m  + \bar\th \G^m d\th,\ \ \ E^{\a}=d\th^{\a},
\label{actpull}
\ee
where $\G^m$ are 32$\times$32 Dirac matrices and the two-form potential $B$, 
given by
\be 
B = - \bar\th \G_{11} \G_m d\th \left(d X^m + {1\over 2} \bar
\th \G^m d\th\right). 
\ee
where $\G_{11} = \G_{0}\G_{1}\ldots \G_{9}$ is the chirality operator 
($\{ \G_{11}, \G^m \} =0$). All the indices are already flat. The NSNS 
fields of type IIB supergravity are the same with $\G_{11}$ replaced by 
$\s_3$.

Let us denote the world-volume coordinates by $\xi^\m, $ $\m = 0, 
1, \ldots, p$ and take the world-volume signature to be $(-+ \cdots +)$. The
world-volume theory is supposed to have global IIA or IIB super-Poincar\'e
symmetry.  This is achieved by constructing it out of manifestly supersymmetric
quantities. The superspace supersymmetry transformations are
\be
\d_\epsilon \th = \epsilon, \qquad \d_\epsilon X^m = \bar\epsilon \G^m
\th\,,
\label{actsusytr}
\ee
and two supersymmetric quantities are $\partial_{\m}\th$ and 
$\Pi_\m^m = \partial_\m X^m - \bar\th \G^m \partial_\m\th$. The 
DBI term of the D-brane action is 
\be
S_{DBI} = - \int d^{p + 1} \s \sqrt{- {\rm {\rm det}} 
(g_{\m\n} + {\cal F}_{\m\n})}\,,
\ee
where $g_{\m\n}$ is the induced world-volume metric 
\be
g_{\m\n} = E_\m{}^m E_\n{}^n\,\eta_{mn}\,,
\ee
and 
\be
{\cal F}=F-B=dA-B\,,
\ee
with A being the U(1) world-volume gauge field. The quantity $\cal F$ 
in which the gauge field appears is required to be supersymmetric.
The variation $\d_{\epsilon}B$ is an exact differential 
form. In fact under the transformations of (\ref{actsusytr})
\be
\d_\epsilon B = -\bar\epsilon \G_{11} \G_m d\th \left(d X^m +
{1\over 2} \bar\th \G^m d\th\right)
+ {1 \over 2} \bar\th \G_{11} \G_m d\th \bar\epsilon \G^m
d\th.
\ee
and then $\d_{\epsilon} {\cal F}=0$ is satisfied for an appropriate 
choice of $\d_{\epsilon} A$, namely: 
\be
\d_\epsilon A =  \bar\epsilon \G_{11} \G_m \th d X^m + {1\over
6} (\bar\epsilon \G_{11} \G_m \th \bar\th \G^m d\th +
\bar\epsilon \G_m \th\bar\th \G_{11} \G^m  d\th).
\ee

As we have seen, the world-volume theory of a D-brane contains 
also a Wess-Zumino type term describing the coupling of the brane 
to RR fields
\be
S_{WZ} = \int \O_{p + 1},
\ee
where $\O_{p + 1}$ is a $(p+1)$-form. To understand the supersymmetry of 
$S_{WZ}$, it is useful to characterize it by a formal $(p+2)$-form
\be  
I_{p+2} = d\O_{p+1}, 
\ee
which is typically a much simpler and more symmetrical expression
than $\O_{p+1}$. It will be entirely constructed from 
the supersymmetry invariants $d\th$, $\Pi^m$ and ${\cal F}$. This implies 
that the variation  $\d_{\epsilon} \O_{p+1}$ must be exact, and therefore 
$S_{WZ}$ is invariant. 

The two terms $S_{DBI}$ and $S_{WZ}$ are then supersymmetric and worldvolume 
diffeomorphism invariant. However, we will not require them to 
be individually kappa symmetry invariant. It is the sum of them what must be invariant. 
Just as for the cases of super $p$-branes we require that, whatever $\d_{\k}\th$ 
is
\be
\d_{\k} X^m = \bar\th \G^m \d_{\k}\th = - \d_{\k} \bar\th \G^m \th,
\label{kappaX}
\ee
Then, in terms of the induced $\gamma$ matrices $\g_\m \equiv 
E_{\m}{}^m\G_m = \Pi_\m^m \G_m$ we have
\be
\d_{\k} g_{\m\n} = - 2\d_{\k}\bar\th (\g_\m \partial_\n + \g_\nu
\partial_\mu)\th,
\label{Gtrans}
\ee
and the kappa variation of B is given by
\be
\d_{\k} B =  -2 \d_{\k} \bar\th \G_{11} \G_m d\th \Pi^m
+ d\Big(-\d_{\k}\bar\th \G_{11} \G_m \th \Pi^m + {1\over 2}
\d_{\k} \bar\th \G_{11} \G_{m} \th \bar \th \G^m d\th
- {1\over 2} \d_{\k} \bar\th \G^m \th \bar\th \G_{11}
\G_m d\th\Big).
\label{Btrans}
\ee
To obtain a simple result for $\d_{\k}{\cal F}$ we can choose the 
transformation of $A$ such that it cancels the closed form in 
(\ref{Btrans}) and then 
\be
\d_{\k} {\cal F} = 2 \d_{\k} \bar\th \G_{11} \G_m d\th \Pi^m.
\ee
In the type IIB case (p odd) one should replace $\G_{11}$ by $\s_3$.

Using eqs. (\ref{Gtrans}) and (\ref{Btrans}) we compute the kappa 
transformation of the DBI term in the action
\be
\d_{\k} {\cal L}_{DBI} = 2\,\sqrt{-det(g\,+\,{\cal F})}\,\d\bar\th\,
\g_{\m}\,\{(g\,-\,{\cal F}\G_{11})^{-1}\}^{\m\n}\,\partial_{\n}\th\,,
\label{actkavarDBI}
\ee
where now, for the type IIB theory, $\G_{11}$ should be replaced by $-\s_3$, 
(the sign appears because it has been moved past $\g_{\m}$). The 
(\ref{actkavarDBI}) can be rewritten in the following form,
\be
\d_{\k}  {\cal L}_{DBI}= 2\d_{\k} \bar\th \g^{(p)} T_{(p)}^\n 
\partial_\n \th,\ \ \ 
\big(\g^{(p)}\big)^2 = 1\,, 
\label{kappaborn}
\ee
where $\g^{(p)}$ and $T_{(p)}^{\n}$ can be represented in terms of  
antisymmetric tensors as follows: 
\br
\g^{(p)}\,=\,\frac{1}{(p+1)!}\,\epsilon^{\m_1\dots\m_{p+1}}\,
\g_{\m_1\dots\m_{p+1}}\,,\ \ \ \ \ \ 
T_{(p)}^{\n}\,=\,\frac{1}{p!}\,\epsilon^{\n_1\dots\n_p\n}\,
T_{\n_1\dots\n_p}\,.\rc
\label{kapparhotp2}
\er
From eqs. (\ref{actkavarDBI}) and (\ref{kappaborn}) it follows that 
$\r^{(p)}\equiv\sqrt{-det(g+{\cal F})}\g^{(p)}$ and $T_{(p)}^{\n}$ have 
to solve the equations
\br
(\r^{(p)})^2\,=\,-det(g+{\cal F})\,,\ \ \ \ \ \ \ 
\r^{(p)}\,\g_{\m}=T_{(p)}^{\n}\,(g - {\cal F}
\G_{11})_{\n\m}\,.\rc
\label{kapparhotp}
\er
The solution to eq. (\ref{kapparhotp}), obtained in \cite{aganagic}, is given in 
terms of the matrix valued one-form ${\Pislash} \equiv \g_\m d\xi^\m = \Pi^m \G_m$ 
and the following sums of differential forms
\br
&&\r = \sum_{p} \r_{p+1}\,,\ \ \ \ \r_{p+1}=\sqrt{-det(g+{\cal F})}\,
{1\over (p+1)!}\g_{\m_1\m_2\dots\m_{p+1}}\,d\xi^{\m_1}\dots d\xi^{\m_{p+1}}\,,
\rc\rc
&&T = \sum_{p} T_p\,,\ \ \ \ T_p={1\over p!}{T_{\n_{1}\n_{2}\dots \n_{p}}} 
d\xi^{\n_{1}}\dots d\xi^{\n_{p}}\,,
\er
where $p$ runs with even or odd values for the IIA or IIB case respectively. The 
solution is then given by
\br
\r_A&=&e^{{\cal F}} S_A (\Pislash)\ \ \ \ \ \ \ \ \
T_A = e^{{\cal F}} C_A (\Pislash)\,, \ \ \ \ \ \ \ (IIA)\nonu\\
\r_B&=&e^{{\cal F}} C_B (\Pislash)\s_1\ \ \ \ \ \
T_B = e^{{\cal F}} S_B (\Pislash)\s_1\,, \ \ \ \ \ (IIB)
\label{acttatb}
\er
where 
\br
S_A(\Pislash)&=&\sum_{l=0}\,(\G_{11})^{l+1}\,\frac{\Pislash^{2l+1}}{(2l+1)!}\,,\ \ 
\ 
C_A(\Pislash)\ =\ \sum_{l=0}\,(\G_{11})^{l+1}\,\frac{\Pislash^{2l}}{(2l)!}\,, \nonu\\
\nonu\\
S_B(\Pislash)&=&\sum_{l=0}\,(\s_3)^{l}\,\frac{\Pislash^{2l+1}}{(2l+1)!}\,,\ \ \ \ \ \ \ \ 
C_B(\Pislash)\ =\ \sum_{l=0}\,(\s_3)^{l+1}\,\frac{\Pislash^{2l}}{(2l)!}\,.
\label{acttatbb}
\er

In order to implement kappa symmetry, we require that the WZ term of the 
action transforms as 
\be
\d_{\k}  {\cal L}_{WZ}= -2\d_{\k} \bar\th T^\n_{(p)} \partial_\n \th\,.
\label{kappawess}
\ee
In this case, the variation of the complete action under the kappa 
transformation has the form
\be
\d ({\cal L}_{DBI}+{\cal L}_{WZ}) = 2\d_{\k} \bar\th (1 - \g^{(p)}) 
T^\n_{(p)} \partial_\n\th. 
\label{kappatotal}
\ee
From eq. (\ref{kapparhotp}) we see that $(\g^{(p)})^2=1$ and then ${1\over 2} 
(1 \pm \gamma^{(p)} )$ are projection operators. The transformation 
\be
\d_{\k} \bar\th = \bar \k (1 + \g^{(p)})
\label{kappabarth}
\ee
makes the complete action kappa symmetry invariant.

We want to find a WZ action such that its kappa variation is the one given 
in (\ref{kappawess}). This variation can be written in terms of the p-form 
$T_p$ as follows:
\be
\d S_{WZ} = 2 (-1)^p\int \d \bar\th T_p d\th = \d \int \O_{p+1},
\ee
Then, the variation of the (p+2)-form $I_{p+2} = d\O_{p+1}$ is given by
\be
\d I_{p+2} = 2(-1)^pd\Big( \d\bar\th T_p d\th\Big) = 
2 (-1)^p\Big( \d d\bar\th T_p d\th -  \d\bar\th dT_p d\th\Big)\,,
\ee
and this equation is solved by
\be
I_{p+2} = (-1)^p d\bar\th T_p d\th\ .
\label{kappadwess}
\ee
Using the results for $T_A$ and $T_B$ given in (\ref{acttatb}) it follows that
\be
d{\cal L}_{WZ}=R e^{{\cal F}}, 
\ee
with, $R=d\bar\th C_A(\Pislash)d\th$\ \  (IIA), or 
$R=-d\bar\th S_B (\Pislash)d\th$\ \  (IIB). This RR field 
strengths satisfy the respective type IIA or IIB supergravity 
equations of motion.

Eqs. (\ref{kappaX}) and (\ref{kappabarth}) give the kappa-symmetry 
transformations of the worldvolume fields, obtained in the 
flat superspace case, namely,
\br
&&\d_{\k} \th=(1 + {({C^{-1}}{\g^{(p)}}C)}^t)\,\k\,=\,
(1 + \G_{\k})\,\k\,, \rc\rc
&&\d_{\k} X^m=\bar\th \G^m \d_{\k}\th = - \d_{\k} \bar\th \G^m \th\,,\rc
\label{ktransfwf}
\er
where C is the charge conjugation matrix ($\bar\th=\th^tC,\ 
\G_m^t=-C^{-1}\G_mC, C^{-1}=C^t=-C$) and the matrix $\G_{\k}$ 
is given by
\be
\G_{\k}\,=\,C{\g^{(p)}}^tC^{-1}\,.
\ee

Given the arbitrary variations $\d Z^M$ of the worldvolume fields $Z^M$, 
let us define 
\be
\d E^{\underline M} \equiv \d Z^M E_M{}^{\underline M}\,. 
\label{kappaevar}
\ee
The transformations (\ref{ktransfwf}) can be rewritten in terms of the 
pull-backs of the super-vielbeins. It follows from eq. (\ref{actpull}) 
that $E_{\a}^{\underline\a}=\d_{\a}^{\underline\a}$ and 
$E_{\m}^{\underline\a}=0$. Then eq. (\ref{kappaevar}) gives
\br
&&\d_\k E^{\underline m}=0\ , \nonu\\
&&\d_\k E^{\underline\a}=(1+ \G_{\k})\k\,.
\er
It is illustrative to obtain explicitely the form of $\G_{\k}$. We will do 
this for the type IIB theory. From eqs. (\ref{acttatb}), (\ref{acttatbb}) 
and (\ref{kapparhotp2}) we see that $\g^{(p)}$ takes the form  
\br
\g^{(p)}\,=\,\frac{(-1)^n}{\sqrt{-det(g+{\cal F})}}\,
&&\sum_{n=0}^{p+1}\,\frac{1}{2^n\,n!\,(p+1-2n)!}\,
\epsilon^{\m_1\dots\m_{p+1}}\,\times\ \ \ \ \ \ \ \ \ \ \rc\rc
&&\times\,{\cal F}_{\m_1\m_2}\dots{\cal F}_{\m_{2n-1}\m_{2n}}
\g_{\m_{2n+1}}\dots\g_{\m_{p+1}}\,\s_3^{\frac{p-3}{2}-n}(i\s_2)\,.\rc
\label{rhop}
\er
We will use here the following identity given in \cite{bergshoeff}:
\be
\epsilon^{\m_1\dots\m_k\l_{k+1}\dots\l_{p+1}}\,\g_{\l_{k+1}\dots\l_{p+1}}\,=\,
(-1)^{\frac{k(k-1)}{2}}\,(p+1-k)!\,\g^{\m_1\dots\m_k}\,\G_{(0)}\,,
\ee
where the matrix $\G_{(0)}$ is defined by
\be
\G_{(0)} = {1\over (p+1)!}\,\epsilon^{\m_1\dots
\m_{(p+1)}}\g_{\m_1\dots \m_{(p+1)}}\,.
\ee
With the form of $\g^{(p)}$ given in (\ref{rhop}) it follows that
\be
\G_{\k}\,=\,\g^{(p)}\,,
\ee
and we get finally the form of $\G_{\k}$ in terms of the bosonic worldvolume 
fields and the pull-backs of the background fields:
\be
\G_{\k}\,=\,{1\over \sqrt{-\det(g+{\cal F})}}\,\,\sum_{n=0}^{p+1} 
\,{1\over 2^n n!}\,\,\g^{\m_1\n_1\dots \m_n\n_n}\,\,{\cal F}_{\m_1\n_1}\dots 
{\cal F}_{\m_n\n_n}\,(-1)^n\,\s_3^{\frac{p-3}{2}-n}(i\s_2)\,\G_{(0)}\,.
\ee
The same expression is valid for type IIA theory if one replaces 
$(-1)^n\,\s_3^{\frac{p-3}{2}-n}(i\s_2)$ by $(\G_{11})^{n+(p+2)/2}$.

\medskip
\subsubsection{D-branes in curved backgrounds}
\medskip

In the generalization to curved gackgrounds all the target space fields 
are functions of the superspace coordinates. The super D-brane action is 
given by
\br
S_{D_p}&=& -T_p\,\int d^{p+1}\xi\, e^{-\phi}\sqrt{-\det (g_{\m\n} + 
\,{\cal F}_{\m\n})}\,+\,T_p\,\int C e^{\cal F}\,,  \rc\rc
g_{\m\n}&=&E_{\m}{}^m E_{\n}{}^n\,g_{mn}\,, \rc\rc
{\cal F}_{\m\n}&=&F_{\m\n}-E_{\m}{}^{\underline M} E_{\n}{}^{\underline N}
B_{\underline{MN}}\,,\rc\rc
C^{(r)}&=&{1\over r!} dZ^{M_1}\cdots dZ^{M_r} C_{M_r\dots M_1}\,=\,
{1\over r!} E^{{\underline M}_1}\cdots E^{{\underline M}_r} 
C_{{\underline M}_r\dots {\underline M}_1}\,,\rc\rc
\label{kappasdbal}
\er
where $g_{\m\n}$ in the induced metric on the world-volume. 

In the case of curved backgrounds the transformations will
depend on the superspace coordinates and they are defined by
\br
&&\d_\k E^a\,(x,\th)=0\ , \nonu\\
&&\d_\k E^{\a}\,(x,\th)=(1+ \G_{\k})\k\,,
\label{kappacurt}
\er
with $\G_{\k}$ satisfying that $(\G_{\k})^2=1$, such that 
${1 \over 2}(1\pm \G_{\k})$ are indeed projectors. 

In refs. \cite{bergshoeff, cederwall} it is proved that 
the above transformations (\ref{kappacurt}) leave the super D-brane 
action (\ref{kappasdbal}) invariant, with the matrix $\G_{\k}$ given by
\be
\G_{\k}\,=\,{1\over \sqrt{-\det(g+{\cal F})}}\,\,\sum_{n=0}^\infty 
\,{1\over 2^n n!}\,\,\g^{\m_1\n_1\dots \m_n\n_n}\,\,{\cal F}_{\m_1\n_1}\dots 
{\cal F}_{\m_n\n_n}\,\,J^{(n)}_{(p)}\,,
\label{kappamatrix}
\ee
where $g$ is the induced 
metric on the worldvolume and  ${\cal F}=F-P[B]$ (P[$\dots$] denoting the 
pullback) and $J^{(n)}_{(p)}$ is the following matrix
\be
J^{(n)}_{(p)} = \cases{(\G_{11})^{n+(p-2)/2}\,\,\G_{(0)} & (IIA)\,,\cr
(-1)^n (\sigma_3)^{n+(p-3)/2}\,\,i\sigma_2 \otimes \G_{(0)} & (IIB)\,,}
\label{kappaj}
\ee
where $\G_{(0)}$ is defined by
\be
\G_{(0)} = {1\over (p+1)!}\,\epsilon^{\m_1\dots
\m_{(p+1)}}\g_{\m_1\dots \m_{(p+1)}}\,.
\label{Gamma0}
\ee
Recall that in the type IIB theory the spinor $\epsilon$ is actually composed
by two Majorana-Weyl spinors $\epsilon_1$ and $\epsilon_2$ of well defined
ten-dimensional chirality, which can be arranged as a two-component vector in
the form
\be
\epsilon\,=\,\pmatrix{\epsilon_1\cr\epsilon_2}\,\,.
\ee
The Pauli matrices appearing in the expression of $J^{(n)}$ act on this
two-dimensional vector. Moreover, in  eqs. (\ref{kappamatrix}) and 
(\ref{Gamma0}) $\g_{\m_1\m_2\dots}$ are
antisymmetrized products of the induced world-volume Dirac matrices
which, in terms of the ten-dimensional constant gamma matrices 
$\G_{\underline{M}}$, are given by
\be
\g_{\m}\,=\,E_{\m}{}^{\underline m}\,\G_{\underline m}\,=\,
\partial_{\m}\,X^m\,E_{m}{}^{\underline m}\,
\G_{\underline m}\,\,,
\label{indgamma}
\ee
where $E_{m}{}^{\underline m}$ is the ten-dimensional vielbein. 

The proof of refs. \cite{bergshoeff, cederwall} asumes that the 
background fields satisfy the corresponding type IIA or IIB superspace 
constraints, which are equivalent to the equations of motion. Thus kappa 
symmetry invariance of D-brane actions requires the brane to move in on-shell 
supergravity backgrounds. We see that the matrix $\G_{\k}$ depends on the 
background fields (the vielbeins and $B$) but also on the worldvolume fields.

As we have seen, it is necessary that the DBI and WZ terms in (\ref{kappasdbal})
are multiplied by the same factor $T_p$ in order to their kappa variations 
cancel each other and the total action remain invariant. In general, the 
coefficient multiplying the WZ term of the action is the charge density $Q_p$,
which determines the strength with which the brane couples to the RR potentials. 
Then, kappa invariance requires that $|Q_p|=T_p$, and thus, the brane must be a 
BPS state. The ambiguity in the sign 
is because we can have a brane or an anti-brane. In all our discussion we have 
considered the case $Q_p=T_p$ . If we had taken $Q_p=-T_p$ then it would appear 
$(1 + \g^{(p)})$ in eq. (\ref{kappatotal})  and then the variation 
$\d_{\k} \th=(1 - \G_{\k})\k$ would leave the action invariant.

\medskip
\subsubsection{Bosonic configurations}
\medskip

It was mentioned that we must gauge fix local kappa symmetry in order 
to remove the extra degrees of freedom in the (GS) formulation of the 
worldvolume theory of the brane. The extra bosonic degrees of freedom 
are removed by choosing the ``static gauge''. In this gauge the world-volume 
general coordinate invariance is used to equate $p + 1$ of the target-space 
coordinates with the world-volume coordinates $(X^\m = \s^\m)$.        

We will be  interested in bosonic configurations ($\th=0$) that preserve a 
fraction of the spacetime supersymmetry. We need the transformation of the 
$\th$ field up to terms linear in $\th$. The supersymmetry and kappa variation 
of $\th$ are  
\be
\d \th = \epsilon + (1 + \G_{\k})\k\,.
\label{kappafix1}
\ee
In general, kappa symmetry and supersymmetry transformations will not leave $\th=0$. 
For any supersymmetry transformation with parameter $\epsilon$ one can 
choose a value of $\k$, such that $\d\th=0$. This is equivalent to gauge 
fix kappa symmetry. Actually, we shall impose a gauge fixing condition of the 
form
\be
{\cal P}\th=0\,,
\label{kappafix2}
\ee
where ${\cal P}$ is a field independent projection operator, 
${\cal P}^2={\cal P}$. Then, the remaining non-vanishing components of 
$\th$ are given by $(1-{\cal P})\th$ and, clearly 
\be
\th={\cal P}\th+(1-{\cal P})\th\,.
\ee
The transformation (\ref{kappafix1}) becomes a global supersymmetry 
transformation. In order to preserve the gauge fixing condition 
(\ref{kappafix2}) is then
\be
\d {\cal P}\th = {\cal P}\epsilon + {\cal P}(1 + \G_{\k})\k = 0\,,
\label{kappafix3}
\ee
which determines $\k=\k(\epsilon)$ and is equivalent to having 
unbroken supersymmetry. The condition of unbroken supersymmetry 
on the remaining non-vanishing components of $\th$ is 
\be
\d (1-{\cal P})\th = (1-{\cal P})\epsilon + 
(1-{\cal P})(1 + \G_{\k})\k(\epsilon)= \epsilon + 
(1 + \G_{\k})\k(\epsilon)=0\,,
\ee
where we have used eq. (\ref{kappafix3}) in the third equality. We can 
project the above condition to the $(1-\G_{\k})$ subspace to obtain the 
condition
\be
\G_{\k} \epsilon = \epsilon\,.
\label{kappafix4}
\ee
No more information can be obtained by projecting to the $(1+\G_{\k})$ 
subspace. We will call (\ref{kappafix4}) ``kappa symmetry condition''. If 
we had considered the case of an anti-brane the transformation of $\th$ 
would be $\d \th = \epsilon + (1 - \G_{\k})\k$ and the condition for unbroken 
supersymmetry $\epsilon + (1 - \G_{\k}(\epsilon))\k=0$, so that, projecting 
onto the $(1+\G_{\k})$ subspace  we obtain the condition
\be
\G_{\k} \epsilon = -\epsilon\ \ \ \ ({\rm anti-brane})\,.
\label{kappafix5}
\ee
The eqs. (\ref{kappafix4}) or (\ref{kappafix5}) can be applied  to determine 
the fraction of space supersymmetry preserved by bosonic membrane configurations. 
The spinor $\epsilon$ is the target space supersymmetry parameter. For brane probes eqs. 
(\ref{kappafix4}, \ref{kappafix5}) are the only supersymmetry conditions that arise. 
However, for supergravity configurations with branes as sources we have the D-brane 
action coupled to supergravity and the conditions above must be complemented 
with the Killing spinor equations of the supergravity theory. Thus the fraction 
of supersymmetry preserved by a D-brane is determined by the number of solutions 
of (\ref{kappafix4}) with $\epsilon$ being Killing spinor solutions of the 
background.

\medskip
\section{Dielectric Branes} \label{secdielbr}
\medskip                                             
\setcounter{equation}{0}

The usual world-volume action for a D$p$-brane can be extended
to the case of N coincident D$p$-branes, where the world-volume
theory involves a U(N) gauge theory \cite{Myers}. The action is 
constructed such that it is consistent with T-duality. The resulting 
action has a variety of potential terms, {\it i.e.,} nonderivative 
interactions, for the nonabelian scalar fields. This action also shows that
D$p$-branes naturally couple to RR potentials of all form degrees,
including both larger and smaller than $p$+1. This mechanism is responsible 
of the fact that D$p$-branes can be ``polarized'' by external fields.

T-duality is most easily implemented in terms of the ten-dimensional 
matrix $E_{\m\n}$ defined by:
\be
E_{\mu\nu}=G_{\mu\nu}-B_{\mu\nu}\ ,
\label{eee}
\ee
where $G_{\mu\nu}$ is the background metric in the 
string-frame and $B_{\mu\nu}$ is the NSNS 2-form potential. 
The action will be formulated in the static gauge, \ie , we use first 
spacetime diffeomorphisms to define the fiducial
world-volume as $x^i=0$ with $i=p+1,\dots,9$, and then with
world-volume diffeomorphisms, the internal coordinates are taken as 
the remaining spacetime coordinates on that surface,
$\s^a=x^a$ with $a=0,1,\dots,p$. The scalar fields $\P^i$, which descibe 
the embedding of the branes in spacetime, are identified with the coordinates 
transverse to the worldvolume through $x^i=2\pi\ls^2\P^i$. In this way the 
scalars and the gauge fields have dimensions of $length^{-1}$.

The approach of \cite{Myers} was to take the D9-brane action as the 
starting point. The D9 fills the entire space and there are no transverse 
directions. Therefore, there are no worldvolume scalar fields. Then a T-duality 
transformation is applied in the $9-p$ directions $x^i$ with $i=p+1,\dots ,9$ 
to produce the corresponding Born-Infeld action for a Dp-brane. The 
resulting action is more easily written in terms of the quantities 
\be
Q^i{}_j\equiv\d^i{}_j+i\l\,[\P^i,\P^k]\,E_{kj}\ ,
\label{myqij}
\ee
where $\l=2\pi\ls^2$ and the $E_{kj}$ are the ones given in (\ref{eee}). 
The final T-dual Born-Infeld action becomes
\be
S_{BI}=-T_p \int d^{p+1}\s\,STr\left(e^{-\p}\sqrt{-det\left(
P\left[E_{ab}+E_{ai}(Q^{-1}-\d)^{ij}E_{jb}\right]+
\,F_{ab}\right)\,det(Q^i{}_j)}
\right)\ ,
\label{biactcom}
\ee
where $P$ denotes the pull-back to the worldvolume of the branes and the 
second index in the expression $(Q^{-1}-\d)^{ij}$ has been raised using 
$E^{ij}$ (rather than $G^{ij}$), where $E^{ij}$ denotes the inverse of 
$E_{ij}$ ($E^{ik}E_{kj}=\d^i{}_j$). 

The Born-Infeld action (\ref{biactcom}) 
is highly nonlinear and is incomplete without a precise prescription for how 
the gauge trace should be implemented. In the above action the symmetrized 
trace prescription suggested by Tseytlin \cite{yet} is adopted. This 
trace is completely symmetric between all nonabelian expressions of the form
$F_{ab}$ and $i[\P^i,\P^j]$. This prescription does not seem to capture the 
full physics of the infrared limit but, on the other hand, in configurations
with supersymmetry, it seems to describe properly the physics of the nonabelian 
Yang-Mills fields and even provides solutions of the full open-string equations 
of motion.

The non-abelian generalization of the Chern-Simons action consistent with T-
duality is given by:
\be
S_{CS}=\m_p\int Tr\left(P\left[e^{i\l\,\hi_\Phi \hi_\Phi}\sum C^{(n)}\,\right]
e^{\cal F}\right)\ ,
\label{csnon}
\ee
where $\cal F$ is given in our conventions by ${\cal F}_{\m\n}=F_{\m\n}-
B_{\m\n}$. 
Apart from the generalization of all worldvolume fields to their abelian  
counterparts and the inclusion of the overall gauge trace, 
$e^{i\l\,\hi_\Phi \hi_\Phi}$ appears inside the 
pull-back. Here $\hi_\P$ denotes the interior product by $\P^i$
regarded as a vector in the transverse space. Acting on forms,
the interior product is an anticommuting operator of form degree --1. 
The interior products in eq. (\ref{csnon}) must act on both
the Neveu-Schwarz two-form and the RR potentials. The exponential makes a 
nontrivial contribution in eq. (\ref{csnon})
because of the nonabelian nature of the displacement vectors
$\P^i$. For example, one has
\be
\hi_\P \hi_\P C^{(2)} = \P^j\P^i\,C^{(2)}_{ij}={1\over2}C^{(2)}_{ij}
\,[\P^j,\P^i]\,.
\label{nonabint}
\ee
It is clear that this action reduces to the form appearing in 
(\ref{fulldpbact}) for the abelian theory of a single Dp-brane. 
The integrand is to be evaluated 
by considering the expression in each set of brackets,
from the innermost around $\sum C^{(n)}e^B$ to the outermost for the
gauge trace. In particular, this means that first $\sum C^{(n)}e^B$
is expanded as a sum of forms in the ten-dimensional spacetime,
and so only a finite set of terms in the exponential will contribute.
Then the exponential $\exp({i\l\,\hi_\P \hi_\P})$ acts on this sum
of forms, and so again only a finite set of terms in this second 
exponential will contribute. As with the Born-Infeld action, the
proposed action (\ref{csnon}) is given with the symmetrized trace 
prescription. 

From the action (\ref{csnon})  it follows that Dp-branes couple not only 
to the RR potential with form degree $n=p+1$ or $n=p-1,p-3,\dots$ that follow 
from the action (\ref{fulldpbact})),  but can also couple 
to the RR potentials with $n=p+3,p+5,\ldots$ through the additional
interactions involving commutators of the nonabelian scalars.

\medskip
\section{Worldvolume Action of the M-theory Five-brane} \label{secPSTact}
\medskip                                             
\setcounter{equation}{0}

The complete Born-Infeld like action for a bosonic five-brane in a background 
of gravitational and antisymmetric gauge fields of $D=11$ supergravity was 
proposed in \cite{PST} and is called the PST action of the M5-brane. The M5 
carries in its worldvolume a second-rank antisymmetric gauge field whose 
field-strength is self-dual (or chiral) in the free field limit. 
When one tries to incorporate the self-duality condition into an action a 
problem arises with preserving manifest Lorentz invariance of the model. The 
lack of manifest Lorentz invariance will result in the lack of manifest general 
coordinate invariance of the worldvolume of the five-brane, which can substantially 
complicate the construction and study  of the complete supersymmetric (and 
$\k$-invariant) super-five-brane action in such a formulation.

In \cite{PST} a $d=6$ worldvolume covariant action of a Born-Infeld type for 
a five-brane with a chiral boson in its worldvolume was proposed. 
In their approach, $d=6$ covariance is achieved by introducing a single auxiliary 
scalar field entering the action in a polynomial way. This auxiliary field 
ensures not only worldvolume covariance but also that all the constraints of 
the model are of the first class, which is important when performing its 
covariant quantization.

The five-brane can be coupled to the antisymmetric gauge field of $D=11$ 
supergravity. When the five-brane couples to the rank-three antisymmetric 
gauge field $C^{(3)}$, the local symmetries require the 
addition to the action of an appropriate Wess-Zumino term. Upon a double 
dimensional reduction of the five-brane worldvolume from $d=6$ to $d=5$ and 
of the target space from $D=11$ to $D=10$, the five-brane action reduces  
to the Born-Infeld action with a Wess-Zumino term for a D4-brane given in 
(\ref{fulldpbact}).

In the PST formalism the worldvolume fields are the three-form field strength 
$F$ and the scalar field $a$ (the PST scalar). The bosonic action is the sum 
of three terms:
\be
S\,=\,T_{M5}\,\int\,d^6\xi\,\Big[\,
{\cal L}_{DBI}\,+\,{\cal L}_{H\tilde H}\,+\,{\cal L}_{WZ}
\,\Big]\,\,,
\label{actPST}
\ee
where the tension of the M5-brane is $T_{M5}\,=\,1/ (2\pi)^5\,\lp^6$. In the
action (\ref{actPST}) the field strength $F=dA_{(2)}$ is combined with the 
pullback 
$P[C^{(3)}]$ of the background potential $C^{(3)}$ to form the field $H$:
\be
H\,=\,F\,-\,P[C^{(3)}]\,\,.
\label{actH}
\ee
Let us now define the field $\tilde H$ as follows:
\be
{\tilde H}^{ij}\,=\,{1\over 3!\,\sqrt{-{\rm det}\,g}}\,
{1\over \sqrt{-(\partial a)^2}}\,
\epsilon^{ijklmn}\,\partial_k\,a\,H_{lmn}\,\,,
\label{acttilH}
\ee
with $g$ being the induced metric on the M5-brane worldvolume. The explicit
form of the three terms of the action is:
\br
{\cal L}_{DBI}&=& -\sqrt{-{\rm det} (g_{ij}\,+\,\tilde H_{ij})}\,\,,\rc\rc
{\cal L}_{H\tilde H}&=&{1\over 24 (\partial  a)^2}\,\,
\epsilon ^{ijkmnr}\,H_{mnr}\,H_{jkl}\,
g^{ls}\partial_i a \,\partial_s a\,=\,H\,\tilde H\,,\rc\rc 
{\cal L}_{WZ}&=&{1\over 6!}\epsilon ^{ijklmn}\,\Bigg[\,
P[C^{(6)}]_{ijklmn}\,+\,10\,H_{ijk}\,P[C^{(3)}]_{lmn}\,\Bigg]\,\,.
\label{actLM5}
\er
The axiliary field $a$ can be eliminated by gauge fixing the corresponding 
local symmetry at the price of the loss of manifest spacetime covariance. 

The gauge transformations of the potentials $C^{(3)}$ and $C^{(6)}$ are 
given by
\be
\delta C^{(3)}\,=\,d\Lambda^{(2)}\,,\ \ \ 
\delta C^{(6)}\,=\,d\Lambda^{(5)}\,+\,\frac{1}{2}\Lambda^{(2)}
\wedge F_{(4)}\,.
\label{M5rr}
\ee
Clearly, $\delta H=0$ provided that $\delta A_{(2)}=\Lambda^{(2)}$ and 
then, the first two terms in the action (\ref{actPST}) are invariant under 
these transformations. The field-strength $F_{(4)}$ satisfies the 
following supergravity equation of motion:
\be
d{}^*F_{(4)}\,=\,-\frac{1}{2}\,F_{(4)}\wedge F_{(4)}\,,
\ee
where ${}^*$ is the Hodge dual in eleven dimensions, and then, we can 
write ${}^*F_{(4)}$ in terms of the potentials $C^{(3)}$ and $C^{(6)}$ 
in the following way:
\be
{}^*F_{(4)}\,=\,dC^{(6)}\,-\,\frac{1}{2}\,C^{(3)}\wedge F_{(4)}\,,
\ee
which is invariant under the gauge transformations given in (\ref{M5rr}). 
The Wess-Zumino term transforms under these transformations as a total 
derivative. Actually,
\be
\delta\left[C^{(6)}+\frac{1}{2}H\wedge C^{(3)}\right]\,=\,
d\left(\Lambda^{(5)}-\frac{1}{2}H\wedge\Lambda^{(2)}\right)\,,
\ee
thus making (\ref{actPST}) gauge invariant.

The form of the action (\ref{actPST}) and the fact that its double dimensional 
reduction results in the action (with Born-Infeld and Wess-Zumino terms) for 
the D4-brane in $D=10$ suggest that it should be possible to construct a $d=6$ 
covariant $\k$-symmetric action for a super-five-brane of M-theory 
analogous to that found for the super D-branes. This action would describe the 
embedding of the five-brane worldvolume into a superfield background of $D=11$ 
supergravity.

Once all the terms of the five-brane action have been completely fixed 
at the bosonic level, one can generalize it to a supersymmetric action by 
replacing background fields with the corresponding superfields in target space 
parametrized by supercoordinates. It has been checked in \cite{PSTSor} that 
the supersymmetric action (\ref{actPST}) posseses also a worldvolume fermionic 
$\k$-symmetry, and hence describes BPS 5-branes configurations preserving half 
the $D=11$ supersymmetry. The corresponding $\k$-matrix is however not uniquely 
defined. The most suitable form of this matrix is the following: 
\be
\G_{\k}=-{\n_m\g^m\over \sqrt{-{\rm det} (g+\tilde H)}}\,\,
\Bigg[\,\g_n t^n\,+\,
{\sqrt{-g}\over 2}\,\g^{np}\,\tilde H_{np}\,
+\,{1\over 5!}\g_{i_1\dots i_5}\,\epsilon^{i_1\dots i_5n}\n_n\,\,
\Bigg]\,\,,
\label{kappaM5}
\ee
where we have defined the following quantities:
\be
\nu_p\,\equiv {\partial_p a\over \sqrt{-(\partial a)^2}},
\,\,\,\,\,\,\,\,\,\,\,\,\,\,\,\,\,\,\,\,
\,\,\,\,\,\,\,\,\,\,\,\,\,\,\,\,\,\,\,\,\,\,
t^m\,\equiv\,{1\over 8}\,
\epsilon^{mn_1n_2p_1p_2q}\,\tilde H_{n_1n_2}\,\tilde H_{p_1p_2}\,\n_q\,\,.
\label{kM5def}
\ee
In eq. (\ref{kappaM5}) $\g_{i_1i_2\dots }$  are antisymmetrized
products of the worldvolume Dirac matrices \break\hfill
$\g_i=\partial_iX^M\,E^{\underline M}_M\,\G_{\underline M}$. 

As common to all super-p-branes the $\k$-symmetry of the brane action 
requires the background superfields to satisfy constraints. In this case 
the $\k$-symmetry is compatible with $D=11$ supergravity constraints which 
put the background on the mass-shell.


\chapter{Baryons from Brane Actions}
\medskip

The interplay between gauge theory and gravity is on the basis of
the Maldacena conjecture \cite{Maldacena}. Indeed, Maldacena has argued
that there is a remarkable duality between classical supergravity in the
near-horizon region and the large $N$ 't Hooft limit of the
$SU(N)$ Yang-Mills theory. In particular, the near-horizon geometry of
parallel D3-branes is equivalent to the space $AdS_5\times S^5$, whose
boundary can be identified with four-dimensional Minkowsky space-time.
In this case one gets the so-called $AdS/CFT$ correspondence between
type IIB superstrings  on $AdS_5\times S^5$ and ${\cal N}=4$
non-abelian super Yang-Mills theory in four dimensions \cite{Gubser,
Witten, Gross}.  Within this context, one can
compute Wilson loop expectation values by considering a fundamental
open string placed on the interior of the $AdS_5$ space and having its
ends on the boundary \cite{Wloop}. This formalism has been used to
extract quark potentials both in the supersymmetric and
non-supersymmetric theory. Moreover, Witten \cite{Wittenbaryon} has
proposed a way to incorporate baryons by means of  a D5-brane wrapped
on the   $S^5$, from which  fundamental strings that end on the
D3-branes emanate \cite{Branbaryon, Imabaryon}. This is the string theory 
counterpart to the gauge theory SU(N) baryon vertex, representing a 
bound state of N external quarks.

The baryon vertex can be described by a worldvolume approach, allowing 
a unified description of the D5-branes and strings.
In ref. \cite{CGS1}, the problem of a D5-brane moving under the
influence of a D3-brane background was considered. The equations of motion of
the static \mbox{D5-brane} are solved if one assumes that the fields
satisfy a certain first-order BPS differential equation that was found
previously by Imamura \cite{Imamura}. This BPS  condition can be
integrated exactly in the near-horizon region of the background
D3-brane geometry and the solutions are spikes of the D5-brane
worldvolume which can be interpreted as bundles of fundamental strings
ending on the D3-branes. Actually, these kinds of spikes are a general
characteristic of the Dirac-Born-Infeld non-linear gauge theory and can
be used to describe strings attached to branes \cite{CM, Gibbons}. 

The BPS condition for the full asymptotically flat D3-brane metric was
analyzed numerically in ref. \cite{CGS1}. As a result of this study one
gets a precise description of the string creation process which takes
place when two D-branes cross each other, \ie\ of  the so-called
Hanany-Witten effect \cite {HW, HWothers}. These results were
generalized in ref. \cite{CGS2} to the case of baryons in confining
gauge theories. Moreover, in ref. \cite{Craps} the BPS differential
equation was obtained as a condition which must be fulfilled in order
to saturate  a lower bound on the energy.

This chapter is based on \cite{Baryon}. Here, we shall study the 
worldvolume dynamics of a
D(8-p)-brane  moving under the action of the gravitational and
Ramond-Ramond fields of a stack of background \mbox{Dp-branes} 
for $p\le 6$.  The D(8-p)-brane is extended along the directions
orthogonal to the worldvolume of the background Dp-branes. For $p=3$
this system is the one analyzed in ref. \cite{CGS1}. By using the
method of ref.  \cite{Craps}, we
will find a BPS first-order differential equation whose solutions also
verify the equation of motion. Remarkably, we will be able to
integrate analytically this BPS equation both in the near-horizon and
asymptotically flat metrics. For $p\le 5$ the solutions we
will find are similar to the ones described in ref. \cite{CGS1}, \ie\
the  D(8-p)-brane worldvolume has spikes which can be interpreted as
flux tubes made up of fundamental strings. Our analytical solution will
allow us to give a detailed description of the shape of these tubes,
their energy and of the process of string creation  from D-brane
worldvolumes.

The organization of this chapter is the following. In section \ref{barwitten} 
we review the baryon vertex construction of \cite{Wittenbaryon}.
In section \ref{barsec2} we formulate our problem. The BPS differential equation is 
obtained in section \ref{barsec3}. In section \ref{barsec4} this BPS condition 
is integrated for $p\le 5$ in the near-horizon region. The $p=6$ system 
requires a special treatment, which is discussed in section \ref{barsec5}. 
The solution of the BPS equation for the full \mbox{Dp-brane} geometry is 
obtained in section \ref{barsec6}. The detailed analysis of this solution 
is performed in appendix 3.A. In section \ref{barsecsym} we analyze the 
supersymmetry preserved by the BPS configurations obtained in section 
\ref{barsec3}. Finally, in section \ref{barsec7}, we summarize our results.

\medskip
\section{The baryon vertex in $SU(N)$ gauge theory} \label{barwitten}
\medskip                                             
\setcounter{equation}{0}

The bayon vertex is a finite energy configuration with $N$ external quarks. 
Finding a baryonic vertex in the ${\cal N}=4$ theory does not mean that
the theory has baryonic particles, or operators.  Baryonic particles
would appear in a theory that has dynamical quark fields
(that is, fields transforming in the fundamental representation of $SU(N)$);
in their absence, we get only a baryonic vertex, a gauge-invariant
coupling of $N$ external charges.
Introducing dynamical quark fields would require breaking some
supersymmetry. It has been argued that Yang-Mills theory with $SU(N)$ 
gauge group has a large $N$ limit as a closed string theory, with an 
effective string coupling constant $\lambda\sim 1/N$. If elementary quarks 
are added, it is believed that there are also open strings (describing 
mesons), with an open string coupling constant $\lambda'\sim 1/\sqrt N$.
Baryons of an $SU(N)$ theory with dynamical quarks are $N$-quark bound
states, which one would expect to have masses of order $N$.  As $N\sim
1/(\lambda')^2$, such states can be interpreted as solitons in the open
string sector. The question is how external quarks in ${\cal N}=4$ super 
Yang-Mills are described in terms of Type IIB on $AdS_5\times S^5$.
With Lorentz signature, the boundary of $AdS_5$ is $S^3\times\RR$,
where $\RR$ is the ``time'' direction; $S^3\times\RR$ is a universal
cover of the conformal compactification of Minkowski space.
External quarks are regarded as endpoints of  strings
in $AdS$ space.  Thus, to compute the energy for a
time-independent configuration with an external quark at a point
$x\in S^3$ and an external antiquark at $y\in S^3$, one considers
configurations in which a string inside
the $AdS$ space  connects the boundary points $x$ and $y$.

The strings in question are elementary Type IIB superstrings
if the external charges are electric charges, \ie , particles in the
fundamental representation of the gauge group $SU(N)$.  External
monopoles would be boundaries of $D$-strings, and external charges
of type $(p,q)$ are boundaries of $(p,q)$ strings.

\begin{figure}
\centerline{\hskip -.8in \epsffile{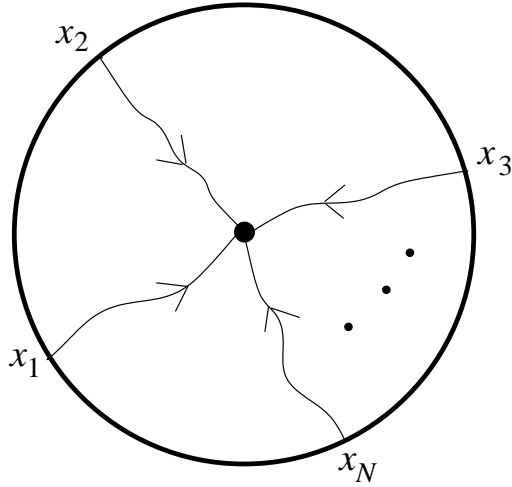}}
\caption{N elementary strings attached to points $x_1,x_2,\dots ,x_N$
on the boundary of AdS space and joining at a baryon vertex in the interior.}
\label{witten1}
\end{figure}

We now want to find a ``baryon'' vertex connecting $N$ external quarks, 
with their color wave functions combined together by an $N^{th}$ order 
antisymmetric tensor of $SU(N)$. For this purpose, we place external quarks at 
boundary points $x_1,x_2,\dots,x_N$.  We consider a configuration in which 
each of the boundary points is the endpoint of an elementary superstring 
in $AdS_5\times S^5$, with all $N$ strings oriented in the same way. We 
want, as in figure 1, to find a ``baryon vertex,'' where these $N$ strings 
can somehow terminate in the interior of $AdS_5\times S^5$.

The baryon vertex is simply a wrapped fivebrane. The reason why the wrapped 
fivebrane is a baryon vertex is the following. In Type IIB superstring theory, 
there is a self-dual five-form field $F_{(5)}$. The $AdS_5\times S^5$ 
compactification which is related to $SU(N)$ gauge theory has $N$ units of 
five-form flux on $S^5$: 
\be
\frac{1}{2\pi}\int_{S^5} F_{(5)}\,=\,N\,.
\label{barvertuno}
\ee
On the $D$-fivebrane worldvolume, there is an $U(1)$ gauge field A.
It couples to $F_{(5)}$ by means of an interaction of the type
\be
\frac{1}{2\pi}\int_{S^5\times\RR}A\wedge F_{(5)}\,.
\ee
Because of this coupling and (\ref{barvertuno}), the $F_{(5)}$ field contributes
$N$ units of $A$-charge.  Since the total charge of a $U(1)$ gauge
field must vanish in a closed universe, there must be $-N$ units
of charge from some other source.

Such a source are elementary strings ending on the fivebrane.
The endpoint of an elementary string that ends on the fivebrane
is electrically charged with respect to the $A$-field, with a charge
that is $+1$ or $-1$ depending on the orientations of the string
and fivebrane.  To cancel the $F_{(5)}$ contribution to the $A$-charge,
we need $N$ strings, all oriented in the same way, ending on the
fivebrane.  The fivebrane is thus a baryon or antibaryon vertex, depending
on its orientation.

We have thus  provided evidence that the gauge theory on $S^3\times\RR$
that is dual to Type IIB on $AdS_5\times S^5$ has the property
that  it is possible to form a gauge-invariant combinations of $N$ quarks,
\ie , of $N$ particles in the fundamental representation of the gauge
group.  This is in agreement with the fact that the gauge group is
believed to be $SU(N)$ (and not $U(N)$, for example).

Let us estimate the energy of the baryon vertex 
in the 't Hooft limit (the string coupling $\l$ going
to zero, and $N$ to infinity, with $\l N$ fixed).
Since the $D$-brane tension is of order $1/\lambda$, that is of order $N$,
and the volume of $S^5$ remains finite for $N\to\infty$,
the baryon vertex for $N$ external electric charges has an energy of
order $N$, as expected for baryons in the large $N$ limit of QCD.  
For $N$ external magnetic
charges, the energy of the baryon vertex is controlled by the tension
of an NS fivebrane, and has an extra factor of $1/\lambda$ or $N$; this
factor seems natural, as  magnetic charges
are boundaries of $D$-strings, whose tension is $1/\lambda$ or $N$ times
the tension of the elementary strings.  In each case, the energy of
the baryon vertex is comparable to the energy of the $N$ strings that
are attached to it, and hence a dynamical study of baryonic states
would involve balancing these two energies.

\setcounter{equation}{0}
\section{Worldvolume  Dynamics} \label{barsec2}
\medskip
The ten-dimensional metric (in the string frame) corresponding to a
stack of $N$ coincident extremal Dp-branes in type II superstring theory
was given in eq. (\ref{strdpbs}). If we parametrize the space transverse to 
the Dp-brane by spherical coordinates, then it takes the following form
\beq
ds^2\,=\,\Bigl[\,f_p(r)\,\Bigr]^{-{1\over 2}}\,\,
(\,-dt^2\,+\,dx_{\parallel}^2\,)\,+\,
\Bigl[\,f_p(r)\,\Bigr]^{{1\over 2}}\,\,
(\,dr^2\,+\,r^2\,d\Omega_{8-p}^2\,)\,\,.
\label{baruno}
\eeq
In eq.  (\ref{baruno}) $x_{\parallel}$ are $p$-dimensional cartesian
coordinates along the branes, $r$ is a radial coordinate
parametrizing the distance to the branes and $d\Omega_{8-p}^2$ is the
line element of an unit $8-p$ sphere. The harmonic function 
$f_p(r)$ appearing in eq. (\ref{baruno}) is given by
\beq
f_p(r)\,=\,a\,+\,\Bigl(\,{R\over r}\Bigr)^{7-p}\,\,,
\label{bardos}
\eeq
where $a=0,1$ and the radius $R$ was given in eq. (\ref{strdbrad}) in terms 
of the number $N$ of branes, the Regge slope $\alpha\,'$ and the string
coupling constant $g_s$. The value $a=1$ in eq. (\ref{bardos}) corresponds 
to the full geometry of
the stack of Dp-branes. Actually, in order to have an asymptotically
flat space-time, we shall restrict ourselves to the case $p<7$. Taking
$a=0$ in the harmonic function $f_p(r)$ is equivalent to approximate
the Dp-brane geometry by the near-horizon metric of the `throat'
region of (\ref{baruno}). 

The Dp-brane solution we are considering is
also characterized by a dilaton field $\phi(r)$ and a 
Ramond-Ramond (RR) $p+2$-form field
$F^{(p+2)}$. The corresponding values are
\bear
e^{-\tilde\phi(r)}\,&=&\,\Bigl[\,f_p(r)\,\Bigr]^{{p-3\over 4}}
\,\,,\rc\rc
F^{(p+2)}_{t\,,\,x_{\parallel}^1\,,\,\cdots\,\,,\,x_{\parallel}^p,\,r}\,&=&\,
{d\over dr}\,\Bigl[\,f_p(r)\,\Bigr]^{-1}\,\,,
\label{barcinco}
\eear
where $\tilde\phi(r)\,=\,\phi(r)\,-\,\phi_{\infty}$ and
$\phi_{\infty}$ is the value of the dilaton field at infinite distance
of the Dp-branes 
(\ie\ $\phi_{\infty}\,=\,\lim_{r\rightarrow\infty}\,\phi(r)$). As was
pointed out in ref. \cite{Maldacena}, in order to trust this type II
supergravity solution, both the curvature in string units and the
dilaton must be small. This fact introduces restrictions in the values
of the radial coordinates for which the correspondence between the
supergravity and gauge theory descriptions is valid (see also 
ref.~\cite{charges}).

Let us now consider a D(8-p)-brane embedded in the transverse
directions of the stack of the background Dp-branes. The dynamics of
this  D(8-p)-brane is determined by its action, which is a sum of a
Dirac-Born-Infeld and a Wess-Zumino term
\be
S\,=\,-T_{8-p}\,\int d^{\,9-p}\s\,e^{-\tilde\phi}\,
\sqrt{-{\rm det}\,(\,g\,+\,F\,)}\,+\,
T_{8-p}\,\int d^{\,9-p}\s\,\, C\wedge e^{{\cal F}}\,,
\label{barseis}
\ee
where $g$ is the induced metric on the worldvolume of the
D(8-p)-brane, $A$  is a worldvolume abelian gauge field and $F=dA$ its
field strength.  The Wess-Zumino term in eq. (\ref{barseis}) couples 
the (7-p)-form potential $C^{(7-p)}$ to $F$. In this particular case 
$C^{(7-p)}$ is related to the pullback of the Hodge dual of the background 
RR (p+2)-form $F^{(p+2)}$ by $^*F^{(p+2)}=dC^{(7-p)}$.
Notice that this RR field acts as a source for the worldvolume gauge
field $A$. The coefficient $T_{8-p}$ appearing in the action
(\ref{barseis}) is the tension of a  D(8-p)-brane, which is given by
\cite{Polchi}
\beq
T_{8-p}\,=\,(2\pi)^{p-8}\,(\,\alpha\,'\,)^{{p-9\over 2}}\,
(\,g_s\,)^{-1}\,\,.
\label{barsiete}
\eeq
Let $\theta^1$, $\theta^2$, $\cdots$, $\theta^{8-p}$ be coordinates
which parametrize the $S^{8-p}$ transverse sphere. The worldvolume
coordinates $\s^{\alpha}$ ($\alpha\,=\,0\,,\,\cdots\,,\,8-p\,)$ will
be taken as
\beq
\s^{\alpha}\,=\,(\,t\,,\,\theta^1\,,\,\cdots\,,\,\theta^{8-p}\,)\,\,.
\label{barocho}
\eeq
We shall assume that the $\theta$'s are spherical angles on $S^{8-p}$
and that $\theta\equiv\theta^{8-p}$ is the polar angle 
($0\le\theta\le\pi$). Therefore, the $S^{8-p}$ line element 
$d\Omega_{8-p}^2$ can be decomposed as
\beq
d\Omega_{8-p}^2\,=\,d\theta^2\,+\,(\,{\rm sin}\,\theta)^{2}\,\,
d\Omega_{7-p}^2\,\,,
\label{barnueve}
\eeq
where $d\Omega_{7-p}^2$, which only depends on 
$\theta^1,\,\cdots\,,\theta^{7-p}$, is the line element of a $S^{7-p}$
sphere. We are going to consider static configurations of the
D(8-p)-brane which only depend on the polar angle $\theta$. Actually, we
shall restrict ourselves to the case in which the only non-vanishing
fields are $r(\theta)$, $x(\theta)$ and $A_0(\theta)$, $x$ being a
direction parallel to the background Dp-branes and $A_0(\theta)$ the component 
of $A$ along the time direction. It is straightforward
to verify that the action for such configurations can be written as
\bear
S\,&=&\,T\,T_{8-p}\,\Omega_{7-p}\,\int d\theta\,\,
({\rm sin}\,\theta)^{7-p}\,\times\rc\rc
&&\times\,
\Bigl[\,-r^{7-p}\,f_p(r)\,
\sqrt{r^2\,+\,r\,'^{\,2}\,+\,{x\,'^{\,2}\over f_p(r)}\,
-\,F_{0,\theta}^2}\,+\,
(-1)^{p+1}\,(7-p)\,A_0\,R^{7-p}\,\Bigr]\,\,,\rc\rc
\label{bardiez}
\eear
where the prime denotes derivative with respect to $\theta$ and 
$T\,=\,\int\,dt$. In eq. (\ref{bardiez}) $\Omega_{7-p}$ is the volume of
the unit (7-p)-sphere, given by
\beq
\Omega_{7-p}\,=\,{2\pi^{{8-p\over 2}}\over 
\Gamma\Bigl(\,{8-p\over 2}\Bigr)}\,\,.
\label{baronce}
\eeq
It is interesting at this point to remember that we are considering
static solutions of the gauge field that only depend on $\theta$. The
electric field for these configurations is
\beq
E\,=\,F_{0,\theta}\,=\,-\partial_{\theta}\,A_0\,\,.
\label{bardoce}
\eeq
Let us now define the displacement field $D_p$ as
\beq
D_p(\,\theta\,)\,\equiv\,{(-1)^{p+1}\over T\,T_{8-p}\,\Omega_{7-p}}\,
{\partial S\over \partial E}\,=\,(-1)^{p+1}
{({\rm sin}\,\theta)^{7-p}\,r^{7-p}\,f_p(r)\,E\over
\sqrt{r^2\,+\,r\,'^{\,2}\,+\,{x\,'^{\,2}\over f_p(r)}\,-\,E^2}}\,\,.
\label{bartrece}
\eeq
The extra factors appearing in the definition (\ref{bartrece}) have been
included for convenience. The Euler-Lagrange equation for $A_0$ implies
an equation for $D_p$ which is easy to find, namely
\beq
{d\over d\theta}\,D_p(\,\theta\,)\,=\,
-(\,7\,-\,p\,)\,R^{7-p}\,(\,{\rm sin }\,\theta\,)^{7-p}\,\,.
\label{barcatorce}
\eeq
Notice that the right-hand side of eq. (\ref{barcatorce}) only depends
on $\theta$. Therefore (\ref{barcatorce}) can be integrated and, as a
result, one can obtain $D_p$ as a function of $\theta$. We shall work
out the explicit form of $D_p(\,\theta\,)$ at the end of this section.
At present we only need to use the fact that $D_p(\,\theta\,)$
satisfies eq. (\ref{barcatorce}). Actually, substituting 
$(\,7\,-\,p\,)\,R^{7-p}\,(\,{\rm sin }\,\theta\,)^{7-p}$ by 
$-\partial_{\theta}\,D_p$ in the Wess-Zumino term of the action and
integrating by parts, one can recast $S$ as
\beq
S\,=\,-T\,U\,\,,
\label{barquince}
\eeq
with $U$ given by
\bear
U\,=\,T_{8-p}\,\Omega_{7-p}\,\int d\theta\,
&&\Biggl[\,\,(-1)^{p+1}\,E\,D_p(\theta)\,+\rc\rc
&&+\,({\rm sin}\,\theta)^{7-p}\,r^{7-p}\,f_p(r)\,
\sqrt{r^2\,+\,r\,'^{\,2}\,+\,{x\,'^{\,2}\over f_p(r)}\,-\,E^2}
\,\,\,\Biggr]\,\,.
\label{bardseis}
\eear
Notice that, as is evident from their relation (\ref{barquince}), $S$ and
$U$ give rise to the same Euler-Lagrange equation. Actually, since we
have eliminated $A_0$ in favor of $D_p$, we can regard $U$ as the
Legendre transform of $S$ and, therefore, $U$ can be considered as an
energy functional for the embedding of the D(8-p)-brane in the
Dp-brane background. The fields $E$ and $D_p$ are related, as is
obvious from eq. (\ref{bartrece}). It is not difficult to invert eq. 
(\ref{bartrece}) and get $E$ in terms of $D_p$. The result is
\beq
E\,=\,(-1)^{p+1}\,
\sqrt{
{r^2\,+\,r\,'^{\,2}\,+\,{x\,'^{\,2}\over f_p(r)}\over
(D_p(\theta))^2\,+
[({\rm sin}\,\theta)^{7-p}\,r^{7-p}\,f_p(r)\,]^2}}
\,\,\,D_p(\theta)\,\,.
\label{bardsiete}
\eeq
Using this relation we can eliminate $E$ from the expression of $U$
\beq
U\,=\,T_{8-p}\,\Omega_{7-p}\,\int d\theta\,
\sqrt{r^2\,+\,r\,'^{\,2}\,+\,{x\,'^{\,2}\over f_p(r)}}\,\,\,
\sqrt{(D_p(\theta))^2\,+
[({\rm sin}\,\theta)^{7-p}\,r^{7-p}\,f_p(r)\,]^2}\,\,.
\label{bardocho}
\eeq
Recall that $D_p(\theta)$ is a known function of $\theta$ (see
below). Therefore eq. (\ref{bardocho}) gives the energy functional $U$ in
terms of $x(\theta)$ and $r(\theta)$. These functions must be
solutions of the Euler-Lagrange equations obtained from $U$. From the
study of the functional $U$ in several situations we will determine
the shape of the D(8-p)-brane embedding in the background geometry. Let
us consider, first of all, the near-horizon approximation, which is
equivalent to taking $a=0$ in the harmonic function (\ref{bardos}). In
this case  $r^{7-p}\,f_p(r)\,=\,R^{7-p}$ and $U$ is given by
\beq
U\,=\,T_{8-p}\,\Omega_{7-p}\,\int d\theta\,
\sqrt{r^2\,+\,r\,'^{\,2}\,+\,
{r^{7-p}\over R^{7-p}}\,\,x\,'^{\,2}}\,\,\,
\sqrt{(D_p(\theta))^2\,+({\rm sin}\,\theta)^{2(7-p)}\,R^{2(7-p)}\,}
\,\,.
\label{bardnueve}
\eeq
By inspecting eq. (\ref{bardnueve}) one easily concludes that $U$
transforms homogeneously under a simultaneous rescaling of the radial
and parallel coordinates of  the form
\beq
(r,x)\,\rightarrow\,(\,\alpha\,r,\,\alpha^{{p-5\over 2}}\,x\,)\,\,,
\label{barveinte}
\eeq
where $\alpha$ is a constant. It follows that the equations of motion
derived from $U$ are invariant under the transformation
(\ref{barveinte}). Moreover, the homogeneous character of $U$ under the
transformation (\ref{barveinte}) implies the following scaling law
\beq
x\,\sim\,{R^{{7-p\over 2}}\over r^{{5-p\over 2}}}\,\,,
\label{barvuno}
\eeq
where the power of $R$ has been determined by imposing dimensional
homogeneity of both sides of (\ref{barvuno}). Eq.  (\ref{barvuno}) is
precisely the holographic UV/IR relation found in ref.
\cite{holography}. According to eq. (\ref{barvuno}), 
for $p<5$, large radial
distances  ($r\rightarrow\infty$) correspond to small values of the
parallel coordinate $x$. For $p=5$, the distance $x$ is insensitive to
changes in $r$, whereas, for $p=6$,  $x$ increases when $r$ grows. The
consequences of this relation for the correspondence between field
theories and near-horizon supergravities have been discussed in ref.
\cite{holography} (see also ref. \cite{holoothers}). 
In our approach we shall see that, indeed, the
$p=6$ case is special and the embedding of the D2-brane in the
D6-brane background geometry has new characteristics, which must be
studied separately.

Let us finish this section by giving the expressions of the
displacement fields $D_p$. The $\theta$-dependence of $D_p$ can be
obtained by integrating the right-hand side of eq. (\ref{barcatorce}).
The results one gets for the different cases are

\bear
D_0(\,\theta\,)\,&=&\,
R^7\,\Bigl[\,{\rm cos}\,\theta\,(\,{\rm sin }^6\theta\,+\,
{6\over 5} \,{\rm sin }^4\,\theta\,+\,{8\over 5} \,
{\rm sin }^2\,\theta\,+\,{16\over 5}\,)\,+\,
{16\over 5}\,(\,2\nu\,-\,1\,)\,\Bigr]\,\,,
\rc\rc
D_1(\,\theta\,)\,&=&\,
R^6\,\Bigl[\,{\rm cos}\,\theta\,(\,{\rm sin }^5\theta\,+\,
{5\over 4} \,{\rm sin }^3\theta\,+\,{15\over 8} \,{\rm sin }\,\theta\,)
\,+\,{15\over 8} \,(\,\pi\nu\,-\,\theta\,)\,\Bigr]\,\,,
\rc\rc
D_2(\,\theta\,)\,&=&\,
R^5\,\Bigl[\,{\rm cos}\,\theta\,(\,{\rm sin }^4\theta\,
+\,{4\over 3} \,{\rm sin }^2\theta\,+\,
{8\over 3}\,)\,+\,{8\over 3}\,(\,2\nu\,-\,1\,)\Bigr]\,\,,
\rc\rc
D_ 3(\,\theta\,)\,&=&\,
R^4\,\Bigl[\,{\rm cos}\,\theta\,(\,{\rm sin }^3\theta\,
+\,{3\over 2} \,{\rm sin }\,\theta\,)\,+\,
{3\over 2} \,(\,\pi\nu\,-\,\theta\,)\,\Bigr]\,\,,\label{barvdos}\\\rc
D_ 4(\,\theta\,)\,&=&\,R^3\,\Bigl[\,{\rm cos}\,\theta\,
(\,{\rm sin}^2\theta\,+\,2\,)\,+\,2\,(\,2\nu\,-\,1\,)\Bigr]\,\,,
\rc\rc
D_ 5(\,\theta\,)\,&=&\,R^2\,\Bigl[\,{\rm cos}\,\theta\,
{\rm sin }\,\theta\,+\,\pi\nu\,-\,\theta\,\Bigr]\,\,,
\rc\rc
D_ 6(\,\theta\,)\,&=&\,R\,\Bigl[\,{\rm cos}\,\theta\,
+\,2\nu\,-\,1\,\Bigr]\,\,,
\nonumber
\eear
where we have parametrized the additive constant of integration by
means of a parameter $\nu$. We have chosen the $\theta$-independent
term in $D_p(\theta)$ in such a way that the value of the displacement
field at $\theta=\pi$ is given by
\beq
D_p(\pi)\,=\,-2\sqrt\pi\,{\Gamma\Bigl(\,{8-p\over 2}\Bigr)\over
\Gamma\Bigl(\,{7-p\over 2}\Bigr)}\,\,R^{7-p}\, (\,1\,-\,\nu\,)\,\,.
\label{barvtres}
\eeq
Eq. (\ref{barvtres}) allows to give a precise meaning to the parameter
$\nu$ \cite{CGS1}. 
In fact, as we shall verify below, there exist solutions for
which $r\rightarrow\infty$ and $x'\rightarrow 0$ when
$\theta\rightarrow\pi$ in a way that simulates a `flux tube' attached
to the D(8-p)-brane. The `tension' (\ie\ the energy per unit radial
distance) for one of these spikes can be obtained from the expression
of $U$ in eq. (\ref{bardocho}). Actually, as $\theta\rightarrow\pi$ one
can check that the terms with $r\,'$ dominate over the other terms in
the first square root in eq. (\ref{bardocho}), whereas the second square
root can be approximated by $|\,D_p(\pi)\,|$. 
As a result \cite{CGS1}, the
tension of the spike is $T_{8-p}\,\Omega_{7-p}\,|\,D_p(\pi)\,|$. Using
the value of $D_p(\pi)$ given in eq. (\ref{barvtres}) and the fact 
that (see eqs. (\ref{strdbrad}), (\ref{barsiete}) and (\ref{baronce})):
\beq
T_{8-p}\,\Omega_{7-p}\,R^{7-p}\,=\,
{NT_f\over 2\sqrt{\pi}}\,
{\Gamma\Bigl(\,{7-p\over 2}\Bigr)\over
\Gamma\Bigl(\,{8-p\over 2}\Bigr)}\,\,,
\label{barvcuatro}
\eeq
we conclude
that the tension of the $\theta=\pi$ spike is  $(1-\nu)\,N\,T_f$,
where $T_f$ is the tension of the fundamental string. This result 
implies that we can interpret the 
$\theta=\pi$ tube as a bundle of $(1-\nu)\,N$ fundamental strings
which, from the gauge theory point of view,  corresponds to a baryon
formed by $(1-\nu)\,N$ quarks. It is clear that, although in the
classical theory $\nu$ is a continuous parameter, upon quantization
$\nu$ should be a multiple of $1/N$ taking values in the range 
$0\le\nu\le1$. 

Let us finally point out that there exist other solutions for which 
$r\rightarrow\infty$ and $x'\rightarrow 0$ when
$\theta\rightarrow 0$. The asymptotic tension for these solutions is 
$T_{8-p}\,\Omega_{7-p}\,|\,D_p(0)\,|$. The value of $D_p(\theta)$ at
$\theta=0$ can be obtained from the values given in eq. (\ref{barvdos}).
The result is
\beq
D_p(0 )\,=\,2\sqrt\pi\,{\Gamma\Bigl(\,{8-p\over 2}\Bigr)\over
\Gamma\Bigl(\,{7-p\over 2}\Bigr)}\,\,R^{7-p}\, \nu\,\,.
\label{barvcinco}
\eeq
Using again eq. (\ref{barvcuatro}), one can verify that 
$T_{8-p}\,\Omega_{7-p}\,|\,D_p(0)\,|\,=\,\nu\,N\,T_f$, which
corresponds to a bundle of $\nu\,N$ fundamental strings. This fact
provides an interpretation of $\nu$ for this second class of
solutions.

\setcounter{equation}{0}
\section{BPS conditions} \label{barsec3}
\medskip
In the remaining of this paper we are going to study solutions of the
equations of motion of the brane probes which are not extended in the
directions parallel to the background branes. This amounts to take
$x\,'=0$ (\ie\ $x(\theta)\,=\,$ constant) in our previous equations.
From the expression (\ref{bardocho}) of $U$ we can get the differential
equation which determines $r$ as a function of $\theta$. Indeed, the
Euler-Lagrange equation derived from $U$ is

\bear
&&{d\over d\theta}\,\Biggr[\,{r'\over \sqrt{r^2\,+\,r\,'^{\,2}}}
\sqrt{(D_p(\theta))^2\,+[({\rm sin}\,\theta)^{7-p}
\,r^{7-p}\,f_p(r)\,]^2}
\,\Biggr]\,=\,\rc\rc\rc
&&=\,{r\over \sqrt{r^2\,+\,r\,'^{\,2}}}\,\,
\sqrt{(D_p(\theta))^2\,+[({\rm sin}\,\theta)^{7-p}\,
r^{7-p}\,f_p(r)\,]^2}
\,+\,\label{barvseis}\\\rc\rc
&&+\,(7-p)a\,\,{\sqrt{r^2\,+\,r\,'^{\,2}}\over r}\,\,
{(\,r\,{\rm sin }\,\theta\,)^{2(7-p)}\,f_p(r)\over
\sqrt{(D_p(\theta))^2\,+[({\rm
sin}\,\theta)^{7-p}\,r^{7-p}\,f_p(r)\,]^2}}\,\,.
\rc\rc
\nonumber
\eear
Trying to obtain a solution for such a complicate second-order
differential equation seems, a priori, hopeless. A possible strategy
to solve eq. (\ref{barvseis}) consists of finding a first integral for
this system. Notice that the integrand of $U$ depends explicitly on
$\theta$ (see eq. (\ref{bardocho})) and, therefore,  there is no first
integral associated to the invariance under shifts of $\theta$ by a
constant. Nevertheless, we will be able to find a first-order equation
such that any function $r(\theta)$ satisfying it is a solution of the
equation  (\ref{barvseis}). This first-order condition is much simpler
than the Euler-Lagrange equation (\ref{barvseis}) and, indeed, we will be
able to solve it analytically, both in the near-horizon ($a=0$) and
asymptotically flat ($a=1$) geometries. For $p=3$ and $a=0$, the first
order equation was found by Imamura 
\cite{Imamura} as a BPS \cite{BPS} condition for the
implementation of supersymmetry in the worldvolume theory of a
D5-brane propagating in a D3 background. This condition was
subsequently extended to ($p=3$, $a=1$)  and ($p=4$, $a=0$) in
refs. \cite{CGS1} and \cite{CGS2} respectively. 
In order to generalize these results,
 we shall  follow here the approach 
of ref. \cite{Craps} (see also ref. \cite{Gomis}),
where the BPS equation was found by requiring the saturation of a
certain bound on the energy functional. 

Following ref. \cite{Craps}, let us define the quantity
\beq
\Delta_p(\,r\,,\,\theta\,)\,\equiv\,
r^{7-p}\,f_p(r)\,\,(\,{\rm sin }\,\theta\,)^{7-p}\,\,.
\label{barvsiete}
\eeq
In terms of $\Delta_p(\,r\,,\,\theta\,)$, we define the function 
$g_p(r, \theta)$ as follows:
\beq
g_p(r, \theta)\,\equiv\,
{\Delta_p(\,r\,,\,\theta\,)\,{\rm sin }\,\theta\,\,+\,
D_p(\,\theta\,)\,{\rm cos }\,\theta\over
\Delta_p(\,r\,,\,\theta\,)\,{\rm cos }\,\theta
\,-\,D_p(\,\theta\,)\,{\rm sin }\,\theta}\,\,.
\label{barvocho}
\eeq
Making use of the function $g_p$, it is easy to verify that the energy
can be put in terms of a square root of a sum of squares. Indeed, it
is a simple exercise to demonstrate that $U$ can be written as
\bear
U\,&=&\,
T_{8-p}\,\Omega_{7-p}\,\int d\theta\,
\,\,\times\rc\rc
&&\times\,\sqrt{{\cal Z}^2\,+\,r^2\,
\Bigl[\,\Delta_p(\,r\,,\,\theta\,)\,
{\rm cos }\,\theta-\,
D_p(\theta)\,{\rm sin }\,\theta\,\Bigr]^2\,\Bigl[\,
{r'\over r}-g_p(r, \theta)\,\Bigr]^2\,\,,
}
\label{barvnueve}
\eear
where ${\cal Z}$ is given by
\beq
{\cal Z}\,=\,r\,
\Bigl[\,\Delta_p(\,r\,,\,\theta\,)\,
{\rm cos }\,\theta-\,
D_p(\theta)\,{\rm sin }\,\theta\,\Bigr]\,
\Bigl[\,1\,+\,{r'\over r}\,g_p(r, \theta)\,\Bigr]\,\,.
\label{bartreinta}
\eeq
In view of eq. (\ref{barvnueve}), it is clear that the energy of the
D(8-p)-brane is bounded as
\beq
U\,\ge\,T_{8-p}\,\Omega_{7-p}\,\int d\theta\,
\Bigl|\,{\cal Z}\,\Bigr|\,\,.
\label{bartuno}
\eeq
This bound is saturated when $r(\theta)$ satisfies the following
first-order differential equation:
\beq
{r'\over r}\,=\,
g_p(r, \theta)\,\,.
\label{bartdos}
\eeq
It is straightforward to prove that any function $r(\theta)$ that
satisfies eq. (\ref{bartdos}) is also a solution of the Euler-Lagrange
equation (\ref{barvseis}). Notice that the condition (\ref{bartdos})
involves the displacement field $D_p(\theta)$ (see eq. (\ref{barvocho})).
The expressions of the $D_p$'s have been given at the end of section 2
(eq. (\ref{barvdos})). However, in order to demonstrate that the
solutions of eq. (\ref{bartdos}) verify the equation of motion
(\ref{barvseis}), we do not need to use the explicit form of 
$D_p(\theta)$. The only property required for this proof is the value
of the derivative of $D_p(\theta)$, given in eq. (\ref{barcatorce}). 
The electric field acquires a very simple expression when the BPS 
(\ref{bartdos}) is satisfied. To see this let us invert eq. (\ref{bartrece}) 
and obtain the electric field $E$ in terms of $D_p(\theta)$ for 
$x^{\prime}\neq 0$. This is given by
\be
E=(-1)^{p+1}\,\frac{\sqrt{r^2+{r^{\prime}}^2}}
{\sqrt{(D_p)^2+(\D_p)^2}}D_p(\theta)\,.
\label{bartdose1}
\ee
Now eq. (\ref{bartdos}) is equivalent to
\be
\frac{\sqrt{r^2+{r^{\prime}}^2}}
{\sqrt{(D_p)^2+(\D_p)^2}}=\frac{r}{\D_p\cos\th-D_p\sin\th}\,.
\label{bartdose2}
\ee
Putting together (\ref{bartdose1}) and (\ref{bartdose2}) we arrive to 
\be
E=(-1)^{p+1}\frac{d}{d\th}(r\cos\th)
\label{barbps}
\ee
As pointed out in ref. \cite{Craps} for 
the $p=3$ case, ${\cal Z}$ can be written as a total derivative:

\beq
{\cal Z}\,=\,
{d\over d\theta}\,\,
\Bigl[\,D_p(\theta)\,r\,{\rm cos }\,\theta\,+\,
\Bigl(\,{a\over 8-p}\,+\,{R^{7-p}\over r^{7-p}}\,\Bigr)
\,\,(\,r\,{\rm sin }\,\theta\,)^{8-p}\,\Bigr]\,\,.
\label{bartseis}
\eeq
It is important to stress the fact that eq. (\ref{bartseis}) can be
proved without using the condition (\ref{bartdos}) or the
Euler-Lagrange equation (\ref{barvseis}) (only eq. (\ref{barcatorce}) has
to be used). Eq. (\ref{bartseis}) implies that the bound (\ref{bartuno})
does not depend on the detailed form of the function $r(\theta)$.
Actually, as a consequence of eq. (\ref{bartseis}), only the boundary
values of $r(\theta)$ matter when one evaluates the right-hand side of
eq. (\ref{bartuno}). The functions $r(\theta)$ which solve eq. (\ref{bartdos})
correspond to those  D(8-p)-brane embeddings which, for given
boundary conditions, have minimal energy. Due to this fact we shall
refer to eq. (\ref{bartdos}) as the BPS condition. 
The search for the solutions of the BPS equation  and their
interpretation will be the subject of the next three sections.

\setcounter{equation}{0}
\section{The near-horizon solution for $p\le 5$} \label{barsec4}
\medskip

In this section we are going to solve the BPS first-order differential
equation in the near-horizon geometry for $p\le 5$. It will become
clear in the process of finding the solution that the $p=6$ case is
singled out (as expected from the holographic relation (\ref{barvuno})).
This $p=6$ case will be discussed separately in section 5.

Let us start by defining the function $\Lambda_p(\,\theta\,)$ by means
of the equation

\beq
\Lambda_p(\,\theta\,)\,\equiv\,{1\over R^{7-p}}\,
\Bigl[\,R^{7-p}\,(\,{\rm sin }\,\theta\,)^{6-p}\,{\rm cos }\,\theta
\,-\,D_p(\,\theta\,)\,\Bigr]\,\,.
\label{bartnueve}
\eeq
The right-hand side of eq. (\ref{bartnueve}) is known and, thus, 
$\Lambda_p(\,\theta\,)$ is a known function of $\theta$ whose explicit
expression can be obtained by substituting the values of
$D_p(\,\theta\,)$, given in eq. (\ref{barvdos}), in 
 (\ref{bartnueve}). Moreover, the BPS condition (\ref{bartdos}) for $a=0$
can be written in terms of $\Lambda_p(\,\theta\,)$ as

\beq
{r'\over r}\,=\,{({\rm sin }\,\theta\,)^{6-p}\,-\,
\Lambda_p(\,\theta\,)\,{\rm cos }\,\theta\over
\Lambda_p(\,\theta\,)\,{\rm sin }\,\theta}\,\,.
\label{barcuarenta}
\eeq
A key point in what follows is that $\Lambda_p(\,\theta\,)$ has a
simple derivative, which can be easily obtained from eq.
(\ref{barcatorce}), namely,

\beq
{d\over d\theta}\,\Lambda_p(\,\theta\,)\,=\,
(\,6\,-\,p)\,(\,{\rm sin }\,\theta\,)^{5-p}\,\,.
\label{barcuno}
\eeq
By inspecting  the right-hand side of this equation we observe that
$p=6$ is a special case. Indeed, when $p\not= 6$,  we can represent the 
$({\rm sin }\,\theta\,)^{6-p}$ term appearing in the BPS condition as

\beq
({\rm sin }\,\theta\,)^{6-p}\,=\,{{\rm sin }\,\theta\over 6-p}\,
\,\,{d\over d\theta}\,\,\Lambda_p(\,\theta\,)\,\,,
\,\,\,\,\,\,\,\,\,\,\,\,\,\,\,\,\,\,\,\,\,\,\,\,\,\,\,\,
(p\not= 6)\,\,.
\label{barcdos}
\eeq
After doing this, the right-hand side of eq. (\ref{barcuarenta}) is
immediately recognized as a logarithmic derivative and, as a
consequence, the BPS condition can be readily integrated. The result
one arrives at is

\beq
r\,(\,\theta\,)\,=\,C\,\,
{\,\Bigl[\,\Lambda_p(\,\theta\,)\,\Bigr]^{{1\over 6-p}}\over
{\rm sin }\,\theta}\,\,,
\label{barctres}
\eeq
where $C$ is a positive constant. For $p<5$ there is a fractional power
of $\Lambda_p(\,\theta\,)$ in the right-hand side of eq.
(\ref{barctres}) and, thus, the solution we have found only makes sense
for those values of $\theta$ such that
$\Lambda_p(\,\theta\,)\,\ge 0$. Moreover, it is clear from eq. 
(\ref{barcuno}) that

\beq
{d\Lambda_p\over d\theta}\,\ge\,0\,\,,
\,\,\,\,\,\,\,\,\,\,\,\,\,\,\,\,\,\,\,\,\,\,\,\,\,\,\,\,
(\,p<6\,,\,\,\,\,0\,\le\,\theta\,\le \,\pi\,)\,\,,
\label{barccuatro}
\eeq
and, therefore, $\Lambda_p(\,\theta\,)$ is a monotonically increasing
function in the interval $0\le\theta\le\pi$. The values of 
$\Lambda_p(\,\theta\,)$ at $\theta\,=\,0,\pi$ are

\bear
\Lambda_p(0)\,&=&-{D_p(0)\over R^{7-p}}\,=\,-\,
2\sqrt\pi\,{\Gamma\Bigl(\,{8-p\over 2}\Bigr)\over
\Gamma\Bigl(\,{7-p\over 2}\Bigr)}\,\, \nu\,\le\,0\,\,,
\rc\rc
\Lambda_p(\pi)\,&=&-{D_p(\pi)\over R^{7-p}}\,=\,
2\sqrt\pi\,{\Gamma\Bigl(\,{8-p\over 2}\Bigr)\over
\Gamma\Bigl(\,{7-p\over 2}\Bigr)}\,\,
(\,1\,- \nu\,)\,\ge\,0\,\,,
\label{barccinco}
\eear
where, in order to establish the last inequalities, we have used the
fact that $0\le\nu\le 1$. From this discussion it follows that there
exists a unique value $\theta_0$ of the polar angle such that
\begin{figure}
\centerline{\hskip -.8in \epsffile{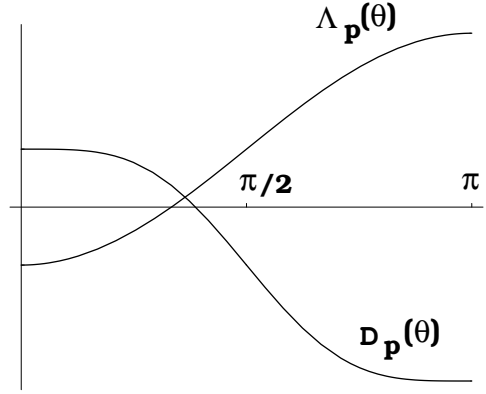}}
\caption{The functions $\Lambda_p(\theta)$ and $D_p(\theta)$ for
$0\le\theta \le\pi$. For illustrative purposes we have plotted these
functions for $p=4$ and $\nu=1/4$.}
\label{barfig1}
\end{figure}

\beq
\Lambda_p\,(\,\theta_{0}\,)\,=\,0\,\,.
\label{barcseis}
\eeq
The form of the function $\Lambda_p\,(\,\theta\,)$ has been
displayed in figure \ref{barfig1}. It is clear that the solution (\ref{barctres}) is
valid for 
$\theta_0\le\theta\le\pi$. Notice that $\theta_0$ depends on $\nu$ and,
actually, is a monotonically increasing function of $\nu$. In fact,
from eq. (\ref{barccinco}) one concludes that $\theta_0\,=\,0$ for 
$\nu\,=\,0$, whereas $\theta_0\,=\,\pi$ for $\nu\,=\,1$.

\begin{figure}
\centerline{\hskip -.8in \epsffile{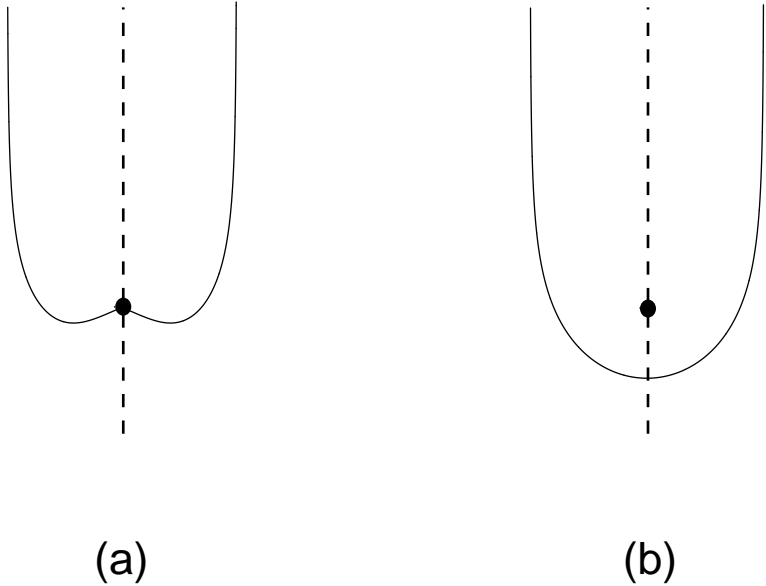}}
\caption{Plot of the near-horizon solution (\ref{barctres}) for
$(p=3,\nu=1/4)$ (a) and $(p=3,\nu=0)$ (b). The discontinuous line
represents the polar axis ($\theta=\pi$ at the top). The dot
corresponds to the origin $r=0$. The background branes are located at
the origin and extend in the directions orthogonal to the plane of the
figure. }
\label{barfig2}
\end{figure}

For $p=5$ the function  appearing in the right-hand side of eq.  
(\ref{barctres}) is
\beq
\Lambda_5(\,\theta\,)\,=\,\theta\,-\,\pi\,\nu\,\,.
\label{barcsiete}
\eeq
Thus, our solution in this case  is
\beq
r\,(\,\theta\,)\,=\,C\,\,
{\,\theta\,-\,\pi\,\nu\over
{\rm sin }\,\theta}\,\,,
\,\,\,\,\,\,\,\,\,\,\,\,\,\,\,\,\,\,\,\,\,\,\,\,\,\,\,\,
(\,p\,=\,5\,)\,\,.
\label{barcocho}
\eeq
As the radial coordinate $r$ must be non-negative, from eq.
(\ref{barcocho}) one immediately concludes that, also in this 
$p=5$ case, $\theta_0\le\theta\le\pi$, where now $\theta_0$ depends
linearly on $\nu$, namely
\beq
\theta_{0}\,=\,\pi\,\nu\,\,,
\,\,\,\,\,\,\,\,\,\,\,\,\,\,\,\,\,\,\,\,\,\,\,\,\,\,\,\,
(\,p\,=\,5\,)\,\,.
\label{barcnueve}
\eeq
The solution we have found coincides with 
the one obtained in refs. \cite{CGS1, CGS2} 
for $p=3,4$. Our result (\ref{barctres}) generalizes these solutions for
any $p<6$ (the solution for $p=6$ will be given in the next section).
By inspecting eq. (\ref{barctres}) it is easy to conclude that, for
$\nu\not=1$, $r(\theta)$ diverges when $\theta\rightarrow\pi$.
Actually, when $\theta\approx\pi$ and $\nu\not=1$, $r(\theta)$ behaves
as
\beq
r(\,\theta\,)\,\approx\,
{C\,\Bigl[\,\Lambda_p(\,\pi\,)\,\Bigr]^{{1\over 6-p}}\over
\pi\,-\,\theta}\,\,,
\,\,\,\,\,\,\,\,\,\,\,\,\,\,\,\,\,\,\,\,\,\,\,\,\,\,\,\,
(\,\nu\,\not=\,1\,)\,\,,
\label{barcincuenta}
\eeq
which corresponds to a ``tube" of radius 
$C\,\Bigl[\,\Lambda_p(\,\pi\,)\,\Bigr]^{{1\over 6-p}}$. We will check
below that the energy of these tubes corresponds to a bundle of 
$(1-\nu)N$ fundamental strings emanating from the D(8-p)-brane. When 
$\nu\not=0$ the critical angle $\theta_0$ is greater than zero and, as
a consequence, eq. (\ref{barctres}) gives $r(\theta_0)\,=\,0$. Actually,
when $\nu\not=0$, the solution has the shape represented in figure 
\ref{barfig2}a.
On the contrary, when $\nu=0$, the function $r(\theta)$ takes a
non-vanishing value at the critical value $\theta_0\,=\,0$. Notice
that, in this last case, both numerator and denominator on the
right-hand side of eq. (\ref{barctres}) vanish. The behaviour of 
$\Lambda_p(\,\theta\,)$ near $\theta=0$ can be obtained by a Taylor
expansion. The successive derivatives of $\Lambda_p(\,\theta\,)$ can be
easily computed from the value of its first derivative, given in eq. 
(\ref{barcuno}).  It is easy to prove that the first non-vanishing
derivative of $\Lambda_p(\,\theta\,)$ at $\theta=0$ is

\beq
{d^{\,6-p}\over d\theta^{\,6-p}}\,\Lambda_p(\,\theta\,)\,\,
\Bigg|_{\theta=0}\,=\,(\,6\,-\,p\,)!\,\,.
\label{barcinuno}
\eeq
Moreover, the $(7-p)^{{\rm th}}$ derivative of 
$\Lambda_p(\,\theta\,)$ at $\theta=0$ vanishes, \ie,

\beq
{d^{\,7-p}\over d\theta^{\,7-p}}\,\Lambda_p(\,\theta\,)\,\,
\Bigg|_{\theta=0}\,=\,0\,\,.
\label{barcidos}
\eeq
Therefore, the expression of $\Lambda_p(\,\theta\,)$ for values of 
$\theta$ close to zero takes the form

\beq
\Lambda_p(\,\theta\,)\,\approx\,\Lambda_p(\,0\,)\,+\,
\theta^{\,6-p}\,+\,c_p\,\theta^{\,8-p}\,+\,
\cdots\,\,,
\label{barcitres}
\eeq
where $c_p$ is a non-vanishing coefficient. Substituting this
expansion in the right-hand side of eq. (\ref{barctres}) for $\nu=0$
leads to the conclusion that
\beq
r(\,0\,)\,=\,C\,,
\,\,\,\,\,\,\,\,\,\,\,\,\,\,\,\,\,\,\,\,\,\,\,\,\,\,\,\,
(\,\nu\,=\,0\,)\,\,,
\label{barcicuatro}
\eeq
\ie\ $r(0)\not=0$ for $\nu=0$, as claimed above. Moreover, $r'(\,0\,)$
vanishes in this case,
\beq
r'(\,0\,)\,=\,0\,,
\,\,\,\,\,\,\,\,\,\,\,\,\,\,\,\,\,\,\,\,\,\,\,\,\,\,\,\,
(\,\nu\,=\,0\,)\,\,.
\label{barcicinco}
\eeq
The shape of the solution for $\nu=0$ has been plotted in figure \ref{barfig2}b.
Notice that the $\nu=0$ tube corresponds to the ordinary baryon with
$N$ quarks.

The solution of the near-horizon BPS condition written in  eq.
(\ref{barctres}) is not the only one. Following 
ref. \cite{CGS1}, we can construct
a new solution, valid for values of $\theta$ in the range 
$0\le\theta\le\theta_0$, in which $\Lambda_p(\,\theta\,)\le 0$, as
follows,

\beq
\tilde r\,(\,\theta\,)\,=\, C\,\,
{\,\Bigl[\,-\Lambda_p(\,\theta\,)\,\Bigr]^{{1\over 6-p}}\over
{\rm sin }\,\theta}\,\,.
\label{barciseis}
\eeq
Eq. (\ref{barciseis}) describes a spike at $\theta\,=\,0$. We shall refer
to the solution (\ref{barciseis}) as a ``lower tube" solution, in
contrast to the ``upper tube" solution of eq. (\ref{barctres}). Actually,
these two types of solutions are related. In order to verify this
fact, let us point out that the functions $D_p(\,\theta\,)$ and 
$\Lambda_p(\,\theta\,)$  change their sign under the transformation 
$\theta\rightarrow\pi-\theta$, $\nu\rightarrow1-\nu$,

\bear
D_p(\,\theta\,;\,\nu\,)\,&=&\,-
D_p(\,\pi\,-\theta\,;\,1\,-\nu\,)\,\,,\rc\rc
\Lambda_p(\,\theta\,;\,\nu\,)\,&=&\,-
\Lambda_p(\,\pi\,-\theta\,;\,1\,-\nu\,)\,\,,
\label{barcisiete}
\eear
and, therefore, the solutions (\ref{barctres}) and (\ref{barciseis}) are
simply related, namely,

\beq
\tilde r\,(\,\theta\,;\,\nu\,)\,=\,
r\,(\,\pi\,-\,\theta\,;\,1\,-\,\nu\,)\,\,.
\label{barciocho}
\eeq
It follows from this relation that the lower flux tubes correspond to 
$\nu N$ quarks. 

Let us now evaluate, following ref. \cite{Craps},  the energy for the
two types of near-horizon solutions we have found. These solutions
saturate the bound (\ref{bartuno}) and, thus, the energy is precisely the
right-hand side of this equation. Moreover, when $r$ is a solution of
eq. (\ref{bartdos}), ${\cal Z }$ can be written as

\beq
{\cal Z }\,=\,r\,[\,\Delta_p(\,r\,,\,\theta\,)\,\,
{\rm cos }\,\theta\,-\,D_p(\,\theta\,)\,
{\rm sin }\,\theta\,]\,\Bigl[\,1\,+\,
\Bigl(\,{r\,'\over r}
\Bigr)^2\,\,\Bigr]\,\,.
\label{barcinueve}
\eeq
It is obvious from this equation that, for our
solutions,  the sign of ${\cal Z }$  is just the sign of 
$\Delta_p(\,r\,,\,\theta\,)\,\,{\rm cos }\,\theta\,-\,D_p(\,\theta\,)\,
{\rm sin }\,\theta$. Moreover, using the definitions of $\Delta_p$ 
({eq.~(\ref{barvsiete})) and $\Lambda_p$ (eq.~(\ref{bartnueve})), one
can prove that 
\beq
\Delta_p(\,r\,,\,\theta\,)\,\,
{\rm cos }\,\theta\,-\,D_p(\,\theta\,)\,
{\rm sin }\,\theta\,=\,a\,r^{7-p}\,
(\,{\rm sin }\,\theta\,)^{7-p}\,{\rm cos }\,\theta\,+\,
R^{7-p}\,\Lambda_p(\,\theta\,)\,{\rm sin }\,\theta\,\,,
\label{barsesenta}
\eeq
where we have used the general harmonic function 
(\ref{bardos}). In the near-horizon case ($a=0$) the first term on the
right-hand side of eq. (\ref{barsesenta}) is absent and, therefore, the
sign of $\Delta_p(\,r\,,\,\theta\,)\,\,{\rm cos }\,
\theta\,-\,D_p(\,\theta\,)\,
{\rm sin }\,\theta$ is just the sign of $\Lambda_p(\theta)$. In the
solutions (\ref{barctres}) and (\ref{barciseis}), the angle $\theta$ can
take values in a range such that $\Lambda_p(\,\theta\,)$ has a well
defined sign. Therefore, we can write

\beq
\int d\theta\,
\Bigl|\,{\cal Z}\,\Bigr|\,=\,
\,\,\Bigl|\,\int d\theta\,{\cal Z}\,\Bigr|\,\,.
\label{bartcuatro}
\eeq
If we now define $Z$ as
\beq
Z\,\equiv\,T_{8-p}\,\Omega_{7-p}\,\int d\theta\,{\cal Z}\,\,,
\label{barttres}
\eeq
it is immediate that the energy of our near-horizon BPS solutions is

\beq
U_{BPS}\,=\,|\,Z\,|\,\,.
\label{bartcinco}
\eeq

In order to evaluate $Z$ for the near-horizon solutions, we make use
of the representation (\ref{bartseis}) of ${\cal Z}$ as a total
derivative. Recall that $\theta$ varies in the range 
$\theta_i\le\theta\le\theta_f$ where $\theta_i=\theta_0$ 
($\theta_i=0$) and $\theta_f=\pi$ 
($\theta_f=\theta_0$) for the upper (lower) tube solution. It follows
that $Z$ can be written as the sum
\footnote{The labels $s$  and $gs$ refer to ``spike" and  ``ground
state", according to the interpretation given to these two
contributions in ref. \cite{Craps}.}, 

\beq
Z\,=\,Z_s\,+\,Z_{gs}\,\,,
\label{bartsiete}
\eeq
where $Z_s$ and $Z_{gs}$ are given by

\bear
Z_s\,&=&T_{8-p}\,\Omega_{7-p}\,D_p(\theta)\,
r(\theta)\,{\rm cos }\,\theta\,\Bigr|_{\theta_i}^{\theta_f}\,\,,\rc\rc
Z_{gs}\,&=&T_{8-p}\,\Omega_{7-p}\,
R^{7-p}\,\,
r(\theta)\, \,\,(\,{\rm sin }\,\theta\,)^{8-p}\,
\Bigr|_{\theta_i}^{\theta_f}\,\,.
\label{bartocho}
\eear
It is clear that $Z$ only depends on the values of $r(\theta)$ at the
boundaries $\theta=\theta_i$ and  $\theta=\theta_f$. In this sense we
can regard $Z$ as a topological quantity which, for fixed boundary
conditions, is invariant under local variations of the fields
\cite{Craps}.

Let us now compute $Z_{gs}$. After substituting the solutions
(\ref{barctres}) and (\ref{barciseis}) of the BPS equations, we get, 

\beq
Z_{gs}\,=\,\,T_{8-p}\,\Omega_{7-p}\,R^{7-p}\,C\,
(\,{\rm sin }\,\theta)^{7-p}\,
\Bigl[\,\pm\Lambda_p(\,\theta\,)\,\Bigr]^{{1\over 6-p}}
\,\,\Biggr|_{\theta_i}^{\theta_f}\,\,,
\label{barsuno}
\eeq
where the $+$ ($-$ ) sign corresponds to the upper (lower) tube
solution. The angles $\theta_i$ and $\theta_f$ can take the values
$0$, $\pi$ and $\theta_0$. For $\theta=0,\pi$ the right-hand side of
eq. (\ref{barsuno}) vanishes due to the $(\,{\rm sin }\,\theta)^{7-p}$
factor, whereas $\theta=\theta_0$ gives also a vanishing contribution
because $\Lambda_p(\,\theta_0\,)=0$ (see eq. (\ref{barcseis})). In
conclusion, we can write
\beq
Z_{gs}\,=\,0\,\,.
\label{barsdos}
\eeq
In the same way one can obtain the values of $Z_s$. One can check in
this case that the contribution of $\theta=\theta_0$ to the right-hand
side of the first equation in (\ref{bartocho}) vanishes and, as a
consequence, only $\theta=0,\pi$ contribute. After using eqs. 
(\ref{barvtres})-(\ref{barvcinco}), one gets the result

\beq
Z_s\,=\,
\cases{(1\,-\,\nu\,)\,N\,T_f\,L\,,&\,\,\,\,\,(upper tube)\,\,,\cr\cr
        -\,\nu\,N\,T_f\,L\,,&\,\,\,\,\,(lower tube)\,\,,\cr}
\label{barstres}
\eeq
where $L=r(\pi)$ ($L=r(0)$) for the upper (lower) tube solution. It is
now evident that $U_{BPS}\,=\,|\,Z_s\,|$ is equal to the energy of a
bundle of $(1-\nu)N$ or $\nu N$ fundamental strings. Notice that this
result is the same as the one found at the end of section 3.2 for the
energy of the $\theta=0$ and $\theta=\pi$ spikes. Let us finally
mention that in ref. \cite{Craps} it has been argued in favor of
interpreting $Z_s$ as a central charge in the worldvolume supersymmetry 
algebra.

\setcounter{equation}{0}
\section{The near-horizon D6-D2  system}  \label{barsec5}
\medskip

In this section we are going to integrate the BPS condition for $p=6$.
Notice that, according to  the definition (\ref{bartnueve}) and the
expression of $D_6(\theta)$ given in eq. (\ref{barvdos}),
$\Lambda_6\,(\,\theta\,)$ is constant, \ie,

\beq
\Lambda_6\,(\,\theta\,)\,=\,1\,-\,2\,\nu\,\,.
\label{barscuatro}
\eeq
For $p=6$ we cannot use eq. (\ref{barcdos}) and, hence, we have to deal
directly with eq. (\ref{barcuarenta}). Actually, eq. (\ref{barcuarenta})
makes sense only  for $\nu\not=1/2$ since
$\Lambda_6\,(\,\theta\,)$, and therefore the denominator of eq.
(\ref{barcuarenta}) vanishes identically for $\nu={1\over 2}$. 
With this restriction\footnote{ The fact that
the $p=6$, $\nu=1/2$ BPS condition  is ill-defined is an artifact of
the near-horizon approximation.}, the $p=6$ BPS condition is

\beq
{r'\over r}\,=\,{1\over 1\,-\,2\nu}\,\,
{1\over {\rm sin }\,\theta}\,-\,
{{\rm cos }\,\theta\over {\rm sin }\,\theta}\,\,.
\label{barscinco}
\eeq
It is not difficult to realize that the right-hand side of eq. 
(\ref{barscinco}) can be written as a total derivative, namely,

\beq
{r'\over r}\,=\,{d\over d\theta}\,\,
\Biggl[\,\,{\rm log}\,\,\Bigl[\,\,
{\Bigl(\,{\rm tan}\,{\theta\over 2}\,\,\Bigr)^{{1\over 1-2\nu}}
\over {\rm sin }\,\theta}\,\,\Bigr]\,\,\Biggr]\,\,.
\label{barsseis}
\eeq
Therefore, the BPS equation can be immediately integrated. It is
convenient to put the result in the form

\beq
r(\,\theta\,)\,=\,A\,\,\,
{\,\,\Bigl[\,
{\rm sin }\,{\theta\over 2}\Bigr]^{{2\nu\over 1\,-\,2\nu}}\over 
\,\,\Bigl[\,
{\rm cos }\,{\theta\over 2}\Bigr]^{{2(\,1-\nu\,)
\over 1\,-\,2\nu}}
}\,\,,
\label{barssiete}
\eeq
where $A$ is a positive constant. The nature of the solution 
(\ref{barssiete}) depends critically on the value of $\nu$.  First of
all, the range of values of $\theta$ is not restricted, \ie\ 
$0\le\theta\le\pi$. If $0\le\nu<1/2$, the 
${\rm cos }\,(\theta/2)$ term present in the denominator of 
(\ref{barssiete}) makes $r\rightarrow\infty$  at $\theta\approx\pi$,
whereas for  $1/2<\nu\le 1$ the solution diverges at 
$\theta\approx 0$. Actually, the solution for 
$0\le\nu<1/2$ is related to the one for $1/2<\nu\le 1$ by means of the
equation

\beq
r(\,\theta\,;\,\nu\,)\,=\,
r(\,\pi\,-\,\theta\,;\,1-\nu\,)\,\,.
\label{barsocho}
\eeq
Eq. (\ref{barsocho}) can be checked easily from the expression
(\ref{barssiete}). Notice that for 
$0<\nu<1/2$ ($1/2<\nu<1$) the D2 brane  passes through the point
$r=0$. This behaviour is similar to the $p<6$ case. However, the
present solution differs substantially  from those studied in section
4. In order to make  this difference manifest, let us introduce the
cylindrical coordinates $(z,\rho)$ as follows:

\beq
z\,=\,-r\,{\rm cos }\,\theta\,\,,
\,\,\,\,\,\,\,\,\,\,\,\,\,\,\,\,\,\,\,\,\,\,\,\,\,\,\,\,
\rho\,=\,r\,{\rm sin }\,\theta\,\,.
\label{barsnueve}
\eeq
The behaviour of the coordinate $\rho$ in the region in which 
$r\rightarrow\infty$ is of special interest to interpret the
asymptotic behaviour of the brane. It is not difficult to find how
$\rho$ depends on $\theta$ for the solution (\ref{barssiete}):

\beq
\rho\,=\,A\,
\Bigl(\,{\rm tan}\,{\theta\over 2}\,\Bigr)^{{1\over 1-2\nu}}\,\,.
\label{barsetenta}
\eeq
Clearly, when $\nu<1/2$, $\rho\rightarrow\infty$ for 
$\theta\rightarrow\pi$, whereas if $\nu>1/2$ the coordinate $\rho$
diverges for  $\theta\approx 0$. This implies that our solution does
not behave as a tube
\footnote{A similar calculation for $p<6$ gives $\rho\rightarrow{\rm
constant}$ at the spikes.}. This fact is more explicit if we rewrite
our solution (\ref{barssiete}) as a function $z=z(\rho)$. After a short
calculation, we get

\beq
z\,=\,{\rho^{2\nu}\over 2\,A^{1-2\nu}}\,\,
\Bigl(\,\rho^{2(1-2\nu)}\,-\,A^{2(1-2\nu)}\,\Bigr)\,\,.
\label{barstuno}
\eeq
When $\rho\rightarrow\infty$, the function $z(\rho)$ behaves
as:
\begin{figure}
\centerline{\hskip -.8in \epsffile{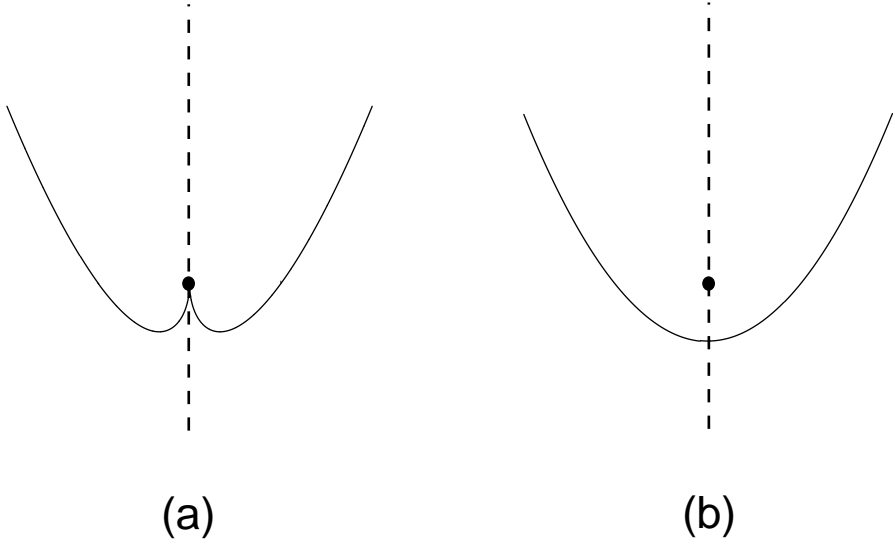}}
\caption{Representation of the D2 embedding for $\nu=1/4$ (a) and 
$\nu=0$ (b). The conventions are the same as in figure \ref{barfig2}.}
\label{barfig3}
\end{figure}

\beq
\lim_{\rho\rightarrow\infty}\,z(\,\rho\,)\,\sim\,
\cases{\rho^{2(1-\nu)}\,\,,&
            \,\,\,\,\,if $\nu\,<\,{1\over 2}\,\,,$\cr\cr
       -\rho^{2\nu}\,\,,&
            \,\,\,\,\,if $\nu\,>\,{1\over 2}$\,\,,\cr
}
\label{barstdos}
\eeq
and thus, as expected, $z\rightarrow+\infty$ ($z\rightarrow-\infty$)
for $\nu<1/2$ ($\nu>1/2$). Moreover, eq. (\ref{barstdos})  implies that
the asymptotic shape of the solution is that of a paraboloid
\footnote{ Strictly speaking, only for $\nu\,=\,0,1$ eq. (\ref{barstuno})
represents a paraboloid.} rather than a tube. We have thus an ``upper
paraboloid" for $\nu<1/2$ and a ``lower paraboloid" for 
$\nu>1/2$. In figure \ref{barfig3} we have plotted the $p=6$ solution for
different values of $\nu$. Notice that for $\nu\,\not=\,0,1$ $z(\rho)$
has an extremum for some value of $\rho\not=0$, while for 
$\nu\,=\,0,1$ the extremum is located at $\rho=0$. This extremum is a
minimum (maximum) for  $\nu<1/2$ ($\nu>1/2$). It is interesting to
point out that, although the solution for $p=6$ differs from the ones
found for
$p\le 5$, the energy $U$ is still given by $|Z_s|$, where $Z_s$ is the
same as in eq. (\ref{barstres}). 

It follows from our previous analysis that $\nu=1/2$ is a critical
point in the behaviour of the D6-D2 system. The BPS condition 
(\ref{barscinco}) is not defined in this case and we have to come back to
the equation of motion (\ref{barvseis}). Fortunately, we will be able to
find the general solution of the field equation in this case. The key
point in this respect is the observation that, for $p=6$ and
$\nu=1/2$, the near-horizon energy density does not depend on
$\theta$ explicitly. Indeed, a simple calculation shows that

\beq
U\,=\,T_2\,\Omega_1\,R\,\,\int d\theta\,
\sqrt{r^2\,+\,r\,'^{\,2}}\,\,,
\,\,\,\,\,\,\,\,\,\,\,\,\,\,\,\,\,\,\,\,\,\,\,\,\
\,\,\,\,\,\,\,\,\,\,\,\,\,\,\,\,\,\,\,\,\,\,\,\,\
(\,\nu\,=\,{1\over 2}\,)\,\,.
\label{barsttres}
\eeq
This $\theta$-independence  implies the ``conservation law"

\beq
r\,'\,{\partial U\over \partial r\,'}\,-\,U\,=\,
{\rm constant}\,\,,
\label{barstcuatro}
\eeq
or, using the explicit form of $U$ given in eq. (\ref{barsttres}),

\beq
{r^2\over \sqrt{r^2\,+\,r\,'\,^2}}\,=\,C\,\,,
\label{barstcinco}
\eeq
where $C\ge 0$ is constant. It is not difficult to integrate the
first-order equation (\ref{barstcinco}). The result is

\begin{figure}
\centerline{\hskip -.8in \epsffile{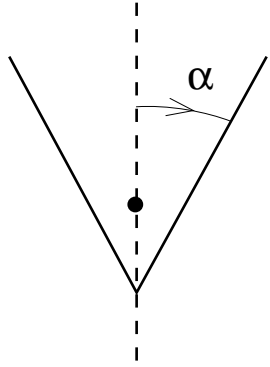}}
\caption{Solution of the equation of motion for $p=6$ and $\nu=1/2$
for the near-horizon metric (eq. (\ref{barstseis})).}
\label{barfig4}
\end{figure}

\beq
r(\,\theta\,)\,=\,{C\over {\rm sin }\,(\,\theta\,+\,\alpha\,)}\,\,,
\label{barstseis}
\eeq
$\alpha$ being a new integration constant. The solution (\ref{barstseis})
represents a cone with opening half-angle $\alpha$ and with its vertex
located at a distance $C/{\rm sin }\,\alpha$ from the origin (see figure
\ref{barfig4}). This fact is neatly shown if we rewrite (\ref{barstseis}) in terms of
the cylindrical coordinates $(\rho, z)$ defined in eq. (\ref{barsnueve}). 
In these coordinates, eq. (\ref{barstseis}) becomes

\beq
\rho\,{\rm cos }\,\alpha\,-\,z\,{\rm sin }\,\alpha\,=\,C\,\,.
\label{barstsiete}
\eeq
It is clear from figure \ref{barfig4} that $\alpha$ can take values in the range 
$-\pi/2\le\alpha\le\pi/2$. When $\alpha=0$ the cone degenerates in the
cylinder $\rho=C$, while for $\alpha\,=\,\pm\pi/2$ the solution 
(\ref{barstsiete}) is the plane $z\,=\,\mp C$. Notice that eq. 
(\ref{barstsiete})  is invariant under the transformation
$\alpha\rightarrow -\alpha$, $z\rightarrow -z$. Thus, for $\alpha>0$ 
($\alpha<0$) $z$ goes to $+\infty$ ($-\infty$) when the angle $\theta$
takes the value $\pi-\alpha$ ($-\alpha$). It is interesting to point
out that, when $\rho\rightarrow\infty$, 
$z$ diverges linearly with $\rho$, which is,
certainly, the  limiting case of the $\nu\not= 1/2$ situation (see eq.
(\ref{barstdos})). This softer behaviour for $\nu= 1/2$ will be also
reflected in the solution for the full Dp-brane metric, as we shall
check in the next section, where we will find the solutions of
the BPS equation for the full Dp-brane  metric (\ref{baruno}). We
will check that for $p=6$ and $\nu=1/2$ this solution is equivalent, in
the near-horizon region, to our general solution (\ref{barstsiete}) with 
$C=0$, \ie\ only those cones passing through the origin correspond to
BPS solutions of the asymptotically flat metric.

\setcounter{equation}{0}
\section{Asymptotically flat background} \label{barsec6}
\medskip

In this section we are going to study the solutions of the BPS
condition beyond the near-horizon approximation. Recall that the full
metric of the Dp-brane background is obtained by taking $a=1$ in the
harmonic function (\ref{bardos}). The full Dp-brane metric is
asymptotically flat and, therefore, one expects that a brane placed
far away from the location of the background branes
is not bent by the gravitational field. We shall concentrate first in
the study of this asymptotic behaviour and, afterwards, we will consider
the complete solution. 

It is convenient for our purposes to rewrite the BPS condition
(\ref{bartdos}) in the coordinates $(z,\rho)$ of eq. (\ref{barsnueve}). In
these coordinates, the D(8-p)-brane embedding is characterized by a
function  $z(\,\rho\,)$. The BPS equation gives  $d\,z/d\,\rho$ in
terms of  $(z,\rho)$. Actually, it is not difficult to relate 
$z(\,\rho\,)$ and its derivative to $r(\theta)$ and $r\,'(\theta)$. By
using only the relation between both coordinate systems one can verify
that

\beq
{dz\over d\rho}\,=\,
{{\rm sin }\,\theta\,-\,{\rm cos }\,\theta\,\,\,{r\,'\over r}\over
{\rm cos }\,\theta\,+\,{\rm sin }\,\theta\,\,\,{r\,'\over r}}\,\,.
\label{barstocho}
\eeq
By substituting the value of
the ratio $r\,'/ r$ given by eq. (\ref{bartdos}) for $a=1$
on the right-hand side of this equation, one arrives
at the following first-order differential equation for the function 
$z(\,\rho\,)$,

\beq
{dz\over d\rho}=\,-
{[\,\rho^2\,+\,z^2]^{{7-p\over 2}}\over 
R^{7-p}\,+\,[\,\rho^2\,+\,z^2]^{{7-p\over 2}}}\,\,\,
{D_p\Bigr(\,{\rm arctan}\,(\,-\rho/z)\,\Bigr)\over
\rho^{7-p}}\,\,.
\label{barstnueve}
\eeq
It is now elementary to evaluate the right-hand side of eq.
(\ref{barstnueve}) in the asymptotic region in which
$\rho\rightarrow\infty$, $z/\rho\rightarrow 0$ 
(and $\theta\rightarrow\pi/2$). In this limit, eq. (\ref{barstnueve})
takes the form

\beq
{dz\over d\rho}\,\sim\,-\,{D_p(\,\pi/2\,)\over
\rho^{7-p}}\,+\,\cdots\,\,,
\label{barochenta}
\eeq
where we have only kept the first term in the expansion in powers of
$R/\rho$. Notice that $z\,'(\rho)\rightarrow 0$ as 
$\rho\rightarrow\infty$, according to our expectations. However, also
in this approach, the behaviour of the $p=6$ case differs from that
corresponding to $p<6$. Indeed, for $p=6$,  $z\,'(\rho)\sim \rho^{-1}$
and $z(\rho)$ diverges logarithmically in the asymptotic region while,
for $p<6$, $z(\rho)$ approaches a constant value as
$\rho\rightarrow\infty$. Let us specify further these two kinds of
behaviours. The coefficient multiplying the power of $\rho$ on the
right-hand side of eq. (\ref{barochenta}) is $D_p(\pi/2)$, which can be
computed from eq. (\ref{barvdos}):

\beq
D_p(\pi/2)\,=\,-2\sqrt\pi\,{\Gamma\Bigl(\,{8-p\over 2}\Bigr)\over
\Gamma\Bigl(\,{7-p\over 2}\Bigr)}\,\,R^{7-p}\, 
(\,{1\over  2}\,-\,\nu\,)\,\,.
\label{barouno}
\eeq
Using this result we can integrate eq. (\ref{barochenta}). For $p<6$, we
get

\beq
z\,(\,\rho\,)\,\sim\,z_{\infty}\,-\,(\,{1\over  2}\,-\,\nu\,)\,
\sqrt\pi\,\,
{\Gamma\Bigl(\,{6-p\over 2}\Bigr)\over
\Gamma\Bigl(\,{7-p\over 2}\Bigr)}\,\,R^{7-p}\, \,
{1\over \rho^{6-p}}\,+\cdots\,\,,
\,\,\,\,\,\,\,\,\,\,\,\,\,\,\,\,\,\,\,\,\,\,
(\,p\not=6\,)\,\,,
\label{barodos}
\eeq
where $z_{\infty}$ is a constant representing the asymptotic value of
$z$. From eq.  (\ref{barodos}) one concludes that the sign of
$z_{\infty}-z$ depends on the sign of 
${1\over  2}\,-\,\nu$. \ie\ if $\nu<1/2$ ($\nu>1/2$) the brane reaches
its asymptotic value of $z$ from below (above). If $p=6$ we get the
expected logarithmic dependence:

\beq
z\,(\,\rho\,)\,\sim\,(\,1\,-\,2\nu\,)\,R\,{\rm log}\,\rho
\,+\,\cdots\,\,.
\label{barotres}
\eeq
Notice that when $\rho\rightarrow\infty$ in eq. (\ref{barotres}),  
$z\rightarrow+\infty$ if $\nu<1/2$ while, on the contrary, 
$z\rightarrow-\infty$ for $\nu>1/2$, in agreement
with our near-horizon analysis of section 5. It is interesting to point
out that, although in this case $z$ diverges as
$\rho\rightarrow\infty$, 
$z/\rho$ still vanishes in this limit, as assumed in the derivation
of eq. (\ref{barochenta}). 

For $\nu=1/2$, the leading term in the asymptotic expansion vanishes
and we have to compute the next-to-leading term. This does not change
significantly the analysis of the $p<6$ case. However, for $p=6$, 
things change drastically when $\nu=1/2$. In fact, one can prove from
eq. (\ref{barstnueve}) that, in this case, $z$ approaches a constant
value $z_{\infty}$ as $\rho\rightarrow\infty$ (and, actually, 
$z-z_{\infty}$ decreases as $\rho^{-1}$ for $\rho>>R$).

Let us now see how one can integrate exactly the BPS condition for the
full metric. The resulting solution must reproduce the asymptotic
behaviour we have just described. For $p=3$, eq. (\ref{barstnueve}) was
integrated numerically in ref. \cite{CGS1}. As we shall show below, our
analytical solution agrees with these numerical results. 
It is more convenient to come back
to our original $(r,\theta)$ coordinates. The basic strategy we will
adopt to integrate the BPS equation is to write it in terms of the
same function $\Lambda_p(\,\theta\,)$ that we have already used to find
the near-horizon solutions. For illustrative purposes, we will do it for
a general value of the parameter $a$ of the harmonic function. After
using the definition (\ref{bartnueve}), eq. (\ref{bartdos}) can be put in
the form

\beq
{r\,'\over r}\,=\,{a\,r^{7-p}\,(\,{\rm sin }\,\theta\,)^{8-p}\,+\,
R^{7-p}\,(\,{\rm sin }\,\theta\,)^{6-p}\,-\,
R^{7-p}\,\Lambda_p(\,\theta\,)\,{\rm cos }\,\theta\over
a\,r^{7-p}\,(\,{\rm sin }\,\theta\,)^{7-p}\,{\rm cos }\,\theta\,+\,
R^{7-p}\,\Lambda_p(\,\theta\,)\,{\rm sin }\,\theta}\,\,.
\label{barocuatro}
\eeq
Notice that eq. (\ref{barocuatro}) for $a=0$ is identical to eq.
(\ref{barcuarenta}). Let us now isolate the terms of (\ref{barocuatro})
depending on $a$ in one of the sides of the equation. Doing this, it is
elementary to prove that eq. (\ref{barocuatro}) can be rewritten as

\beq
a\,r^{7-p}\,(\,{\rm sin }\,\theta\,)^{7-p}\,
{d\over d\theta}\,\,(\,r\,{\rm cos }\,\theta\,)\,=\,
R^{7-p}\,\Bigl[\,r\,(\,{\rm sin }\,\theta\,)^{6-p}\,-\,
\Lambda_p(\,\theta\,)\,
{d\over d\theta}\,\,(\,r\,{\rm sin}\,\theta\,)\,
\Bigr]\,\,.
\label{barocinco}
\eeq
On the right-hand side of this expression we recognize a term
containing $({\rm sin }\,\theta\,)^{6-p}$, whose representation as a
derivative of the function $\Lambda_p(\,\theta\,)$ was crucial to
integrate the near-horizon BPS condition for  $p\not= 6$. Actually, 
assuming that $p\not= 6$ and using eq. (\ref{barcdos}), our differential
equation becomes

\bear
&&a\,r^{7-p}\,(\,{\rm sin }\,\theta\,)^{7-p}\,
{d\over d\theta}\,\,(\,r\,{\rm cos }\,\theta\,)\,=\,\rc\rc
&&={R^{7-p}\over 6-p}\,\Bigl[\,r\,{\rm sin }\,\theta\,\,
{d\over d\theta}\,\,\Lambda_p(\,\theta\,)-\,
(6-p)\,\Lambda_p(\,\theta\,)\,
{d\over d\theta}\,\,(\,r\,{\rm sin}\,\theta\,)\,
\Bigr]\,\,,
\,\,\,\,\,\,\,\,\,\,\,\,\,\,\,\,\,\,\,\,\,\,
(\,p\not=6\,)\,\,.\rc
\label{baroseis}
\eear
Taking $a=0$ we recover the solutions (\ref{barctres}) and
(\ref{barciseis}). From now on we shall put  $a=1$ in all our expressions. 
In this case we can pass the 
$r^{7-p}\,(\,{\rm sin }\,\theta\,)^{7-p}$ factor in the left-hand side
 of eq.  (\ref{baroseis}) to the right-hand side and, remarkably, this
equation can be written as

\beq
{d\over d\theta}\,\,(\,r\,{\rm cos }\,\theta\,)\,=\,
{R^{7-p}\over 6-p}\,\,\,{d\over d\theta}\,\,
\Biggl[\,{\Lambda_p(\,\theta\,)\over 
(\,r\,{\rm sin }\,\theta\,)^{6-p}}\,\Biggr]\,\,,
\,\,\,\,\,\,\,\,\,\,\,\,\,\,\,\,\,\,\,\,\,\,\,\,\,\,\,\,\,\,\,\,
(\,p\not=6\,)\,\,.
\label{barosiete}
\eeq
The integration of eq. (\ref{barosiete}) is now immediate:

\beq
r\,{\rm cos }\,\theta\,=\,{R^{7-p}\over 6-p}\,\,
{\Lambda_p(\,\theta\,)\over 
(\,r\,{\rm sin }\,\theta\,)^{6-p}}
\,-\,z_{\infty}\,\,,
\,\,\,\,\,\,\,\,\,\,\,\,\,\,\,\,\,\,\,\,\,\,\,\,\,\,\,\,\,\,\,\,
(\,p\not=6\,)\,\,,
\label{baroocho}
\eeq
where $z_{\infty}$ is a constant of integration which can be
identified with the asymptotic value of $z$ introduced in eq. 
(\ref{barodos}). Actually, in terms of the coordinates $(z,\rho)$, our
solution (\ref{baroocho}) can be written as

\beq
z\,=\,z_{\infty}\,-\,{R^{7-p}\over 6-p}\,\,
\,\,{\Lambda_p
\Bigr(\,{\rm arctan}\,(\,-\rho/z)\,\Bigr)\over
\rho^{6-p}}\,\,,
\,\,\,\,\,\,\,\,\,\,\,\,\,\,\,\,\,\,\,\,\,\,\,\,\,\,\,\,\,\,\,\,
(\,p\not=6\,)\,\,.
\label{baronueve}
\eeq
Eqs. (\ref{baroocho}) and (\ref{baronueve}) give the analytical solution of
the  BPS differential equation for $p\not=6$.  It is
easy to verify that eq. (\ref{baronueve}) for $\rho\rightarrow\infty$ is
reduced to eq. (\ref{barodos}). Moreover, by derivating 
both sides of eq.  (\ref{baronueve}) with respect to
$\rho$, one can extract the value of
$dz/d\rho$. The result in terms of $(z,\rho)$ is precisely the one
written in (\ref{barstnueve}), \ie\ the function $z(\rho)$ defined by 
(\ref{baronueve}) is a solution of the BPS condition in cylindrical
coordinates.

\begin{figure}
\centerline{\epsffile{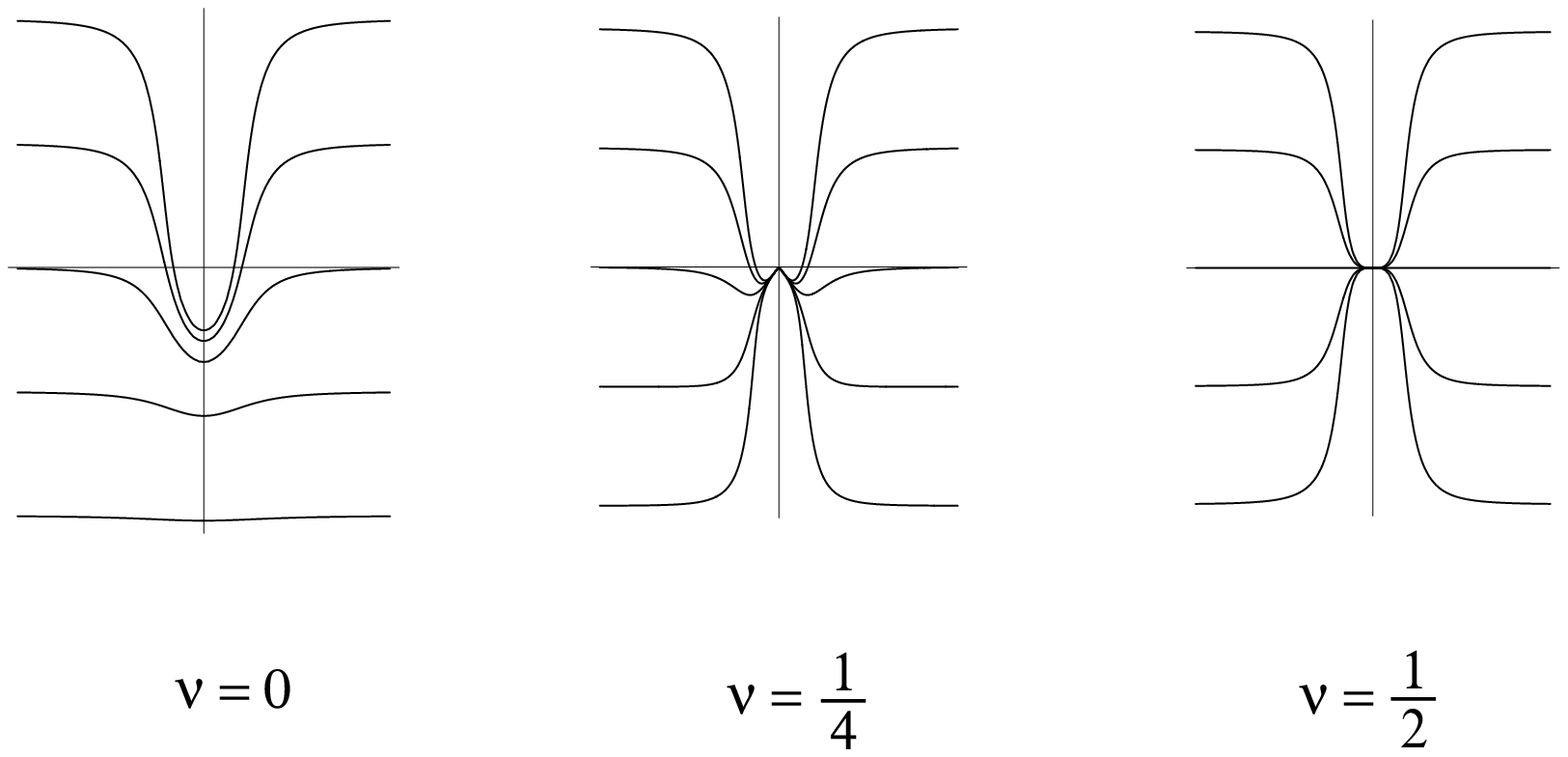}}
\caption{Solutions of the BPS differential equation for the
asymptotically flat metric (eq.~(\ref{baronueve})) for $p=3$ and
several values of $\nu$ and $z_{\infty}$. }
\label{barfig5}
\end{figure}

Let us now study  the analytical solution we have found
and, specially, how it interpolates between the near-horizon and
asymptotic regions. Notice, first of all,  that 
eqs.  (\ref{baroocho})  and (\ref{baronueve}) give $r(\theta)$ and $z(\rho)$
in implicit form. In particular, eq. (\ref{baroocho}) determines
$r(\theta)$ as the root of a polynomial of degree $7-p$ which, for
$p\ge 3$, can be solved algebraically. However, the implicit equations 
(\ref{baroocho})  and (\ref{baronueve}) are much easier to handle and,
actually, can be used to plot the embeddings. In figure \ref{barfig5} we have
represented these embeddings for different values of $\nu$ and 
$z_{\infty}$. Moreover, by studying eqs. (\ref{baroocho})  and
(\ref{baronueve}) we can characterize the general features of our BPS
solutions. The detailed analysis of these solutions is performed in
appendix A. It follows from this analysis that our solution
(\ref{baronueve}) coincides with the near-horizon embeddings in a region
close to the origin. Actually, as can be seen from the plots of figure
\ref{barfig5}, the integral of the BPS differential equation for the full metric
contains cylindrical regions which resemble closely  the tubes found
in the near-horizon analysis. It is not difficult to relate the
parameter $z_{\infty}$ of eq. (\ref{baronueve}) to the constant $C$
appearing in the lower and upper tube solutions (eqs. (\ref{barctres}) and
(\ref{barciseis})). For large $|\,z_{\infty}\,|$ it can be seen that both
types of solutions are approximately equal if we are sufficiently near 
the origin and if the following relation,

\beq
C^{6-p}\,=\,{R^{7-p}\over (\,6\,-\,p\,)\,|\,z_{\infty}\,|}\,\,,
\label{barnoventa}
\eeq
is satisfied.

For $0<\nu<1$ the solution reaches the origin $r=0$ by means of a
tube, which is an upper or lower tube depending on the sign of 
$z_{\infty}$ (see figure \ref{barfig5}). For $|\,z_{\infty}\,|\rightarrow\infty$
we will verify that this tubes can be regarded as bundles of
fundamental strings, exactly in the same way as in the near-horizon
case. 

The behaviour of the $\nu=0$ solution is different from that
we have just described for $0<\nu<1$\footnote{The solution for $\nu=1$
can be reduced to that for $\nu=0$ (see  eq. (\ref{barapauno})).}. When 
$z_{\infty}\rightarrow-\infty$, the D(8-p)-brane is nearly flat. As we
increase $z_{\infty}$, an upper tube develops and for 
$z_{\infty}\rightarrow+\infty$ a bundle of $N$ fundamental strings,
connecting the Dp and D(8-p) branes, is created. As pointed out in
ref. \cite{CGS1}, this solution provides a concrete realization of the
so-called Hanany-Witten effect \cite{HW}.  

Let us now study the energy of our solutions. It is clear from the
plots of figure \ref{barfig5} that $r$ is not a single-valued function of
$\theta$. It is thus more convenient to use the $(z,\rho)$
coordinates in order to have a global description of the energy
functional. From eq. (\ref{bardocho}) it is straightforward to obtain 
the expression of $U$ as an integral over $\rho$. The result can be
put in the form

\beq
U\,=\,T_{8-p}\,\Omega_{7-p}\,\,\int d\rho\,\,
\sqrt{\Bigl(\,\,\Delta_p\,-\,D_p\,{dz\over d\rho}\,\,\Bigr)^2\,+\,
\Bigl(\,\,D_p\,+\,\Delta_p\,{dz\over d\rho}\,\,\Bigr)^2}\,\,.
\label{barnuno}
\eeq

In eq. (\ref{barnuno}) $D_p$ depends on $\rho$ and $z(\rho)$ as in eq. 
(\ref{barstnueve}) and $\Delta_p$ is the function defined in eq.
(\ref{barvsiete}). It is important to stress that eq.~(\ref{barnuno}) gives
the energy  for any embedding $z(\rho)$. Notice that on the right-hand
side of eq. (\ref{barnuno}) we have the sum of two squares. The second of
these two terms vanishes when the BPS condition 
$dz/d\rho=-D_p/\Delta_p$ (see eq. (\ref{barstnueve})) holds. It is
clear from these considerations that, if we define

\beq
 X\,\equiv\,\Delta_p\,-\,D_p\,{dz\over d\rho}\,\,,
\label{barextrauno}
\eeq
we obtain the following lower bound for the energy of any embedding:

\beq
U\,\ge\,T_{8-p}\,\Omega_{7-p}\,\,\int d\rho\,\,
\Bigl|\,X\,\Bigr|\,\,.
\label{barextrados}
\eeq
The bound (\ref{barextrados}) is saturated precisely when the embedding
satisfies the first-order BPS equation (\ref{barstnueve}). A remarkable 
aspect of the function $X$ appearing in the right-hand side of eq. 
(\ref{barextrados}) is that it can be put as a total derivative. Indeed,
one can verify that

\beq
X\,=\,
{d\over d \rho}\,\,\Biggl[\,-z\,D_p\,+\,
\Bigl(\,{1\over 8-p}\,+\,
{R^{7-p}\over (\,\rho^2\,+\,z^2\,)^{{7-p\over 2}}}\,\Bigr)\,\,
\rho^{8-p}\,\,\Biggr]\,\,.
\label{barextratres}
\eeq
Eq. (\ref{barextratres}) is the analog in these coordinates of eq.
(\ref{bartseis}). As it happened with eq. (\ref{bartseis}), in order to
prove eq. (\ref{barextratres}) one only has to use eq. (\ref{barcatorce})
and, therefore, (\ref{barextratres}) is valid for any function $z(\rho)$. 
An important consequence of eq. (\ref{barextratres}) is that the bound 
(\ref{barextrados}) only depends on the boundary conditions of the 
embedding at infinity. Thus, one can say that the BPS embeddings are
those that minimize the energy for a given value of $z(\rho)$ at  
$\rho\rightarrow\infty$. From eq. (\ref{barnuno}) it is very easy to obtain
the energy $U_{BPS}$ of one of such BPS solutions. The result one
arrives at is

\beq
U_{BPS}\,=\,-\,T_{8-p}\,\Omega_{7-p}\,\,\int d\rho\,\,
\Bigl[\,{dz\over d\rho}\,+\, 
\Bigl(\,{dz\over d\rho}\,\Bigr)^{-1}\,\Bigr]\,\,D_p\,\,.
\label{barntres}
\eeq
Apart from being simple, eq. (\ref{barntres}) is specially suited for our
purposes. Notice, first of all, that the integrand on the right-hand
side of (\ref{barntres}) is always non-negative due to eq.
(\ref{barstnueve}). Secondly, we can use (\ref{barntres}) to evaluate the
energy of a tubular portion of the brane. Indeed, 
for large $|z_{\infty}|$, 
$\Bigr|{dz\over d\rho}\Bigl|$ is large in one of these
tubes and the argument  of $D_p$
is almost constant and equal to $\pi$ (0) for an upper (lower) tube. 
Neglecting the term containing $\Bigl(\,{dz\over d\rho}\,\Bigr)^{-1}$
on the right-hand side of eq. (\ref{barntres}), and defining the length
of the tube as

\beq
L_{tube}\,=\,\int_{tube}\,d\rho\,\Bigr|{dz\over d\rho}\Bigl|\,\,,
\label{barncuatro}
\eeq
we get the following values for the energy of the tubes

\beq
U_{tube}\,=\,
\cases{-\,T_{8-p}\,\Omega_{7-p}\,D_p(\,\pi\,)\,L_{tube}\,\,,&
                   \,\,\,\,\,\,(upper tube)\,\,,\cr\cr
        T_{8-p}\,\Omega_{7-p}\,D_p(\,0\,)\,L_{tube}\,\,,&
                    \,\,\,\,\,\,(lower tube)\,\,.}
\label{barncinco}
\eeq

Taking into account eqs. (\ref{barvtres}) and (\ref{barvcinco}),  and using
eq. (\ref{barvcuatro}), one can easily show from the result in eq. 
(\ref{barncinco}) that $U_{tube}$ coincides with $|Z_s|$, where the
values of $Z_s$ are given in eq. (\ref{barstres}). This confirms our
interpretation of the tubes as bundles of $(1-\nu)N$ and $\nu N$
fundamental strings. 

It is not difficult to calculate the energy of the whole brane.
Actually, by means of eq.~(\ref{barextratres}) we
can perform the integral appearing in the right-hand side of eq.
(\ref{barntres}). As we are calculating the energy of an infinite brane,
this integral is divergent. In order to regulate this divergence, let
us introduce a cutoff $\rho_c$, in such a way that the integral 
in eq. (\ref{barntres}) is
performed between $\rho=0$ and $\rho=\rho_c$. It is not difficult to
check from the properties of our solutions that the contribution of 
the lower limit $\rho=0$ is zero and, thus, only the value at the
cutoff contributes.  After this process, 
$U_{BPS}$ can be put as

\beq
U_{BPS}\,=\,-T_{8-p}\,\Omega_{7-p}\,\,
\Biggl[\,z\,D_p\,-\,
\Bigl(\,{1\over 8-p}\,+\,
{R^{7-p}\over (\,\rho^2\,+\,z^2\,)^{{7-p\over 2}}}\,\Bigr)\,\,
\rho^{8-p}\,\,\Biggr]_{\rho\,=\,\rho_c}\,\,\,.
\label{barnsiete}
\eeq
We have now to take $\rho_c\rightarrow\infty$ and, therefore, 
$z(\,\rho_c\,)\rightarrow z_{\infty}$. Notice that, in this limit, the
argument of the function $D_p$ is $\pi/2$. Using eq. (\ref{barouno}), 
$U_{BPS}$ can be put as a sum of two terms, namely,

\beq
U_{BPS}\,=\,\Bigl(\,{1\over 2}\,-\,\nu\,\Bigr)\,N\,T_f\,z_{\infty}\,+\,
T_{8-p}\,\Omega_{7-p}\,\,\Biggl[\,{\rho_c^{8-p}\over 8-p}\,+\,
R^{7-p}\,\rho_c\,\Biggr]\,\,.
\label{barnocho}
\eeq
Notice that the first term on the right-hand side of eq. (\ref{barnocho})
is finite and depends linearly on $z_{\infty}$, while the second term
diverges when $\rho_c\rightarrow\infty$ and is independent of 
$z_{\infty}$. According to  ref. \cite{Craps}, one can give the
following interpretation to this divergence. Let us consider a
D(8-p)-brane embedded in the metric  (\ref{baruno}) along the plane $z=0$
or, equivalently, such that its 
worldvolume is determined by the equation  
$\theta=\pi/2$. Notice that in this configuration the brane is not bent
at all and, for this reason, it will be referred to as the ``ground
state" of the brane. Let
$g_{gs}$ be the induced metric on the worldvolume of the D(8-p)-brane
for this ground state configuration. Putting the worldvolume gauge
fields to zero  and substituting $g$ by $g_{gs}$ in eq. 
(\ref{barseis}), we get the action of the ground state

\beq
S_{gs}\,=\,-T_{8-p}\,\int d^{9-p}\,\xi\,
e^{-\tilde\phi}\,\,
\sqrt{-{\rm det}\,\,(\,g_{gs}\,)}\,\,.
\label{barnnueve}
\eeq

The energy $E_{gs}$ of the ground state is obtained from $S_{gs}$ as

\beq
E_{gs}\,=\,-{S_{gs}\over T}\,\,,
\label{barcien}
\eeq
where $T=\int dt$. Using the metric given in eq. (\ref{baruno})  and the
value of the dilaton field displayed in eq. (\ref{barcinco}),
the calculation of  $E_{gs}$  is a simple exercise. By comparing the
result of this computation with the right-hand side of eq.
(\ref{barnocho}), one discovers that $E_{gs}$ is equal to the divergent
contribution to $U_{BPS}$. Therefore, one can subtract $E_{gs}$ from 
$U_{BPS}$ and define a renormalized energy $U_{ren}$ as
\beq
U_{ren}\,\equiv\,U_{BPS}\,-\,E_{gs}\,\,.
\label{barctuno}
\eeq
It follows from eq. (\ref{barnocho}) that $U_{ren}$ is given by

\beq
U_{ren}\,=\,
\Bigl(\,{1\over 2}\,-\,\nu\,\Bigr)\,N\,T_f\,z_{\infty}\,\,.
\label{barctdos}
\eeq
By derivating $U_{ren}$ with respect to $z_{\infty}$, we learn that
there is a net constant force acting on the D(8-p)-brane, which is
equal to $(\,{1\over 2}\,-\,\nu\,\Bigr)\,N\,T_f$, \ie\ equivalent to
the tension of $(\,{1\over 2}\,-\,\nu\,\Bigr)\,N$ fundamental strings.
In ref. \cite{CGS1} this force was interpreted as a consequence of the
fact that, due to the $p+2$-form flux captured by the D(8-p)-brane, the
latter is endowed with an effective charge equal to 
$(\,{1\over 2}\,-\,\nu\,\Bigr)\,N$ units.

Let us now integrate the BPS
differential equation for $p=6$. Using the value of $\Lambda_6(\theta)$
in eq. (\ref{barocinco}), we get (for $a=1$)

\beq
r\,{\rm sin }\,\theta\,\,
{d\over d\theta}\,\,(\,r\,{\rm cos}\,\theta\,)\,=\,R\,
\Bigl[\,r\,+\,(\,2\nu\,-\,1\,)\,
{d\over d\theta}\,\,(\,r\,{\rm sin}\,\theta\,)\,\Bigr]\,\,,
\,\,\,\,\,\,\,\,\,\,\,\,\,\,\,\,\,\,\,\,\,\,\,\,\,\,\,\,\,\,\,\,
(\,p=6\,)\,\,.
\label{barcttres}
\eeq
This equation can be recast in the form

\beq
{d\over d\theta}\,\,(\,r\,{\rm cos}\,\theta\,)\,=\,R\,
{d\over d\theta}\,\,\Biggl[\,{\rm log}\,\Bigl[\,
{\rm tan}\,\bigl(\,{\theta\over 2}\,\bigr)\,
(\,r\,{\rm sin}\,\theta\,)^{2\nu-1}\,\Bigr]\,\Biggr]
\,\,,\,\,\,\,\,\,\,\,\,\,\,\,\,\,\,\,\,\,\,\,\,\,\,\,\,\,\,\,\,\,\,\,
(\,p=6\,)\,\,,
\label{barctcuatro}
\eeq
and, therefore, its integration is immediate. It is more interesting
to write the result in cylindrical coordinates. After a short
calculation one gets

\beq
\Bigl(\,z\,+\,\sqrt{\rho^2\,+\,z^2}\,\,\Bigr)\,
\rho^{2(\nu-1)}\,=\,e^{{k-z\over R}}\,\,
\,\,,\,\,\,\,\,\,\,\,\,\,\,\,\,\,\,\,\,\,\,\,\,\,\,\,\,\,\,\,\,\,\,\,
(\,p=6\,)\,\,,
\label{barctcinco}
\eeq
where $k$ is a constant of integration. It is easy to check that the
function $z(\rho)$ parametrized by eq. (\ref{barctcinco}) has the
asymptotic behaviour displayed in eq. (\ref{barotres}). It follows that,
as expected, $z(\rho)$ does not reach a constant value for $p=6$ and 
$\nu\not=1/2$. The profiles of the $\nu \not=1/2$ curves are similar to
the near-horizon ones, plotted in figure 3. The only difference is
that, instead of the behaviour written in eq. (\ref{barstdos}), 
$|\,z(\rho)\,|$ grows logarithmically as $\rho\rightarrow\infty$.
Moreover, similarly to what happens to the $p\le 5$ solution, 
eq. (\ref{barctcinco}) is invariant under the transformation

\beq
\rho\rightarrow\rho\,,
\,\,\,\,\,\,\,\,\,\,\,\,\,
z\rightarrow -z\,,
\,\,\,\,\,\,\,\,\,\,\,\,\,
k\rightarrow -k\,,
\,\,\,\,\,\,\,\,\,\,\,\,\,
\nu\rightarrow 1 -\nu\,\,.
\label{barctseis}
\eeq

It is also easy to check that the function $z(\rho)$ defined by eq. 
(\ref{barctcinco})  coincides, in the near-horizon region, with the one
written down in eq. (\ref{barstuno}) if the constants $A$ and $k$ are
identified as follows:

\beq
A^{2\nu-1}\,=\,e^{{k\over R}}\,\,,
\,\,\,\,\,\,\,\,\,\,\,\,\,\,\,\,\,\,\,\,\,\,\,\,\,\,\,\,\,\,\,\,\,\,
(\,\nu\,\not=\,{1\over 2}\,)\,\,.
\label{barctsiete}
\eeq

\begin{figure}
\centerline{\epsffile{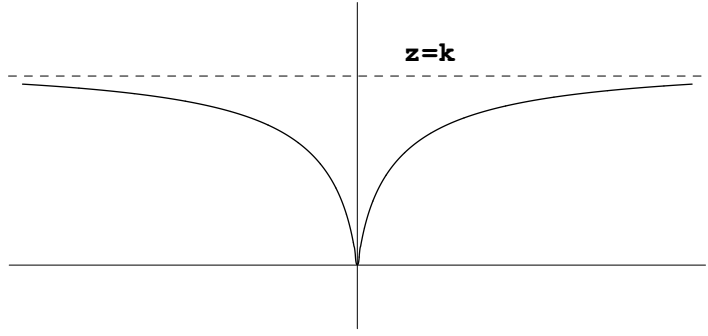}}
\caption{BPS embedding of the D2-brane in the full D6-brane geometry
for $\nu=1/2$. }
\label{barfig6}
\end{figure}

For $\nu\,=\,1/2$ the behaviour of the solution changes drastically.
Indeed, in this case, one can solve eq. (\ref{barctcinco}) for $\rho$ and
get $\rho=\rho(z)$. The result is

\beq
\rho\,=\,
{z\over {\rm sinh}\,{k-z\over R}}\,\,,
\,\,\,\,\,\,\,\,\,\,\,\,\,\,\,\,\,\,\,\,\,\,\,\,\,\,\,\,\,\,\,\,\,\,
(\,\nu \,=\,{1\over 2}\,,\,p=6\,)\,\,.
\label{barctocho}
\eeq
The function (\ref{barctocho}) has been represented graphically in figure
\ref{barfig6}. From this plot one notices that
$\lim_{\rho\rightarrow\infty}\,\,z(\rho)=k$. Actually, one can prove
using eq. (\ref{barctocho}) that for 
$k\ge 0$ ($k\le 0$) the variable $z$ takes values in the range 
$0\le z\le k$ ($k\le z\le 0$). Thus, in this ($p=6$, $\nu=1/2$) case,
the constant $k$ plays the same role as $z_{\infty}$ did for 
$p\le 5$. In particular, for $k=0$ the solution (\ref{barctocho}) is
equivalent to the equation $z=0$, which is precisely the same solution
we found for $p\le 5$, $\nu=1/2$ and $z_{\infty}=0$, 
\ie\ one obtains the ``ground state" solution in both cases.

Let us finally point out that, by linearizing eq. (\ref{barctocho}) in
$\rho$ and $z$, one gets the equation of a cone passing through the
origin and with half-angle equal to

\beq
{\rm tan}\,\alpha\,=\,{1\over {\rm sinh}\,\,\Bigl(\,
{k\over R}\,\Bigr)}\,\,.
\label{barctnueve}
\eeq
Therefore, with the  identification (\ref{barctnueve}), we get a perfect
agreement with the near-horizon solutions (\ref{barstsiete}) with $C=0$.

\medskip
\section{Supersymmetry} \label{barsecsym}
\medskip                                             
\setcounter{equation}{0}

In this section we will see that the configurations satisfying the BPS bound 
(\ref{barvseis}) preserve 1/4 of the target space supersymmetry. The number of 
supersymmetries preserved by the probe is the number of independent solutions 
of the equation (\ref{kappafix4})
\be
(1-\G_{\k})\epsilon=0\,,
\label{barsymuno}
\ee
where $\epsilon$ are the Killing spinors of the background. The equation 
above is a projection condition, as we have seen in the derivation of 
kappa-symmetric D-brane actions in the previous chapter. The matrix $\G_{\k}$ 
depends on the geometry of the background and on the worldvolume fields. 
Then (\ref{barsymuno}) is a local projection condition giving different 
projectors for different points on the brane worldvolume. 
We have been considering configurations in which the transverse scalar and the 
gauge field depend only on the polar coordinate $(r(\th), A_0(\th))$ so that 
the projector in (\ref{barsymuno}) will be a function of $\th$. As we will see 
the Killing spinor will also depend on $\th$. Thus, if we want some 
supresymmetries to be preserved globally we must achieve some configuration 
of the worldvolume fields such that eq. (\ref{barsymuno}) is reduced to a 
$\th$-independent projection condition on a constant spinor. 

Our BI field-strength has only the $F_{0\th}$ component non-vanishing and 
the expansion of $\G_{\k}$ given in (\ref{kappamatrix}) terminates at linear 
order:
\be
\G_{\k}=\frac{1}{\sqrt{-det(g+{\cal F})}}\,\left[ J_{(8-p)}^{(0)}+
\g^{0\th}{\cal F}_{0\th}J_{(8-p)}^{(1)}\right]\,,
\ee
where the matrices $J_{(p)}^{(n)}$ were defined in (\ref{kappaj}) and the induced 
gamma matrices are
\be
\g_{\m}=\partial_{\m}X^mE_{m}{}^{\underline m}\G_{\underline m}\,.
\ee
with $\G_{\underline m}$ the ten-dimensional constant Dirac matrices and 
$E_m{}^{\underline m}$ the ten-dimensional vielbeins. The index $0$ denotes 
the component along the time direction. For the geometry (\ref{baruno}) and 
in the static gauge (\ref{barocho}) we get
\br
&&\g_0=f_p^{-1/4}\G_{\underline 0}\,, \nonu\\ \nonu\\
&&\g_{\th^i}=f_p^{1/4}r \sin\th\,e_{\th^i}{}^{\underline \th^i} 
\G_{\underline 0}\,, \nonu\\ \nonu\\
&&\g_{\th}=f_p^{1/4}(r\G_{\underline \th}+r^{\prime}\G_{\underline r})\,,
\er
where $e_{\th^i}{}^{\underline \th^i}$ are the vielbeins of the unit $S^{7-p}$ 
sphere. Let us consider first the odd $p$ case (type IIB case). In this case 
$\G_{\k}$ is given by
\be
\G_{\k}=\frac{\s_3^{\frac{5-p}{2}}}{\sqrt{r^2+{r^{\prime}}^2-F_{0\th}^2}}\,
\left[r\G_{\underline {0\th}}\s_3+r^{\prime}\G_{\underline {0r}}
\s_3+F_{0\th}\s_3\right](-1)^{(7-p)}\G_*(i\s_2)\,,
\label{barsymdos}
\ee
where we have used the definition $\G_{*} \equiv \G_{\underline {\th^1\dots 
\th^{7-p}}}$.

The Killing spinors of Dp-brane backgrounds have been given in 
(\ref{strkisp}):
\be
\epsilon\,=\,g(r)\,e^{\frac{\th}{2}\G_{{\underline {r\th}}}}\prod_{i=1}^{7-p}
e^{-\frac{\th_i}{2}\G_{\underline {\th^i\th^{i+1}}}}\,\epsilon_0\,,
\label{barsymtres}
\ee
with $\epsilon_0$ a constant spinor. Both $\epsilon$ and $\epsilon_0$ satisfy 
the following projection condition:
\be 
\G_{\underline {01\dots p}}(i\s_2)\s_3^{\frac{p-3}{2}}\epsilon=\epsilon\,,
\label{barsymcuatro}
\ee
where the indices $1,\dots ,p$ are used to denote the components along 
directions $x^1,\dots ,x^p$ respectively. This projection is the 1/2 
supersymmetry breaking condition corresponding to 
extremal (BPS) D-brane backgrounds. It corresponds to a D1 or a D3 for 
$p=1$ or $p=3$
and an anti-D5 for $p=5$ ($\ol {D5}$). This choice is made so that the 
supersymmetry preserved by the probe is the same for the three cases. In fact, 
as we will see, this is the supersymmetry preserving condition imposed by 
fundamental strings lying along the radial direction.
The spinors (\ref{barsymtres}) can be reexpressed in the form
\be
\epsilon=e^{\frac{\th}{2}\G_{\underline {r\th}}}\,\hat\epsilon\,,
\label{barsymcinco}
\ee
where $\hat\epsilon$ is the $\th$-independent part of $\epsilon$ 
($\hat\epsilon=
g(r)\,\prod_{i=1}^{7-p}e^{-\frac{\th_i}{2}\G_{\underline {\th^i\th^{i+1}}}}
\,\epsilon_0$). Eq. (\ref{barsymcinco}) can be interpreted as a rotation 
in the $(r,\th)$ plane with angle $\th$ of the Killing spinor $\hat\epsilon$ 
at the pole ($\th=0$). In the type IIB case the Killing spinor has possitive 
chirality: 
\be 
\G_{11}\epsilon=\epsilon\,,\ \ \ \ \ \G_{11}=\G^{\underline {01\dots pr
\th^1\dots \th^{7-p}\th}}\,.
\label{barsymseis}
\ee
One can combine the chirality condition and the background projection 
condition (\ref{barsymcuatro}) to obtain the relation
\be
\G_{*}\epsilon=(-1)^{(7-p)}\G_{\underline {r\th}}(i\s_2)
\s_3^{\frac{p-3}{2}}\epsilon\,.
\ee
This can be substituted in (\ref{barsymdos}) and the $\k$-symmetry projection 
can be rewritten as
\be
\left[-r\G_{\underline {0r}}\s_3+r^{\prime}\G_{\underline {0\th}}
\s_3+F_{0\th}\G_{\underline {r\th}}\right]
\epsilon=-\sqrt{r^2+{r^{\prime}}^2-F_{0\th}^2}\,\epsilon
\label{barsymsiete}
\ee
We have seen in eq. (\ref{barbps}) that the BPS condition can be recast in the 
form 
\be
F_{0\th}=\frac{d}{d\th}(r\cos\th)\,.
\label{barsymocho}
\ee
Substituting this value of $F_{0\th}$ in eq. (\ref{barsymsiete}) and using 
the form of the Killing spinor given in (\ref{barsymcinco}) we get
\be
\left[-re^{-\th \G_{\underline {r\th}}}\G_{\underline {0r}}\s_3+
r^{\prime}e^{-\th \G_{\underline {r\th}}}\G_{\underline {0\th}}\s_3+
(r^{\prime}\cos\th -r\sin\th)\G_{\underline{r\th}}\right]
\hat\epsilon=-(r\cos\th+r^{\prime}\sin\th)\,\hat\epsilon\,.
\label{barsymnueve}
\ee
If we consider eq. (\ref{barsymnueve}) for the particular case $\th=0$ we 
get the condition
\be
\left[-r\G_{\underline {0r}}\s_3+
r^{\prime}\G_{\underline {0\th}}\s_3+
r^{\prime}\G_{r\th}\right]
\hat\epsilon=-r\,\hat\epsilon\,,
\label{barsymdiez}
\ee
which is equivalent to impose the following condition on the spinor 
$\hat\epsilon$:
\be
(1-\G_{\underline {0r}}\s_3)\hat\epsilon=0\,,
\label{barsymonce}
\ee
showing the supersymmetry breaking condition of the brane probe at the pole. 
One can see that this condition implies (\ref{barsymnueve}) also for 
$\th\neq 0$. We then conclude that the $\k$-symmetry condition 
$\G_{\k}\epsilon=\epsilon$ is equivalent to eq. (\ref{barsymonce}) for the 
BPS configuration (\ref{barsymocho}). Since the matrix $\G_{\underline {0r}}$ 
commutes with the pairs of matrices appearing in the definition of 
$\hat\epsilon$ the condition (\ref{barsymonce}) is equivalent to
\be
(1-\G_{\underline {0r}}\s_3)\epsilon_0=0\,.
\label{barsymdoce}
\ee
This is a coordinate independent condition over the constant spinor 
$\epsilon_0$, which shows that the brane probe breaks 1/2 of the 
supersymmetries preserved by the background (\ref{barsymcuatro}). 
Eq. (\ref{barsymdoce}) is precisely the 
projection condition for a fundamental string stretching along the radial 
direction, as expected from the interpretation of the energy in 
(\ref{bartseis}) as the one corresponding to a bundle of fundamental strings.

For the type IIA case the Killing spinor has the same form as in 
(\ref{barsymtres}) but the background projection is now given by:
\be 
\G_{\underline{01\dots p}}(\G_{11})^{\frac{p-2}{2}}\epsilon=-\epsilon\,.
\label{barsymtrece}
\ee
The condition $\G_{\k}\epsilon=\epsilon$ takes the form
\be
\left[-r\G_{\underline {0r}}\G_{11}+r^{\prime}\G_{\underline {0\th}}
\G_{11}-F_{0\th}\G_{\underline {r\th}}\right]
\epsilon=-\sqrt{r^2+{r^{\prime}}^2-F_{0\th}^2}\,\epsilon\,,
\label{barsymcatorce}
\ee
and is the same as (\ref{barsymsiete}) except or the sign in the $F_{0\th}$ 
term. However, in this case, the BPS condition is $F_{0\th}=-
\frac{d}{d\th}(r\cos\th)$ and then we arrive at (\ref{barsymnueve}) again, but 
with $\G_{11}$ instead of $\s_3$. The resulting condition is the supersymmetry 
of a string in the radial direction in type IIA theory:
\be
(1-\G_{\underline {0r}}\G_{11})\epsilon_0=0\,.
\label{barsymquince}
\ee

Of course, all the reasoning above could be inverted. That is, in 
(\ref{barsymsiete}), with $F_{0\th}$ unknown, we could have imposed eq. 
(\ref{barsymdoce}), which means that we require the solution to have 
an spike which can be interpreted as a bundle of fundamental strings 
in the radial direction. This would fix 
the value of $F_{0\th}$ to the one given in (\ref{barsymocho}).

\setcounter{equation}{0}
\section{Discussion} \label{barsec7}
\medskip

In this chapter we have studied the embedding of a D(8-p)-brane in the
background geometry of a stack of coincident Dp-branes. This embedding
is governed by the worldvolume action of the D(8-p)-brane 
(eq. (\ref{baruno})), which determines the equation of motion
(\ref{barvseis}). By using a BPS argument we have found a bound for the
energy of the system such that those embeddings which saturate it 
are also a solution of the equation of motion. This equation of motion
(eq.~(\ref{bartdos})) is a first-order differential equation which,
amazingly, can be solved analytically both in the near-horizon and
asymptotically flat geometries. 

The solutions of the BPS equations give the deformation of the
D(8-p)-brane under the influence of the gravitational and RR fields
created by the background branes. Generically, these solutions contain
tubes connecting the D(8-p)-brane to the background branes which,
after analyzing its energy, can be interpreted as bundles of
fundamental strings. From the point of view of the gauge theory defined
on the worldvolume of the background branes, these tubes represent
baryonic multiquark states. The relation between the  BPS  condition and
supersymmetry was also discussed. It was shown that the BPS differential 
equation is precisely the requirement one must impose
to the brane embedding in order to preserve one half of the supersymmetry 
of the background.  

The baryonic configurations discussed in this chapter can be generalized 
to M-theory. In \cite{kappa} there were found configurations of a baryonic 
M5-brane  probe in the background of a stack of N M5-branes. These 
configurations saturate the BPS bound and are $1/4$ supersymmetric. In the 
near-horizon limit  the geometry becomes $AdS_7\times S^4$ and the brane 
construction provides  a worldvolume realization of the baryon string 
vertex of the $(2,0)$-supersymmetric six-dimensional conformal field theory 
on coincident M5-branes \cite{Ali}. For the full M5-background, these 
configurations are a realization of the Hanany-Witten effect in M-theory.


\setcounter{equation}{0}
\section{Appendix 3A}                                              
\medskip                                         
\renewcommand{\theequation}{\rm{A}.\arabic{equation}}

In this appendix we are going to carry out an study of the solution
found in section \ref{barsec6} of the BPS differential equation for the
asymptotically flat metric for $p\le 5$ (eqs. (\ref{baroocho}) and
(\ref{baronueve})). 

The first thing we want to
point out in this respect is that, due to the property of
$\Lambda_p$ displayed in eq. (\ref{barcisiete}),  our solution 
(\ref{baronueve}) is invariant under the transformation

\beq
\rho\rightarrow\rho\,,
\,\,\,\,\,\,\,\,\,\,\,\,\,
z\rightarrow -z\,,
\,\,\,\,\,\,\,\,\,\,\,\,\,
z_{\infty}\rightarrow -z_{\infty}\,,
\,\,\,\,\,\,\,\,\,\,\,\,\,
\nu\rightarrow 1 -\nu\,\,.
\label{barapauno}
\eeq
Due to this invariance we can restrict ourselves to the case 
$0\le\nu\le 1/2$. The solutions outside this range of $\nu$ can be
obtained by performing a reflection with respect to the $z=0$ axis.
Therefore,  unless otherwise stated, we will assume in what follows
that   $\nu\le 1/2$. Moreover, it is more convenient for our purposes
to rewrite eq. (\ref{baronueve}) in the equivalent form

\beq
(\,z_{\infty}\,-\,z\,)\,\rho^{6-p}\,=\,
{R^{7-p}\over 6-p}\,\,\Lambda_p\,
\Bigr(\,{\rm arctan}\,(\,-\rho/ z)\,\Bigr)\,\,.
\label{barapados}
\eeq
We will start our analysis of the solution by studying its cut with
the $\rho=0$ axis.  By looking
at eq. (\ref{barapados}) it follows immediately that $\Lambda_p$ must
vanish for $\rho=0$. Thus, the angle $\theta$ at which the solution
reaches the $\rho=0$ axis must be $\theta\,=\,\theta_0$, $\theta_0$
being the same angle appearing in  the near-horizon solution (see eq. 
(\ref{barcseis})). Following eq.  (\ref{barsnueve}), 
$\rho\,=\,r(\theta)\,{\rm sin}\,\theta$ and, therefore, 
if $\rho=0$ for $\theta=\theta_0$ one must have 
$r(\theta_0)\,{\rm sin}\,\theta_0\,=\,0$. There are two
possibilities to fulfill this equation. If, first of all, $\nu\not=0$,
the angle $\theta_0$ is non-vanishing, which means that, necessarily, 
$r(\theta_0)=0$, \ie\ the solution reaches the origin $r=0$ at an angle 
$\theta_0$. Notice that, as it should occur near  $r=0$,  this is
precisely what happens in the near-horizon solution. The second
possibility is  $\nu=0$. In this case $\theta_0=0$ and the
vanishing of $\rho$ does not require $r=0$. Again,  this is in agreement
with the near-horizon solution. Actually, the distance $r(0)$ at which
the 
$\nu=0$ solution cuts the $\rho=0$ axis can be determined, as in
section \ref{barsec4}, by expanding $\Lambda_p(\theta)$ around $\theta_0=0$. Using
eq. (\ref{barcitres}) one gets that $r(0)$ must satisfy

\beq
\bigl[\,r(0)\,\bigr]^{6-p}\,\,\,
\bigl[\,r(0)\,+\,z_{\infty}\,\bigr]\,=\,
{R^{7-p}\over 6-p}\,\,,
\,\,\,\,\,\,\,\,\,\,\,\,\,\,\,\,\,\,\,\,\,\,\,\,\,\,
(\,\nu\,=\,0\,)\,\,.
\label{barapatres}
\eeq
As the right-hand side of this equation is positive, it follows that 
$r(0)\,>\,-z_{\infty}$. When this condition is satisfied, the
left-hand side of eq. (\ref{barapatres}) is a monotonically increasing
function of $r(0)$ and, therefore, there exists a unique solution for
$r(0)$.

Let us now consider the cut with the $z=0$ axis. We shall denote
the corresponding value of the  $\rho$ coordinate by 
$\rho_0$, \ie\ 
$\rho_0\,=\,\rho(\,z=0\,)$. If $\rho_0\not= 0$, the  value
of the angle  $\theta$ for $z=0$ is $\theta=\pi/2$ and, after
evaluating the function $\Lambda_p$ for this value of $\theta$, eq. 
(\ref{barapados}) gives

\beq
\rho_0^{6-p}\,=\,2\sqrt{\pi}\,\,
{\Gamma\Bigl(\,{8-p\over 2}\Bigr)\over
\Gamma\Bigl(\,{7-p\over 2}\Bigr)}\,\,\,
({1\over 2}\,-\,\nu\,)\,\,\,{R^{7-p}\over z_{\infty}}\,\,.
\label{barapacuatro}
\eeq

Notice that for $\nu\not=1/2$ the right-hand side of eq. 
(\ref{barapacuatro}) only makes sense for $z_{\infty}>0$. Therefore, only
for $z_{\infty}>0$  the solution for $\nu\not=1/2$ cuts the $z=0$ axis
at a finite non-vanishing value of the coordinate $\rho$. For 
$\nu=1/2$ and $z_{\infty}\not=0$ eq. (\ref{barapacuatro}) has no solution
for $\rho_0\not= 0$. All these features appear in the plots of figure
\ref{barfig5}.  

From eq. (\ref{barapados}) one can extract the range of allowed values
of the coordinate $z$. Indeed, it follows from eq. (\ref{barapados})
that the signs of $z_{\infty}-z$ and $\Lambda_p$ are the same. On the
other hand, we know that $\Lambda_p(\theta)$ is positive for 
$\theta>\theta_0$ and negative for $\theta<\theta_0$. Thus, when 
$\theta>\theta_0$ ($\theta<\theta_0$) one must have 
$z<z_{\infty}$ ($z>z_{\infty}$),  while for $\theta=\theta_0$ either
$\rho=0$ or else $z=z_{\infty}$. As 
$\theta_0\le\pi/2$ for $\nu\le 1/2$, one can easily see that these
results imply that 
$z\le 0$ for $z_{\infty}\le 0$,  whereas $z$ can be positive or
negative  if $z_{\infty}> 0$. 

It is easy to prove that $z$, as a function of $\rho$, must have a
unique extremum. For $\nu< 1/2$ this extremum is actually a minimum,
as we are going to verify soon. From eq. (\ref{barstnueve}) it is clear
that $dz/d\rho$ vanishes if and only if $D_p$ is zero. Recall that the
Gauss' law (eq. (\ref{barcatorce})) implies that $D_p$ is a monotonically
decreasing function of $\theta$ (see figure \ref{barfig1}). Moreover, it follows
from eqs. (\ref{barvtres}) and (\ref{barvcinco}) that $D_p(0)\ge 0$ and
$D_p(\pi)\le 0$. Thus, it must necessarily exist a unique value
$\theta_m$ of $\theta$ such that

\beq
D_p(\,\theta_m\,)\,=\,0\,\,.
\label{barapacinco}
\eeq
Clearly, at the point of the curve $z(\rho)$ at which $\theta=\theta_m$
the derivative $dz/d\rho$ is zero.  It is interesting to compare
$\theta_m$ with the angle
$\theta_0$ for which $\Lambda_p$ is zero. By substituting 
$\theta=\theta_0$ in the equation which relates 
$D_p(\,\theta\,)$ and $\Lambda_p(\,\theta\,)$ (eq. (\ref{bartnueve})),
one gets

\beq
D_p(\,\theta_0\,)\,=\,R^{7-p}\,
(\,{\rm sin }\,\theta_0\,)^{6-p}\,\,
{\rm cos }\,\theta_0\,\,.
\label{barapaseis}
\eeq

For $\nu\le 1/2$ one has $\theta_0\,\le\,\pi/2$ and, thus, eq. 
(\ref{barapaseis}) gives $D_p(\,\theta_0\,)\,\ge\,0$. The monotonic
character of $D_p(\,\theta\,)$ implies that 
$0\,\le\,\theta_0\,\le\,\theta_m\,\le\,\pi/2\,$. Notice that 
$\theta_0\,=\,\theta_m$ if $\nu=0$ ($\theta_0\,=\,\theta_m\,=\,0$) or 
$\nu=1/2$ ($\theta_0\,=\,\theta_m\,=\,\pi/2$).  The coordinate $z$ of
the extremum is 
$z_m\,=\,-r(\theta_m)\,{\rm cos }\,\theta_m$. As 
$\theta_m\le \pi/2$, one must have $z_m\le 0$. Moreover, since 
$\theta_m\ge \theta_0$, then $\Lambda_p (\theta_m)\ge 0$ and 
eq. (\ref{barapados}) gives $z_m\le z_{\infty}$. By using the value of 
$\Lambda_p$ at $\theta=\theta_m$, which is

\beq
\Lambda_p (\theta_m)\,=\,(\,{\rm sin }\,\theta_m\,)^{6-p}\,\,
{\rm cos }\,\theta_m\,\,,
\label{barapasiete}
\eeq
(see eq. (\ref{bartnueve})) one can obtain an expression which determines
$z_m$, namely,

\beq
|\,z_m\,|^{6-p}\,\,(\,z_{\infty}\,-\,z_m\,)\,=\,
{R^{7-p}\over 6-p}\,\,
\Bigr(\,{\rm cos }\,\theta_m\,\Bigr)^{7-p}\,\,.
\label{barapaocho}
\eeq

It is not difficult to verify now that, when $\nu<1/2$, the extremum
at $\theta=\theta_m$ is a minimum. In order to prove  it, one must
evaluate  $d^2z/d\rho^2$. This can be done by derivating eq. 
(\ref{barstnueve}). After putting 
$\rho\,=\rho_m\,=\,r(\theta_m)\,{\rm sin }\,\theta_m$, and using the
fact that $dz/d\rho$ vanishes for $\rho\,=\rho_m$, one arrives at

\beq
{d^2\,z\over d\rho^2}\Big|_{\rho\,=\,\rho_m}
\,=\,-
{(\,7-\,p\,)\,\,R^{7-p}\over (\,\rho_m^{2}\,+\,z_m^2\,)\,\,
\Bigl[\,R^{7-p}\,+\,
(\,\rho_m^{2}\,+\,z_m^2\,)^{{7-p\over 2}}\,\Bigl]}\,\,\,z_m\,\,.
\label{barapanueve}
\eeq
When $\nu<1/2$, one has $z_m<0$ and, as a consequence, the right-hand
side of eq. (\ref{barapanueve}) is strictly positive. Therefore $z(\rho)$
has a minimum for $\rho\,=\,\rho_m$ as claimed. Notice that,  when
$\nu=1/2$, \ie\ $\theta_m=\pi/2$, the right-hand side of eq. 
(\ref{barapaocho}) vanishes and, thus, $z_m=0$ (the value $z=z_{\infty}$
is reached asymptotically). 

Eq. (\ref{barapaocho}) can be solved immediately if $z_{\infty}=0$.
Indeed, in this case, one has

\beq
|\,z_m\,|^{7-p}\,=\,
{R^{7-p}\over 6-p}\,\,
\Bigr(\,{\rm cos }\,\theta_m\,\Bigr)^{7-p}\,\,,
\,\,\,\,\,\,\,\,\,\,\,\,\,\,\,\,\,\,
(\,z_{\infty}\,=\,0\,)\,\,.
\label{barapadiez}
\eeq
For general values of $z_{\infty}$ we cannot give the general solution
of eq. (\ref{barapaocho}).  However, one can extract valuable information
from this equation. An important point concerning eq. (\ref{barapaocho}) 
is the fact that its right-hand side does not depend on $z_{\infty}$
(it only depends on $p$ and $\nu$) and, therefore, it remains constant
when we change the asymptotic value of the $z$ coordinate. This allows
us to study the behaviour of the minimum in different limiting
situations. Let us suppose, first of all, that
$z_{\infty}\,\rightarrow\,+\infty$ for
$\nu<1/2$. As $z_m<0$, the difference
$z_{\infty}-z_{m}\rightarrow\,+\infty$. In order to keep the
right-hand side of eq. (\ref{barapaocho}) finite, one must have 
$z_m\rightarrow\,0^{-}$. Actually $z_m$ should behave as

\beq
z_m\,\,\sim\,\,-{1\over (\,z_{\infty}\,)^{{1\over 6-p}}}\,\,,
\,\,\,\,\,\,\,\,\,\,\,\,\,\,\,\,\,\,
(\,z_{\infty}\,\rightarrow\,+\infty\,,\,\nu< {1\over 2}\,)\,\,.
\label{barapaonce}
\eeq
On the other hand if
$z_{\infty}\,\rightarrow\,-\infty$, again for $\nu<1/2$, the only
possibility to maintain 
$|\,z_m\,|^{6-p}\,\,(\,z_{\infty}\,-\,z_m\,)$ constant is that
$z_{\infty}\,-\,z_m\rightarrow\,0^{+}$ (recall that 
$z_m\le z_{\infty}$ and, thus, $z_m\rightarrow\,0$ is impossible if 
$z_{\infty}\,\rightarrow\,-\infty$). The actual behaviour of 
$z_{\infty}\,-\,z_m$ in this limit is

\beq
z_{\infty}\,-\,z_m
\,\,\sim\,\,{1\over (\,z_{\infty}\,)^{ 6-p}}\,\,,
\,\,\,\,\,\,\,\,\,\,\,\,\,\,\,\,\,\,
(\,z_{\infty}\,\rightarrow\,-\infty\,,\,\nu< {1\over 2}\,)\,\,.
\label{barapadoce}
\eeq

The $\nu=0$ embedding presents some characteristics which make it
different from the $\nu\not= 0$ ones. We have already shown that the 
$\nu=0$  solution cuts the $\rho=0$ axis at a coordinate $z=-r(0)$,
with $r(0)$ determined by eq. (\ref{barapatres}). Moreover, for 
$\nu=0$ the function $\Lambda_p$ on the right-hand side of 
eq. (\ref{barapados}) is  non-negative for all $\rho\not=0$,
which implies that  $z\le z_{\infty}$ for these solutions. Recall
that, in this case, $z(\rho)$ has a minimum at $\rho=0$. For 
$z_{\infty}\,\rightarrow -\infty$, due to eq. (\ref{barapadoce}),
$z(\rho)$ is approximately a constant function, \ie\ 
$z(\rho)\,\approx\,z_{\infty}$. If, on the contrary, 
$z_{\infty}\,\rightarrow +\infty$, an ``upper tube" connecting the
point $z=-r(0)\rightarrow 0$ (see eq. (\ref{barapaonce})) 
and the asymptotic region is developed (see figure \ref{barfig5}).

$\nu=1/2$ is also a special case. Indeed, for $\nu=1/2$ the angle
$\theta_0$ is equal to $\pi/2$ and, therefore, the function 
$\Lambda_p$ in eq. (\ref{barapados}) takes positive (negative) values for 
$z>0$ ($z<0$). By studying the signs of both sides of eq. 
(\ref{barapados}) one easily concludes that for $z_{\infty}>0$ one must
have $0\le z<z_{\infty}$, whereas for $z_{\infty}<0$ the coordinate
$z$ takes values on the range $z_{\infty}<z\le 0$ (see figure \ref{barfig5}). When 
$z_{\infty}= 0$, the analysis of eq. (\ref{barapados}) leads to the
conclusion that the solution is the hyperplane $z=0$.

It is also possible to determine the range of possible values that 
the coordinate  $\theta$ can take. Actually $\theta$ has generically an
extremal value, which we shall denote by $\theta_*$. The existence of 
$\theta_*$ can be proved by computing the derivative $d\theta/dr$ for
our solution. A short calculation  proves that this derivative
vanishes if the coordinate $z$ takes the value

\beq
z_{*}\,=\,{6-p\over 7-p}\,\,z_{\infty}\,\,.
\label{barapatrece}
\eeq
Alternatively, one could determine $z_{*}$ as the point at which the
denominator of the BPS condition (\ref{bartdos}) vanishes. By computing
the second derivative $d^2\theta/dr^2$ at $\theta=\theta_*$, one can
check that $\theta_*$ is a minimum (maximum) if $z_{\infty}<0$ 
($z_{\infty}>0$)\footnote{For $\nu=0$ and $z_{\infty}<0$, the value
(\ref{barapatrece}) is not reached (recall that $z\le z_{\infty}$ when
$\nu=0$). In this case the coordinate $\theta$ varies monotonically
along the embedding. }. It is also interesting to point out that for
$z=z_{*}$ the function ${\cal Z}$ of eq. (\ref{barcinueve}) changes its
sign. This is consistent with the extremal nature of $\theta_*$.


\medskip
\chapter{Flux Stabilization}
\medskip

\renewcommand{\theequation}{{\rm\thesection.\arabic{equation}}}

In a recent paper \cite{Bachas},  Bachas, Douglas and Schweigert have
shown how D-branes on a group manifold  are stabilized against shrinking
(see also ref. \cite{Pavel}). The concrete model studied in ref.
\cite{Bachas} was the motion of a D2-brane in the geometry of the $SU(2)$
group manifold. Topologically, $SU(2)$ is equivalent to a three-sphere $S^3$ 
and the D2-brane is embedded in this $S^3$ along a two-sphere $S^2$ which, in a 
system of spherical coordinates, is placed at constant latitude angle $\th$.
The D2-brane dynamics is determined by the Dirac-Born-Infeld action, in which a
worldvolume gauge field is switched on. An essential ingredient in this
analysis is the quantization condition of the worldvolume flux. 

By using the quantization condition one can easily find the form of the 
worldvolume gauge field strength in terms of a quantization integer, and 
the corresponding value of the energy of the D2-brane. The minimum of this 
energy determines the embedding of the brane in the group manifold, which
occurs at a finite set of latitude angles $\th$. It turns out that the
static configurations found by this method are stable under small
perturbations and exactly match those obtained by considering open strings 
on group  manifolds \cite{KS,KO}.

This chapter is organized as follows. In section \ref{secgrmn} 
we review shortly the dynamics of branes in group manifolds.
In section \ref{flbachas} we discuss the flux stabilization in a 
NS5-brane background, which was studied in \cite{Bachas}. Section 
\ref{flRR} is devoted to the analysis of flux stabilization of branes wrapped 
on spheres in the geometry of RR 
backgrounds and a quantization condition for the electric components 
of the worldvolume gauge field is proposed. This section is based 
on \cite{Flux1}. In sections \ref{flncbs} and \ref{flns5dpbs} the 
analysis is extended to backgrounds created by bound 
states of branes. Both sections are based on \cite{Flux2}. The 
generalization to M-theory of the quantization condition given in 
\cite{Bachas} is proposed in section \ref{flM2} and it is used to describe 
the stabilization of an M5-brane in the background created by an stack of 
M5-branes. In appendix 4.A we collect the functions which determine the 
location of the wrapped brane configurations in the transverse sphere. In 
appendix 4.B it is shown how to obtain wrapped D3-branes in terms of 
D-strings polarized \'a la Myers. Finally, in section \ref{flconcl}, 
we comment about the results.

\setcounter{equation}{0}
\section{Strings in group manifolds and D-branes} \label{secgrmn}
\medskip

Strings moving on a group manifold are described  by the WZW model, 
which is a solvable theory \cite{Gawedzki,all}. The action is given by
\be
S = \,\int_{\S} d^2z\,L^{sm}\,+\,\int_B\,L_{WZ}\,,
\label{flWZWm}
\ee
where $L^{sm}=Tr\,(g^{-1}\partial g g^{-1}\bar\partial g)$ 
is the sigma model part of the action and $L_{WZ}=\o^{(3)}$ 
is the Wess-Zumino term. Here, the field $g(z,\bar z)$ is a map from a closed 
orientable Riemann surface to the group $\bf G$, $k\in\ZZ_+$ is the level of 
the current algebra, $\S$ is the worldsheet manifold, $B$ is a 
three-dimensional manifold whose boundary is the worldsheet $\S$ and 
$\o^{(3)}=\frac{1}{3}Tr\left((g^{-1}\,dg)^3\right)$. 

We denote by $\bf g$ to the Lie algebra corresponding to $\bf G$, and we 
choose a basis of generators $\{T_a\}$ for $\bf g$. The invariant metric on 
$\bf g$ has components 
$G_{ab}\equiv\langle T_a,T_b\rangle$.  The exact conformal invariance of 
this model is based, as is well known, on its infinite-dimensional symmetry 
group ${\bf G}_L(z)\times{\bf G}_R(\bar z)$. The symmetry acts on a group 
element $g(z.\bar z)$ as:
\be
g\rightarrow h_L(z)\,g\,h_R(\bar z)\,.
\label{flels}
\ee
If the worldsheet has a boundary, there is a relation between the left-moving 
and right-moving modes, and the ${\bf G}_L\times{\bf G}_R$ symmetry is broken. 
However, one 
can still preserve the diagonal symmetry $G$, say the symmetry
\be
g\rightarrow hgh^{-1}\,,
\label{flelsb}
\ee
corresponding to $h_L=h_R^{-1}=h$ in (\ref{flels}). The presence of this 
symmetry constrains the boundary conditions that can be placed on $g$. 
Allowing $g$(boundary)$=f$ for some $f\in{\bf G}$ we must allow also 
$g$(boundary)$=hfh^{-1}$ for every $h\in{\bf G}$. This means that $g$ on 
the boundary takes values in the conjugacy class containing $f$. Then, each 
possible boundary condition corresponds to a
D-brane wrapped on a (twisted) conjugacy class of the group \cite{Ale,FFFS,Sonia}.

When the worldsheet $\S$ has boundaries, it cannot be the boundary of 
a three-dimensional manifold, since the boundary of a boundary is the empty 
set. To define 
the WZ term in (\ref{flWZWm}) for this case, one should fill the holes in the 
worldsheet by adding discs, and extend the mapping from the worldsheet into 
the group manifold to these discs. One further demands that the whole disc $D$ 
is mapped into a region (which will be also referred to as $D$) inside the conjugacy 
class in which the corresponding boundary lies. Then we take $B$ to be a  
three-manifold bounded by the union $\S\cup D$, which now has no boundaries.

The resulting action should preserve the symmetry (\ref{flels}) in the bulk 
of the worldsheet which becomes (\ref{flelsb}) on the boundary. If we take 
$C(f)$ to be the conjugacy class containing $D$ and the group element $f\in\bf G$
\be
C(f)\,\equiv\,\{\,hfh^{-1}\,|\ h\in {\bf G}\}\,,
\ee
then, to keep the symmetry (\ref{flelsb}) one has to modify the action by adding 
to it an integral over the disc $D$ of some two form $\o^{(2)}$, defined on 
$C(f)$ and the proper action is now of the form
\be
S = \,\int_{\S} d^2z\,L^{sm}\,+\,\int_B\,L_{WZ}\,-
\,\int_D\,\o^{(2)}\,,
\label{flWZWm2}
\ee
where $d\o^{(2)}=\o^{(3)}$ on the conjugacy class $C(f)$. 

The modified action (\ref{flWZWm2}) is invariant, by construction, under 
continuous deformations of $D$ inside $C(f)$. However, if one compares the 
value of the action for $D$ and $D^{\prime}$, two different choices of embedding 
the disc in $C(f)$ with the same boundary, $D^{\prime}$ may not be a continuos 
deformation 
of $D$ in $C(f)$. In that case the two ways to evaluate the action (\ref{flWZWm2}) 
do not agree. There is no natural choice between the two embeddings and if 
one chooses $D$ and $D^{\prime}$ such that their union covers the whole $C(f)$ the 
difference between the action $S_D$ (the value of (\ref{flWZWm2}) with embedding 
$D$) and $S_{D^{\prime}}$ with embedding $D^{\prime}$ is
\be
\D S\,=\,\int_{\hat B}\,\o^{(3)}\,-\,\int_{C(f)}\,\o^{(2)}\,,
\label{flds}
\ee
where $\hat B$ is a manifold whose boundary is $C(f)$. 
Nevertheless, the quantum theory is still well defined if 
\be
\D S=2\pi n\,,\ \ \ \ \ n\in\ZZ\,.
\ee

The affine symmetry ${\bf G}(z)\times{\bf G}(\bar z)$ guarantees the exact 
conformal invariance of the model and is characterised by the conserved 
currents 
\be
J(z)=-\partial g g^{-1}\,,\qquad\qquad \bar J(\bar z)=g^{-1}\bar\partial g\,,
\label{flbcurrents}
\ee
which are the natural dynamical variables. If we
introduce the mode expansions for the two currents
\be
J_a(z) = \sum_{n\in\ZZ}\,J_{an}\,z^{-n-1}\,,\qquad\qquad
\bar J_a(\bar z) = \sum_{n\in\ZZ}\,\bar J_{an}\,\bar z^{-n-1}\,,
\ee
we can write down the affine Lie algebra that they satisfy
\br
[J_{an},J_{bm}] &=&{f_{ab}}^c J_{c\,n+m}\,+\,nk\,\d_{n+m,0}\,G_{ab}\,,\rc 
\left[{\bar J}_{an},{\bar J}_{bm}\right]&=&{f_{ab}}^c \bar J_{c\,n+m}\,+
\,nk\,\d_{n+m,0}G_{ab}\,,\rc 
\left[J_{an},\bar J_{bm}\right] &=&0\,\,.
\er
The relative sign in the definition (\ref{flbcurrents}) of the two chiral 
currents ensures that the two currents satisfy the same chiral algebra.

D-branes can be defined in two ways: they can be described either in terms
of boundary conditions of open strings or as boundary states in the 
closed string sector. Boundary conditions were initially derived from
the vanishing of the boundary term in the action of the free string
(or, more generally, in the case of a flat background).  A Neumann
boundary condition imposed in a certain direction ensures that there
is no flow of momentum through the boundary of the string worldsheet,
\ie ,
\be
\partial_n X^{||}|_{\partial\S} = 0\,.
\ee
Here, the subscript $n$ indicates the direction normal to the boundary
of the worldsheet $\partial\S$, whereas the superscript $||$ denotes the
direction parallel to the D-brane worldvolume. On the contrary, a
Dirichlet boundary condition in a given direction constrains the
boundary of the string worldsheet to lie within the hyperplane normal
to that direction.  This implies the following condition
\be
\partial_t X^{\perp}|_{\partial\S} = 0\,,
\ee
where the subscript $t$ refers to the direction tangent to the
boundary of the string worldsheet and $\perp$ denotes the component of 
the background field normal to the worldvolume of the D-brane.

Let us first consider the case of strings in a flat string background, as 
a previous step to study strings propagating in group manifolds. The bosonic 
fields $X^{\m}$ 
decompose into a left and a right-moving part
\be
X^{\m}(\s,\t) = X^{\m}_L(\s_+)\,+\,X^{\m}_R(\s_-)\,, 
\ee
where $\s_+=\t+\s$ and $\s_-=\t-\s$. The mode expansion is
\be
\partial_+ X^{\m}_L = \sum_{n\in\ZZ}\a^{\m}_n\,e^{-in\,\s_+}\,,\qquad 
\partial_- X^{\m}_R = \sum_{n\in\ZZ} \tilde\a^{\m}_n\,e^{-in\,\s_-}\,, 
\ee
with the standard notation $\partial_{\pm} = \partial_{\t}\pm\partial_{\s}$.  The
canonical commutation relations for the modes read
\br
\left[\a^{\m}_n,\a^{\n}_m\right]&=&n\,\d_{m+n,0}\,\eta^{\m\n}\,,\rc
\left[{\tilde\a}^{\m}_n,{\tilde\a}^{\n}_m\right]&=&n\,\d_{m+n,0}\,\eta^{\m\n}\,,\rc 
\left[\a^{\m}_n,{\tilde\a}^{\n}_m\right] &=&0\,.
\er

In the closed string picture the boundary of the worldsheet is taken to be 
at $\tau =0$. A Neumann boundary condition in a given direction say, $X^{\m}$,
is therefore given by the condition
\be
\partial_{\t} X^{\m}|_{\t=0} = 0\,.
\ee
On the contrary, a Dirichlet boundary condition in the same direction reads 
\be
\partial_{\s} X^{\m}|_{\t=0} = 0\,.
\ee
One can easily work out these conditions in terms of the modes and
obtain
\br
\a^{\m}_n + {\tilde\a}^{\m}_{-n}&=&0\,,\hspace{1cm}{\rm (N)}\rc
\a^{\m}_n - {\tilde\a}^{\m}_{-n}&=&0\,,\hspace{1cm}{\rm (D)}\,.
\er
The above conditions on the modes are not to be taken as
operatorial relations on the quantum fields since they do not 
yield automorphisms of the oscillator algebra. Rather, they are only satisfied
on certain boundary states $|B\rangle$, which are characterised
by the property that they satisfy a certain boundary condition,
namely,
\br
(\a^{\m}_n + \tilde\a^{\m}_{-n})|B\rangle &=&0\,,\hspace{1cm}({\rm N})\rc
(\a^{\m}_n - \tilde\a^{\m}_{-n})|B\rangle &=&0\,,\hspace{1cm}({\rm D})
\er
in a given direction $X^{\m}$. A boundary state $|B\rangle$ is specified by a 
complete set of boundary conditions, for all the directions of the target 
manifold. The conditions above can be recast in the form
\br
\left(\partial X^{\m}dz\,-\,\bar\partial X^{\m}\,d\bar z\right)\,|B\rangle\,=\,0\,,
\hspace{1cm}({\rm N})\,,\rc
\left(\partial X^{\m}dz\,+\,\bar\partial X^{\m}d\bar z\right)\,|B\rangle\,=\,0\,\,,
\hspace{1cm}({\rm D})\,.
\er
The most important requirement that a boundary state must satisfy is
conformal invariance.  In the case of a bosonic theory this means that
the holomorphic and antiholomorphic sectors satisfy the following
condition at the boundary
\be
\left(T (z) - \bar T (\bar z)\right)|B\rangle = 0\,,
\ee
where, in this case, $T = \partial X\cdot\partial X$ and 
$T = \bar\partial X\cdot\bar\partial X$. Any boundary state 
characterised by an arbitrary combination of Neumann and Dirichlet conditions 
satisfies the above condition and therefore preserves conformal invariance.

In a group manifold the currents $-J_a(z)$ and $\bar J_a(\bar z)$ are 
the nonabelian analogues of the currents $\partial X(z)$ and 
$\bar\partial X(\bar z)$ in flat space and one can immediately
generalise the Neumann and Dirichlet conditions for free fields to 
the following gluing conditions \cite{Sonia}:
\br
(J_{a}(z)\,dz\,+\,\bar J_{a}(\bar z)\,d\bar z)|B\rangle &=0\,,\ \ \ ({\rm N})\,,\rc
(J_{a}(z)\,dz\,-\,\bar J_{a}(\bar z)\,d\bar z)|B\rangle &=0\,,\ \ \ ({\rm D})\,.
\er
In terms of the modes of the currents we get:
\br
(J_{an} - \bar J_{a~-n})|B\rangle &=0\,,\ \ \ ({\rm N})\,,\rc
(J_{an} + \bar J_{a~-n})|B\rangle &=0\,,\ \ \ ({\rm D})\,.
\er
Since left multiplication is generated by the currents $J^a$ while right 
multiplication is generated by 
$\bar J^a$, the invariance of the boundary condition $g\in C(f)$ under 
$g\rightarrow hgh^{-1}$ implies that the currents satisfy
\be
J_a=\bar J_a\,,
\ee
which confirms their interpretation as Dirichlet boundary conditions.

\medskip
\subsubsection{The SU(2) group}
\medskip

The group $SU(2)$ has the topology of a sphere $S^3$. A general group 
element $U$ can be parameterized in the form $U=e^{(i\vec\psi\cdot\vec\s)}$, 
where $\vec\psi\equiv(\psi_1,\psi_2,\psi_3)$ forms a vector of length $\th$ 
pointing in the direction $(\th^1,\th^2)$ and $\s_i$ are the Pauli matrices, 
which realize the algebra of $su(2)$ (see \cite{Sonia2,KKZ}). The sigma model 
metric and the Wess-Zumino three-form $\omega^{(3)}$ are given by:
\br
ds^2&=&\frac{k}{2\pi}\,\left[d\th^2\,+\,(\sin\th)^2\,
\left((d\th^1)^2\,+\,(\sin\th^1)^2\,(d\th^2)^2\right)\right]\,,\rc\rc
\omega^{(3)}&=&\frac{k}{\pi}\,(\sin\th)^2\,\sin\th^1\,d\th\wedge d\th^1
\wedge d\th^2\,,
\er
where $k$ is the level. The line element is proportional to the line 
element of the unit sphere $S^3$ parametrized by the angles $\th, \th^1, \th^2$
and $\epsilon_{(3)}$ is its volume element. 

In the case of ${\bf G}=SU(2)$, conjugacy classes are parameterized by a 
single angle $\th$, $0\le\th\le\pi$, corresponding to the choice 
$f=e^{i\th\s_3}$. The conjugacy classes are thus $S^2$ spheres at constant 
angle $\th$, parametrized by the angles $\th^1$ and $\th^2$. Therefore, a 
boundary condition which preserves (\ref{flelsb}) is 
\be
g(boundary)\in C_{\th}\,.
\ee
The form taken by  $\o^{(2)}$ on the conjugacy class $C_{\th}$ 
is 
\be
\o^{(2)}\,=\,\frac{k}{4\pi}\,\sin 2\th\,\epsilon_{(2)}\,,
\ee 
where $\epsilon_{(2)}$ is the volume element of the unit $S^2$.
This gives for the change in the action 
for two topologically different embeddings in (\ref{flds})
\be
\D S\,=\,2k\th,
\ee
and the quantum theory is well defined if the conjugacy classes on which a 
boundary state can live are quantized, and thus, the corresponding $\th$ 
must satisfy
\be
\th=\pi\frac{n}{k}\,,
\label{angles}
\ee
where $n$ in an integer satisfying $0\le n\le k$. The conjugacy classes  
parametrized by $\th$ correspond to D2-branes wrapped on $S^2$ spheres 
at constant polar angle. There are two point-like conjugacy classes $e$ 
and $-e$ corresponding to 
$\th=0$ and $\th=\pi$ which correspond to D-particles. Thus, there are $k+1$ 
different boundary conditions preserving the symmetry $g\rightarrow ghg^{-1}$ 
which one can impose to the WZW sigma model on the $SU(2)$ group manifold at 
level $k$.

\setcounter{equation}{0}
\section{Flux stabilization in a NS5-brane background} \label{flbachas}
\medskip

Let us consider a D2-brane in the background created by a stack of N parallel 
NS5-branes. In the near-horizon region the metric and dilaton are given by
\br
ds^2&=&-dt^2\,+\,dx_{||}^2\,+\,\frac{N\,\a^{\prime}}{r^2}\,dr^2\,+\,
N\,\a^{\prime}\,d\O_3^2\,,\rc\rc
e^{-\p}&=&\frac{r}{\sqrt{N\,\a^{\prime}}}\,.
\label{flbdos}
\er
The solution is charged under the NSNS 3-form $H$, given by:
\be
H  \equiv {dB} = 2\,N\,\a^{\prime}\,(\sin\th)^2\,\sin\th^1\,
d\th\wedge d\th^1\wedge d\th^2\,.
\label{flbtres}
\ee
The $S^3$ sphere transverse to the NS5 gives the SU(2) manifold if we identify 
the number of branes $N$ with the level $k$ of the associated current algebra.
We have seen that D-branes stay on integer conjugacy classes, which correspond 
to $S^2$ spheres at constant polar angle $\th$. Therefore we will search 
for brane configurations in which the D2-brane is embedded in the $S^3$ of 
the background at constant angle $\th$, which corresponds to the D2 wrapped 
in the $S^2$ sphere spanned by $(\th^1, \th^2)$.
We can choose a gauge in which the NS 2-form B is proportional to the volume 
form of the 2-sphere,
\be
B\,=\,-N\a^{\prime} C_5(\th)\,\epsilon_{(2)}\,,
\label{flbB}
\ee
where the function $C_5(\th)$ is defined as $C_5(\th)\equiv\sin\th\,
\cos\th\,-\,\th$ and its derivative is given by
\be
\frac{d}{d\th}C_5(\th)=-2\,(\sin\th)^2
\label{flbcinco}
\ee

The action for the D2 is the DBI action but there is no coupling to RR fields, 
and we are left with the Born-Infled part of the action
\be
{\cal L}_{BI}=-T_2\,\int\,d^3\s\,e^{-\p}\,\sqrt{-det(g+{\cal F})}\,,
\label{flbseis}
\ee
where $T_2$ is the tension of the brane and $\cal F$ is the modified field 
strength ${\cal F}=F-B$.

In order to find stable $S^2$-wrapped configurations of the D2-brane
probe, we need to switch on a non-vanishing worldvolume field which
could prevent the collapse to one of the poles of the $S^2$. The value 
of this worldvolume field is determined by some quantization condition 
which can be obtained by coupling the D2-brane to a F1-string. 

\begin{figure}
\centerline{\hskip -.8in \epsffile{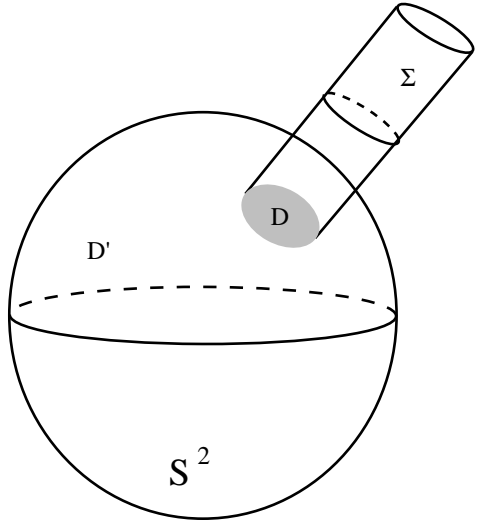}}
\caption{An F1-string with worldvolume $\S$ having its boundary on
the worldvolume of a D-brane. If $\S$ is attached to a submanifold
of the brane worldvolume with the topology of $S^2$, there are two
possible disks $D$ and $D'$ on the $S^2$ whose boundary is $\partial
\S$.}
\label{flfig3}
\end{figure}

Let us consider an open F1-string with worldvolume given by a
two-manifold $\S$ whose boundary $\partial\S$ 
lies on the worldvolume of an
D2-brane. Clearly, $\partial\S$ is also the boundary
of some disk $D$ on the worldvolume of the D2-brane. Let $\hat\S$ be a
three-manifold whose boundaries are $\S$ and $D$, \ie\ 
$\partial\hat\S\,=\,\S\,+\,D$. The coupling of the F1
to the supergravity background and to the D2-brane is described by an
action of the form

\be
S_{int}\,[\,\hat\S, D\,]\,=\,
T_f\,\left[\int_{\S}\,B\,+\,\int_{\partial D}\,A\,\right]\,\,,
\label{flbotres}
\ee
where $T_f$ is the tension of the fundamental string, given by:
\be
T_f\,=\,{1\over 2\pi\,\a^{\prime}}\,\,.
\label{flbocuatro}
\ee
In a topologically trivial situation, if we represent $H$ as 
$dB$ and $F=dA$,  the above action is equivalent to
\be
S_{int}\,=\,T_f\,\left[\int_{\hat\S}\,H\,+
\,\int_{D}\,{\cal F}\right]\,\,,
\label{flbocinco}
\ee
where we have used that
\be
\int_{\S}\,B\,=\,\int_{\S\cup D}\,B\,-\,\int_{D}\,B\,=
\,\int_{\hat\S}\,H\,-\,\int_{D}\,B\,.
\ee

We shall regard eq. (\ref{flbocinco}) as the definition of the interaction
term of the F1-string. Notice that, in general, $\hat\S$ and
$D$ are not uniquely defined. To illustrate this point let us consider
the case in which we attach the F1-string to a D2-brane worldvolume which
has some  submanifold with the topology of $S^2$. This is precisely the
situation in which we are interested in. As illustrated in figure \ref{flfig3}, 
we have two possible choices for the disk in eq. (\ref{flbotres}) namely,
we can choose the ``internal" disk $D$ or the ``external" disk
$D'$. Changing $D\rightarrow D'$, the manifold $\hat\S$ changes to 
$\hat\S'$, with $\partial \hat\S'\,=\,\S\,+\,D'$ and, in
general, $S_{int}$ also changes. However, in the quantum-mechanical
theory, the action appears in a complex exponential of the form 
$\exp[\,i\,S_{int}\,]$. Thus, we should require that
\be
e^{iS_{int}\,[\,\hat\S, D\,]}\,=\,
e^{iS_{int}\,[\,\hat\S', D'\,]}\,\,.
\label{flboseis}
\ee
The condition (\ref{flboseis}) is clearly equivalent to
\be
S_{int}\,[\,\hat\S', D'\,]\,-\,S_{int}\,[\,\hat\S, D\,]\,=\,
2\pi n\,\,,
\label{flbosiete}
\ee
with $n\in\ZZ$. The left-hand side of eq. (\ref{flbosiete}) can be
straightforwardly computed from eq. (\ref{flbotres}). Actually, 
if $\hat{\cal B}$ is the 3-ball bounded by $D'\cup(-D)\,=\,S^2$, 
then $\hat\S^{\prime}-\hat\S=\hat{\cal B}$ and one has
\be
S_{int}\,[\,\hat\S^{\prime}, D'\,]\,-\,S_{int}\,[\,\hat\S, D\,]\,=\,
T_f\,\left[\int_{\hat\S^{\prime}-\hat\S}\,H\,+\,
\int_{D'-D}\,{\cal F}\right]\,=\,
T_f\,\left[\int_{\hat{\cal B}}\,H\,+\,
\int_{\partial\hat{\cal B}}\,{\cal F}\right]\,\,.
\label{flboocho}
\ee
Using this result in eq. (\ref{flbosiete}), we get the condition
\be
\int_{\hat{\cal B}}\,H\,+\,
\int_{\partial\hat{\cal B}}\,{\cal F}\,=\,{2\pi n\over T_f}\,,
\,\,\,\,\,\,\,\,\,\,\,\,\,\,\,\,\,
n\in\ZZ\,\,.
\label{flbonueve}
\ee
If $H$ can be represented as $dB$ on $\hat{\cal B}$, the
first integral on the left-hand side of eq. (\ref{flbonueve})
can be written as an integral of $B$ over 
$\partial\hat{\cal B}\,=\,S^2$. Our parametrization of $B$ 
(eq. (\ref{flbB})) is certainly non-singular if we  are
outside of the poles of the $S^2$. If this is the case, we get the
quantization condition
\be
\int_{S^2}\,F\,=\,{2\pi n\over T_f}\,\,,
\,\,\,\,\,\,\,\,\,\,\,\,\,\,\,\,\,
n\in\ZZ\,\,.
\label{flbqc}
\ee

One can find a solution of the quantization condition (\ref{flbqc}) if 
one takes $F$ as
\be
F\,=\,\frac{1}{2T_f}\,n\,\epsilon_{(2)}\,=\,\pi\a^{\prime}\,n\,\epsilon_{(2)}\,\,,
\label{flbsiete}
\ee
where $\epsilon_{(2)}$ is the volume form of the $S^2$ sphere, 
$\epsilon_{(2)}\,=\,(\sin\th^1)^2\,d\th^1\wedge d\th^2$, and then
\be
\int_{S^2}\,\epsilon^{(2)}\,=\,4\pi\,\,.
\label{flbocho}
\ee
This gives the following expression for ${\cal F}$:
\be
{\cal F}_{12}\,=\,N\a^{\prime}\left[\sin\th\cos\th-\th+\frac{\pi n}{N}\right]\,
(\sin\th)^2\,=\,N\a^{\prime}{\cal C}_{5,n}(\th)\,(\sin\th)^2\,\,,
\label{flbnueve}
\ee
with the function ${\cal C}_{5,n}(\th)$ defined by
\be
{\cal C}_{5,n}(\th)\equiv\sin\th\cos\th-\th+\frac{\pi n}{N}\,.
\ee

With the value of ${\cal F}_{12}$ given in (\ref{flbsiete}) one gets the action
\be
S\,=\,-\,T_2\,N\,\a^{\prime}\,\int\,dtd\th^1d\th^2\,e^{-\p}\,(\sin\th)^2\,
\sqrt{(\sin\th)^4\,+\,({\cal C}_{5,n}(\th))^2}\,.
\label{flbdiez}
\ee
The Hamiltonian is obtained by performing a Legendre transformation, and 
if we integrate over the angles $\th^1$ and $\th^2$ the result is
\be
H\,=\,T_2\,N\,\a^{\prime}\,e^{-\p}\,4\pi\,
\sqrt{(\sin\th)^4\,+\,({\cal C}_{5,n}(\th))^2}
\label{flbonce}
\ee
The energy has a unique minimum for the angle $\bar\th$ which solves the 
equation
\be
\L_{5,n}(\bar\th)\,=\,0\,,
\label{flbtrece}
\ee
where $\L_{5,n}(\th)$ is the function defined by
\be
\L_{5,n}(\th)\,\equiv\,\sin\th\cos\th\,-\,{\cal C}_{5,n}(\th)\,=
\,\th\,-\,\frac{\pi n}{N}\,.
\label{flbdoce}
\ee
The equation (\ref{flbtrece}) is solved for the angle
\be
\bar\th=\frac{\pi n}{N}\,.
\label{flbcatorce}
\ee
These are the same angles as the ones given in in (\ref{angles}) if we identify  
the number of branes in the stack, $N$, with the level of the algebra, $k$.
The limiting values $n=0$ and $n=N$ correspond to the D2-brane collapsed at the poles 
of the $S^2$. The corresponding value of the energy is given by
\be
E_n(\th)\,=\,T_2\,N\,\a^{\prime}\,e^{-\p}\,4\pi\,
\sin(\frac{\pi n}{N})\,.
\label{flbquince}
\ee
This result agrees with the one obtained by the CFT approach. We have seen 
that the underlying CFT imposes quantization conditions on the allowed 
conjugacy classes of the group manifold, which can be interpreted  
geometrically in terms of the embedding of the D-brane 
worldvolume in the group manifold. To each conjugacy class one associates a 
Cardy boundary state \cite{Cardy} of the WZW model. The mass of the D-brane
configuration can be obtained, in this approach, by computing \`a la
Polchinski \cite{Polchi} the matrix element between Cardy states of the
string theory cylinder diagram and the result can be compared with  the one
obtained in a gravitational field theory. Apart from a finite shift in the 
level of the current algebra, the mass obtained in this way is
exactly the same as the one given in (\ref{flbquince}), computed with the 
Born-Infeld action and the quantization condition (\ref{flbqc}). For other 
aspects of the open string approach and of the flux quantization condition 
(\ref{flbqc}) see ref. \cite{all}.

The agreement between the Born-Infeld and CFT approaches for the system
of ref. \cite{Bachas} is quite remarkable. For this reason the
generalization of this result to other backgrounds and brane probes is
very interesting. The natural generalization to consider is a
Ramond-Ramond (RR) background. This case was studied in ref.
\cite{PR}, where it was shown that the brane probe must be partially
wrapped on some angular directions and extended along the radial
coordinate.

\setcounter{equation}{0}
\section{Wrapped branes in Ramond-Ramond backgrounds} \label{flRR}
\medskip

Following the analysis of ref. \cite{PR}, in this section we  shall study 
the motion of a D(8-p)-brane in the background of a stack
of parallel Dp-branes. The external region of the Dp-brane metric has
$SO(9-p)$ rotational symmetry, which is manifest when a system of
spherical coordinates is chosen. In this system of coordinates a
transverse $S^{8-p}$ sphere is naturally defined and the constant latitude
condition on the $S^{8-p}$ determines a $S^{7-p}$ sphere. We shall embed
the D(8-p)-brane in this background in such a way that it is wrapped on
this $\,S^{7-p}\subset S^{8-p}$ constant latitude sphere and extended
along the radial direction. Therefore, as in section \ref{flbachas},
the brane configuration is characterized by an angle $\th$, which
parametrizes  the latitude of the  $S^{7-p}$. 

In order to analyze this Dp-D(8-p) system by means of the Born-Infeld
action, we shall establish first some quantization condition which,
contrary to (\ref{flbqc}),  will now involve the electric components of the
worldvolume gauge field. By using this quantization rule we shall find a
finite set of stable brane configurations characterized by some angles
$\th$ which generalize the ones found in section \ref{flbachas}. The
energy of these configurations will be also computed and, from this
result, we shall conclude that semiclassically our D(8-p)-brane
configurations can be regarded as a bound state of fundamental strings.
On the other hand, we will find a first order BPS differential equation
whose fulfillment implies the saturation of an energy bound and whose
constant $\th$ solutions are precisely our wrapped configurations.
This BPS equation is the one \cite{Imamura}  satisfied by the baryon
vertex \cite{Wittenbaryon}, which will allow us to interpret our
configurations as a kind of short distance limit (in the radial
direction) of the baryonic branes \cite{Baryon,CGS1,CGS2,Craps}.

The ten-dimensional metric corresponding to a stack of $N$ coincident
extremal Dp-branes was given in section \ref{sugrasol} and in the 
near-horizon region is
\be
ds^2\,=\,\Bigl[\,{r\over R}\,\Bigr]^{{7-p\over 2}}\,\,
(\,-dt^2\,+\,dx_{\parallel}^2\,)\,+\,
\Bigl[\,{R\over r}\,\Bigr]^{{7-p\over 2}}\,\,
(\,dr^2\,+\,r^2\,d\O_{8-p}^2\,)\,\,,
\label{fl1dos}                                                                                                                                                                                                                                                                                                          
\ee
where $x_{\parallel}$ represent $p$ cartesian
coordinates along the branes, $r$ is a radial coordinate
parametrizing the distance to the branes and $d\O_{8-p}^2$ is the
line element of an unit $8-p$ sphere. We have written the metric in
the string frame. The radius $R$ is given in eq. (\ref{strdbrad}) and 
the values of the dilaton field $\phi(r)$ and of the Ramond-Ramond (RR) 
(8-p)-form field strength $F^{(8-p)}$ are given in (\ref{strdpbe}) and 
respectively (\ref{strdbF8pf}).

Let $\th^1$, $\th^2$, $\dots$, $\th^{8-p}$ be coordinates
which parametrize the $S^{8-p}$ transverse sphere. 
We shall assume that the $\th$'s are spherical angles on $S^{8-p}$
and that $\th\equiv\th^{8-p}$ is the polar angle 
($0\le\th\le\pi$). Therefore, the $S^{8-p}$ line element 
$d\O_{8-p}^2$ can be decomposed as:
\be
d\O_{8-p}^2\,=\,d\th^2\,+\,(\,{\rm sin}\,\th)^{2}\,\,
d\O_{7-p}^2\,\,.
\label{fl1cinco}
\ee
In these coordinates it is not difficult to find a potential for the
RR gauge field. Indeed, let us  define the function $C_p(\th)$ 
as the solution of the differential
equation
\be
{d\over d\th}\, C_p(\th)\,=\,-(7-p)\,({\rm sin}\,\th)^{7-p}\,\,,
\label{fl1seis}
\ee
with the initial condition
\be
C_p(0)\,=\,0\,\,.
\label{fl1siete}
\ee
It is clear that one can find by elementary integration a unique
solution to the problem of eqs. (\ref{fl1seis}) and (\ref{fl1siete}). Thus 
$C_p(\th)$ can be considered as a known function of the polar angle
$\th$. In terms of $C_p(\th)$, the RR potential $C^{(7-p)}$
can be represented as
\be
C^{(7-p)}\,=\,-R^{(7-p)}\,C_p(\th)\,\,\epsilon_{(7-p)}\,\,.
\label{fl1ocho}
\ee
By using eq. (\ref{fl1seis}) it can be easily verified that
\footnote{For simplicity, we choose the orientation
of the transverse $S^{8-p}$  sphere such that 
$\epsilon_{(8-p)}=(\sin\th)^{7-p}\,d\th\wedge\epsilon_{(7-p)}$.} 
\be
F^{(8-p)}\,=\,d\,C^{(7-p)}\,\,.
\label{fl1nueve}
\ee
Let us now consider a D(8-p)-brane embedded along the transverse
directions of the stack of  Dp-branes. The 
action of such a brane probe is  the sum of a
Dirac-Born-Infeld and a Wess-Zumino term
\be
S\,=\,-T_{8-p}\,\int d^{\,9-p}\s\,e^{-\tilde\p}\,
\sqrt{-{\rm det}\,(\,g\,+\,F\,)}\,+\,
T_{8-p}\,\int \,\, F\wedge\,C^{(7-p)}\,\,,
\label{fl1diez}
\ee
where $g$ is the induced metric on the worldvolume of the D(8-p)-brane
and $F$ is a worldvolume abelian gauge field strength. The coefficient 
$T_{8-p}$ in eq. (\ref{fl1diez}) is the tension of the D(8-p)-brane,
given by
\be
T_{8-p}\,=\,(2\pi)^{p-8}\,(\,\a\,'\,)^{{p-9\over 2}}\,
(\,g_s\,)^{-1}\,\,.
\label{fl1once}
\ee
The worldvolume
coordinates $\s^{\a}$ ($\a\,=\,0\,,\,\dots\,,\,8-p\,)$ will
be taken as
\be
\s^{\a}\,=\,(\,t\,,\,r\,,\,
\th^1\,,\,\dots\,,\,\th^{7-p}\,\,)\,\,.
\label{fl1doce}
\ee
\begin{figure}
\centerline{\hskip -.8in \epsffile{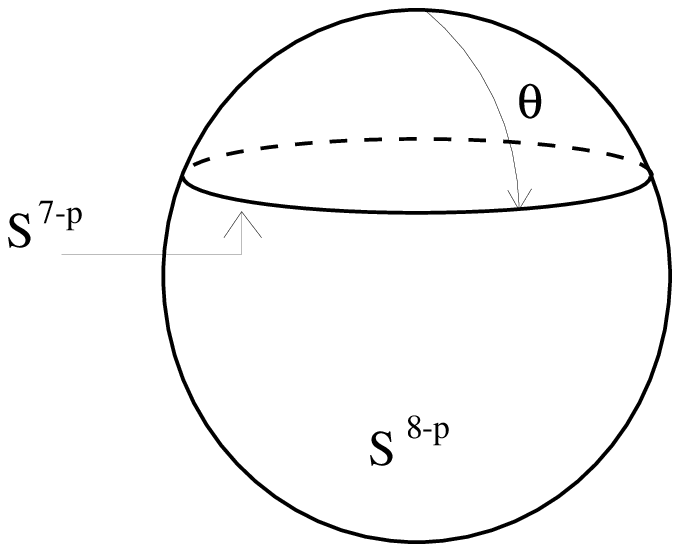}}
\caption{The points of the  $S^{8-p}$ sphere with the same polar angle
 $\th$ define a  $S^{7-p}$ sphere. The angle $\th$ represents the
latitude on $S^{8-p}$,  measured from one of its poles.
}
\label{flfig1}
\end{figure}

With this choice the embedding of the brane probe is described by a
function $\th=\th(\s^{\a})$. Notice that the
hypersurface $\th\,=\,{\rm constant}$ defines a $S^{7-p}$ sphere on
the transverse $S^{8-p}$ (see figure \ref{flfig1}). These configurations with
constant polar angle represent a D(8-p)-brane wrapped on a $S^{7-p}$
sphere and extended along the radial direction. These are the kind of
configurations we want to study in this paper. Actually, we will
consider first a more general situation in which the polar angle
depends only on the radial coordinate, \ie\ when
$\th= \th(r)$. It is a rather simple exercise to compute the
induced metric $g$ in this case. Moreover, by inspecting the form of
the RR potential $C^{(7-p)}$ in eq. (\ref{fl1ocho}) and the Wess-Zumino
term in the action, one easily concludes that this term acts as a source
for the worldvolume electric field $F_{0,r}$ and, thus, it is natural
to assume that $F_{0,r}$ is different from zero. If we  take this
component of $F$ as the only non-vanishing one, the action can be
written as
\be
S\,=\,
\int_{S^{7-p}}\,d^{7-p}\th\,\,
\int\,drdt\,\,{\cal L}(\th, F)\,\,,
\label{fl1trece}
\ee
where the lagrangian density ${\cal L}(\th, F)$ is given by
\be
{\cal L}(\th, F)\,=\,-\,T_{8-p}\,R^{7-p}\,\sqrt{\hat g}\,\,\Bigl[\,
({\rm sin}\,\th)^{7-p}
\,\,\sqrt{1\,+\,r^2\,\th\,'^{\,2}\,-\,F_{0,r}^2}\,+\,
F_{0,r}\,C_p(\th)\,\Bigr]\,\,.
\label{fl1catorce}
\ee
In eq. (\ref{fl1catorce}) $\hat g$ is the determinant of the metric of the
$S^{7-p}$ and $\th\,'$ denotes $d\th/d r$. 

\medskip
\subsection{Quantization condition}
\medskip

The equation  of motion of the gauge field, derived from the
lagrangian density of eq. (\ref{fl1catorce}), implies that
\be
{\partial {\cal L}\over \partial F_{0,r}}\,=\,
{\rm constant}\,\,.
\label{fl1quince}
\ee
In order to determine the value of the constant on the right-hand side
of eq. (\ref{fl1quince}) let us follow the procedure of ref. \cite{PR}
and couple the D-brane to a Neveu-Schwarz (NS) Kalb-Ramond field $B$. As
is well-known, this coupling can be performed by substituting $F$ by
$F-B$ in ${\cal L}$, \ie\ by doing 
${\cal L}(\th, F)\,\rightarrow {\cal L}(\th, F-B)$ in eq.
(\ref{fl1catorce}). At  first order in $B$, this substitution generates a
coupling of the D-brane to the NS field $B$ of the form
\be
\int_{S^{7-p}}\,\,d^{7-p}\th\,\,\,\int\,drdt\,\,
{\partial \,{\cal L}\over\partial F_{0,r}}\,\,B_{0,r}\,\,,
\label{fl1dseis}
\ee
where we have assumed that only the $B_{0,r}$ component of the $B$
field is turned on. 

We shall regard eq. (\ref{fl1dseis}) as the interaction energy of a
fundamental string source in the presence of the D-brane. This source 
is extended along the radial direction and, thus, it is quite natural
to require that the coefficient of the $B$ field, integrated over 
$S^{7-p}$, be an integer
multiple of the fundamental string tension, namely,
\be
\int_{S^{7-p}}\,\,d^{7-p}\th\,\,\,
{\partial \,{\cal L}\over\partial F_{0,r}}\,=\, n\,T_{f}\,\,,
\label{fl1dsiete}
\ee
with $n\in\ZZ$.  Eq. (\ref{fl1dsiete}) is the quantization condition we
were looking for in these RR backgrounds and will play in our analysis
a role similar to the one played in section \ref{flbachas} by the flux
quantization condition (eq. (\ref{flbqc})). Notice that eq.  (\ref{fl1dsiete})
constraints the electric components of $F$, whereas eq. (\ref{flbqc})
involves the magnetic worldvolume field\footnote{It is interesting to
point out that the left-hand side of eq. (\ref{fl1dsiete}) can be written
in terms of the integral over the $S^{7-p}$ sphere of the worldvolume 
Hodge dual of $\partial \,{\cal L}/\partial 
F_{\a, \b}$.}. Thus, our quantization rule
is a kind of electric-magnetic dual of the one used in section \ref{flbachas}. 
This has a nice interpretation in the case in which $p$
is odd, which corresponds to the type IIB theory. Indeed, it is known in
this case  that the electric-magnetic
worldvolume duality corresponds to the S-duality of the
background \cite{EMduality}. In particular, when
$p=5$, the D5 background can be converted, by means of an S-duality
transformation,  into a NS5 one, which is precisely the type of
geometry considered in section \ref{flbachas}. 

By using the explicit form of the lagrangian density (eq.
(\ref{strdbrad})), the left-hand side of our quantization condition can
be easily calculated:
\be
\int_{S^{7-p}}\,\,d^{7-p}\th\,\,\,
{\partial \,{\cal L}\over\partial F_{0,r}}\,=\,
T_{8-p}\,\O_{7-p}\,R^{7-p}\,\Biggl[\,
{F_{0,r}\,\,({\rm sin}\,\th)^{7-p}\,\over 
\sqrt{1\,+\,r^2\,\th\,'^{\,2}\, -\,F_{0,r}^2}}\,\,-\,\,
C_p(\th)\,\Biggr]\,\,,
\label{fl1docho}
\ee
where $\O_{7-p}$ is the volume of
the unit $(7-p)$-sphere, given by:
\be
\O_{7-p}\,=\,{2\pi^{{8-p\over 2}}\over 
\G\Bigl(\,{8-p\over 2}\Bigr)}\,\,.
\label{fl1dnueve}
\ee

By using eqs. (\ref{fl1docho}) and (\ref{fl1dsiete}) one can obtain 
$F_{0,r}$ as a function of $\th(r)$ and the integer $n$. Let us
show how this can be done. First of all, by using eqs. 
(\ref{fl1once}), (\ref{fl1dnueve}) and (\ref{strdbrad}) it is straightforward to
compute the global coefficient appearing on the right-hand side of eq. 
(\ref{fl1docho}), namely,
\be
T_{8-p}\,\O_{7-p}\,R^{7-p}\,=\,
{NT_f\over 2\sqrt{\pi}}\,
{\G\Bigl(\,{7-p\over 2}\Bigr)\over
\G\Bigl(\,{8-p\over 2}\Bigr)}\,\,. 
\label{fl1veinte}
\ee
Secondly, 
let us define the function ${\cal C}_{p,n}(\th)$ as:
\be
{\cal C}_{p,n}(\th)\,=\,C_p(\th)\,+\,2\,\sqrt{\pi}\,\,
{\G\Bigl(\,{8-p\over 2}\Bigr)\over
\G\Bigl(\,{7-p\over 2}\Bigr)}\,\,
{n\over N}\,\,.
\label{fl1vuno}
\ee
Notice that ${\cal C}_{p,n}(\th)$ satisfies the same differential
equation as  $C_p(\th)$ (eq. (\ref{fl1seis})) with different initial
condition. Moreover,  by inspecting eqs. (\ref{fl1dsiete}), (\ref{fl1docho})
and  (\ref{fl1veinte}), one easily concludes that $F_{0,r}$ can be put in
terms of ${\cal C}_{p,n}(\th)$. The corresponding expression is
\be
F_{0,r}\,=\,\sqrt{\,
{1\,+\,r^2\,\th\,'^{\,2}\over 
{\cal C}_{p,n}(\th)^2\,+\,
({\rm sin}\,\th)^{2(7-p)}}}\,\,{\cal C}_{p,n}(\th)\,\,.
\label{fl1vdos}
\ee
Let us now evaluate the energy of the system. By performing a Legendre
transformation, we can represent the hamiltonian $H$ of the
D(8-p)-brane as
\be
H\,=\,\int_{S^{7-p}}\,\,d^{7-p}\th\,\,\,
\int dr\,\Big[\, F_{0,r}\,{\partial \,{\cal L}\over\partial
F_{0,r}}\,-\, {\cal L}\,\Big]\,\,.
\label{fl1vtres}
\ee
By using (\ref{fl1vdos}) one can eliminate $F_{0,r}$ from the expression
of $H$. One gets
\be
H\,=\,T_{8-p}\,\O_{7-p}\,R^{7-p}\,\int dr\,
\sqrt{1\,+\,r^2\,\th\,'^{\,2}}\,\,
\sqrt{{\cal C}_{p,n}(\th)^2\,+\,
({\rm sin}\,\th)^{2(7-p)}}\,\,. 
\label{fl1vcuatro}
\ee
It is now simple to find the constant $\th$ configurations which
minimize the energy. Indeed, we only have to require the vanishing of 
$\partial\,H/\partial\th$ for $\th^{'}=0$. Taking into account
that ${\cal C}_{p,n}(\th)$ satisfies eq. (\ref{fl1seis}), we arrive at
\be
{\partial\,H\over \partial\th}\,\,\,\Biggr|_{\th^{'}=0}\,\,\,=
\,\,\,(7-p)\,\,T_{8-p}\,\O_{7-p}\,R^{7-p}\,\,\,
{({\rm sin}\,\th)^{7-p}\,
[\,({\rm sin}\,\th)^{6-p}\,{\rm cos}\,\th\,-\,
{\cal C}_{p,n}(\th)\,\,]\over
\sqrt{{\cal C}_{p,n}(\th)^2\,+\,
({\rm sin}\,\th)^{2(7-p)}}}\,\,.
\label{fl1vcinco}
\ee
Moreover, if we define the functions\footnote{The functions
$\L_{p,n}(\th)$ for different values of $p$ have been listed in
appendix A. These functions are the same as the ones defined in (\ref{bartnueve}) 
for the near-horizon analisys of the baryon vertex if one identifies $\n=n/N$}
\be
\L_{p,n}(\th)\,\equiv\,
({\rm sin}\,\th)^{6-p}\,{\rm cos}\,\th\,-\,
{\cal C}_{p,n}(\th)\,\,,
\label{fl1vseis}
\ee
it is clear by looking at the right-hand side of eq. (\ref{fl1vcinco}) 
 that the energy is minimized either when $\th=0,\pi$ (\ie\ when 
$\sin\th=0$) or when $\th=\bar\th_{p,n}$, where 
$\bar\th_{p,n}$ is determined by the condition
\be
\L_{p,n}(\bar\th_{p,n})\,=\,0\,\,.
\label{fl1vsiete}
\ee
The solutions $\th=0,\pi$ correspond to singular configurations in
which the D(8-p)-brane collapses at the poles of the $S^{7-p}$ sphere.
For this reason we shall concentrate on the analysis of the 
$\th=\bar\th_{p,n}$ configurations. 

It was shown in eq. (\ref{barcuno}) that when 
$p< 6$, $\L_{p,n}(\th)$  is a monotonically increasing function in the
interval $0<\th<\pi$. In what follows we shall
restrict ourselves to the case $p\le 5$, and therefore the function 
$\L_{p,n}(\th)$ vanishes 
for $0<\th <\pi$ if and only if $\L_{p,n}(0)\,<0$ 
and $\L_{p,n}(\pi)\,>\,0$ (see figure \ref{barfig1}). These values are 
given in eq. (\ref{barccinco}) we conclude
that the first condition occurs when $n>0$, whereas 
$\L_{p,n}(\pi)\,>\,0$ if $n<N$. It follows that there exists only
one solution $\bar\th_{p,n}\in (0,\pi)$ of eq. (\ref{fl1vsiete}) for
each
$n$ in the interval $0<n<N$. Then, we have found exactly $N-1$ 
angles which correspond to nonsingular wrappings of the D(8-p)-brane
on a $S^{7-p}$ sphere. Notice that for $n=0$ ($n=N$) the solution of
eq. (\ref{fl1vsiete}) is $\bar\th_{p,0}=0$ ($\bar\th_{p,N}=\pi$) 
(see eq. (\ref{barccinco})). Therefore, we can identify
these $n=0,N$ cases with the singular configurations previously
found. In general, when $n$ is varied from $n=0$ to $n=N$ the angle 
 $\bar\th_{p,n}$ increases from $0$ to $\pi$ (\ie\ from one of the
poles of the $S^{8-p}$ sphere to the other). 

It is not difficult to find the energy of these wrapped configurations.
Actually we only need to substitute $\th=\bar\th_{p,n}$ in 
eq. (\ref{fl1vcuatro}). Taking into account (see eqs. (\ref{fl1vseis}) and
(\ref{fl1vsiete})) that
\be
{\cal C}_{p,n}(\bar\th_{p,n})\,=\,
(\,\sin\bar\th_{p,n}\,)^{6-p}\,\cos\bar\th_{p,n}\,\,,
\label{fl1ttres}
\ee
one easily finds that the energy of these solutions can be written as
\be
H_{p,n}\,=\,\int\,dr\,{\cal E}_{p,n}\,\,,
\label{fl1tcuatro}
\ee
where the constant energy density ${\cal E}_{p,n}$ is given by
\be
{\cal E}_{p,n}\,=\,
{NT_f\over 2\sqrt{\pi}}\,
{\G\Bigl(\,{7-p\over 2}\Bigr)\over
\G\Bigl(\,{8-p\over 2}\Bigr)}\,\,
({\rm sin}\,\bar\theta_{p,n})^{6-p}\,\,.
\label{fl1tcinco}
\ee
Similarly, by substituting eq. (\ref{fl1ttres}) in eq. (\ref{fl1vdos}), we
can get the worldvolume electric field for our configurations, namely,
\be
\bar F_{0,r}\,=\,{\rm cos}\,\bar\th_{p,n}\,\,. 
\label{fl1tseis}
\ee
Let us now analyze some particular cases of our equations. First of
all, we shall consider the $p=5$ case, \ie\ a D3-brane wrapped on a
two-sphere under the action of a D5-brane background. The function
$\L_{5,n}(\th)$ is (see eq. (\ref{barcsiete}))
\be
\L_{5,n}(\th)\,=\,\th\,-\,{n\over N}\,\pi\,\,,
\label{fl1tsiete}
\ee
and the equation $\L_{5,n}(\th)=0$ is trivially solved by the
angles
\be
\bar\th_{5,n}\,=\,{n\over N}\,\pi\,\,.
\label{fl1tocho}
\ee
Notice that the set of angles in eq. (\ref{fl1tocho}) is the same as that
of section \ref{flbachas}. Using this result in eq. (\ref{fl1tcinco}) we get the
following energy density:
\be
{\cal E}_{5,n}\,=\,{NT_f\over \pi}\,
{\rm sin}\,\Big[\,{n\over N}\,\pi\,\Big]\,\,,
\label{fl1tnueve}
\ee
which is very similar to the result found in section \ref{flbachas}. Next,
let us take $p=4$, which corresponds to a D4-brane wrapped on a
three-sphere in the background of a stack of D4-branes. The corresponding 
$\L_{p,n}(\th)$ function is
\be
\L_{4,n}(\th)\,=\,-2\,\Big[\,{\rm cos}\,\th\,+\,
2\,{n\over N}\,-\,1\,\Big]\,\,,
\label{fl1cuarenta}
\ee
and the solutions of eq. (\ref{fl1vsiete}) in this case are easily found,
namely,
\be
{\rm cos}\,\bar\th_{4,n}\,=\,1\,-\,2\,{n\over N}\,\,.
\label{fl1cuno}
\ee
The corresponding energy density takes the form
\be
{\cal E}_{4,n}\,=\,{n(N-n)\over N}\,\,T_f\,\,.
\label{fl1cdos}
\ee
Notice that, in this D4-D4 case, the energy density of eq.
(\ref{fl1cdos}) is a rational fraction of the fundamental string tension.

For general $p$ the equation $\L_{p,n}(\th)\,=\,0$ is much
difficult to solve analytically. In order to illustrate this point let
us write down the equation to solve in the physically interesting case
$p=3$:
\be
\bar\th_{3,n}\,-\,\cos\bar\th_{3,n}\,\sin\,\bar\th_{3,n}\,=\,
{n\over N}\,\pi\,\,.
\label{fl1ctres}
\ee
Despite of the fact that we are not able to find the analytical
solution of the equation $\L_{p,n}(\th)\,=\,0$ for $p\le 3$,
we can get some insight on the nature of our solutions from some general
considerations. First of all, it is interesting to point out the
following property of the functions $\L_{p,n}(\th)$ 
(see eq. (\ref{barcisiete})):
\be
\L_{p,n}(\th)\,=\,-\L_{p,N-n}(\pi-\th)\,\,.
\label{fl1extrauno}
\ee
Eq. (\ref{fl1extrauno}) can be proved either from the definition of the 
$\L_{p,n}(\th)$'s or from their expressions listed in appendix A.
It follows from this equation that our set of angles 
$\bar\th_{p,n}$ satisfy
\be
\bar\th_{p,N-n}\,=\,\pi\,-\,\bar\th_{p,n}\,\,.
\label{fl1extrados}
\ee
By using (\ref{fl1extrados}) in the expression of the energy density 
${\cal E}_{p,n}$ (eq. (\ref{fl1tcinco})), one immediately gets the
following periodicity relation
\be
{\cal E}_{p,N-n}\,=\,{\cal E}_{p,n}\,\,.
\label{fl1extratres}
\ee

Another interesting piece of information can be obtained by considering
the semiclassical $N\rightarrow\infty$ limit. Notice that 
$\L_{p,n}$ depends on $n$ and $N$ through their ratio $n/N$ (see
eqs. (\ref{fl1vseis})) and  (\ref{fl1vuno})). Then, taking $N\rightarrow\infty$
with fixed $n$ is equivalent to make $n\rightarrow 0$ for finite $N$. We
have already argued that if $n\rightarrow 0 (\n\rightarrow 0)$ the angle 
$\bar\th_{p,n}\rightarrow 0$. In order to solve the equation 
$\L_{p,n}(\th)\,=\,0$ for small $\th$, let us expand 
$\L_{p,n}(\th)$ in Taylor series around $\th=0$. It turns
out (see eq. (\ref{barcinuno})) that the first non-vanishing derivative of
$\L_{p,n}(\th)$ at $\th=0$ is the $(6-p)^{th}$ one and near $\th=0$ we can 
write
\be
\L_{p,n}(\th)\,\approx\,\L_{p,n}(0)\,+\,\th^{6-p}\,
+\,\dots\,\,.
\label{fl1ccuatro}
\ee
It follows immediately that for $N\rightarrow\infty$ the value of 
$\bar\th_{p,n}$ is given by
\be
\big(\,\bar\th_{p,n}\,\big)^{6-p}\,\approx\,-\L_{p,n}(0)\,\,.
\label{fl1ccinco}
\ee
Taking into account eq. (\ref{barccinco}) and the general expression of
the energy density (eq. (\ref{fl1tcinco})), we can easily verify that
\be
\lim_{N\rightarrow\infty}\,{\cal E}_{p,n}\,=\,n\,T_f\,\,,
\label{fl1cseis}
\ee
a fact which can be verified directly for $p=4,5$ from our analytical
expressions of the energy density (eqs. (\ref{fl1tnueve}) and
(\ref{fl1cdos})).  It is now clear from eq. (\ref{fl1cseis}) that our
configurations can be interpreted as bound states of $n$ fundamental
strings. Actually, one can prove quite generally that the following
inequality holds
\be
{\cal E}_{p,n}\,\le\,n\,T_f\,\,,
\label{fl1extracuatro}
\ee
which shows that the formation of our bound states is energetically
favored. This is an indication of their stability, which we will
verify directly in section \ref{flncstability}.

In order to prove (\ref{fl1extracuatro}), it is  very useful again to
consider the dependence of the energy on $1/N$. Notice that for 
$1/N\rightarrow 0$ both sides of eq. (\ref{fl1extracuatro}) are equal (see
eq. (\ref{fl1cseis})). The energy ${\cal E}_{p,n}$ depends on $1/N$
both explicitly and implicitly  (through $\bar\th_{p,n}$). If we
consider $1/N$ as a continuous variable, then one has
\be
{d\over d\Big({1\over N}\Big)}\,\,\bar\th_{p,n}\,\,=\,\,
{nNT_f\over (6-p)\,{\cal E}_{p,n}}\,\,\sin\bar\th_{p,n}\,\,.
\label{fl1extracinco}
\ee
Eq. (\ref{fl1extracinco}) is obtained by differentiating eq. 
(\ref{fl1vsiete}) and using eqs. (\ref{barcuno}) and (\ref{barccinco}) (the
latter determines the explicit dependence of $\L_{p,n}$ on 
$1/N$). We are now ready to demonstrate (\ref{fl1extracuatro}). For this
purpose let us consider the quantity $({\cal E}_{p,n}-n\,T_f)/N$, which
we will regard as a function of $1/N$. We must prove that this quantity
is always less or equal than zero. Clearly,  eq. (\ref{fl1cseis}) implies
that  $({\cal E}_{p,n}-n\,T_f)/N\rightarrow 0$ for
$1/N\rightarrow 0$. Moreover, by using (\ref{fl1extracinco}) it is
straightforward to compute the derivative
\be
{d\over d\Big({1\over N}\Big)}\,\,\Bigg[\,
{{\cal E}_{p,n}-n\,T_f\over N}\,\Bigg]\,=\,-n\,T_f\,\,
(\,1\,-\,\cos\bar\th_{p,n}\,)\,\,,
\label{fl1extraseis}
\ee
which vanishes for $N\rightarrow \infty$ and is always negative  for
finite $N$ and $0<n<N$. Thus, it follows that $({\cal E}_{p,n}-n\,T_f)/N$
is negative for finite $N$ and, necessarily, eq. (\ref{fl1extracuatro})
holds. 

As a further check of (\ref{fl1extracuatro}) one can compute the first
correction to ${\cal E}_{p,n}-n\,T_f$ for finite $N$. By Taylor
expanding ${\cal E}_{p,n}$ in powers of $1/N$, and using eq. 
(\ref{fl1extracinco}), one can prove that
\be
{\cal E}_{p,n}\,\,-\,\,n\,T_f\,\approx\,\,-
\,{6-p\over 2(8-p)}\,\,n\,T_f\,\,\,\Big(\,
{\cal C}_{p,n}(0)\,\Big)^{{2\over 6-p}}\,\,+\,\dots\,\,,
\label{fl1extrasiete}
\ee
where ${\cal C}_{p,n}(0)$, which is of order $1/N$, has been   given in
eq. (\ref{barccinco}).

In order to confirm the interpretation of our results given above, let us study
the supersymmetry preserved by our brane probe configurations. In general, 
the number of supersymmetries preserved by a D-brane is the number of
independent solutions of the equation 
$\G_{\k}\,\epsilon\,=\,\epsilon$,
where $\epsilon$ is a Killing spinor of the background and $\G_{\k}$
is the $\k$-symmetry matrix. The Killing spinors $\epsilon$ associated to a D-brane 
background satisfy the projection condition
\be
(i\s_2)\,\s_3^{\frac{p-3}{2}}\,
\G_{\underline{x^0\dots x^p}}\,\,\epsilon\,\,=\,\,
\epsilon\,,
\ee
where $\G_{\underline {x^{\m_1}x^{\m_2}}\dots}$ are antisymmetrized products
of ten-dimensional constant gamma matrices. The $\k$-symmetry matrix of the 
D(8-p)-brane probe is given by
\be
\G_{\k}\,=\,{i\s_2\,\s_3^{\frac{p-3}{2}}\over 
\sqrt{1\,-\,F_{0,r}^2}}\,
\Big[\,F_{0,r}\,-\,\G_{\underline {x^0r}}\s_3\,\Big]\,
\G_{*}\,\,,
\label{fl1csiete}
\ee
where $\G_{*}\,=\,\G_{\underline {\th^1\dots\th^{7-p}}}$.
The condition $\G_{\k}\,\epsilon\,=\,\epsilon$ is:
\be
{1\over \sin\th}\,\Big[\,\cos\th\,+\,
\G_{\underline {x^0r}}\,\s_3\,\Big]\,
\G_{\underline {\th r}}\,\epsilon\,=\,
\epsilon\,\,.
\label{fl1cocho}
\ee
Notice that, in order to derive (\ref{fl1cocho}) we have used that
$F_{0,r}=\cos\th$ in eq. (\ref{fl1csiete}). Moreover, 
introducing the $\theta$-dependence of the spinors, \ie\ 
$\epsilon\,=\,\exp[-{\th\over 2}\,\G_{\underline {\th r}}\,]\,
\hat\epsilon$, with $\hat\epsilon$ independent of $\th$, 
we get the following condition on $\hat\epsilon$:
\be
\G_{\underline {x^0r}}\,\s_3\hat\epsilon\,=\,\hat\epsilon\,\,,
\label{fl1cnueve}
\ee
which can be rewritten as
\be
\G_{\underline {x^0r}}\,\s_3
\epsilon_{{\,\big |}_{\th=0}}=\,\epsilon_{{\,\big |}_{\th=0}}\,\,,
\label{fl1sesenta}
\ee
which certainly corresponds to a system of fundamental strings in the radial
direction. Notice that the point $\th=0$ can be regarded as the ``center of
mass" of the expanded fundamental strings.

\medskip
\subsection{BPS configurations and the baryon vertex}
\medskip

In this section we shall show that the wrapped configurations found
above solve a BPS differential equation. With this purpose in mind,
let us now come back to the more general situation in which the angle
$\th$ depends on the radial coordinate $r$. The hamiltonian for a
general function $\theta(r)$ was given in eq. (\ref{fl1vcuatro}). By
means of a simple calculation it can be verified that this hamiltonian
can be written as
\be
H\,=\,T_{8-p}\,\O_{7-p}\,R^{7-p}\,\int dr\,
\sqrt{\,{\cal Z}^2\,+\,{\cal Y}^2}\,\,,
\label{fl1csiete2}
\ee
where, for any function $\th(r)$, ${\cal Z}$ is a total derivative:
\be
{\cal Z}\,=\,{d\over dr}\,\Big[\,r\Big(\,
({\rm sin}\,\th)^{6-p}\,-\,\L_{p,n}(\th\,
)\,{\rm cos}\,\th\, \,\Big)\,\Big]\,\,,
\label{fl1cocho2}
\ee
and ${\cal Y}$ is given by
\be
{\cal Y}\,=\,{\rm sin}\,\th\,\L_{p,n}(\th\,)\,
-\,r\th\,'\,\Big[\,({\rm sin}\,\th)^{6-p}\,-\,
\L_{p,n}(\th\,)\,{\rm cos}\,\th\,\Big]\,\,.
\label{fl1cnueve2}
\ee
It follows from eq. (\ref{fl1csiete2}) that $H$ is bounded as
\be
H\,\ge\,T_{8-p}\,\O_{7-p}\,R^{7-p}\,\int dr\,
\big |\,{\cal Z}\,\big |\,\,.
\label{fl1cincuenta}
\ee
Since ${\cal Z}$ is a total derivative, the bound on the right-hand
side of eq. (\ref{fl1cincuenta}) only depends on the boundary values of 
$\th(r)$. This implies that any $\th(r)$ saturating the bound is
also a solution of the equations of motion. This saturation of the
bound clearly occurs when ${\cal Y}\,=\,0$ or, taking into account eq. 
(\ref{fl1cnueve2}), when  $\th(r)$ satisfies the following first-order
differential equation:
\be
\th\,'\,=\,{1\over r}\, \,\,
{{\rm sin}\,\th\,\L_{p,n}(\th\,)\over
({\rm sin}\,\th)^{6-p}\,-\,\L_{p,n}(\th\,)
\,{\rm cos}\,\th}\,\,.
\label{fl1ciuno}
\ee
It is straightforward to verify directly that any 
solution $\th(r)$ of eq. (\ref{fl1ciuno}) also solves the second-order
differential equations of motion derived from the hamiltonian of eq. 
(\ref{fl1vcuatro}). Moreover, by using eq. (\ref{fl1ciuno}) to evaluate the
right-hand side of eq. (\ref{fl1vdos}), one can demonstrate that the BPS
differential equation is equivalent to the following relation between
the electric field $F_{0,r}$ and $\th(r)$:
\be
F_{0,r}\,=\,\partial_r\,(\,r\,{\rm cos}\,\th\,)\,=\,
{\rm cos}\,\th\,-\,r\th\,'\,{\rm sin}\,\th\,\,.
\label{fl1cidos}
\ee
Notice now that eq. (\ref{fl1ciuno}) admits solutions with 
$\th={\rm constant}$ if and only if $\th\,=\,0\,,\,\pi$ 
or when $\th$ is a zero of 
$\L_{p,n}(\th\,)$. Thus our wrapped configurations are
certainly solutions of the BPS differential equation. As a
confirmation of this fact, let us point out that, for constant 
$\th$, the electric field of eq. (\ref{fl1cidos}) reduces to the value
displayed in eq. (\ref{fl1tseis}). 

Eq. (\ref{fl1ciuno}) was first proposed (for $p=3$) in section \ref{barsec4} to
describe the baryon vertex (see also refs. \cite{Baryon,CGS1,CGS2,Craps})
\footnote{In these studies of the baryon vertex a different choice of
worldvolume coordinates is performed. Instead of taking  these
coordinates as in eq. (\ref{fl1doce}), one takes 
$\s^{\a}\,=\,(\,t\,,\,
\th^1\,,\,\dots\,,\,\th^{7-p}\,,\,\th\,\,)$ and the embedding
of the D(8-p)-brane is described by a function
$r\,=\,r(\s^{\a})$.}. In section \ref{barsecsym} it was verified, by
looking at the $\k$-symmetry of the brane probe, that the condition 
(\ref{fl1cidos}) is enough to preserve $1/4$ of the bulk supersymmetry.
Actually, in section \ref{barsec4} it was obtained the general solution of 
the BPS differential equation (\ref{fl1ciuno}). In implicit form this 
solution, given in eq. (\ref{barctres}), is
\be
{[\,\L_{p,n}(\th\,)\,]^{{1\over 6-p}}\over
{\rm sin}\,\th}\,=\,C\,r\,\,,
\label{fl1citres}
\ee
\begin{figure}
\centerline{\hskip -.8in \epsffile{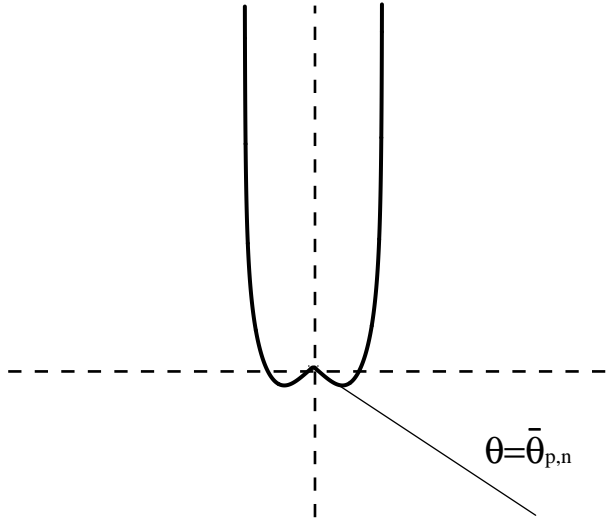}}
\caption{Representation of a typical solution of the BPS equation
(\ref{fl1citres}) for $C\not= 0$. In this plot $r$ and $\th$ are the
polar coordinates of the plane of the figure. We have also plotted the 
$\th\,=\,\bar\th_{p,n}$ curve, which  is the solution of
(\ref{fl1citres}) for $C= 0$.}
\label{flfig2}
\end{figure}
where $C$ is a constant. Our constant angle solutions
$\th\,=\,\bar\th_{p,n}$ can be obtained from eq.~(\ref{fl1citres})
by taking $C=0$, whereas the baryon vertex solutions correspond to
$C\not= 0$. A glance at eq.~(\ref{fl1citres}) reveals that, by
consistency, $\th$ must be restricted to take values in an interval
such that the function $\L_{p,n}(\th\,)$ has a fixed sign. If,
for example, $\th\in(0,\bar\th_{p,n})$, then 
$\L_{p,n}(\th\,)<0$ and, by redefining the phase of $C$, we
get a consistent solution in which $r$ is a non-negative real number.
Similarly, we could have $\th\in(\bar\th_{p,n},\pi)$ since 
$\L_{p,n}(\th\,)>0$ for these values. In both cases 
$\bar\th_{p,n}$ is a limiting angle. Actually, for 
$0<n<N$ one immediately infers from eq. (\ref{fl1citres}) that 
$\bar\th_{p,n}$ is the angle reached when $r\rightarrow 0$. The
baryon vertex solutions behave \cite{Baryon,CGS1,CGS2,Craps} as a bundle of
fundamental strings in the asymptotic region $r\rightarrow \infty$ (see
figure \ref{flfig2}). The number of fundamental strings is precisely $n$ for the
solution with 
$\th\in(0,\bar\th_{p,n})$ (and $N-n$ when 
$\th\in(\bar\th_{p,n},\pi)\,$). Notice that $r\rightarrow
\infty$ when $\th=0$ ($\th=\pi$) for the solution with $n$
($N-n$) fundamental strings, whereas in the opposite limit 
$r\rightarrow 0$ the solution displayed in eq. (\ref{fl1citres}) is
equivalent to our $\th\,=\,\bar\th_{p,n}$  wrapped
configuration. This is quite suggestive and implies that one can 
regard our constant angle configurations as a short distance limit 
(in the radial direction) of the baryon vertex solutions.

\medskip
\subsection{Fluctuations and stability} \label{flncstability}
\medskip
We are now going to study fluctuations around the static
configurations found above. Let us parametrize these fluctuations as
follows: 
\be
\th\,=\,\bar\th_{p,n}\,+\,\xi\,\,,
\,\,\,\,\,\,\,\,\,\,\,\,\,\,\,\,\,\,\,\,
F_{0,r}\,=\,\cos\bar\th_{p,n}\,+\,f\,\,,
\label{fl1cicuatro}
\ee
where $\xi$ and $f$ are small quantities which depend on the worldvolume
coordinates $\sigma^{\a}$. We are going to prove in this section
that the $\th=\bar\th_{p,n}$ solution is stable under the
perturbation of eq. (\ref{fl1cicuatro}). In order to achieve this goal we
must go back to the action written in eq. (\ref{fl2ocho}).  We shall
evaluate this action for an angle $\theta$ and an electric field as in
eq. (\ref{fl1cicuatro}). Let us represent the perturbation $f$
by means of a potential as $f\,=\,\partial_0a_r\,-\,\partial_ra_0$.
We shall choose a gauge in which the components $a_{\hat i\,}$ of the
potential  along the sphere $S^{7-p}$ vanish. 
Then we see that, for consistency, we  must include in our
perturbation the components of the gauge field  strength of the type 
$F_{\hat  i\,,r}\,=\,\partial_{\hat i\,}a_r$ and
$F_{0,\hat  i}\,=\,-\partial_{\hat i\,}a_0$. Under these circumstances
it is not difficult to compute the lagrangian density for the action
(\ref{fl2ocho}) up to second order in $\xi$, $f$,  $F_{\hat  i\,,r}$
and $F_{0,\hat  i}$.
After some calculation one gets
\br
{\cal L}\,&=&\,-\sqrt {\hat g}\,R^{7-p}\,T_{8-p}\,
\L_{p,n}(0)\,f\,+\,\sqrt {\hat g}\,R^{7-p}\,T_{8-p}\,
(\,\sin\bar\th\,)^{6-p}\,\times\rc\rc
&\times&{1\over 2}\Bigg\{\,
R^{7-p}\,r^{p-5}\,(\,\partial_0\xi\,)^2\,-\, 
r^2\,(\,\partial_r\xi\,)^2\,-\,(\,\partial_{\hat i\,}
\xi\,)^2\,+\,\rc\rc  
&+&{R^{p-7}\,r^{5-p}\over
(\sin\bar\th)^2}\,\big[\,\Bigg({R\over r}\Bigg)^{7-p}\,F_{0,\,\hat
i}^2\,-\, F_{\hat i,\,r}^2\,\big]\,+\,
\,(7-p)\xi^2\,+\,{f^2\over (\sin\bar\theta)^2}\,+\,
2(7-p)\,\,{f\xi\over \sin\bar\th}\,\,\Bigg\}\,\,,\rc
\label{fl1cicinco}
\er
where, to simplify the notation, we have written 
$\bar\th$ instead of $\bar\th_{p,n}\,$, $\hat g_{\hat i\,\hat j}$
represents the metric of the $S^{7-p}$ sphere and we have denoted
\be
(\,\partial_{\hat i\,}\xi\,)^2\,=\,\hat
g^{\hat i\,\hat j}\partial_{\hat i\,}\xi\,
\partial_{\hat j\,}\xi\,\,,
\,\,\,\,\,\,\,\,\,\,\,\,\,\,\,
F_{\hat i,\,r}^2\,=\,\hat g^{\hat i\hat j}F_{\hat i,\,r}\,F_{\hat j,\,r}
\,\,,
\,\,\,\,\,\,\,\,\,\,\,\,\,\,\,
F_{0\,,\hat i}^2\,=\,\hat g^{\hat i\hat j}F_{0\,,\hat i}
\,F_{0\,,\hat j}
\,\,.
\label{fl1ciseis}
\ee
In eq. (\ref{fl1cicinco}) we have dropped the zero-order term. 
Moreover, the first term on the right-hand side of eq.
(\ref{fl1cicinco}) is a first-order term which, however, does not
contribute to the equations of motion. In fact, by computing the
variation of the action with respect to $a_0$, $a_r$ and $\xi$ we get
the following set of equations:
\br
&&\partial_r\,\Big[\,{f\over \sin\bar\th}\,
+\,(7-p)\,\xi\,\Big]\,+\,
{1\over r^2\,\sqrt{\hat g}}\,
\partial_{\hat i\,}\,\Big[\,\sqrt{\hat g}\,\hat g^{\hat i\,\hat j}\,
{F_{0,\hat j}\over \sin\bar\th}\,\Big]\,
\,=\,0\,\,,\rc\rc
&&r^{p-5}\,R^{7-p}\partial_{0}\,\Big[\,{f\over \sin\bar\th}\,
+\,(7-p)\,\xi\,\Big]\,-\,{1\over \sqrt{\hat g}}\,
\partial_{\hat i\,}\,\Big[\,\sqrt{\hat g}\,\hat g^{\hat i\,\hat j}\,
{F_{\hat j,\,r}\over \sin\bar\theta}\,\Big]\,=\,0\,\,,\rc\rc
&&R^{7-p}r^{p-5}\,\partial_0^2\xi\,-\,
\partial_r(r^2\partial_r\xi)\,-\,\nabla^2_{S^{(7-p)}}\,\xi\,+\,
(p-7)\,\Big[\,\xi\,+\,{f\over \sin\bar\th}\,\Big]\,=\,0\,\,.\rc
\label{fl1cisiete}
\er
The first equation in (\ref{fl1cisiete}) is nothing but the Gauss law.
Moreover,  if we further fix the gauge to $a_0=0$ 
(\ie\ $f\,=\,\partial_0a_r$, 
$F_{\hat i,\,r}\,=\,\partial_{\hat i\,}a_r$ and 
$F_{0,\hat  i}\,=\,0$), the second equation  in (\ref{fl1cisiete}) can be
written as
\be
r^{p-5}\,R^{7-p}\,\Big[\,{\partial_{0}^2a_r\over \sin\bar\theta}\,+\,
(7-p)\,\partial_{0}\xi\,\Big]\,-\,{1\over \sin\bar\theta}\,
\nabla^2_{S^{(7-p)}}\,a_r\,=\,0\,\,,
\label{fl1ciocho}
\ee
where $\nabla^2_{S^{(7-p)}}$ is the laplacian operator on the 
$S^{(7-p)}$ sphere. In order to continue with our analysis,
let us now expand $a_r$ and $\xi$ in spherical harmonics of
$S^{(7-p)}$
\br
&&a_r(\,t,r,\th^1,\dots,\th^{7-p}\,)\,=\,
\sum_{l\ge 0, m}\,Y_{l,m}(\,\th^1,\dots,\th^{7-p}\,)\,
\a_{l,m}(\,t,r\,)\,\,,\rc\rc
&&
\xi(\,t,r,\th^1,\dots,\th^{7-p}\,)\,=\,
\sum_{l\ge 0, m}\,Y_{l,m}(\,\th^1,\dots,\th^{7-p}\,)\,
\zeta_{l,m}(\,t,r\,)\,\,.\rc
\label{fl1cinueve}
\er
The spherical harmonics $Y_{l,m}$ are well-defined functions on 
$S^{(7-p)}$ which are eigenfunctions of the laplacian on the sphere,
namely,
\be
\nabla^2_{S^{(7-p)}}\,Y_{l,m}\,=\,
-l(l+6-p)\,Y_{l,m}\,\,.
\label{fl1sesenta2}
\ee
By plugging the mode expansion (\ref{fl1cinueve}) into the equations of
motion (\ref{fl1cisiete}) and (\ref{fl1ciocho}), and using eq.
(\ref{fl1sesenta2}), we can obtain some equations for 
$\a_{l,m}(\,t,r\,)$ and $\zeta_{l,m}(\,t,r\,)$. Actually, if we
define
\be
\eta_{l,m}\,\equiv\,
{\partial_{0}\a_{l,m}\over\sin\bar\th}\,+\,(7-p)\,\zeta_{l,m}
\,\,,
\label{fl1suno}
\ee
then, the Gauss law in this $a_0\,=\,a_{\hat i}\,=\,0$ gauge can be
simply written as
\be
\partial_{r}\,\eta_{l,m}\,=\,0\,\,,
\label{fl1sdos}
\ee
whereas the other two equations of motion give rise to
\br
&&R^{7-p}\,r^{p-5}\,\partial_0\,\Big[\,
{\partial_{0}\a_{l,m}\over\sin\bar\th}\,+\,(7-p)\,\zeta_{l,m}
\,\Big]\,+\,l(l+6-p)\, {\a_{l,m}\over\sin\bar\th}\,=\,0\,\,,
\rc\rc
&&R^{7-p}\,r^{p-5}\,\partial_0^2\zeta_{l,m}\,-\,
\partial_r(\,r^2\,\partial_r\,\zeta_{l,m})\,
+\,l(l+6-p)\,\zeta_{l,m}\,+\rc\rc
&&+\,(p-7)\,\Big[\,\zeta_{l,m}\,+
{\partial_{0}\a_{l,m}\over\sin\bar\th}\,\Big]\,=\,0\,\,.
\rc
\label{fl1stres}
\er
Let us analyze first eqs. (\ref{fl1sdos}) and (\ref{fl1stres}) for 
$l=0$. From the first equation in (\ref{fl1stres}) it follows that
\be
\partial_0\eta_{0,m}\,=\,0\,\,.
\label{fl1scuatro}
\ee
Thus,  as $\partial_r\eta_{0,m}\,=\,0$ (see eq. (\ref{fl1sdos})), one
concludes that
\be
\eta_{0,m}\,=\,{\rm constant}\,\,.
\label{fl1scinco}
\ee
By using this result and the definition of $\eta_{l,m}$ given in eq. 
(\ref{fl1suno}), we can express $\partial_0\,\a_{0,m}$ in terms of 
$\zeta_{0,m}$ and the additive constant appearing in eq. 
(\ref{fl1scinco}). By substituting this relation in the second equation
in (\ref{fl1stres}), we get
\be
R^{7-p}\,r^{p-5}\,\partial_0^2\zeta_{0,m}\,-\,
\partial_r(\,r^2\,\partial_r\,\zeta_{0,m})\,+\,
(6-p)(7-p)\,\zeta_{0,m}\,=\,{\rm constant}\,\,.
\label{fl1sseis}
\ee
It is interesting to rewrite eq. (\ref{fl1sseis}) in the following form.
First of all, we define the wave operator ${\cal O}_p$ that acts on any
function $\psi$ as
\be
{\cal O}_p\,\psi\equiv\,R^{7-p}\,r^{p-5}\,\partial_0^2\,\psi\,-\,
\partial_r\,(\,r^2\,\partial_r\psi\,)\,\,.
\label{fl1ssiete}
\ee
Then, if $m^2_0$ is given by
\be
m^2_0\,=\,(6-p)(7-p)\,\,,
\label{fl1socho}
\ee
eq. (\ref{fl1sseis}) can be written as
\be
\Big(\,\,{\cal O}_p\,+\,m^2_0\,\,\Big)\,\zeta_{0,m}\,
=\,{\rm constant}\,\,,
\label{fl1snueve}
\ee
which means that $\zeta_{0,m}$ is a massive mode with mass $m_0$.
Notice that, as $p<6$, $m^2_0$ is strictly positive.

For a general value of $l>0$ the equations of motion can be
conveniently expressed in terms of the variables  $\eta_{l,m}$ and
$\zeta_{l,m}$. Indeed, by differentiating with respect to the time the
first equation (\ref{fl1stres}), and using the definition (\ref{fl1suno}),
we can put them in terms of  $\eta_{l,m}$ and
$\zeta_{l,m}$. Actually, if we define the mass matrix ${\cal M}_p$ as
\be
{\cal M}_p\,=\,\pmatrix{
l\,(l+6-p)\,+\,(7-p)\,(6-p)&&p-7\cr\cr
(p-7)\,l\,(l+6-p)&&l\,(l+6-p)}\,\,,
\label{fl1setenta}
\ee
the equations of motion can be written as
\be
\Big(\,\,{\cal O}_p\,+\,{\cal M}_p\,\,\Big)\,
\pmatrix{\zeta_{l,m}\cr\eta_{l,m}}
=\,0\,\,,
\label{fl1stuno}
\ee
where ${\cal O}_p$ is the wave operator defined in eq. (\ref{fl1ssiete}).
In order to check that our wrapped configurations are stable, we must
verify that the eigenvalues of the matrix ${\cal M}_p$ are
non-negative. After a simple calculation one can show that these
eigenvalues are
\be
m_l^2\,=\,\cases{(l+6-p)\,(l+7-p)&for $l=0,1,\dots \,\,,$\cr\cr
              l\,(l-1)&for $l=1,2,\dots\,\,,$\cr}
\label{fl1stdos}
\ee
where we have already included the $l=0$ case. Eq. (\ref{fl1stdos})
proves that there are not negative mass  modes in the spectrum of
small oscillations for $p<6$, which demonstrates that, as claimed, our
static solutions are stable.

\setcounter{equation}{0}
\section{Flux quantization in the (D(p-2), Dp) background} \label{flncbs}
\medskip

The string frame metric $ds^2$ and the dilaton $\p$ generated by a stack of 
(D(p-2), Dp) bound states $(p\ge 2)$ are given in section \ref{strsecbs} 
(see eqs. (\ref{strncbs1})-(\ref{strncbs5})). The Dp-brane of the background 
extends along the directions $x^0\dots x^p$, whereas the  D(p-2)-brane 
component lies along $x^0\dots x^{p-2}$ and $N$ 
denotes the number of branes of the stack. The components of the Hodge 
dual fields ${}^*\,F^{(p)}$ and ${}^*\,F^{(p+2)}$ along the directions 
transverse to the bound state are given in eq. (\ref{strncbsdfs}) and they can 
be represented by means of two RR potentials 
$C^{(9-p)}$ and $C^{(7-p)}$ which are, respectively, a $(9-p)$-form and 
a $(7-p)$-form. In order to write the relevant components of these potentials,
let us parametrize the $S^{8-p}$ transverse sphere by means of the spherical
angles $\th^1$, $\th^2$, $\dots$, $\th^{8-p}$ and let 
$\th\equiv\th^{8-p}$ be the polar angle measured from one of the poles of
the sphere ($0\le\th\le\pi$). Then, the $S^{8-p}$ line element 
$d\O_{8-p}^2$ can be decomposed as:
$d\O_{8-p}^2\,=\,d\th^2\,+\,(\,{\rm sin}\,\th)^{2}\,\,
d\O_{7-p}^2$, where $d\O_{7-p}^2$ is the metric of the constant
latitude $(7-p)$-sphere. In terms of the functions  $C_p(\th)$ defined in 
(\ref{fl1seis}) and (\ref{fl1siete}) the components of the RR potentials 
in which we are interested in are
\br
C^{(7-p)}_{\th^1,\dots,\th^{7-p}}
&=&-\,\cos\varphi\,R^{7-p}\,C_p(\th)\,
\sqrt{\hat g^{(7-p)}}\,\,, \rc
C^{(9-p)}_{x^{p-1},x^p,\th^1,\dots, \th^{7-p}}
&=&-\,\sin\varphi\,R^{7-p}\,
h_p\,f_p^{-1}\,C_p(\th)\,
\sqrt{\hat g^{(7-p)}}\,\,.
\label{fl2siete}
\er
In eq. (\ref{fl2siete}) $\hat g^{(7-p)}$ is the determinant of the metric of the 
unit $S^{7-p}$ sphere. 

Let us now place a D(10-p)-brane probe in the (D(p-2), Dp) geometry. The action
of such a brane probe is the sum of a Dirac-Born-Infeld and a Wess-Zumino term,
namely,
\br
S\,=\,
-T_{10-p}\,\int d^{\,11-p}\xi\,e^{-\tilde\phi}\,
\sqrt{-{\rm det}\,(\,g\,+\,{\cal F}\,)}\,+\,
T_{10-p}\int\Bigg[\,\,C^{(9-p)}\,\wedge\,{\cal F}+\,
{1\over 2}\, C^{(7-p)}\,\wedge\,
{\cal F}\wedge\,{\cal F}\,\Bigg],\rc
\label{fl2ocho}
\er
where $g$ is the induced metric on the worldvolume of the brane probe, 
$T_{10-p}$ is the tension of the  D(10-p)-brane and 
${\cal F}=F-B$, with $F$ being a $U(1)$ worldvolume gauge field and $B$ the
NSNS gauge potential (actually its pullback to the probe worldvolume).  We want
to find stable configurations in which the probe is partially wrapped on the 
$S^{7-p}$ constant latitude sphere. From the analysis performed in section
\ref{flRR} for RR backgrounds, it follows that we must extend the probe along
the radial coordinate and switch on an electric worldvolume field along this
direction. Moreover, our background has a $B$ field with non-zero components
along the 
$x^{p-1}x^{p}$ plane. Then, in order to capture the flux of the $B$ field, we
must also extend our D(10-p)-brane probe along the $x^{p-1}x^{p}$ directions.
Therefore, the natural set of worldvolume coordinates 
$\s^{\a}$ ($\a=0,\dots,10-p$) is
$\s^{\a}\,=\,(t,x^{p-1},x^{p},r,\th^1,\dots,\th^{7-p})$,  where
$t\equiv x^{0}$.
Moreover, we will adopt the following ansatz for the field ${\cal F}$:
\be
{\cal F}\,=\,F_{0,r}\,dt\wedge dr\,
+\,{\cal  F}_{p-1,p}\,dx^{p-1}\wedge dx^{p}\,\,.
\label{fl2nueve}
\ee
Notice that ${\cal  F}_{p-1,p}$ gets a contribution from the pullback of $B$,
namely,
\be
{\cal  F}_{p-1,p}\,=\,  F_{p-1,p}\,-\,h_pf_p^{-1}\tan\varphi\,\,.
\label{fl2diez}
\ee
It is interesting to point out that the components of ${\cal  F}$ in eq. 
(\ref{fl2nueve}) are precisely those which couple to the RR potentials 
(\ref{fl2siete}) in the Wess-Zumino term of the action. With the choice of
worldvolume coordinates we have made above, the embedding of the brane probe in
the transverse space is encoded in the dependence of the polar angle $\th$ on
the  $\s^{\a}$'s. Although we are interested in configurations in which
$\th$ is constant, we will consider first the more general situation in
which $\th=\th(r)$. It is easy to compute the lagrangian in this case.
One gets
\be
L\,=\,\int dx^{p-1}dx^p\int_{S^{7-p}}\,d^{7-p}\th\,
\sqrt{\hat g}\,\int\,drdt\,{\cal L}(\th, F)\,\,,
\label{fl2once}
\ee
where $\hat g\equiv \hat g^{(7-p)}$ and the lagrangian density 
${\cal L}(\th, F)$ is
\br
{\cal L}(\theta, F)\,=\,-T_{10-p}\,R^{7-p}&&\Big[\,
(\sin\th)^{7-p}\,
\sqrt{h_pf_p^{-1}+h_p^{-1}{\cal  F}_{p-1,p}^2}\,\,
\sqrt{1+r^2\th'^2-F_{0,r}^2}\,+\,\rc\rc
&&+\,\cos\varphi F_{p-1,p}\,F_{0,r}\,C_p(\th)\,
\Big]\,\,.
\label{fl2doce}
\er
We want to find solutions of the equations of motion derived from 
${\cal L}(\th, F)$ in which both the angle $\theta$ and the worldvolume
gauge field are constant. The equation of motion for
$\th$ with $\th=\bar\th={\rm constant}$ reduces to
\be
\cos\bar\th\,\sqrt{1-F_{0,r}^2}\,
\sqrt{h_pf_p^{-1}+h_p^{-1}{\cal  F}_{p-1,p}^2}\,-\,
\sin\bar\th\,\cos\varphi\,F_{0,r}\,F_{p-1,p}\,=\,0\,\,.
\label{fl2trece}
\ee
If $F_{0,r}$ and $F_{p-1,p}$ are constant, eq. (\ref{fl2trece}) is only consistent
when its left-hand side is independent of $r$. However, the square root
involving ${\cal  F}_{p-1,p}$ does depend on $r$ in general. Actually, after a
simple calculation one can verify that
\be
h_pf_p^{-1}+h_p^{-1}{\cal F}_{p-1,p}^2\,=
\,\cos^2\varphi\,F_{p-1,p}^2\,+
\,f_p^{-1}\,\Big(\,F_{p-1,p}\sin\varphi\,-\,
{1\over \cos\varphi}\Big)^{2}\,\,.
\label{fl2catorce}
\ee
By inspecting the right-hand side of eq. (\ref{fl2catorce}) one immediately
concludes that it is only independent of $r$ when $F_{p-1,p}$ takes the value
\be
F_{p-1,p}\,=\,{1\over \sin\varphi\cos\varphi}
\,=\,2\csc (2\varphi)\,\,.
\label{fl2quince}
\ee
Plugging back this value of $F_{p-1,p}$ into eq. (\ref{fl2trece}), one gets that 
$F_{0,r}\,=\,\cos\bar\th$. In order to determine the allowed values of 
$\bar\th$, and therefore of $F_{0,r}$, we need to impose a quantization
condition. Let us consider again a configuration in which $\th=\th(r)$
and assume that $F_{p-1,p}$ is given by eq. (\ref{fl2quince}). Moreover, let us
introduce a quantization volume ${\cal V}$ in the $x^{p-1}x^{p}$ plane which
corresponds to one unit of flux, namely,
\be
\int_{{\cal V}}\,dx^{p-1}\,dx^{p}\,F_{p-1,p}\,=\,
{2\pi\over T_f}\,\,,
\label{fl2dseis}
\ee
where $T_{f}\,=\,( 2\pi\a\,'\,)^{-1}\,$. 
By using the constant value of $F_{p-1,p}$ written in eq. (\ref{fl2quince}), one
gets that ${\cal V}$ is given by
\be
{\cal V}\,=\,2\pi^2\a'\sin(2\varphi)\,.
\label{fl2dsiete}
\ee
We now determine $F_{0,r}$ by imposing the quantization condition of 
(\ref{fl1dsiete}) on the volume ${\cal V}$, \ie,
\be
\int_{{\cal V}}\,dx^{p-1}\,dx^{p}\,
\int_{S^{7-p}}\,\,d^{7-p}\th\,\sqrt{\hat g}\,
{\partial \,{\cal L}\over\partial F_{0,r}}\,=\, n\,T_{f}\,\,,
\label{fl2docho}
\ee
with $n\in\ZZ$. By using the explicit form of ${\cal L}$ and $F_{p-1,p}$, one
can easily compute the left-hand side of the quantization condition 
(\ref{fl2docho}):
\br
\int_{{\cal V}}\,dx^{p-1}\,dx^{p}\,
\int_{S^{7-p}}\,\,d^{7-p}\th\,\sqrt{\hat g}\,
{\partial \,{\cal L}\over\partial F_{0,r}}\,=\, 
{T_{10-p}\,\O_{7-p}\,R^{7-p}\,{\cal V}\over \sin\varphi}
\Biggl[\,
{F_{0,r}\,\,({\rm sin}\,\th)^{7-p}\,\over 
\sqrt{1\,+\,r^2\,\th\,'^{\,2}\, -\,F_{0,r}^2}}\,\,-\,\,
C_p(\theta)\,\Biggr]\,\,,\rc
\label{fl2dnueve}
\er
where $\O_{7-p}$ is the volume of the unit $(7-p)$-sphere and we have
assumed that $F_{0,r}$ does not depend on $\th^1,\dots\th^{7-p}$. It is
not difficult now to find $F_{0,r}$ as a function of $\th(r)$ and the
quantization integer $n$. Let us notice that the global 
coefficient of the right-hand side of (\ref{fl2dnueve}) is
\be
{T_{10-p}\,\O_{7-p}\,R^{7-p}\,{\cal V}\over \sin\varphi}
\,=\,
{NT_f\over 2\sqrt{\pi}}\,
{\G\Bigl(\,{7-p\over 2}\Bigr)\over
\G\Bigl(\,{8-p\over 2}\Bigr)}\,\,. 
\label{fl2veinte}
\ee
In terms of the functions ${\cal C}_{p,n}(\th)$ defined in (\ref{fl1vuno}) 
the electric field is given by
\be
F_{0,r}\,=\,\sqrt{\,
{1\,+\,r^2\,\th\,'^{\,2}\over 
{\cal C}_{p,n}(\th)^2\,+\,
({\rm sin}\,\th)^{2(7-p)}}}\,\,{\cal C}_{p,n}(\th)\,\,.
\label{fl2vdos}
\ee
Once $F_{0,r}$ is known, we can obtain the hamiltonian $H$ by means of a
Legendre transformation:
\be
H\,=\,\int\,dx^{p-1}\,dx^{p}\,
\int_{S^{7-p}}\,\,d^{7-p}\th\,\,\sqrt{\hat g}\,
\int dr\,\Big[\, F_{0,r}\,{\partial \,{\cal L}\over\partial
F_{0,r}}\,-\, {\cal L}\,\Big]\,\,.
\label{fl2vtres}
\ee
By using eqs. (\ref{fl2doce}) and (\ref{fl2vdos}), one easily obtains the 
following expression of $H$:
\be
H\,=\,{T_{10-p}\,\O_{7-p}\,R^{7-p}\over \sin\varphi}\,
\int\,dx^{p-1}\,dx^{p}\,\int dr\,
\sqrt{1\,+\,r^2\,\th\,'^{\,2}}\,\,
\sqrt{\,({\rm sin}\,\th)^{2(7-p)}+\,
\Big({\cal C}_{p,n}(\theta)\Big)^2}\,\,. 
\label{fl2vcuatro}
\ee
The constant $\th$ solutions of the equation of motion are those which
minimize $H$ for $\th\,'=0$. The vanishing of $\partial H/\partial \th$ 
for $\th\,'=0$ occurs when $\th=\bar\th_{p,n}$, where $\bar\th_{p,n}$ 
is determined by the condition
\be
\L_{p,n}(\bar\th_{p,n})\,=\,0\,\,,
\label{fl2vseis}
\ee
where the functions $\L_{p,n}(\th)$ are the ones defined in (\ref{fl1vseis}).
In section \ref{flRR} it was proved that there exists a unique solution  
$\bar\th_{p,n}$ in the interval $[0,\pi]$ for $p\le 5$ and $0\le n\le N$ and 
for all values of $p\le 5$, $\bar\th_{p,0}=0$ and $\bar\th_{p,N}=\pi$, which
correspond to singular configurations in which the brane probe collapses at one
of the poles of the $S^{7-p}$ sphere. Excluding these points, there are exactly
$N-1$ angles  which minimize the energy. The corresponding electric field is
$F_{0,r}=\cos\bar\th_{p,n}$ and if we integrate $x^{p-1}$
and $x^{p}$ in eq. (\ref{fl2vcuatro}) over the quantization volume ${\cal V}$, 
we obtain  the energy $H_{p,n}$ of these solutions on the volume ${\cal V}$, 
which can be written as
\be
H_{p,n}\,=\,\int\,dr\,{\cal E}_{p,n}\,\,,
\label{fl2vsiete}
\ee
where the constant energy density ${\cal E}_{p,n}$ is given by
\be
{\cal E}_{p,n}\,=\,
{NT_f\over 2\sqrt{\pi}}\,
{\G\Bigl(\,{7-p\over 2}\Bigr)\over
\G\Bigl(\,{8-p\over 2}\Bigr)}\,\,
({\rm sin}\,\bar\theta_{p,n})^{6-p}\,\,.
\label{fl2vocho}
\ee
The expression of ${\cal E}_{p,n}$  in (\ref{fl2vocho}) is the same as that 
found in (\ref{fl1tcinco}) for a Dp-brane background. It was argued in 
section \ref{flRR} that ${\cal E}_{p,n}$ can be interpreted as the energy 
density of a bound state of $n$ fundamental strings since ${\cal E}_{p,n}\le 
n T_f$ and ${\cal E}_{p,n}\rightarrow n T_f$ in the semiclassical limit 
$N\rightarrow\infty$. Thus, we are led to propose that the states we have 
found are, in fact, a bound state of polarized fundamental strings stretched 
along the radial direction and distributed over the $x^{p-1}x^p$ plane in 
such a way that there are $n$ fundamental strings in the volume ${\cal V}$. 
Notice that ${\cal V}\rightarrow 0$ when  $\varphi\rightarrow 0$ and, thus, 
the bound state becomes point-like in the $x^{p-1}x^p$ directions as 
$\varphi\rightarrow 0$. This fact is in agreement with the results of section 
\ref{flRR}, since, in this limit, the (D(p-2), Dp) background becomes the 
Dp-brane geometry. 

Let us study
the supersymmetry preserved by our brane probe configurations. For simplicity, 
we shall restrict ourselves to the analysis of the $p=3$ case, \ie\ for the 
(D1, D3) background. The Killing spinors in this case have the form
\be
\epsilon\,=\,e^{{\a\over 2}\,\G_{\underline{x^{2}x^{3}}}\,
\s_3}\,\,\tilde\epsilon\,\,,
\label{fl2vnueve}
\ee
where $\tilde\epsilon$ is a spinor which satisfies 
$(i\s_2)\,
\G_{\underline{x^0\dots x^3}}\,\,\tilde\epsilon\,\,=\,\,
\tilde\epsilon$ and $\a$ is given by: 
\be
\sin\a\,=\,f_3^{-{1\over 2}}\,h_3^{{1\over 2}}\sin\varphi\,\,,
\,\,\,\,\,\,\,\,\,\,\,\,\,\,\,\,\,\,\,\,\,\,
\cos\a\,=\,h_3^{{1\over 2}}\cos\varphi\,\,.
\label{fl2treinta}
\ee
The $\k$-symmetry matrix of the D7-brane probe can be
put as
\be
\G_{\k}\,=\,{i\s_2\over 
\sqrt{1\,-\,F_{0,r}^2}}\,
\Big[\,F_{0,r}\,-\,\G_{\underline {x^0r}}\s_3\,\Big]\,
e^{-\eta\G_{\underline {x^2x^3}}\s_3}\,\G_{*}\,\,,
\label{fl2tuno}
\ee
where now $\G_{*}\,=\,\G_{\underline {\th^1\dots\th^4}}$ and $\eta$
is
\be
\sin\eta\,=\,{f_3^{-{1\over 2}}\,h_3^{{1\over 2}}\over 
\sqrt{h_3f_3^{-1}+h_3^{-1}{\cal F}_{2,3 }^2}
}\,\,,
\,\,\,\,\,\,\,\,\,\,\,\,\,\,\,\,\,\,\,\,\,\,
\cos\eta\,=\,{{\cal F}_{2,3 }\,h_3^{-{1\over 2}}\over 
\sqrt{h_3f_3^{-1}+h_3^{-1}{\cal F}_{2,3 }^2}
}\,\,.
\label{fl2tdos}
\ee
For our configurations, in which $F_{2,3 }$ is given by eq. (\ref{fl2quince}), 
the angles $\a$  and $\eta$ of eqs. (\ref{fl2treinta}) and (\ref{fl2tdos})
are equal, and the
$\G_{\k}\,\epsilon\,=\,\epsilon$ condition becomes
\be
{1\over \sin\th}\,\Big[\,\cos\th\,+\,
\G_{\underline {x^0r}}\,\s_3\,\Big]\,
\G_{\underline {\th r}}\,\tilde\epsilon\,=\,
\tilde\epsilon\,\,,
\label{fl2ttres}
\ee
where we have used that $F_{0,r}=\cos\th$. The $\th$-dependence of the spinors, 
is $\tilde\epsilon\,=\,\exp[-{\th\over 2}\,\G_{\underline {\th r}}\,]\,
\hat\epsilon$, with $\hat\epsilon$ independent of $\th$, 
we get the same condition on $\hat\epsilon$ as in eq. (\ref{fl1cnueve}).  
The supersymmetry preserved by the brane probe
\be
\G_{\underline {x^0r}}\,\s_3
\epsilon_{{\,\big |}_{\th=0}}=\,\epsilon_{{\,\big |}_{\th=0}}\,\,,
\label{fl2tcinco}
\ee
corresponds again to a system of fundamental strings in the radial
direction with the point $\th=0$ being the ``center of
mass" of the expanded fundamental strings.

\setcounter{equation}{0}
\section{Flux quantization in the (NS5, Dp) background} \label{flns5dpbs}
\medskip

Let us now consider the background given in section \ref{strsecNS5Dp} 
(eqs. (\ref{strns5dpbs1})-(\ref{strns5dpbs6})) generated by a stack of $N$ bound
states of NS5-branes and Dp-branes for $1\le p\le 5$. The NS5-branes of this 
background extend along the $tx^1\dots x^5$ coordinates, whereas the Dp-branes 
lie along $tx^1\cdots x^p$ and are smeared in the $x^{p+1}\dots x^5$ coordinates. 
The integers $l$ and $m$ represent, respectively, the number of NS5-branes in 
the bound state and the number of Dp-branes in a $(5-p)$-dimensional volume 
${\cal V}_p\,=\,(2\pi \sqrt{\a'})^{5-p}$ in the  $x^{p+1}\dots x^5$
directions (see section \ref{strsecNS5Dp}). The near-horizon  harmonic function 
$H_{(l,m)}(r)$ is defined as
\be
H_{(l,m)}(r)\,=\,{R^2_{(l,m)}\over r^2}\,\,,
\label{fl2tseis}
\ee
with $R^2_{(l,m)}\,=\,N\,\Bigl[\,\mu_{(l,m)}\,\Bigr]^{{1\over 2}}
\,\a'$ and $\m_{(l,m)}\,=\,l^2\,+\,m^2\,g_s^2$.

We shall choose spherical coordinates, and
we will represent the $S^3$ line element as 
$d\O_{3}^2\,=\,d\th^2\,+\,(\,{\rm sin}\,\th)^{2}\,\,
d\O_{2}^2$. The NSNS potential $B$ is given by
\be
B\,=\,-l\,N\a'\,C_5(\th)\,\epsilon_{(2)}\,\,,
\label{fl2cuarenta}
\ee
and the relevant components of the two RR potentials $C^{(7-p)}$ and 
$C^{(5-p)}$ are
\br 
&&C^{(7-p)}_{x^{p+1},\dots, x^{5},\th^{1},\th^{2}}\,=\,
-m\,N\a'\,C_5(\th)\,\sqrt{\hat g^{(2)}}\,\,,\rc\rc
&&C^{(5-p)}_{x^{p+1},\dots ,x^{5}}\,=\,
{lm\over \mu_{(l,m)}}\,
\Big(\,\Bigl[\,H_{(l,m)}(r)\,\bigr]^{{1\over 2}}\,-\,
\bigl[\,H_{(l,m)}(r)\,\bigr]^{-{1\over 2}}
\,\Big)\,h_{(l,m)}(r)\,\,.
\label{fl2cuno}
\er
In eqs. (\ref{fl2cuarenta}) and (\ref{fl2cuno})
$C_5(\th)$ is the function defined in eq. (\ref{fl1seis}) for $p=5$, namely 
$C_5(\th)=\sin \theta\cos\th\,-\,\th$,  $\epsilon_{(2)}$ is the volume
element of the constant latitude sphere $S^2$ and $\hat g^{(2)}$ the
determinant of its metric. Moreover, to simplify the equations that follow  we
shall take from now on $g_s=1$ (the dependence on $g_s$ can be easily
restored).

By inspecting the form of the NSNS and RR potentials is eqs. (\ref{fl2cuarenta})
and (\ref{fl2cuno}) one easily realizes that, in order to get flux-stabilized
configurations, one must consider a D(8-p)-brane probe wrapping the $S^2$ and
extended along $r,x^{p+1},\dots,x^{5}$. The action of such a probe is
\br
S\,&=&\,
-T_{8-p}\,\int d^{\,9-p}\s\,e^{-\tilde\phi}\,
\sqrt{-{\rm det}\,(\,g\,+\,{\cal F}\,)}\,+\,\rc\rc
&&+\,T_{8-p}\,\int\Big[\,C^{(7-p)}\wedge {\cal F}\,+\,
{1\over 2}\,C^{(5-p)}\wedge {\cal F} \wedge {\cal F}\,\Big]\,\,.
\label{fl2cdos}
\er
We shall take in (\ref{fl2cdos}) the following  set of worldvolume 
coordinates 
$\s^{\a}=(t,x^{p+1},\dots,x^{5},r,\th^1,\th^{2})$ and 
we will consider configurations of the brane probe in which $\th$ is a
function of $r$.

\medskip
\subsubsection{Quantization conditions}
\medskip

To determine the worldvolume gauge field $F$,  we first impose 
the flux quantization condition
\be
\int_{S^2}\,F\,=\,{2\pi n_1\over T_f}\,\,,
\,\,\,\,\,\,\,\,\,\,\,\,\,\,\,\,\,\,
n_1\in\ZZ\,\,.
\label{fl2ctres}
\ee
Eq. (\ref{fl2ctres}) can be easily solved, and its solution fixes the magnetic
components of $F$. Actually, if we assume that the electric worldvolume field
has only components along the radial direction, one can write the solution of 
(\ref{fl2ctres}) as 
$F\,=\,\pi n_1\a'\epsilon_{(2)}\,+\,F_{0,r}dt\wedge dr$,
which is equivalent to the following expression of ${\cal F}$
\be
{\cal F}\,=\,f_{12}(\th)\epsilon_{(2)}\,+\,F_{0,r}dt\wedge dr\,\,,
\label{fl2ccuatro}
\ee
with $f_{12}(\th)$ being
\be
f_{12}(\th)\,\equiv\,lN\a'C_5(\th)\,+\,\pi n_1\a'\,\,.
\label{fl2ccinco}
\ee
Using eq. (\ref{fl2ccuatro}) in (\ref{fl2cdos}) one finds that the 
lagrangian of the system is 
\be
L\,=\,\int dx^{p+1}\dots dx^5\int_{S^2}\,d^2\th\,
\sqrt{\hat g}\,\int\,drdt\,{\cal L}(\th, F)\,\,,
\label{fl2cseis}
\ee
where $\hat g\equiv \hat g^{(2)}$ and the lagrangian density is given by
\br
{\cal L}(\th, F)\,&=&\,-T_{8-p}\,\,\Bigg[\,
\sqrt{r^4\,\Bigl[\,H_{(l,m)}(r)\,\Bigr]^{{3\over 2}}\,(\sin\th)^4
\,+\,h_{(l,m)}(r)\,f_{12}(\th)^2}\,\times\rc\rc
&&\times\,\sqrt{\Bigl[\,H_{(l,m)}(r)\,\Bigr]^{{1\over 2}}
(1+r^2\th'^2)\,-\,h_{(l,m)}(r)\,F_{0,r}^2}\,+\,\rc\rc
&&+\,(mN\alpha'C_5(\th)\,-\,C^{(5-p)} f_{12}(\th))\,F_{0,r}
\,\Bigg]\,\,.
\label{fl2csiete}
\er
In eq.  (\ref{fl2csiete}) we have suppressed the indices of the RR potential 
$C^{(5-p)}$. We now impose the following quantization condition 
(see eq. (\ref{fl2docho}))
\be
\int_{{\cal V}_p}\,\,
dx^{p+1}\dots dx^5\int_{S^2}\,d^2\th\,
\sqrt{\hat g}\,\,{\partial {\cal L}\over 
\partial F_{0,r}}\,=\,n_2\,T_f\,\,,
\label{fl2cocho}
\ee
where $n_2\in\ZZ$ and ${\cal V}_p$ is the quantization volume defined before
eq. (\ref{fl2tseis}) (\ie\ ${\cal V}_p=(2\pi \sqrt{\a'})^{5-p}$). 
Eq. (\ref{fl2cocho}) allows to obtain
$F_{0,r}$ as a function of $\th(r)$ and of the quantization integers $n_1$
and $n_2$.  By means of a Legendre transformation one can get the form of
the hamiltonian of the system. After some calculation one arrives at
\br
&&H\,=\,T_{8-p}\O_2\,\int dx^{p+1}\dots dx^5
\int dr\sqrt{1\,+\,r^2\th'^2}\,\times\,\label{fl2cincuenta}\\\rc
&&\times\,\sqrt{R^4_{(l,m)}\,(\sin\th)^4\,+\,
[\mu_{(l,m)}]^{-1}\big[\,\big(lf_{12}(\th)+m\Pi(\th)\big)^2+
H_{(l,m)}(r)\big(mf_{12}(\th)-l\Pi(\th)\big)^2\,\big]}\,\,,
\nonumber
\er
where $\Pi(\th)$ is the function
\be
\Pi(\th)\,\equiv\,mN\a'C_5(\th)\,+\,
\pi n_2\a'\,\,.
\label{fl2ciuno}
\ee
By inspecting the right-hand side of eq. (\ref{fl2cincuenta}) one immediately
reaches the conclusion that  there exist configurations with  constant $\theta$
which minimize the energy only when $mf_{12}(\th)\,=\,l\Pi(\th)$. By
looking at eqs. (\ref{fl2ccinco}) and (\ref{fl2ciuno}) it is immediate to verify that
this condition is equivalent to $mn_1\,=\,ln_2$. Since $l$ and $m$
are coprime, one must have  $n_1\,=\,l\,n\,$, $n_2\,=\,mn\,$ with $n\in\ZZ$. 
Then, our two quantization integers  $n_1$ and $n_2$ are not independent and 
$f_{12}(\th)$ and $\Pi(\th)$ are given in terms of $n$ by
\be
f_{12}(\th)\,=\,lN\a'\,{\cal C}_{5,n}(\th)\,,
\,\,\,\,\,\,\,\,\,\,\,\,\,\,\,\,\,\,\,
\Pi(\th)\,=\,mN\a'\,{\cal C}_{5,n}(\th)\,\,,
\label{fl2cidos}
\ee
where ${\cal C}_{5,n}(\th)$ is the function defined in eq. (\ref{fl1vuno}) for
$p=5$. By using eq. (\ref{fl2cidos}) in eq. (\ref{fl2cincuenta}), one gets the
following expression of $H$: 
\br
H\,=\,T_{8-p}\,\O_2\,R_{(l,m)}^2\,
\int dx^{p+1}\dots dx^5\int\,\, dr
\,\times\,\rc\rc\,
\times\,
\sqrt{1\,+\,r^2\th'^2}\,\,
\sqrt{(\sin\th)^4\,+\,
\Big(\,{\cal C}_{5,n}(\th)\Big)^2}\,\,.
\label{fl2citres}
\er
By comparing the right-hand side of eq. (\ref{fl2citres}) with that of eq. 
(\ref{fl2vcuatro}) one immediately realizes that the constant angles which 
minimize the energy are the solutions of eq. (\ref{fl2vseis}) 
(or (\ref{fl1vsiete})) for $p=5$, \ie\
$\th\,=\,\bar\th_{5,n}\,=\,{n\over N}\,\pi$ with $0\le n\le N$. 
The electric field $F_{0,r}$ which we must have in
the  worldvolume in order to wrap the D(8-p)-brane at
$\th\,=\,\bar\th_{5,n}$ is easily obtained from eq. (\ref{fl2cocho}). After
a short calculation one gets that, for a general value of $g_s$, $F_{0,r}$ is
given by
\be
\bar F_{0,r}\,=\,{mg_s\over 
\sqrt{l^2+m^2g_s^2}}\,\,
\cos\,\Big[\,{n\over N}\,\pi\Big]\,\,.
\label{fl2cicuatro}
\ee
Let 
$H_n^{(l,m)}$ be the energy of our configurations and 
${\cal E}_n^{(l,m)}$ the corresponding energy density, whose integral over
$x^{p+1}\cdots x^5 r$ gives $H_n^{(l,m)}$. After a short calculation one easily
proves that
\be
{\cal E}_n^{(l,m)}\,=\,
{NT_{6-p}(m,l)\over \pi}\,
\sin\Big[\,{n\over N}\,\pi\Big]\,\,,
\label{fl2cicinco}
\ee
where, for an arbitrary value of $g_s$, $T_{6-p}(m,l)$ is given by:
\be
T_{6-p}(m,l)\,=\,
{1\over (2\pi)^{6-p}\,(\a')^{{7-p\over 2}} g_s}\,\,
\sqrt{l^2\,+\,m^2\,g_s^2}\,\,.
\label{fl2ciseis}
\ee
$T_{6-p}(m,l)$  is the tension of a bound state of fundamental strings and
D(6-p)-branes \cite{FDp}. In such a (F, D(6-p))-brane state, $l$ is the number of
D(6-p)-branes, whereas $m$ parametrizes the number of fundamental strings.
Indeed, one can check that $T_{6-p}(0,l)=lT_{6-p}$ and, on the other hand, 
$T_{6-p}(m,0){\cal V}_p=mT_f$, which means that there are $m$ fundamental
strings in the $(5-p)$-dimensional volume ${\cal V}_p$. These strings are
stretched in the radial direction and  smeared in the 
$x^{p+1}\dots x^5$ coordinates. This interpretation of $T_{6-p}(m,l)$ 
suggests that our configurations with $\th\,=\,\bar\th_{5,n}$ are bound
states of (F, D(6-p))-branes. Indeed, since 
${\cal E}_n^{(l,m)}\rightarrow n T_{6-p}(m,l)$ as $N\rightarrow\infty$, the
number of (F, D(6-p))-branes which form our bound state is precisely the
quantization integer $n$. Moreover, we can determine the supersymmetry
preserved by our configuration. This analysis is similar to the one carried out
at the end of section \ref{flncbs}. Let us present the result of this study for the 
(NS5, D3) background, which corresponds to taking $p=3$ in our general
expressions. If $\epsilon$ denotes a Killing spinor of the background, only 
those $\epsilon$ which satisfy
\be
\Big[\,\cos\a\,\G_{\underline{x^0r}}\s_3\,+\,
\sin\a\,\G_{\underline{x^0x^4x^5r}}(i\s_2)\,\Big]\,
\epsilon_{{\,\big |}_{\th=0}}\,=\,\epsilon_{{\,\big |}_{\th=0}}\,,
\label{fl2cisiete}
\ee
generate a supersymmetry transformation which leaves our  configuration
invariant. In eq. (\ref{fl2cisiete}) $\a$ is given by
\be
\sin\a\,=\,{l\over [\mu_{(l,m)}]^{1/2}}\,
H_{(l,m)}^{1/4}\,h_{(l,m)}^{1/2}\,\,,
\,\,\,\,\,\,\,\,\,\,\,\,\,\,\,\,\,\,\,\,\,\,\,
\cos\a\,=\,{mg_s\over [\mu_{(l,m)}]^{1/2}}\,
H_{(l,m)}^{-1/4}\,h_{(l,m)}^{1/2}\,\,.
\label{fl2ciocho}
\ee
The supersymmetry projection (\ref{fl2cisiete}) certainly corresponds to that of a 
(F, D3) bound state of the type described above, with $\a$ being the mixing
angle.

\medskip
\subsection{Stability}
\medskip

The static configurations of the D(8-p)-branes studied above are stable
under small perturbations, as one can check following the same steps
as in section \ref{flncstability}. We shall consider for simplicity the 
case of $p=5$. First of all, we parametrize the angle
fluctuations as
\be
\th\,=\,\bar\th_{5,n}\,+\,\xi\,\,,
\label{fl1ctvnueve}
\ee
whereas the gauge field fluctuates as:
\be
{\cal F}\,=\,\big[\,f_{12}(\th)\,+\,g\,\big]\,\epsilon_{(2)}\,+\,
\big[\,\bar F_{0,r}\,+\,f\,\big]\,dt\wedge dr\,\,.
\label{fl1cttreinta}
\ee
The angle fluctuation $\xi$ and the electric (magnetic) field
fluctuation $f$ ($g$) are supposed to be small and only terms up to
second order are retained in the lagrangian. The corresponding
equations of motion involve now the wave operator 
${\cal O}_5^{(l,m)}$, which acts on any function $\psi$  as
\be
{\cal O}_5^{(l,m)}\,\psi\,\equiv\,R^2_{(p,q)}\,\partial_0^2\,\psi\,-\,
\partial_r(r^2\partial_r\psi)\,\,.
\label{fl1cttuno}
\ee
Notice that ${\cal O}_5^{(l,m)}$ is obtained from ${\cal O}_5$ in eq. 
(\ref{fl1ssiete}) by means of the substitution 
$R\rightarrow R_{(p,q)}$. Let us combine $\xi$, $f$ and $g$ into the
field $\eta$, defined as
\br
\eta&\equiv&{1\over l^2\,H_{(l,m)}(r)\,\,+\,m^2\sin^2\bar\th_{5,n}}
\,\,\times\rc\rc
&&\times\Big[\,[\mu_{(l,m)}]^{{1\over 2}}\,
\big(\,m\sin\bar\th_{5,n}\,f\,+\,{l\over r^2}\,g\,\big)\,+\,
2m^2\sin^2\bar\th_{5,n}\,\xi\,\Big]\,\,,
\label{fl1cttdos}
\er
and let us expand $\xi$ and $\eta$ is spherical harmonics of $S^2$. If 
$\zeta_{j,n}$ and $\eta_{j,n}$ denote their modes respectively, one
can prove after some calculation that the equations of motion for 
$\zeta_{j,k}$ and $\eta_{j,k}$ can be written as
\be
\Big(\,{\cal O}_5^{(l,m)}\,+\,{\cal M}_5\,\Big)\,
\pmatrix{\zeta_{j,k}\cr\eta_{j,k}}\,=\,0\,\,. 
\label{fl1ctttres}
\ee
In eq. (\ref{fl1ctttres}) ${\cal M}_5$ is the matrix defined in eq. 
(\ref{fl1setenta}), whose eigenvalues, as proved in section 2.3, are always
non-negative. There is also a decoupled mode $\sigma$, whose
expression in terms of $\xi$, $f$ and $g$ is
\br
\s&\equiv&
{\sin\bar\th_{5,n}
\over r[\,
l^2\,H_{(l,m)}(r)\,\,+\,m^2\sin^2\bar\th_{5,n}\,]}
\,\,\times\rc\rc
&&\times\Big[\,[\mu_{(l,m)}]^{{1\over 2}}\,
\big(\,m\sin\bar\th_{5,n}\,g\,-\,l\,R^2_{(l,m)}\,f\,\big)\,-\,
2lm\,R^2_{(l,m)}\sin\bar\th_{5,n}\,\xi\,\Big]\,\,.
\label{fl1cttcuatro}
\er
The equation of motion of $\s$ can be written as
\be
\Big(\,{\cal O}_5^{(l,m)}\,+\,j(j+1)\,\Big)\,\s_{j,k}\,=\,0\,\,,
\label{fl1cttcinco}
\ee
where $\s_{l,m}$ are the modes of the expansion of $\s$ in
$S^2$-spherical harmonics. It is evident from eq. (\ref{fl1cttcinco})
that the mass eigenvalues of  $\s_{l,m}$ are non-negative, which
confirms that the configurations around which we are expanding are
stable.

\setcounter{equation}{0}
\section{Flux stabilization of  M5-branes} \label{secflM5}
\medskip

Our purpose in this section is to study the mechanism of flux
stabilization in M-theory. We shall consider, in particular, a M5-brane
probe in a M5-brane background. By using the Pasti-Sorokin-Tonin (PST)
action for the M5-brane probe, discussed in section \ref{secPSTact}, 
we shall look for static
configurations in which the probe is wrapped on a three-sphere. After
establishing a flux quantization condition similar to (\ref{flbqc}), we
shall find these configurations and we will show that they closely
resemble  those found for the D4-D4 system. Actually, our states can be
interpreted semiclassically as BPS bound states of M2-branes and they are
related to the short distance limit of the baryonic vertex of
M-theory \cite{Ali,kappa}.

We shall consider the solution of
the equations of motion of eleven dimensional supergravity associated 
to a stack of $N$ parallel M5-branes, given in section \ref{strsecmtbs}. 
If we decompose the $S^4$ line element $d\O_4^2$ as in eq. (\ref{fl1cinco}) 
and use the same orientation conventions as in the preceding sections, 
one can readily check that $C^{(3)}$ can be taken as
\be
C^{(3)}\,=\,-R^3\,C_4(\th)\epsilon_{(3 )}\,\,,
\label{fl1stseis}
\ee
where $C_4(\th)$ is the function
defined in  eqs. (\ref{fl1seis}) and (\ref{fl1siete}), namely, 
$C_4(\th)\,=\,\cos\th\sin^2\theta\,+\,2(\cos\th\,-\,1)$.

We will put in this background a probe M5-brane, whose action will be
given by the PST formalism (\ref{actPST}). The fields of this
formalism include a three-form field strength $F$, whose potential is a
two-form field $A$ (\ie\ $F=dA$) and a scalar field $a$ (the PST scalar).
In the PST action the field strength $F$ is be combined with (the pullback of) 
the background potential $C^{(3)}$ to form the field $H\,=\,F\,-\,C^{(3)}$.

We will extend our M5-brane probe along the  directions transverse to
the M5-branes of the background and along one of the directions
parallel to them. Without loss of generality we will take the latter to
be the $x^5$ direction. Accordingly, our worldvolume coordinates
$\s^{\a}$ will be taken to be
\be
\s^{\a}\,=\,(\,t,r,x^5,\th^1,\th^2,\th^3)\,\,,
\label{fl1ouno}
\ee
and the embedding of the M5-brane probe is determined by a function
$\th\,=\,\th(\,\s^{\a}\,)$. As in the case of the RR
background, we shall mainly look for solutions with 
$\th\,=\,{\rm constant}$, which represent a M5-brane wrapped on a
three-sphere and extended along the $r$ and $x^5$ directions. 

As discussed in section \ref{secPSTact}, the scalar  $a$ is an auxiliary
field which can be eliminated from the action by fixing its  gauge
symmetry. The price one must pay for this gauge fixing is the loss of
manifest covariance. A particularly convenient choice for $a$  is
\be
a\,=\,x^5\,\,.
\label{fl1odos}
\ee
In this gauge the components of the worldvolume potential $A$ with
$x^5$ as one of its indices can be gauge fixed to zero.
Moreover, if we consider configurations of $A$ and of the embedding angle
$\theta$ which are independent of $x^5$, one readily realizes that the
components of the three-forms $F$ and $H$ along $x^5$ also vanish and,
as  a consequence, only the square root term of the PST action 
(\ref{actPST}) is non-vanishing. As we will verify soon this
constitutes a great simplification.

\medskip
\subsection{Quantization condition and M5-brane configurations} \label{flM2}
\medskip

In order to find stable $S^3$-wrapped configurations of the M5-brane
probe, we need to switch on a non-vanishing worldvolume field which
prevents the collapse to one of the poles of the $S^3$. As in
section \ref{flbachas} (and ref. \cite{Bachas}) the value of this worldvolume 
field is determined by some quantization condition. This condition follows 
from an analysis similar to the one in section \ref{flbachas} by coupling an 
M5-brane to an open M2-brane. 

The worldvolume of an open M2-brane is given by a three-manifold $\S$ whose 
boundary $\partial\S$ lies on the worldvolume of an M5-brane. For simplicity 
we shall consider the  case in which  $\partial\S$ has only one component and 
we get the representation given by figure \ref{flfig3}, with the $S^2$ changed 
by an $S^3$. Again, $\partial\Sigma$ is also the boundary of some disk $D$ on 
the worldvolume of the M5-brane and $\hat\S$ is a four-manifold with $\partial
\hat\S\,=\,\S\,+\,D$. The coupling of the brane to the supergravity background 
and to the M5-brane is described by an
action of the form

\be
S_{int}\,[\,\hat\S, D\,]\,=\,
T_{M2}\,\int_{\hat\S}\,F^{(4)}\,+\,T_{M2}\,\int_{D}\,H\,\,,
\label{fl1otres}
\ee
where $T_{M2}$ is the tension of the M2-brane, given by
\be
T_{M2}\,=\,{1\over (2\pi)^2\,\lp^3}\,\,.
\label{fl1ocuatro}
\ee
In a topologically trivial situation, if we represent $F^{(4)}$ as 
$dC^{(3)}$ and $F=dA$,  the above action reduces to the more familiar
expression
\beq
S_{int}\,=\,T_{M2}\,\int_{\Sigma}\,C^{(3)}\,+
\,T_{M2}\,\int_{\partial D}\,A\,\,.
\label{fl1ocinco}
\eeq

Let us consider the case in which we attach the M2-brane to a 
M5-brane worldvolume which has some  submanifold with the topology of $S^3$. 
The two possible choices for the disk in eq. (\ref{fl1otres}), the 
internal disk $D$ or the external disk $D'$, change $\hat\S\rightarrow
\hat\S'$, with $\partial \hat\Sigma'\,=\,\Sigma\,+\,D'$. We require that

\be
S_{int}\,[\,\hat\S', D'\,]\,-\,S_{int}\,[\,\hat\S, D\,]\,=\,
2\pi n\,\,,
\label{fl1osiete}
\ee
with $n\in\ZZ$. The left-hand side of eq. (\ref{fl1osiete}) is:
\be
S_{int}\,[\,\hat\S', D'\,]\,-\,S_{int}\,[\,\hat\S, D\,]\,=\,
T_{M2}\,\int_{\hat{\cal B}}\,F^{(4)}\,+\,T_{M2}\,
\int_{\partial\hat{\cal B}}\,H\,\,.
\label{fl1oocho}
\ee
where $\hat{\cal B}$ is the 4-ball bounded by $D'\cup(-D)\,=\,S^3$, 
and we get the condition
\be
\int_{\hat{\cal B}}\,F^{(4)}\,+\,
\int_{\partial\hat{\cal B}}\,H\,=\,{2\pi n\over T_{M2}}\,,
\,\,\,\,\,\,\,\,\,\,\,\,\,\,\,\,\,
n\in\ZZ\,\,.
\label{fl1onueve}
\ee
If $F^{(4)}$ can be represented as $dC^{(3)}$ on $\hat{\cal B}$, the
first integral on the left-hand side of eq. (\ref{fl1onueve})
can be written as an integral of $C^{(3)}$ over 
$\partial\hat{\cal B}\,=\,S^3$. Our parametrization of $C^{(3)}$ 
(eq. (\ref{fl1stseis})) is certainly non-singular if we  are
outside of the poles of the $S^4$. If this is  the case we get the
quantization condition
\be
\int_{S^3}\,F\,=\,{2\pi n\over T_{M2}}\,\,,
\,\,\,\,\,\,\,\,\,\,\,\,\,\,\,\,\,
n\in\ZZ\,\,.
\label{fl1noventa}
\ee
Eq. (\ref{fl1noventa}), which is the M-theory analogue of eq. 
(\ref{flbqc}), is the quantization condition we were looking for. It is 
very simple  to obtain a solution of this equation. Let us take
$F$ proportional to the volume element $\epsilon_{(3)}$ of  the $S^3$.
Taking into account that the volume of the unit three-sphere is
$\Omega_3\,=\,2\pi^2$ (see eq. (\ref{fl1dnueve})), we can write down
immediately the following solution of eq. (\ref{fl1noventa}):
\be
F\,=\,{n\over \pi T_{M2}}\,\,\epsilon_{(3)}\,\,.
\label{fl1nuno}
\ee
We can put this solution in a more convenient form if we use the
following relation between the M2-brane tension and the radius $R$
\be
T_{M2}\,=\,{N\over 4\pi R^3}\,\,, 
\label{fl1ndos}
\ee
which follows from eqs. (\ref{strm5rad}) and (\ref{fl1ocuatro}). By using 
eq. (\ref{fl1ndos}), one can rewrite eq. (\ref{fl1nuno}) as
\be
F\,=\,4R^3\,{n\over N}\,\epsilon_{(3)}\,\,.
\label{fl1ntres}
\ee
We can  use the ansatz (\ref{fl1ntres}) and the potential $C^{(3)}$ of
eq. (\ref{fl1stseis}) to compute the three-form field $H$ of eq. 
(\ref{actH}). It turns out that the result for $H$ can be written
in terms of the function ${\cal C}_{4,n}(\theta)$ defined in eq. 
(\ref{fl1vuno}). One gets
\be
H\,=\,R^3\,{\cal C}_{4,n}(\theta)\,\epsilon_{(3)}\,\,.
\label{fl1ncuatro}
\ee

Let us now assume that the angle $\theta$ characterizing the M5-brane
embedding only depends on the radial coordinate $r$ and, as
before, let us denote by $\theta'$ to the derivative $d\theta/dr$. As
was mentioned above, in the $a=x^5$ gauge and for this kind of
embedding, only the first term of the PST action (\ref{actPST}) is
non vanishing and, as a consequence, all the dependence on $H$ of this
action comes through the field $\tilde H$ defined in eq.
(\ref{acttilH}). Actually, the only non-vanishing component of
$\tilde H$ is
\be
\tilde H_{0r}\,=\,-{i\over (\sin\theta)^3}\,
\sqrt{{R\over r}}\,\sqrt{1\,+\,r^2\theta'^2}\,\,
{\cal C}_{4,n}\,(\theta)\,\,.
\label{fl1ncinco}
\ee
After a simple calculation one can obtain the induced metric $g$ and,
using eq. (\ref{fl1ncinco}), the lagrangian density of the M5-brane. The
result is
\be
{\cal L}\,=\,-T_{M5}\,R^3\,\sqrt{\hat g}\,\,
\sqrt{1\,+\,r^2\theta'^2}\,
\sqrt{\,(\sin\theta)^6\,+\,(\,{\cal C}_{4,n}(\theta)\,)^2}\,\,,
\label{fl1nseis}
\ee
where $\hat g$ is the determinant of the metric of a unit 3-sphere. 
Notice the close similarity of this result and the hamiltonian density
of eq. (\ref{fl1vcuatro}) for $p=4$, \ie\ for the D4-D4 system. Indeed,
it is immediate to check that the solutions with constant $\theta$ are
the same in both systems, \ie\ $\theta=\bar\theta_{4,n}$ with $0<n<N$,
where  $\bar\theta_{4,n}$ is given in eq. (\ref{fl1cuno}) (for $n=0,N$
we have the singular solutions with $\theta=0,\pi$). This result is
quite natural since the D4-D4 system can be obtained from the M5-M5
one by means of a double dimensional reduction along the $x^5$
direction. The energy density for these solutions can be easily
obtained from the lagrangian (\ref{fl1nseis}). One gets
\be
{\cal E}_n^{M5}\,=\,{n(N-n)\over N}\,T_{M2}\,\,,
\label{fl1nsiete}
\ee
which, again, closely resembles   the D4-D4 energy of eq.
(\ref{fl1cdos}). In particular ${\cal E}_n^{M5}\rightarrow n\,T_{M2}$ as
$N\rightarrow\infty$, which implies that, semiclassically, our
configurations can be regarded as bound states of M2-branes. Moreover,
one can check that eq. (\ref{fl1ciuno}) with $p=4$ is a BPS condition for
the M5-brane system. The integration of this equation can be read from
eq.  (\ref{fl1citres}) and represents a baryonic vertex in M-theory
\cite{kappa}, $n$ being the number of M2-branes which form the baryon at
$r\rightarrow\infty$. The $\theta=\bar\theta_{4,n}$ solution can be
obtained as the $r\rightarrow 0$ limit of the M-theory baryon, in
complete analogy with the analysis at the end of section 2.2.

\medskip
\subsection{Fluctuations and stability}
\medskip

We will now perturb our static solution in order to check its
stability. We must allow the angle $\theta$ to
deviate from $\bar\theta_{4,n}$ and the worldvolume field strength
$F$ to vary from the value displayed in eq. (\ref{fl1ntres}). The best
way to find out which components of $F$ must be included in the 
perturbation is to choose a gauge. As  $F$ in eq. (\ref{fl1ntres}) has
only components along the sphere $S^3$, one can represent it by means
of a potential $\bar A_{\hat i\,\hat j}$ which also has component only
on $S^3$ (in what follows indices along $S^3$  will be denoted with a
hat). Accordingly, the perturbation of $F$ will be parametrized as a
fluctuation of the $S^3$-components of the potential $A$. Thus, we put
\beq
\theta\,=\,\bar\theta_{4,n}\,+\,\xi\,\,,
\,\,\,\,\,\,\,\,\,\,\,\,\,\,\,\,\,\,\,\,
A_{\hat i\,\hat j}\,=\,\bar A_{\hat i\,\hat j}\,
+\,\alpha_{\hat i\,\hat j}\,\,,
\label{fl1nocho}
\eeq
where $\xi$ and $\alpha_{\hat i\,\hat j}$ are small. For simplicity 
we shall assume that $\xi$ and the $\alpha_{\hat i\,\hat j}$'s do
not depend on $x^5$. Using the parametrization of $A$ in eq. 
(\ref{fl1nocho}), it is clear that the $S^3$-components of the three-form
field $H$ can be written as
\beq
H_{\hat i\,\hat j\,\hat k}\,=\,R^3\,
[\,{\cal C}_{4,n}(\theta)\,+\,f\,]\,\,
\,\,\,{\epsilon_{\hat i\,\hat j\,\hat k}\over \sqrt{\hat g}}\,\,,
\label{fl1nnueve}
\eeq
where $f$ can be put in terms of derivatives of the type 
$\partial_{\hat i}\,\alpha_{\hat j\,\hat k}$. In eq. (\ref{fl1nnueve})
$\hat g$ is the determinant of the metric of the $S^3$ and we are
using the convention 
$\epsilon^{\hat1\, \hat2\,\hat 3\,}\,=\,
\epsilon_{\hat 1\, \hat 2\,\hat 3\,}/{\hat g}\,=\,1$. As 
$\alpha_{\hat i\,\hat j}$ in (\ref{fl1nocho}) depends on $t$ and $r$, it
follows that we have now non-zero components 
$H_{0\hat i\,\hat j}\,=\,\partial_0\,\alpha_{\hat i\,\hat j}$ 
and $H_{r\hat i\,\hat j}\,=\,\partial_r\,\alpha_{\hat i\,\hat j}$.
Thus, in the gauge (\ref{fl1odos}), the non-vanishing components of
$\tilde H$ are $\tilde H_{0r}$, 
$\tilde H_{0\hat i}$ and $\tilde H_{r\hat i}$. To the relevant order,
these components take the values
\bear
\tilde H_{0r}&=&-i\,\sqrt{{R\over r}}\,\cot\bar\theta\,+\,
\,{i\over (\sin\bar \theta)^2}\,
\,\sqrt{{R\over r}}\,\Big(\,3\xi\,-\,{f\over \sin\bar\theta}
\,\Big)\,
-3i\,\,{\cos\bar\theta \over (\sin\bar\theta)^3}\,\,
\sqrt{{R\over r}}\,\Big(\,2\xi^2\,-\,\xi\,{f\over \sin\bar\theta}
\,\Big)\,\,+\rc\rc
&&\,+\,{i\over 2}\,R^2\,\cot\bar\theta\Bigg[\,\sqrt{{R^3\over r^3}}\,
(\partial_t\xi)^2\,-\,\sqrt{{r^3\over R^3}}\,
 (\partial_r\xi)^2\,\Bigg]\,+\,{i\over 2}\, 
\sqrt{{R\over r}}{\cos\bar\theta \over
(\sin\bar\theta)^3}\,  (\partial_{\hat i}\xi)^2\,\,, \rc\rc
\tilde H_{0\hat i}&=&{i\over 2R\sin\bar\theta}\, \,
\sqrt{{r^3\over R^3}}\,\,\hat g_{\hat i\,\hat j}\,\,
{\epsilon^{\hat j\,\hat l\,\hat m}\,\over \sqrt{\hat g}}\,\,
H_ {r\,\hat l\,\hat m}\,\,,\rc\rc
\tilde H_{r\hat i}&=&{i\over 2R\sin\bar\theta}\, 
\sqrt{{R^3\over r^3}}\,\,\hat g_{\hat i\,\hat j}\,\,
{\epsilon^{\hat j\,\hat l\,\hat m}\,\over \sqrt{\hat g}}\,\,
H_ {0\,\hat l\,\hat m}\,\,,
\label{fl1cien}
\eear            
with $\bar\theta\equiv \bar\theta_{4,n}$. Using these results we can
compute the lagrangian for the fluctuations. After some calculation
one arrives at
\bear
{\cal L}\,&=&\,-\,\sqrt{\hat g}\,R^3\,T_{M5}\,\cos\bar\theta\,f\,+\,
\sqrt{\hat g}\,R^3\,T_{M5}\,\big(\,\sin\bar\theta\,\big)^2
\,\,\times\rc\rc
&&\times\,{1\over 2}\,\, \Bigg[\,R^3r^{-1}(\partial_t\xi)^2\,-\,
r^2\,(\partial_r\xi)^2\,-\,(\partial_{\hat i}\xi)^2\,
+\,{1\over 2R^3r(\sin\bar\theta)^2}\,(H_{0\hat j\,\hat k})^2\,-\,\rc\rc
&&-{r^2\over 2R^6(\sin\bar\theta)^2}\,(H_{r\hat j\,\hat k})^2\,-\,
6\xi^2\,-\,{f^2\over (\sin\bar\theta)^2}\,
+\,6\,{f\xi\over \sin\bar\theta}\,\Bigg]\,\,,\rc
\label{fl1ctuno}
\eear
where $(H_{0\hat j\,\hat k})^2$ and $(H_{r\hat j\,\hat k})^2$ are
contractions with the metric of the $S^3$. In eq. (\ref{fl1ctuno}) we have
kept terms up to second order and we have dropped the zero-order term. 

The analysis of the equations of motion derived from eq.
(\ref{fl1ctuno}) is similar to the one performed in section 2.3. For
this reason we will skip the details and will give directly the final
result. Let us expand $f$ and $\xi$ is spherical harmonics of $S^3$ as
in eq. (\ref{fl1cinueve}) and let  $f_{l,m}(t,r)$ and $\zeta_{l,m}(t,r)$
denote their modes respectively. The equations of motion of these
modes can be written as
\beq
\Bigg(R^3r^{-1}\partial_0^2\,-\,\partial_r\,r^2\,\partial_r\,
+\,{\cal M}_4\,\Bigg)\,
\pmatrix{\zeta_{l,m}\cr\cr
         {f_{l,m}\over \sin\bar\theta}}\,=\,0\,\,,
\label{fl1ctdos}
\eeq
where the mass matrix ${\cal M}_4$ is the same as
that corresponding the  D4-D4 system (\ie\ the one of eq. 
(\ref{fl1setenta}) for $p=4$). Notice that the wave operator on the
left-hand side of eq. (\ref{fl1ctdos}) is formally the same as 
${\cal O}_4$ in eq. (\ref{fl1ssiete}) (although the radius $R$ is not the
same quantity in both cases). Thus, the eigenvalues of the mass matrix
are non-negative and, actually, the same as in the D4-D4 system.
Therefore our static M-theory configurations are indeed stable.

\setcounter{equation}{0}
\section{Discussion}  \label{flconcl}
\medskip

In this chapter we studied certain  configurations of branes  which
are partially wrapped on spheres and also extended, in the case of 
non-threshold bound state backgrounds, in some directions parallel to the 
bound state. These spheres are placed on the transverse region of the 
supergravity background, and their positions, characterized by a polar 
angle which measures their latitude in a system of spherical coordinates, 
are quantized and given by a very specific set of values. It was checked 
that our configurations are stable by analyzing their behaviour under small 
fluctuations and, by studying their energy, it was concluded that they can be 
regarded as a bound state of strings or, in the cases of the M5-M5 system, 
M2-branes, and of the (NS5,Dp) bound states, (F,D(6-p))-branes. We have 
verified this fact explicitly in appendix 4B for the case of a wrapped 
D3-brane in the background of a NS5-brane. Indeed, we have proved that, 
by embedding a D1-brane in a fuzzy two-sphere in the NS5-brane background, 
one obtains exactly the same energies and allowed polar angles as for a 
wrapped D3-brane in the same geometry. Clearly, a similar description of 
all the cases studied here would be desirable and would help to understand
more precisely the r\^ole of noncommutative geometry in the formation of
these bound states. In this sense it is interesting to point out that
the polarization of multiple fundamental strings in a RR background was
studied in ref. \cite{Schiappa}

Contrary to ref. \cite{Bachas}, the problems treated here do not have a
CFT description  to compare with. Thus, we do not know to what
extent we can trust our Born-Infeld results. However, one could argue
that it was followed the same methodology as in ref. \cite{Bachas} and,
actually, the configurations obtained can be connected to the ones in
\cite{Bachas} by string dualities. Moreover, the BPS nature of these 
configurations makes one reasonably confident of the correctness of the  
conclusions. 

The presence of a non-trivial supergravity background is of crucial
importance in the analysis. Indeed, these backgrounds induce worldvolume
gauge fields on the brane probes, which prevent their collapse. The
stabilization mechanisms found here are a generalization of the one
described in refs. \cite{Bachas, Pavel}, and are based on a series of
quantization rules which determine the values of the worldvolume gauge
fields.

It seems that the general
rule to find flux-stabilized configurations in a non-threshold bound state
background is to consider probes which are also extended in the directions
parallel to  the bound state in such a way that the probe could capture the flux
of the background gauge fields. However, nothing guarantees that the
corresponding configurations are  free of pathologies. To illustrate this fact,
let us consider the case of the background generated by  a (F,Dp) bound state. 
The string frame metric and dilaton for this bound state are (\ref{strcinueve})
\br
ds^2&=&f_p^{-1/2}\,h_p^{-1/2}
\Big[\,-(\,dx^0\,)^2\,+\,(\,dx^{1}\,)^2\,+\,
\,h_p\,\Big((\,dx^{2}\,)^2\,+\,\dots\,+\,\,(\,dx^{p}\,)^2\Big)\,\Big]\,+\rc\rc
&&+\,f_p^{1/2}\, h_p^{-1/2}\,
\Big[\,dr^2\,+\,r^2\,d\O_{8-p}^2\,\Big]\,\,,\rc\rc
e^{\tilde\p}&=&f_p^{{3-p\over 4}}\,\,h_p^{{p-5\over 4}}\,\,,
\label{fl2cinueve}
\er
while the $B$ field is 
$B=\sin\varphi\,\,f_p^{-1}\,\,dx^{0}\wedge dx^{1}$ and the RR potentials are
\br
C^{(7-p)}_{\th^1,\dots,\th^{7-p}}
&=&-\,\cos\varphi\,R^{7-p}\,C_p(\th)\,
\sqrt{\hat g^{(7-p)}}\,\,, \rc\rc
C^{(9-p)}_{x^{0},x^1,\th^1,\dots, \th^{7-p}}
&=&-\,\sin\varphi\,\cos\varphi\,R^{7-p}\,
f_p^{-1}\,C_p(\th)\,
\sqrt{\hat g^{(7-p)}}\,\,.
\er
According to our rule we should place a D(10-p)-brane 
probe extended along
$(t,x^{1},x^{2},r,\\\th^1,\cdots,\th^{7-p})$.
Moreover, we will adopt the ansatz
${\cal F}\,=\,{\cal F}_{0,1}\,dt\wedge dx^1\,
+\,  F_{2,r}\,dx^{2}\wedge dx^{r}$ for the gauge field, 
with ${\cal  F}_{0,1}\,=\,  F_{0,1}\,-\,f_p^{-1}\sin\varphi$ with constant
values of  $F_{0,1}$ and $F_{2,r}$. Following the same steps as in our previous
examples, we obtain that there exist  constant $\theta$ configurations if
$F_{0,1}\,=\,-\cos^2\varphi/\sin\varphi$. This is an overcritical field 
which makes negative the argument of the square root of the Born-Infeld term of
the action and, as a consequence, the corresponding value of $F_{2,r}$
is imaginary, namely  $F_{2,r}=-i\cos\th$. These configurations are clearly
unacceptable.

\section{Appendix 4A}                                 
\renewcommand{\theequation}{\rm{A}.\arabic{equation}}  
\medskip 
\setcounter{equation}{0}

In this appendix we collect the expressions of the functions 
$\L_{p,n}(\th)$ for $0\le p\le 5$. They are
\br
\L_{0,n}(\th)&=&-{2\over 5}\,\Big[\,\cos\th\,\Big(\,
3\sin^4\th\,+\,4\sin^2\th\,+\,8\,\Big)\,+\,8\,\Big(\,
2\,{n\over N}\,-\,1\Big)\,\Big]\,\,,\rc\rc
\L_{1,n}(\th)&=&-{5\over 4}\,\Big[\,\cos\th\,\Big(\,
\sin^3\th\,+\,{3\over 2}\,\sin\th\,\Big)\,+\,
{3\over 2}\,\Big(\,{n\over N}\,\pi\,-\,\th\,\Big)\,\Big]\,\,, \rc\rc
\L_{2,n}(\th)&=&-{4\over 3}\,\Big[\,\cos\th\,\Big(\,
\sin^2\theta\,+\,2\,\Big)\,+\,2\,\Big(\,
2\,{n\over N}\,-\,1\Big)\,\Big]\,\,,\rc\rc
\L_{3,n}(\th)&=&-{3\over 2}\,\Big[\,\cos\th\,\sin\th\,+\,
{n\over N}\,\pi\,-\,\th\,\Big]\,\,,\rc\rc
\L_{4,n}(\th)&=&- 2\,\Big[\,\cos\th\,+\,
2\,{n\over N}\,-\,1\Big]\,\,,\rc\rc
\L_{5,n}(\th)&=&\th\,-\,{n\over N}\,\pi\,\,.
\label{fl1apauno}
\er
The functions ${\cal C}_{p,n}(\th)$ and $ C_p(\th)$ can be
easily obtained from (\ref{fl1apauno}) by using their relation with the 
$\L_{p,n}(\th)$'s (see eqs. (\ref{fl1vseis}) and 
(\ref{fl1vuno})).

\section{Appendix 4B}                                 
\medskip                                             
\renewcommand{\theequation}{\rm{B}.\arabic{equation}}  
\setcounter{equation}{0}

In this appendix we will show how one can represent the wrapped branes
studied in the main text as a bound state of strings. We will make use
of the Myers polarization mechanism \cite{Myers}, in which the strings
are embedded in a noncommutative space. Actually, it will only be considered a
particular case of those analyzed in sects. \ref{flbachas}-\ref{secflM5}, 
namely the one of
section \ref{flns5dpbs} with $l=1$, $m=0$, \ie\ the D3-brane in the
background of the NS5-brane. For convenience we will choose a new set
of coordinates to parametrize the space transverse to the NS5. Instead
of using the radial coordinate $r$ and the three angles $\theta^1$, 
 $\th^2$ and  $\th$ (see eq. (\ref{fl1cinco})), we will work with
four cartesian coordinates $z,x^1, x^2, x^3$, which, in terms of the
spherical coordinates, are given by 
\br
z&=&r\cos\theta\,\,,\cr
x^1&=&r\sin\th\cos\th^2\,\,,\cr
x^2&=&r\sin\th\sin\th^2\cos\th^1\,\,,\cr
x^2&=&r\sin\th\sin\th^2\sin\th^1\,\,.
\label{fl1apbuno}
\er
Conversely, $r$ and $\th$ can be put in terms of the new
coordinates as follows:
\br
r\,&=&\,
\sqrt{(z)^2\,+\,(x^1)^2\,\,+(x^2)^2\,\,+(x^3)^2\,\,}\,\,,\cr\cr\cr
\tan\th&=&{\sqrt{(x^1)^2\,\,+(x^2)^2\,\,+(x^3)^2\,\,}\over z}\,\,.
\label{fl1apbdos}
\er
In what follows some of our expressions will contain $r$ and $\th$.
It should be understood that they are given by the functions of 
$(\,z,x^i\,)$ written in eq. (\ref{fl1apbdos}). The near-horizon metric
and the dilaton for a stack of N NS5-branes are (see eq. (\ref{flbdos}) 
or (\ref{strns5dpbs2}) with $l=1, m=0$)
\br
ds^2&=&-dt^2\,+\,dx_{\parallel}^2\,+\,
{N\a'\over r^2}\,\,
\Big(\,(dz)^2\,+\,(dx^1)^2\,+\,(dx^2)^2\,+\,(dx^3)^2\,\Big)\,\,,\rc\rc
e^{-\phi}\,&=&\,{r\over \sqrt{N\a'}}\,\,.
\label{fl1apbtres}
\er
Moreover, the non-vanishing components of the $B$ field in the new 
coordinates can be obtained from eq. (\ref{flbB}). They are 
\be
B_{x^ix^j}\,=\,-N\a'\,{C_5(\th)\over r^3\sin^3\th}\,\,
\epsilon_{ijk}\,x^k\,\,.
\label{fl1apbcuatro}
\ee

According to our analysis of section \ref{flns5dpbs}, the wrapped D3-brane in this
background can be described as a bound state of D1-branes. Thus it is
clear that we must consider a system of $n$ D1-branes, moving in the
space transverse to the stack of $N$ Neveu-Schwarz fivebranes. We will
employ a static gauge where the two worldsheet coordinates will be
identified with $t$ and $z$. The Myers proposal for the action of this
system  was given in eq. (\ref{biactcom}). The worldsheet gauge field 
strength $F_{ab}$ is zero in our case since there is no source term for it. 
The indices $a,b,\dots$ correspond to directions
parallel to the worldsheet (\ie\ to $t$ and $z$), whereas $i,j\dots$
refer to directions transverse to the D1-brane probe. 

Let us now make the standard identification between the transverse
coordinates $x^i$ and the scalar fields $\p^i$, namely,
\be
x^i\,=\,\l\,\phi^i\,\,,
\label{fl1apbocho}
\ee
where $\l=2\pi\a^{\prime}$. Notice that, after the identification 
(\ref{fl1apbocho}), the $x^i$'s become noncommutative coordinates 
represented by matrices, since the transverse scalars $\p^i$ are 
matrices taking values in the adjoint representation of U(n). Actually,
as in ref. \cite{Myers}, we will make the following ansatz for the scalar
fields:
\be
\phi^i\,=\,{f\over 2}\,\a^i\,\,,
\label{fl1apbnueve}
\ee
where $f$ is a c-number to be determined and
the $\alpha^i$'s are $n\times n$ matrices corresponding to the
$n$-dimensional irreducible representation of $su(2)$:
\be
[\,\a^i\,,\,\a^j\,]\,=\,2i\epsilon_{ijk}\,\a^k\,\,.
\label{fl1apbdiez}
\ee
As the quadratic Casimir of the $n$-dimensional irreducible
representation of $su(2)$ is $n^2-1$, we can write
\be
(\a^1)^2\,\,+(\a^2)^2\,\,+(\a^3)^2\,=\,
(n^2\,-\,1)\,I_n\,\,,
\label{fl1apbonce}
\ee
where $I_n$ is the $n\times n$ unit matrix. By using eqs.
(\ref{fl1apbocho}) and (\ref{fl1apbnueve}) in (\ref{fl1apbonce}), we get
\be
(x^1)^2\,\,+(x^2)^2\,\,+(x^3)^2\,=\,{\l^2\,f^2\over 4}\,
(n^2\,-\,1)\,I_n\,\,,
\label{fl1apbdoce}
\ee
which shows that, with our ansatz, the $x^i$'s are coordinates of a
fuzzy two-sphere of radius $\l\,f\,\sqrt{n^2-1}/2$. On the other
hand, if we treat the $x^i$'s as commutative coordinates, it is easy
to conclude from eqs. (\ref{fl1apbuno}) and (\ref{fl1apbdos}) that the
left-hand side of (\ref{fl1apbdoce}) is just $(r\sin\th)^2$. In view of
this, when the  $x^i$'s  are non-commutative  we should identify the
expression written in eq. (\ref{fl1apbdoce}) with 
$(r\sin\th)^2\,I_n$. Thus, we put
\be
{f\over 2}\,=\,{r\sin\th\over \l\sqrt{n^2\,-\,1}}\,\,.
\label{fl1apbtrece}
\ee
Notice that, as can be immediately inferred from eq. (\ref{fl1apbdos}),
$r$ and $\th$ depend on the $x^i$'s through the sum
$\sum_i\,(x^i)^2$, which is proportional to the $su(2)$ quadratic
Casimir. Then, as matrices, $r$ and $\theta$ are  multiple of  the
unit matrix and, thus, we can consider them as commutative
coordinates. This, in particular,  means that the  elements of the
metric tensor $G_{\m\n}$ are also commutative, whereas, on the
contrary, the components of the $B$ field have a non-trivial matrix
structure. By substituting our ansatz in eqs. (\ref{fl1apbtres}) and 
(\ref{fl1apbcuatro}), we get for the transverse components of the 
$E_{\m\n}$ tensor, defined in (\ref{eee}), the following expression:
\be
E_{ij}\,=\,{N\a'\over r^2}\,\Big[\,
\d^i_{\,\,j}\,\,+\,\,{1\over\sqrt{n^2\,-\,1}}\,\,\,
{C_5(\th)\over \sin^2\th}\,\,
\epsilon_{ijk}\,\a^k\,\Big]\,\,.
\label{fl1apbcatorce}
\ee
The quantities $Q^i_{\,\,j}$, defined in eq. (\ref{myqij}), can be
readily obtained from eq. (\ref{fl1apbcatorce}), namely, 
\be
Q^i_{\,\,j}\,=\,\Big(\,1\,+\,{N\over \pi}\,
{C_5(\th)\over \sqrt{n^2\,-\,1}}\,\Big)\,\d^i_{\,\,j}\,-\,
{N\over \pi}\,{C_5(\th)\over (n^2\,-\,1)^{3/2}}
\,\a^j\,\a^i\,-\,{N\over \pi}\,
{\sin^2\th\over n^2-1}\,\,\epsilon_{ijk}\,\a^k\,\,.
\label{fl1apbquince}
\ee
In order to compute the pullback appearing in the first determinant of
the right-hand side of eq. (\ref{biactcom}), we need to characterize
the precise embedding of the D1-brane in the transverse
non-commutative space. Actually, it is straightforward to write our
ansatz for the $x^i$'s as
\be
x^i\,=\,z\,\,{\tan\th\over  \sqrt{n^2\,-\,1}}\,\,\,\a^i\,\,.
\label{fl1apbdseis}
\ee
Moreover, the kind of configurations we are looking for have constant
$\th$ angle. Thus, eq.  (\ref{fl1apbdseis}) shows that, in this case,
the $x^i$'s are linear functions of the worldsheet coordinate $z$. By
using this result it is immediate to find the expression of the first
determinant in (\ref{biactcom}). One gets
\be
-{\rm det}\Big(
P\big[E_{ab}+E_{ai}\,(\,Q^{-1}-\d\,)^{ij}\,E_{jb}\big]\,
\Big)\,=\,{N\alpha'\over r^2}\,+\,
{\tan^2\th\over n^2-1}\,\,\,
\a^i\,\Big[Q^{-1}\Big]_{ij}\,\a^j\,\,,
\label{fl1apbdsiete}
\ee
where $Q^{-1}$ satisfies 
$Q^{ij}\Big[Q^{-1}\Big]_{jk}\,=\,\d^i_{\,\,k}$ with $Q^{ij}$ 
being:
\be
Q^{ij}\,=\,E^{ij}\,+\,i\l\,\,[\,\phi^i\,,\,\phi^j\,]\,\,.
\label{fl1apbdocho}
\ee
As expected on general grounds, a system of D-strings can model a
D3-brane only when the number $n$ of D-strings is very large. Thus, if we
want to make contact with our results of section 4, we should consider
the limit in which $n\rightarrow\infty$  and keep
only the leading terms in the $1/n$ expansion. Therefore, it this
clear that, in this limit, we can replace $n^2-1$ by $n^2$ in all 
our previous expressions. Moreover, as argued in ref. \cite{Myers}, the
leading term in a symmetrized trace of $\a$'s of the form 
${\rm STr}\big(\,(\a^i\a^i)^m\,\big)$ is $n(n^2)^m$. Then, at
leading order in $1/n$, one can make the following replacement inside
a symmetrized trace:
\be
\a^i\a^i\,\,\rightarrow\,\,n^2\,I_n\,\,.
\label{fl1apbdnueve}
\ee
With this substitution the calculation of the action (\ref{biactcom})
drastically simplifies. So, for example, by using (\ref{fl1apbquince}),
one can check that, in the second term under the square root of 
(\ref{biactcom}), we should make the substitution
\be
{\rm det} \Big(\,Q^i_{\,\,j}\,\Big)\,\rightarrow\,
\Bigg(\,{N\over \pi n}\,\Bigg)^2\,\,
\Bigg[\,\big(\sin\th\big)^4\,+\,
\big(\,C_5(\th)\,+\,{\pi n\over N}\,\big)^2\,\Bigg]\,I_n\,\,.
\label{fl1apbveinte}
\ee
Moreover, as 
${\cal C}_{5,n}(\th)\,=\,C_5(\th)\,+\,{\pi n\over N}$, eq. 
(\ref{fl1apbveinte}) is equivalent to:
\be
{\rm det} \Big(\,Q^i_{\,\,j}\,\Big)\,\rightarrow\,
\Bigg(\,{N\over \pi n}\,\Bigg)^2\,\,
\Bigg[\,\big(\sin\th\big)^4\,+\,
\big(\,{\cal C}_{5,n}(\th)\,\big)^2\,\Bigg]\,I_n\,\,.
\label{fl1apbvuno}
\ee

We must now perform the substitution (\ref{fl1apbdnueve}) on the
right-hand side of eq. (\ref{fl1apbdsiete}). First of all, we must
invert the matrix of eq. (\ref{fl1apbdocho}). Actually, it is not difficult
to obtain the expression of $E^{ij}$. After some calculation one gets
\be
E^{ij}\,=\,{r^2\over N\a'}\,\,
{\sin^4\th\over\sin^4\th\,+\,\Big(\,C_5(\th)\,)^2}\,\,
\Bigg[\,\d^i_{\,\,j}\,\,+\,\,
{\Big(\,C_5(\th)\,)^2\over n^2\sin^4\th}\,\a^i\,\a^j
\,\,-\,\,{C_5(\th)\over n\sin^2\th}\,\,
\epsilon_{ijk}\,\a^k\,\Bigg]\,\,.
\label{fl1apbvdos}
\ee
Plugging this result on the right-hand side of eq. (\ref{fl1apbdocho}),
and adding the commutator of the scalar fields, one immediately
obtains $Q^{ij}$. By inverting this last matrix one arrives at the
following expression of $[Q^{-1}]_{ij}$:
\be
[Q^{-1}]_{ij}\,=\,{N\a'\over r^2}\,\,\,
{\sin^4\th\,+\,\Big(\,C_5(\th)\,)^2\over 
(1+a^2)\sin^4\th}\,\,\Bigg[\,\d^i_{\,\,j}\,\,
+\,\,{a^2-b\over n^2(1+b)}\,\a^i\,\a^j\,\,+\,\,
{a\over n}\,\epsilon_{ijk}\,\a^k\,\Bigg]\,\,,
\label{fl1apbvtres}
\ee
where, at leading order, $a$ and $b$ are given by
\be
a={N\over \pi n\sin^2\th}\,\,\Big[\,\sin^4\th\,+\,
C_5(\th)\,{\cal C}_{5,n}(\th)\,\Big]\,\,,
\,\,\,\,\,\,\,\,\,\,\,\,\,\,\,\,\,
b\,=\,{\Big(\,C_5(\th)\,)^2\over\sin^4\th}\,\,.
\label{fl1apbvcuatro}
\ee
By contracting $[Q^{-1}]_{ij}$ with $\a^i\a^j$ and applying
the substitution (\ref{fl1apbdnueve}), one gets a remarkably simple result:
\be
\a^i\,\Big[Q^{-1}\Big]_{ij}\,\a^j\,\rightarrow\,
n^2\,{N\a'\over r^2}\,I_n\,\,.
\label{fl1apbvcinco}
\ee
By using eq. (\ref{fl1apbvcinco}), one immediately concludes that we should
make the following substitution:
\be
-{\rm det}\Big(
P\big[E_{ab}+E_{ai}\,(\,Q^{-1}-\delta\,)^{ij}\,E_{jb}\big]\,
\Big)\,\,\rightarrow\,
{N\a'\over r^2\cos^2\th}\,I_n\,\,.
\label{fl1apbvseis}
\ee
It is now straightforward to find the action of the D1-branes in the
large $n$ limit. Indeed, by using eqs. (\ref{fl1apbvuno}) and
(\ref{fl1apbvseis}), one gets
\be
S_{D1}\,=\,-T_1\,\int dtdz\,{N\over \pi\cos\th}\,\,
\sqrt{\big(\sin\th\big)^4\,+\,
\big(\,{\cal C}_{5,n}(\th)\,\big)^2}\,\,.
\label{fl1apbvsiete}
\ee
From eq. (\ref{fl1apbvsiete}) one can immediately obtain the hamiltonian
of the D-strings. In order to compare this result with the one
corresponding to the wrapped D3-brane, let us change the worldsheet
coordinate from $z$ to $r\,=\,z/\cos\th$. Recalling that $\th$
is constant for the configurations under study  and using that 
$T_1/\pi\,=\,4\pi\alpha'\,T_3\,=\,T_3\O_2\,\a'$, we get the
following hamiltonian:
\be
H\,=\,T_3\O_2\,N\a'\,\int dr\,
\sqrt{\big(\sin\th\big)^4\,+\,
\big(\,{\cal C}_{5,n}(\th)\,\big)^2}\,\,,
\label{fl1apbvocho}
\ee
which, indeed,  is the same as in the one in eq. (\ref{fl2citres}) for
$l=1, m=0$. Notice that $n$, which in our present approach is  the number
of D-strings, corresponds to the quantization integer of the D3-brane
worldvolume gauge field. It follows that the minimal energy
configurations occur for
$\th\,=\,\pi n/N$ and its energy density is the one written in eq. 
(\ref{fl2cicinco}).  This agreement shows that our ansatz represents 
D-strings growing up into a D3-brane configuration of the type studied
in the main text. 

Let us finally point out that the same ansatz of eqs.
(\ref{fl1apbnueve})  and (\ref{fl1apbtrece}) can be used to describe the
configurations in which D0-branes expand into a D2-brane in the NS5
background of eqs. (\ref{fl1apbtres}) and (\ref{fl1apbcuatro}). In this
case, which corresponds to the situation analyzed in section \ref{flbachas},
the D2-branes are located at fixed $r$ and one only has to compute the
determinant of the matrix (\ref{fl1apbquince}) in the D0-brane action. By
using eq.  (\ref{fl1apbvuno}) one easily finds the same hamiltonian and
minimal energy configurations as those of section \ref{flbachas}.


\chapter{Giant Gravitons}
\medskip

\renewcommand{\theequation}{{\rm\thesection.\arabic{equation}}}

One of the most interesting things we have recently learnt from string theory is
the fact that a system can increase its size with increasing momentum. There are
several manifestations of this phenomenon, which is opposite to the standard
field theory intuition, in several contexts related to string theory such as  the
infrared ultraviolet connection \cite{holography} or the noncommutative geometry
\cite{NCG}.

In ref. \cite{GST} McGreevy, Susskind and Toumbas found another example of the
growth in size with energy. These authors considered a massless particle moving
in a spacetime of the form $AdS_m\times S^{p+2}$ and discovered that there
exists a configuration in which an expanded brane (the giant graviton) has
exactly the same quantum numbers as the point-like particle. This expanded brane
wraps the spherical part of the spacetime and is stabilized against shrinking
by the flux of the Ramond-Ramond (RR) gauge field. The size of the giant graviton
increases with its angular momentum and, as the radius of the brane cannot be
greater than the radius of the spacetime, one gets that there exists an upper
bound for the momentum of the brane. This fact is a realization of the
so-called stringy exclusion principle. Moreover, in refs. \cite{GMT, HHI}  was
proved that the giant gravitons of ref. \cite{GST} are BPS configurations which
preserve the same supersymmetry as the point-like graviton. It was also shown in
\cite{GMT, HHI} that there also exist gravitons expanded into the $AdS$ part of
the spacetime which, however, do not have an upper bound on their angular
momentum due to the non-compact nature of the $AdS$ spacetime. 

The overall physical picture one gets from these results is that for high
momenta the linearized approximation to supergravity breaks down and one is
forced to introduce interactions in order to describe the dynamics of the 
massless modes  of the theory. An effective procedure to represent these
interactions is to assume that the massless particles polarize and become a
brane. The mechanism responsible of this polarization is the Myers
dielectric brane effect discussed in section \ref{secdielbr}. 

The blowing up of gravitons into branes can take place in backgrounds different
from $AdS_m\times S^{p+2}$. Indeed, in ref. \cite{DTV} it was found that there
are giant graviton configurations of D(6-p)-branes moving in the near-horizon
geometry of a dilatonic background created by a stack of Dp-branes. For other
aspects of the expanded graviton solutions see refs. \cite{DJM}-\cite{KM}.

In this chapter we will find giant graviton solutions for probes which move in
the geometry created by a stack of non-threshold bound states of the type 
(D(p-2), Dp) for $2\le p \le 6$, (which was given in section \ref{strsecbsnc}), 
and we will extend this analysis to M-theory backgrounds generated by a stack of 
non-threshold bound states of the type (M2,M5), given in section \ref{strsecmtbs}. 
As we have seen, these backgrounds are both $1/2$ supersymmetric. The (D(p-2), Dp) 
solutions are characterized by the presence of a non-zero Kalb-Ramond field 
$B$, directed along the Dp-brane, from the Neveu-Schwarz sector of the
superstring, together with the corresponding RR fields. We will place in this
background a brane probe in such a way that it can capture both the RR flux
(as in the Dp backgrounds) and the flux of the $B$ field. This last requirement
implies that we must extend our probe along two directions parallel to the
background and, actually, our probe will be a D(8-p)-brane wrapped on an
$S^{6-p}$ sphere transverse to the background and extended along a
(noncommutative) plane parallel to it. In the M-theory case the background is 
characterized by a self-dual three-form $C$ and the brane probe is an M5 wrapped 
on a $S^2$ sphere and extended along a plane  parallel to the background such 
that the flux of the $C$-field is captured. We will verify that, by switching on a
particular worldvolume gauge field, one can find configurations of the
D(8-p) or M5-brane which behave as a massless particle.  We will check that the
energy  of these giant graviton configurations is exactly the same as that of a
massless particle moving in the metric of the background.
Generically, the brane falls into the center of the gravitational potential
along a trajectory which will be determined. We will also study the supersymmetry
projection introduced by the brane and we will show that it breaks completely
the supersymmetry of  the background, exactly in the same way 
as  a wave which propagates with the velocity of the
center of mass of the brane. Thus, also from the point of view of supersymmetry,
the expanded brane mimics the massless particle. 
 
The chapter is organized as follows. In section \ref{DMF} we will review the theory of 
electric dipoles moving in a magnetic field as an example of system for 
which physics of increasing momentum is governed by increasingly large distances. 
When the theory is defined on a 
2-sphere there is a bound on the angular momentum when the ends of the dipole 
separate to the antipodes of the sphere. In section \ref{ADSGG} we will analyse 
the gravitons found by McGreevy, 
Susskind and Toumbas \cite{GST} and we will show that they are BPS states preserving 
the same supersymmetries as the point-like graviton. In section \ref{NCGG} we 
find giant graviton configurations in the background of a stack of non-threshold 
bound states of $Dp$ and $D(p-2)$ branes. Finally, in \ref{MTGG}, we extend these 
results 
to $M$-theory backgrounds with a self-dual three-form potential field.


\setcounter{equation}{0}
\section{Dipoles in Magnetic Fields } \label{DMF}

As we mentioned in the introduction physics in non-commutative 
spaces is an example of system whose size grows with 
increasing momentum.
In this section we briefly review the dipole analogy for NC-field
theory \cite{jabbari} \cite{bigatti}.
We begin with a pair of unit charges of opposite sign
moving on a plane with a constant magnetic field $B$.
The Lagrangian is
\be
{\cal L} ={m \over 2} \left(\dot x_1^2 +\dot x_2^2 \right)
+{B\over 2} \e_{ij} \left(\dot x_1^i x_1^j - \dot x_2^i x_2^j \right)
-{K \over2}(x_1 -x_2)^2.
\label{dip1}
\ee
Let us suppose that the mass is so small so that the first term in
Eq.(\ref{dip1}) can be ignored. Let us also introduce center of mass and
relative coordinates
\br
X&=&(x_1 +x_2)/2 \rc\rc
\D &=& (x_1 -x_2)/2.
\label{dip2}
\er
The Lagrangian becomes
\be
{\cal L}=B \e_{ij}\dot X^i \D^j -2K\D^2.
\label{dip3}
\ee
From Eq.(\ref{dip3}) we see that $\D$ plays the role of canonical 
momentum conjugate to $X$
\be
P_i=B\e_{ij}\D^j.
\label{dip4}
\ee
Thus when the dipole is moving with momentum $P$ in some direction
it is stretched to a size
\be
|\D| =|P|/B.
\label{dip5}
\ee
in the perpendicular direction.

Now suppose the dipole is moving on the surface of a sphere of
radius $R$. Assume also that the sphere has a magnetic flux $N$. Then
\be
\frac{1}{2\,\pi}\,\int_{S^2}\,F\,=\,\frac{1}{2\,\pi}\,
\int_{S^2}\,B\,R^2\sqrt{g^{(2)}}\,d\th\,d\p\,=
\,\frac{1}{2 \pi}\,B\,R^2\,\O_2\,=\,N\,.
\label{dip6}
\ee
where $\th,\p$ are the angles parametrizing the sphere 
and $F\,=\,B\,\sqrt{g^{(2)}}\,d\th\,\wedge\,d\p$. Then
\be
2 \pi N= \Omega_2 B R^2.
\label{dip7}
\ee
We can get a rough idea of what happens
by just saying that when the momentum of the dipole $P$ is about $2BR$
the dipole will be as big as the sphere as $\D\,=\,2\,R$ (that is, 
the ends of the dipole are at the poles of the sphere). At this point the angular
momentum is the maximum value
\be
L=PR \sim BR^2,
\label{dip8}
\ee
and this is of order the total magnetic flux $N$.

We will now do a more precise analysis and see that the maximum angular
momentum is exactly $N$. Parameterize the sphere by two angles $\theta,\phi$.
The angle $\phi$ measures angular distance from the equator. It is
$\pm \pi/2 $ at the poles. The azimuthal angle $\theta$ goes from
$0$ to $2\pi$. We work in a gauge in which the $\theta$ component
of the vector potential is non-zero. It is given by
\be
A_{\theta} = N\,{1-\sin{\phi}\over 2R \cos{\phi}}
\label{dip9}
\ee
For a unit charged point particle moving on the sphere the term coupling the
velocity to the vector potential is
\be
{\cal L}_A =A_{\th}\,v_{\th}\,=\,A_{\theta} R \cos{\phi}\,\dot{\theta}\,
=\,N R{1-\sin{\phi}\over 2R }
\,\dot{\theta}.
\label{dip10}
\ee

Now consider a dipole with its center of mass moving on the
equator. The positive charge is at position $(\theta,\phi)$ and the
negative charge is at $(\theta,-\phi)$. For the motion we
consider $\phi $ is time independent. Eq.(\ref{dip10}) becomes
\be
{\cal L}_A =N({1-\sin{\phi} \over 2})
\dot{\theta} - N({1 + \sin{\phi} \over 2})
\dot{\theta}
\label{dip11}
\ee
or
\be
{\cal L}_A ={-N\sin{\phi}}
\,\dot{\theta}.
\label{dip12}
\ee

Again we want to consider a slow-moving dipole whose mass is 
so small that its kinetic term may be ignored 
compared to the coupling to the 
magnetic field, i.e. $mR << N$.

The total Lagrangian is
\be
{\cal L}=  - N \sin{\p}\,\dot{\th}
\ee
and the angular momentum is
\be
L = - N\sin{\phi}.
\ee
The angular momentum will reach its maximum when $\phi = \pi/2$ at
which point
\be
|L_{max}|= N.
\ee
The fact that the angular momentum of a single field quantum in 
non-commutative field
theory is
bounded by $N$ is well known in the context of non-commutative field theory
on a sphere \cite{madore}. Here we see that it is a large distance 
effect.


\medskip
\setcounter{equation}{0}
\section{Giant Gravitons in $AdS \times S$ spaces} \label{ADSGG}
\medskip

In this  section we generalize the results of the previous paragraph
 and consider the cases of (test) brane configurations in background 
spacetimes of the form $AdS_{m}\times S^n$. There are three different 
backgrounds of this type, the $M2$ and $M5$ branes of $D=11$ supergravity 
and the $D3$ branes in the type $IIB$ supergravity. We will find giant 
graviton configurations for the three cases simultaneously. This three
 cases correspond to $(m,n)\,=\,(4,7),(7,4)$ and $(5,5)$ respectively.
The full line element for the metric on $AdS_{m}\times S^n$ takes the
form $ds^2=ds^2_{AdS}+ds^2_{sph}$. If we use global coordinates on $AdS_m$, 
the metric is given by
\be
ds^2_{\rm AdS}=-\left(1+{u^2\over L^2}\right)dt^2+{du^2\over
1+{u^2\over L^2}}+u^2 d\O_{m-2}^2\ ,
\label{ads1}
\ee
with $L=\frac{2}{n-3}R$, where $L$ is the radius of AdS and $R$ is the 
radius of the sphere. The metric for the $S^n$ is 
\be
d\,s_{sph}^2=\,R^2\,d\O_n^2
\ee
where $d\O_n$ is the line element of the unit $n$-sphere
\be
d\O_{n}^2=(d\chi^1)^2+\sin^2\chi^1\left((d\chi^2)^2+\sin^2\chi^2\left(\ 
\dots
+\sin^2\chi_{n-1}(d\chi^{n})^2\right)\right).
\ee
Let us parametrize $S^n$ using cartesian coordinates $X^1,.....,X^{n+1}$ so that
\vspace{.5cm}
\br
X^{1}\,&=&\,R\,\cos\,\chi^1 \cr
X^2\,&=&\,R\,\sin\,\chi^1\,\cos\,\chi^2 \cr
X^3\,&=&\,R\,\sin\,\chi^1\,\sin\,\chi^2\,\cos\,\chi^3 \cr
&\vdots&\cr
X^{n}\,&=&\,R\,\sin\,\chi^1\,\sin\,\chi^2\dots\,\sin\,\chi^{n-1}\,\cos\,\chi^n \cr
X^{n+1}\,&=&\,R\,\sin\,\chi^1\,\sin\,\chi^2\dots\,\sin\,\chi^{n-1}\,\sin\,\chi^n\, ,
\label{ads2}
\er
where $0\le \chi^1 ,\chi^2 ,\dots,\chi^{n-1}\,\le \pi$ and $\chi^n$ is the
azimuthal angle ($0\le \chi^n \le 2\,\pi$). Then
\be
(X^1)^2 + (X^2)^2 + (X^3)^2 + (X^4)^2 +\,\dots\,+ (X^{n+1})^2 = R^{2}.
\label{ads3}
\ee

We will embed the brane probe on the $S^{n-2}$ surface parametrized by 
the angles $\chi^3 ,\chi^4 ,\dots,\chi^n$, such that the brane is 
allowed to move in the $X^1,X^2$ plane and its size depends on its 
location in this plane according to 
\be
r\,=\,R\,\sin\chi^1\sin\chi^2\,=\,R\,\r,
\ee
where we have defined 
\be
\r\,\equiv\,\sin\chi^1\,\sin\chi^2.
\label{ads4}
\ee
We see that the size is at its maximum value, $r=R$, when 
$\cos \chi^1=\cos \chi^2=0$, or equivalently, when the 
brane is at the origin ($X^1=X^2=0$). 
From eq.(\ref{ads2}) we see that
\be
(X^1)^2 + (X^2)^2\,=\,R^2 - r^2\,=\,R^2 (1- \r^2),
\ee
and then, the membrane can move around a circle of constant size 
in the $(X_1,X_2)$ plane. We can change to polar coordinates. Let 
us set
\br
X^1 &=& R\,\sqrt{1 - \r^2}\,\cos\p \cr
X^2 &=& R\,\sqrt{1 - \r^2}\,\sin\p.
\label{ads5}
\er
where $\r$ was defined in eq. (\ref{ads4}).

In terms of the coordinates 
$\r, \p, \chi^3\,\dots\,\chi^n$, the
metric on the n-sphere becomes
\be
ds^2_{sph} = R^2\left(\,{1 \over (1 - \r^2)}d\r^2 + (1 - \r^2)d\p^2 +
\r^2d\O_{n-2}^2\,\right),
\label{ads6}
\ee
where $d\O_{n-2}^2$ is the metric of a unit $(n-2)$--sphere. If we rename as 
$\th^1,\dots,\th^{n-2}$ the angles $\chi^3,\dots,\chi^n$ 
parametrizing the $(n-2)$-sphere and with the coordinates chosen above,
 we explicitly write the  components of the $RR$ $(n-1)$-form potential on the $S^n$
 as
\be
C^{(n-1)}_{\p\th^1\dots\th^{n-2}}=\b_n R^{n-1} \r^{n-1}\,
\sin^{n-3}\th_1\cdots\sin\th_{n-3}
\equiv \b_n R^{n-1} \r^{n-1}\,\sqrt{{\hat g}^{(n-2)}}\ ,
\label{ads7}
\ee
where $\sqrt{{\hat g}^{(n-2)}}$ is the volume element of the unit 
$(n-2)$-sphere. The constant $\b_n$ is a sign ($\b_4=+1=\b_5$,
while $\b_7=-1$) so that the four-form field strength appears with 
a positive coefficient in both of the M-theory backgrounds. This 
means that we are considering a background of branes 
(with positive charge) and 
not of {\it anti}-branes.

Now we put an $(n-2)$-brane probe in this background wrapped on 
the $S^{n-2}$. The action for the $(n-2)$-brane is
\be
S_{n-2}\,=\,-T_{n-2}\int d^{n-1}\xi\ \sqrt{-g}+T_{n-2}\int P[C^{(n-1)}]
\label{ads8}
\ee
where $g_{\a\b}$ is the pull-back of the spacetime metric to the
world-volume, \ie ,
\be
g_{\a\b}=\pa_{\a}X^M\pa_{\b}X^N\,G_{MN},
\label{ads9}
\ee
and $P[C^{(n-1)}]$ denotes the pull-back of the ($n-1$)-form potential. 
Aditional bosonic world-volume fields as the self-dual three form on 
the $M5$ and the $U(1)$ gauge field on the $D3$-brane are consistently 
set to zero in the equations of motion, as there is no source term 
for these fields in the action. 

As in refs. \cite{GST}\,\cite{GMT} we look for stable solutions where 
the $(n-2)$-brane has expanded on the $S^n$ to a sphere of fixed size 
$\r$. We choose the static gauge, \ie, world-volume coordinates 
$\xi^{\a}$ will be taken as
\be
\xi^{\alpha}\,=\,(t,\th^1,\dots,\th^{n-2})\,\,.
\label{ads10}
\ee
Let us consider a trial solution of the form
\be
\r=\r(t)\ ,\qquad\p=\p(t)\ ,\qquad
u=0,
\label{ads11}
\ee
which corresponds to a spherical $(n-2)$-brane of radius $R\,\r$ 
moving in the $(\r,\p)$ plane inside the $S^n$.
For this embedding the pull-back of the metric is
\begin{displaymath}
g_{ij} = \left(\begin{array}{cc}
          -1\,+\,R^2\,(1\,-\,\r^2)\,\dot{\p}^2 & 0 \\
          0 & R^2\,\r^2 \,\hat g_{ij}^{(n-2)}
          \end{array}\right)
\label{ads12}
\end{displaymath}
where $(\hat g^{(n-2)})_{ij}$ denotes the metric on the unit $(n-2)$-sphere.
Substituting the trial solution (\ref{ads11}) into the world-volume action 
(\ref{ads8}) and integrating over the angular coordinates, yields the
following Lagrangian 
\br
L_{n-2}=\,T_{n-2}\,\O_{n-2}\,R^{n-2}\,\left[-\r^{n-2}\,\sqrt{\,1
-\,R^2\frac{\dot\r^2}{(1-\r^2)}\,-\,R^2 
\,(1-\r^2)\,{\dot\p}^2}\,+
\,R\,\r^{n-1} \,\dot\p\right]\ .\rc
\label{ads13}
\er
As indicated in ref. \cite{GMT} note that with the sign of the $RR$-
potential, discussed after eq. (\ref{ads7}), the second term
in eq. (\ref{ads13}) would be negative for $n=7$. Nevertheless we 
choose it to be positive, which corresponds to the brane probe having the
opposite charge, that is, we are considering an {\it anti}-M5-brane 
brane expanding into $S^7$. We would obtain the same Lagrangian if we 
considered an $M5$ in the background of an {\it anti}-M5. It is also 
worth mention that this sign can be absorved by a redefinition of 
$\dot\p$ which corresponds to the brane turning in the inverse direction.

The quantization of the $RR$ $n$-form flux on the $S^n$ implies that
\be
T_{n-2}\,\O_{n-2}\,R^{n-1}\,=\,N\, .
\label{ads14}
\ee
If we use this equation in (\ref{ads13}) we get the Lagrangian
\be
L_{n-2}=\frac{N}{R}\left[-\r^{n-2}\,
\sqrt{1\,-\,R^2\,(1-\r^2)\,{\dot\p}^2\,-\,
R^2\,\frac{{\dot\r}^2}{1-\r^2}}\,+\,
R\,\r^{n-1}\,{\dot\p}\right]\ .
\label{ads15}
\ee
Let us study the dynamics of the system. The momentum conjugate to 
$\r$ and $\p$ are
\br
P_{\r}&=&{\partial L_{n-2}\over \partial \dot \rho}\,\equiv\,
T_{n-2}\,\O_{n-2}\,R^{n-1}\,\pi_{\r}\,=\,N
\,\pi_{\r}\,,\rc\rc
P_{\phi}&=&{\partial L_{n-2}\over \partial \dot \phi}\,\equiv\,
T_{n-2}\,\O_{n-2}\,R^{n-1}\,\pi_{\phi}\,=\,N\,\pi_{\p}\,,
\label{ads16}
\er
with the reduced momenta $\pi_{\r}$ and $\pi_{\p}$ given by
\br
\pi_{\r}&=&\,R\,{\r^{n-2}\over 1- \r^2}\,\,{\dot\r\over
\sqrt{1\,-\,R^{2}\frac{{\dot\r}^2}{1-\r^2}-\,R^2\,(1-\r^2)\,{\dot\p}^2}}\,\,,\rc\rc
\pi_{\p}&=&\,R\,(1-\r^2)\,\r^{n-2}\,\,{\dot\p\over
\sqrt{1\,-\,R^{2}\frac{{\dot\r}^2}{1-\r^2}-\,R^2(1-\r^2)\,{\dot\p}^2}}
\,\,+\,\,\r^{n-1}\,\,.\rc
\label{ads17}
\er
The Hamiltonian is given by
\be
H\,=\,\dot\r\,P_{\r}\,
+\,\dot\p P_{\p}\,-\,L_{n-2}\,\equiv\,
\,T_{n-2}\,\O_{n-2}\,R^{n-1}\,h\,=\,N\,h\,\,,
\label{ads18}
\ee
in terms of the reduced Hamiltonian $h$. From (\ref{ads15}) we get 
\be
h\,=\,\frac{1}{R}\,\Bigg[\,\r^{2(n-2)}\,+\,
(1-\r^2)\,\pi_{\r}^2\,+\,
{\Big(\pi_{\p}-\r^{n-1}\Big)^2\over 1-\r^2}
\,\,\Bigg]^{{1\over 2}}\,\,.
\label{ads19}
\ee
The first and last terms within the square brackets can be combined 
to a sum of squares, and 
the Hamiltonian can be rewritten as
\be
h\,=\,\frac{1}{R}\,\Bigg[\,\pi_{\p}^{2}\,+\,
(1-\r^2)\,\pi_{\r}^2\,+\,
{\Big(\pi_{\p}\,\r-\r^{n-2}\Big)^2\over 1-\r^2}
\,\,\Bigg]^{{1\over 2}}\,\,.
\label{ads20}
\ee
The eq. (\ref{ads14}) is crucial in achieving this result and 
to the existence of giant graviton configurations.

Now we try to obtain solutions with fixed size, that is, with $\r$ 
constant. From eq. (\ref{ads17}) we see that
\be
\pi_{\r}\,=\,0\,\,.
\label{asd21}
\ee
Then, from the hamiltonian equation of motion 
$\dot\pi_{\r}\,=\,-\pa h\,/\,\pa \r=0$ implies that we 
have constant $\r$ configurations at the 
extrema of h as a function on $\r$ 
($h$ can be regarded as the potential that 
determines the equilibrium radius). The equation 
for the extrema is
\be
\frac{\partial h}{\partial \r}\,\sim\,
(\pi_{\p}\,\r-\r^{n-2})\,
\left((n-3)\r^{n-1}-(n-2)\r^{n-3}+\pi_{\p}\right)\,=0\, . 
\ee
and is solved either when $(\pi_{\p}\,\r-\r^{n-2})\,=\,0$ 
or $\left((n-3)\r^{n-1}-(n-2)\r^{n-3}+\pi_{\p}\right)=0$. 
In Figure \ref{fig1}, we represent the behaviour of $H$ 
versus $\r$ for $\pi_{\p}\le1$ and $n$ even, 
and we see that there are two degenerate minima at 
\be
\r=0\,\,,
\label{ads22}
\ee
and
\be
\pi_{\p}\,=\,\r^{n-3}\,\,.
\label{ads23}
\ee
which are the two possible solutions of 
$(\pi_{\p}\,\r-\r^{n-2})\,=\,0$ for the three 
cases of interest ($n$ equal to 7, 5 or 4), and one 
intermediate maximum which solves $\left((n-3)\r^{n-1}-(n-2)
\r^{n-3}+\pi_{\p}\right)=0$. Both minima 
describe stable configurations with constant $\r$.
 
The Hamiltonian (\ref{ads20}) is independent of $\p$ 
and then the momentum $\pi_{\p}$ is a constant of motion, 
thus the eq. (\ref{ads23}) only makes sense for 
$\r\,=\,$constant. At either of the 
minima, the energy evaluates to
\be
H\,=\,\frac{P_{\p}}{R}\,=\,N\,\frac{\pi_{\p}}{R}
\label{ads24}
\ee
\begin{figure}
\centerline{\epsffile{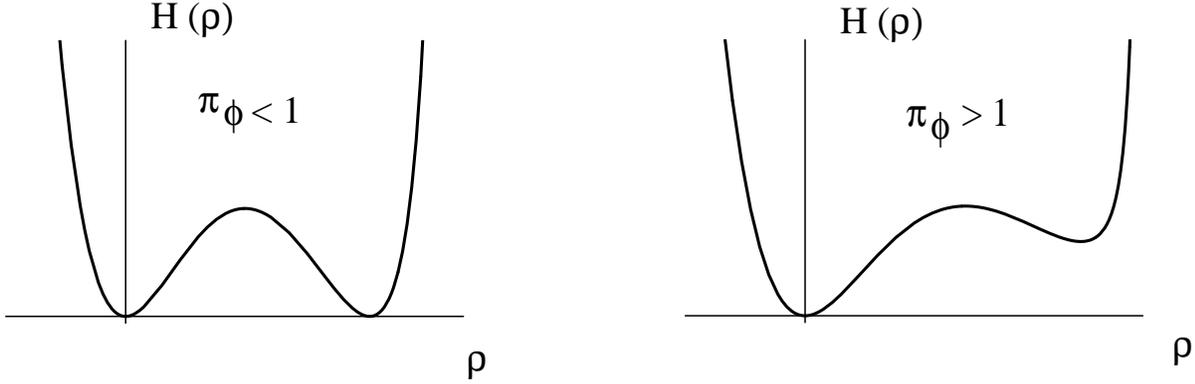}}
\caption{Energy of expanded $(n-2)$-brane, $n$ even,
as a function of its radius. For
$\pi_{\p}=P_{\p}/N\le 1$ (left figure) there are two degenerate minima. For
$\pi_{\p}>1$ the second minimum acquires higher energy and eventually disappears
completely.}
\label{fig1}
\end{figure}
The minimum at $\r\,=\,\pi_{\p}^{1/n-3}$ cannot exist at $\pi_{\p}>1$ 
as $\r\le 1$ (see the definition of $\r$ in eq. (\ref{ads4})). In 
Figure \ref{fig1} we also see that as $\pi_{\p}$ grows beyond 
$\pi_{\p}=1$ the minimum at $\r>0$ first lifts above the one at $\r=0$ 
before disappearing for
\be
\pi_{\p}>2\left({n-2\over n-1}\right)^{n-1\over2}\ .
\ee
Results are similar for the case 
of odd n. 

Otherwise, if we invert the 
relation (\ref{ads17}) between $\pi_{\p}$ and $\dot\p$ we find
\be
\dot\p\,=\,\frac{1}{R}
{\pi_{\p}-\r^{n-1}\over 1-\rho^2}\,\,
{\Bigg[\,1\,-\,R^{2}
{\dot\r^2\over 1-\r^2}\,\Bigg]^{{1\over 2}}\over
\Bigg[\,\pi_{\p}^2\,+\,
{\big(\,\pi_{\p}\rho\,-\,\r^{n-2}\,\big)^2
\over 1-\r^2}\,\Bigg]^{{1\over 2}}}\,\,.
\label{ads25}
\ee
If we substitute in this equation $\dot\r\,=\,0$ and one 
of the conditions (\ref{ads22}) or (\ref{ads23}) one gets
$\dot\p\,=\,\frac{1}{R}$, \ie , the velocity of the center 
of mass of the brane, $\dot\p\,R$, equals the speed of light. Also, 
the fact that the brane can increase in size up to $\r=1$ implies 
that the angular momentum, $P_{\p}$, has a maximum value
\be
P_{\p}^{max}\,=\,N
\ee
when the brane has its maximal size.

The movement of the center of mass of the brane probe 
in the full ($m$+$n$)-dimensional
background is along a null trajectory, since
\be
ds^2\,=\,(G_{t\,t}\,+\,G_{\p\,\p}\,\dot\p^2)\,dt^2\,=
\,\left(-1\,+\,R^2\,(1\,-\,\r^2)\,\dot\p^2\right)\,dt^2=0
\ee
when evaluated for $\r=0$ and $\dot\p=1/R$.
This is the expected result for a massless `point-like'
graviton.
However, note that in the expanded configurations, the motion
of each element of the sphere is along a timelike trajectory, with
$ds^2=-\r^2\,dt^2$. 

It was found that the quantum numbers of a $(n-2)$-brane  moving 
on the $n$-sphere of a $AdS_m\times S^n$ space-time are exactly the same 
as those of a massless particle moving along the same trajectory as 
the center of mass of the brane. In the next section we will show 
that all of these configurations 
are supersymmetric. In principle, one could expect this 
since, from the point of view of $m$-dimensional supergravity in the
AdS space, the stable brane configurations correspond to massive states with
$M = P_\p/R$. The motion on the $\S^n$ means that these states are
also charged under a $U(1)$ subgroup of the $SO(n+1)$ gauge symmetry in
the reduced supergravity theory. With the appropriate normalizations,
the charge is $Q= P_\p/R$, and hence one finds that these
configurations satisfy the appropriate BPS bound ($Q\,=\,M$).

It is important to specify that we have chosen global coordinates 
in AdS (see eq. (\ref{ads1})) and we have put the brane probe 
at the center of AdS ($u=0$). Then, the brane probe  does not 
move in the radial direction and the energy in global coordinates 
is equal (in units of the radius) to the angular momentum, as we 
have seen in the last paragraph. We could instead, have used 
Poincar\'e coordinates and, in this case, there would be an r-dependent 
$\sqrt{-g_{t\,t}}$ factor in eq. (\ref{ads20}), where r is the 
radial direction. Due to it, the massless particle, or equivalently 
the expanded brane, initially at rest in the radial direction 
will fall to $r=0$.

\subsubsection{Supersymmetry for Gravitons}

In this section we will show that the configurations studied in the 
previous paragraph are in fact BPS states preserving the same 
supersymmetries as the point-like graviton. We will do this for 
the case of D3-branes propagating in a 10-dimensional
type IIB background compactified on AdS$_5 \times S^5$. The 
other two cases follow in the same line.
Ten-dimensional, Type IIB supergravity is described by the vielbein, a 
complex
Weyl gravitino, a real four-form  $C^{(4)}_{MNPQ}$ with self-dual 
field
strength  $F^{(5)}_{MNPQR}$, a complex two-form $C^{(2)}_{MN}$, a complex 
spinor $\Lambda$
and a complex scalar $\Phi$. In the AdS$_5 \times S^5$ background, we
have $\Phi=  C^{(2)}_{MN} = \Lambda =\Psi_M=0$.
The five-form field strength is 
$F= \frac{4}{R}[\varepsilon({\rm AdS}_5)+\varepsilon(S^5)]$.
Given that the scalar and two-form vanish in this background, the
supersymmetry variation of the complex spinor automatically vanishes,
$\delta\Lambda=0$. Hence to examine the background supersymmetries, the
only nontrivial variation which needs to be considered is that of the 
gravitino.
For the given background, the variation of the
gravitino takes the form
\be
\delta \Psi_M = {D}_M \e - \frac{i}{1920}{\Gamma_M}^{PQRST}F_{PQRST}\e\ ,
\label{adss1}
\ee
in terms of ten-dimensional gamma matrices. 
Demanding $\delta\Psi_M=0$ leads to the Killing spinor equations
\be
{D}_M \e -\frac{i}{4}(\G^{t\,r\,\a^1\,\a^2\,\a^3}
+ \G^{\r\,\p\,\chi^1 \,\chi^2\,\chi^3})\G_M \e =0\ .
\label{adss2}
\ee
We recall that here $ \e$ is a complex Weyl spinor satisfying $\G_{11}
\e = \e$, \ where  
$\G_{11}\equiv\G^{t\,r\,\a^1\,\a^2\,\a^3\,\r\,\p\chi^1\,\chi^2\,\chi^3}$.
The Killing spinor condition can be rewritten then
as
\br
&&{D}_\m \e -\frac{i}{2\,R}\g_{AdS}\G_\m \e =0
\ \ \ ,\ \ \ \g_{AdS}=\G^{t\,r\,\a^1\,\a^2\,\a^3}
\equiv \g\ , \nonumber\\
&&{D}_m \e -\frac{i}{2\,R}\g_{5}
\G_m \e =0 \ \ \ \ \ ,\ \ \ \ \g_{_5}
=\G^{\r\,\p\chi^1\,\chi^2\,\chi^3}\ ,
\label{adss3}
\er
on AdS$_5$ and $ S^5$, respectively.
The Killing spinor solutions are now
\br
\e &=& e^{\frac{i}{2}\,\b\,\g_5 \G^{\underline \r}} e^{ \frac{i}{2} \p
\g_5 \G^{\underline \p}} 
e^{- \frac{i}{2} \chi^1 \G^{\underline{\chi^1\,\r}}}
e^{- \frac{i}{2} \chi^2 \G^{\underline{\chi^2\,\chi^1}}}
e^{- \frac{i}{2} \chi^3 \G^{\underline{\chi^3\,\chi^2}}}
\,\,\times \nonumber\\
&&\quad\times\,\,e^{i\frac{\a}{2}\G^{\underline r}\,\g} 
e^{-i\frac{t}{2L}\G^{\underline t}\,\g}
e^{\frac{\a^1}{2}\G^{\underline{\a^1\,r}}}
e^{\frac{\a^2}{2}\G^{\underline{\a^2\,\a^1}}}
e^{\frac{\a^3}{2}\G^{\underline{\a^3\,\a^2}}}
\,\e_0\ ,
\label{adss4} 
\er
where $\b\,=\,$arcsin$\,\r,\ \sinh\alpha = r/R$ and $\e_0$ is an arbitrary constant 
spinor. Hence we have a maximally 
supersymmetric
solution of the type IIB supergravity equations with 32 residual 
supersymmetries.

For our D3-brane configurations, residual supersymmetries
are determined by imposing the kappa-symmetry constraint $\G_{\k} \e=\pm\e$ 
on the background Killing spinors. In our $D3$-brane configurations the 
world-volume gauge fields vanish and the matrix of $\k$- symmetry (\ref{kappamatrix}) 
is given by:
\be
\G_{\k}\,=\,\frac{1}{4!\,\sqrt{|g|}}\epsilon^{\m_1\dots \m_4}\,
\g_{\m_1\dots \m_4}\,(i\s_2)\,
=-\frac{i}{4!}\,\epsilon^{i_1\dots i_4} \partial_{i_1}X^{M_1}
\cdots\partial_{i_4}X^{M_4}\G_{M_1\dots M_4}\ .
\label{adss11}
\ee
For the giant gravitons on $S^5$, we consider the embedding of a
D3-brane as in eqs. (\ref{ads10}) and (\ref{ads11}).
With this configuration, one finds
\be
\G_{\k}\,=\,-\frac{i}{\r}\,(\G^{\underline{t\,\chi^1\,\chi^2\,\chi^3}} -
\sqrt{1\,-\,\r^2}\,\,\G^{\underline {\p\,\chi^1\,\chi^2\,\chi^3}} )\ ,
\label{adss12}
\ee
where we have substituted $\dot\p \,=\,1/R$. The condition $\G_{\k}\e=\e$ 
becomes 
\be
\left[ \G^{\underline{t\,\chi^1\,\chi^2\,\chi^3}} -\,\sqrt{1\,-\,\r^2}\,
\,\G^{\underline{\p\,\chi^1\,\chi^2\,\chi^3}} + \r \right]\e =0\ ,
\label{adss13}
\ee
or equivalently, after pulling out a factor of 
$\G^{\underline{\p\,\chi^1\,\chi^2\,\chi^3}}$,
\be
(e^{-i \th\,\g_5\,\G^{{\underline \r}}} + \G^{\underline{t \p}})\e=0\ .
\label{adss14}
\ee
Substituting the Killing spinors of eq. (\ref{adss4}) 
we see that $\G^{\underline{t \p}}$ moves through the factor 
$e^{\frac{i}{2} \th \g_5 \G^{\underline \r}}$ such that
\be
\G^{\underline{t\,\p}}\,e^{-\frac{i}{2}\,\b\,\g_5\,\G^{\underline \r}}
\,=\,e^{-\frac{i}{2}\,\b\,\g_5 \G^{\underline \r}}\,\G^{\underline{t\,\p}}\,.
\label{adss15}
\ee
and then, we can write eq. (\ref{adss14}) as 
\br
(e^{-i\ \b\,\g_5\,\G^{\underline \r}} + \G^{\underline{t \p}})\,\,\e 
\,=\,
e^{-\frac{i}{2}\,\b\,\g_5\,\G^{\underline \r}}\,(\,1\,
+\,\G^{\underline{t \p}}\,)\,
e^{ \frac{i}{2}\,\p\,\g_5\,\G^{\underline \p}}
e^{- \frac{i}{2} \chi^1 \G^{\underline{\chi^{1}\r}}}\,\times
\rc
\rc
\times \ 
e^{- \frac{i}{2} \chi^2 \G^{\underline{\chi^2\,\chi^1}}}
e^{- \frac{i}{2} \chi_3 \G^{\underline{\chi_3\,\chi^2}}}
e^{i\frac{\a}{2} \G^{\underline r}\g_5} 
e^{-i\frac{t}{2L}\G^{\underline{t\,\g}}}
e^{\frac{\a^1}{2}\G^{\underline{\a^1\,r}}}
e^{\frac{\a^2}{2}\G^{\underline{\a^2\,\a^1}}}
e^{\frac{\a^3}{2}\G^{\underline{\a^3\,\a^2}}}
\ \e_0\ .\nonu \\ 
\vspace{.5cm}
\label{adss16}
\er
The remaining exponentials commute with $\G^{\underline{t \phi}}$ except the 
one with $\G^{\underline r}\,\g_5$. However, the test brane configurations 
are at $u\,=\,R\,$sinh$ \a\,=\,0$ and this factor reduces to the identity 
when evaluated on the test brane world-volume. Thus, eq. (\ref{adss16}) is 
equivalent to the condition
\be
\G_{\underline{t\,\p}}\,\e_0\,=\,\e_0\,\,.
\label{adss17}
\ee
The matrix $\G_{\underline{t\,\phi}}$ is a projector since 
$(\G_{\underline{t\,\phi}})^2=1$ and $tr(\G_{\underline{t\,\phi}})=0$, and then  
the $D3$-brane preserves half of the supersymmetries of type $IIB$ ten-
dimensional supergravity and the giant gravitons in 
$AdS\times S$ are BPS states. Furthermore eq. (\ref{adss17}) is 
exactly the supersymmetry preserved by a point-like particle 
moving along the $\p$ direction in the ten-dimensional spacetime.


\medskip
\setcounter{equation}{0}
\section{Giant gravitons in (D(p-2),Dp)  backgrounds} \label{NCGG}
\medskip

In these section  giant graviton configurations in more 
general backgrounds than $AdS\times S$ spaces will be obtained. These backgrounds 
include 
$Dp$-brane backgrounds as a limiting case. As we will see, the main difference 
among these configurations and the ones in $AdS\times S$ spaces is that 
they are no longer BPS states, as the supersymmetries preserved by 
the probe are different than the ones preserved by the background. 
Nevertheless, the way supersymmetry is broken is the same for this 
configurations and for a point-like graviton with the same quantum numbers. 
The supergravity background that will considered is the one generated  by a stack of
$N$  non-threshold bound states of Dp and D(p-2) branes for 
$2\le p \le 6$ given is section \ref{strsecbsnc}.

As before we will use the particular coordinate system
of eq. (\ref{ads5}) for the unit $8-p$ sphere $S^{8-p}$. 
In terms of $(\rho,\phi)$, 
the metric of $S^{8-p}$ takes the form:
\beq
d\Omega_{8-p}^2\,=\,{1\over 1-\rho^2}\,d\rho^2\,+\,
(1\,-\,\rho^2\,)\,d\phi^2\,+\,\rho^2\,d\Omega_{6-p}^2\,\,,
\label{ncocho}
\eeq
where $d\Omega_{6-p}^2$ is the metric of a unit $6-p$ sphere.

The Hodge duals of the RR field strengths (\ref{strncbs5}) can be easily computed 
from eqs. (\ref{strncbs5}) and (\ref{strncbs1}). The $(10-p)$-form ${}^*\,F^{(p)}$ 
and the $(8-p)$-form ${}^*\,F^{(p+2)}$ is a $(8-p)$-form have  the following 
components 
in the coordinate system of (\ref{ads5}):
\bear
{}^*\,F^{(p)}_{x^{p-1},x^p,\rho,\phi,\theta^1\cdots, \theta^{6-p}}
&=&(-1)^{p+1}\,(7-p)\,\sin\varphi\,R^{7-p}\,\rho^{6-p}\,
h_p\,f_p^{-1}\,\sqrt{\hat g^{(6-p)}}\,\,,\rc\rc
{}^*\,F^{(p+2)}_{\rho,\phi,\theta^1\cdots,\theta^{6-p}}
&=&(-1)^{p+1}\,(7-p)\,\cos\varphi\,R^{7-p}\,\rho^{6-p}\,
\,\sqrt{\hat g^{(6-p)}}\,\,,
\label{ncnueve}
\eear
where $\theta^1,\cdots, \theta^{6-p}$ are coordinates of the unit $6-p$ sphere 
and $ \hat g^{(6-p)}$ is the determinant of the $S^{6-p}$ metric. These forms 
satisfy the equations:
\beq
d{}^*\,F^{(p)}\,=\,H\,\wedge\,{}^*\,F^{(p+2)}\,\,,
\,\,\,\,\,\,\,\,\,\,\,\,\,\,\,\,\,\,
d{}^*\,F^{(p+2)}\,=\,0\,\,.
\label{ncdiez}
\eeq
Then, one can represent ${}^*\,F^{(p)}$ and ${}^*\,F^{(p+2)}$ in terms of
three potentials as follows:
\bear
{}^*\,F^{(p)}&=&dC^{(9-p)}\,-\,H\,\wedge\,C^{(7-p)}\,\,,\rc
{}^*\,F^{(p+2)}&=&dC^{(7-p)}\,-\,H\,\wedge\,C^{(5-p)}\,\,,
\label{nconce}
\eear
where $H=dB$. In eq. (\ref{nconce}) $C^{(r)}$ is an $r$-form. Actually only for
$p=3$ the term $H\,\wedge\,C^{(5-p)}$ in the second of these equations gives a
non-vanishing contribution. By a direct calculation one can check that the other
two potentials $C^{(7-p)}$ and  $C^{(9-p)}$ have the components:
\bear
C^{(7-p)}_{\phi,\theta^1\cdots,\theta^{6-p}}
&=&(-1)^{p+1}\,\cos\varphi\,R^{7-p}\,\rho^{7-p}\,
\sqrt{\hat g^{(6-p)}}\,\,, \rc\rc
C^{(9-p)}_{x^{p-1},x^p,\phi,\theta^1\cdots, \theta^{6-p}}
&=&(-1)^{p+1}\,\sin\varphi\,R^{7-p}\,\rho^{7-p}\,
h_p\,f_p^{-1}\,\sqrt{\hat g^{(6-p)}}\,\,.
\label{ncdoce}
\eear
Apart from the ones displayed in eqs. (\ref{strncbs5}) and (\ref{ncnueve}), 
the RR field strengths and their duals have another components. Actually, due
to the identification of $F^{(p)}$ and ${}^*F^{(10-p)}$, the forms appearing in
these equations are not all different. Using this fact it is not difficult to
obtain all the components of the  RR gauge forms. From this result one can
check that eq. (\ref{ncdiez}) is satisfied and, thus, the representation of 
${}^*F^{(p)}$ and ${}^*F^{(p+2)}$ in terms of the different potentials appearing
on the right-hand side of eq. (\ref{nconce}) holds (although these potentials
have another components in addition to the ones written in eq. (\ref{ncdoce})). 
Let us specify this for $p=3$. In this case one can easily check that $F^{(3)}$
and $F^{(5)}$ are given by:
\bear
F^{(3)}\,&=&\,\sin\varphi\,\partial_rf_3^{-1}\,
dx^0\wedge dx^1\wedge dr\,\,,\rc\rc
F^{(5)}\,&=&\,\cos\varphi\,\Big[\,h_3\partial_rf_3^{-1}
dx^0\wedge \cdots\wedge dx^{3}\wedge dr\,+\,
4R^4\rho^3d\rho\wedge d\phi\wedge\epsilon_{(3)}\,\Big]\,\,.
\label{nctrece}
\eear
Moreover, ${}^*F^{(3)}$ and ${}^*F^{(5)}=F^{(5)}$ can be represented in terms of
three RR potentials $C^{(6)}$, $C^{(4)}$ and $C^{(2)}$ as in eq. 
(\ref{nconce}) with:
\bear
C^{(6)}&=&\sin\varphi\, R^4\,\rho^4\,h_3f_3^{-1}\,
dx^2\wedge dx^3 \wedge d\phi \wedge \epsilon_{(3)}\,\,, \rc\rc 
C^{(4)}&=&\cos\varphi\,\Big[\,h_3f_3^{-1}dx^0\wedge \cdots\wedge dx^{3}\,+\,
R^4\rho^4d\phi\wedge\epsilon_{(3)}\,\Big]\,\,,\rc\rc
C^{(2)}&=&-\sin\varphi f_3^{-1} dx^0\wedge dx^1\,\,,
\label{nccatorce}
\eear
where $\epsilon_{(3)}$ is the volume form of the unit $S^3$ 
(with this choice of $C^{(2)}$ one has $F^{(3)}\,=\,-dC^{(2)}$).  One
can treat similarly the other cases.

\medskip
\subsubsection{The brane probe}
\medskip

Let us now embed a D(8-p)-brane in the near-horizon region of the (D(p-2), Dp)
geometry. In this region $r$ is small and one can approximate the harmonic
function $f_p$ appearing in the supergravity solution as:
\beq
f_p\,\approx\,{R^{7-p}\over r^{7-p}}\,\,.
\label{ncquince}
\eeq
The D(8-p)-brane probe we will be dealing with wraps the (6-p) transverse
sphere and  extends along  the $x^{p-1}x^p$ directions. The response of the
probe to the background is determined by its action $S$, which was written in 
eq. (\ref{fulldpbact}). For an D(8-p)-brane the Dirac-Born-Infeld  term $S_{DBI}$ is:
\beq
S_{DBI}\,=\,
-T_{8-p}\,\int d^{\,9-p}\s\,e^{-\tilde\phi_D}\,
\sqrt{-{\rm det}\,(\,g\,+\,{\cal F}\,)}\,\,,
\label{ncdsiete}
\eeq
where $g$ is the induced worldvolume metric,  $T_{8-p}$ is the tension of the
D(8-p)-brane:
\beq
T_{8-p}\,=\,(2\pi)^{p-8}\,(\,\alpha\,'\,)^{{p-9\over 2}}\,
(\,g_s\,)^{-1}\,\,,
\label{ncdocho}
\eeq
and, if $P[\cdots]$ denotes the pull-back to the worldvolume of a bulk field, 
${\cal F}$ is given by:
\beq
{\cal F}\,=\,F\,-\,P[B]\,=\,dA\,-\,P[B]\,\,,
\label{ncdnueve}
\eeq
with $F$ being the $U(1)$ worldvolume gauge field strength and $A$ its
potential. The Wess-Zumino term of the action $S_{WZ}$ couples the probe to the
RR potentials of the background. For the brane probe configuration we are
considering and the (D(p-2), Dp) background, $S_{WZ}$ is given by:
\beq
S_{WZ}\,=\,T_{8-p}\int\Bigg[\,\,P[C^{(9-p)}]\,+\,
{\cal F}\wedge P[C^{(7-p)}]\,\,\Bigg]\,\,.
\label{ncveinte}
\eeq

The worldvolume coordinates $\s^{\alpha}$ ($\alpha=0,\cdots,8-p$) will be
taken as:
\beq
\s^{\alpha}\,=\,(t,x^{p-1},x^{p},\theta^1,\cdots,\theta^{6-p})\,\,.
\label{ncvuno}
\eeq
As we will confirm soon, the set of coordinates (\ref{ncvuno}) is quite
convenient to study the kind of configurations we are interested in. These
configurations are embeddings of the D(8-p)-brane which, in our system of
coordinates, are described by functions of the type:
\beq
r\,=\,r(t)\,\,,
\,\,\,\,\,\,\,\,\,\,\,\,\,
\rho\,=\,\rho(t)\,\,,
\,\,\,\,\,\,\,\,\,\,\,\,\,
\phi\,=\,\phi(t)\,\,.
\label{ncvdos}
\eeq
Let us now evaluate the action for the ansatz (\ref{ncvdos}). We shall begin by
studying the Wess-Zumino term. In this term only the components of $C^{(7-p)}$
and $C^{(9-p)}$ written in eq. (\ref{ncdoce}) contribute. Actually, it is easy
to see that the pull-back of the RR potential $C^{(7-p)}$ is coupled to the 
$x^{p-1}x^{p}$ component of ${\cal F}$. Assuming that ${\cal
F}_{x^{p-1},x^{p}}$ is independent of the angles $\theta^1\cdots\theta^{6-p}$,
one gets that $S_{WZ}$ can be written as:
\beq
S_{WZ}\,=\,T_{8-p}\,\Omega_{6-p}\,R^{7-p}\,\cos\varphi\,
\int dt\,dx^{p-1}\,dx^{p}\,
\rho^{7-p}\,(-1)^{p+1}\,\dot\phi\,
\Big[\,{\cal F}_{x^{p-1},x^{p}}\,+\,h_pf_p^{-1}\tan\varphi\,\Big]\,\,,
\label{ncvtres}
\eeq
where $\dot\phi=d\phi/dt$ and $\Omega_{6-p}$ is the volume of the $S^{6-p}$
sphere, given by:
\beq
\Omega_{6-p}\,=\,{2\pi^{{7-p\over 2}}\over 
\Gamma\Bigl(\,{7-p\over 2}\Bigr)}\,\,.
\label{ncvcuatro}
\eeq
Notice that for the ansatz (\ref{ncvdos}) the scalars $r$, $\rho$ and $\phi$ do
not depend on the coordinates $x^{p-1}$ and $x^{p}$ and, therefore, 
$P[B]$ has only non-zero components along the directions
$x^{p-1}x^{p}$. By using the actual value of the $B$ field for the 
(D(p-2), Dp) background in the definition of ${\cal F}$ (eq. (\ref{ncdnueve})),
one immediately gets that the term inside the square brackets on the 
right-hand side of eq. (\ref{ncvtres}) is:
\beq
{\cal F}_{x^{p-1},x^{p}}\,+\,h_pf_p^{-1}\tan\varphi\,=\,F_{x^{p-1},x^{p}}\,\,.
\label{ncvcinco}
\eeq
In what follows we will assume that the only non-zero component of the
worldvolume gauge field is $F_{x^{p-1},x^{p}}$ and we will denote from now on 
${\cal F}_{x^{p-1},x^{p}}$ and $F_{x^{p-1},x^{p}}$ simply by ${\cal F}$ and $F$
respectively. Then, the total action can be written as:
\beq
S\,=\,\int\,dt\,dx^{p-1}\,dx^p\,{\cal L}\,\,,
\label{ncvseis}
\eeq
where the lagrangian density ${\cal L}$ is given by:
\bear
{\cal L}&=&T_{8-p}\,\Omega_{6-p}\,R^{7-p}\,\times\rc\rc
&&\times\Bigg[\,-\rho^{6-p}\lambda_1\,
\sqrt{r^{-2}\,f_p^{-1}\,\,-\,r^{-2}\dot r^2\,-\,
{\dot\rho^2\over 1-\rho^2}\,-\,
(1-\rho^2)\,\dot\phi^2}\,+\,\lambda_2\,(-1)^{p+1}\rho^{7-p}\,\dot\phi
\,\,\Bigg]\,\,.\rc\rc
\label{ncvsiete}
\eear
In eq. (\ref{ncvsiete}) we have introduced the functions $\lambda_1$ and 
$\lambda_2$, which are defined as:
\beq
\lambda_1\,=\,\sqrt{h_pf_p^{-1}\,+\,{\cal F}^2\,h_p^{-1}}\,\,,
\,\,\,\,\,\,\,\,\,\,\,\,\,\,\,\,\,\,\,\,\,\,\,\,\,\,
\lambda_2\,=\,F\cos\varphi\,\,.
\label{ncvocho}
\eeq
Before starting the analysis of the lagrangian density (\ref{ncvsiete}), let us
discuss how the brane probe is extended along the $x^{p-1}x^{p}$ directions.
The main motivation to extend the probe along these directions is to allow the
worldvolume of the brane to capture the flux of the $B$ field of the
background. It is thus natural to characterize the spreading of the
D(8-p)-brane in the $x^{p-1}x^{p}$ plane by means of the flux of the $F$ field.
Accordingly we will extend our brane along the $x^{p-1}\,x^p$ directions in such
a way that there are $N'$ units of worldvolume flux. \ie :
\beq
\int\,\, dx^{p-1}\,dx^p\,F\,=\,{2\pi\over T_f}\,\,N'\,\,,
\label{ncvnueve}
\eeq
with $T_f=(2\pi\alpha')^{-1}$ being the fundamental string tension. Clearly,
given $F$ (which will be determined below), eq. (\ref{ncvnueve}) gives the volume
occupied by the brane probe in the noncommutative plane in terms of the flux
number $N'$.

Let us now resume our study of the dynamics of the system by performing a
canonical hamiltonian analysis of the lagrangian density  (\ref{ncvsiete}). First
of all, for simplicity, let us absorb the sign $(-1)^{p+1}$ sign of the
Wess-Zumino term of ${\cal L}$ by redefining
$\dot\phi$ if necessary. The density of momenta associated to ${\cal L}$  are:
\bear
{\cal P}_r&=&{\partial {\cal L}\over \partial \dot r}\,\equiv\,
T_{8-p}\,\Omega_{6-p}\,R^{7-p}\,\lambda_1\,\pi_r\,\,,\rc\rc
{\cal P}_{\rho}&=&{\partial {\cal L}\over \partial \dot \rho}\,\equiv\,
T_{8-p}\,\Omega_{6-p}\,R^{7-p}\,\lambda_1\,\pi_{\rho}\,\,,\rc\rc
{\cal P}_{\phi}&=&{\partial {\cal L}\over \partial \dot \phi}\,\equiv\,
T_{8-p}\,\Omega_{6-p}\,R^{7-p}\,\lambda_1\,\pi_{\phi}\,\,,
\label{nctreinta}
\eear
where we have defined the reduced momenta $\pi_r$, $\pi_{\rho}$ and 
$\pi_{\phi}$. By using the explicit value of  of ${\cal L}$,
given in eq. (\ref{ncvsiete}), we get:
\bear
 \pi_r&=&{\rho^{6-p}\over r^2}\,\,{\dot r\over
\sqrt{r^{-2}\,f_p^{-1}\,\,-\,r^{-2}\dot r^2\,-\,
{\dot\rho^2\over 1-\rho^2}\,-\,(1-\rho^2)\,\dot\phi^2}}\,\,,\rc\rc
\pi_{\rho}&=&{\rho^{6-p}\over 1- \rho^2}\,\,{\dot \rho\over
\sqrt{r^{-2}\,f_p^{-1}\,\,-\,r^{-2}\dot r^2\,-\,
{\dot\rho^2\over 1-\rho^2}\,-\,(1-\rho^2)\,\dot\phi^2}}\,\,,\rc\rc
 \pi_{\phi}&=&(1-\rho^2)\rho^{6-p}\,\,{\dot \phi\over
\sqrt{r^{-2}\,f_p^{-1}\,\,-\,r^{-2}\dot r^2\,-\,
{\dot\rho^2\over 1-\rho^2}\,-\,(1-\rho^2)\,\dot\phi^2}}
\,\,+\,\,\Lambda\,\rho^{7-p}\,\,,\rc
\label{nctuno}
\eear
where, in the third of these expressions we have introduced the quantity
$\Lambda$, defined as:
\beq
\Lambda\,=\,{\lambda_2\over \lambda_1}\,\,.
\label{nctdos}
\eeq
The hamiltonian density of the system can be obtained in the standard way,
namely:
\beq
{\cal H}\,=\,\dot r\,{\cal P}_r\,+\,\dot\rho\,{\cal P}_{\rho}\,
+\,\dot\phi{\cal P}_{\phi}\,-\,{\cal L}\,\equiv\,
\,T_{8-p}\,\Omega_{6-p}\,R^{7-p}\,\lambda_1\,h\,\,,
\label{ncttres}
\eeq
where, in analogy with what we have done for the momenta, we have defined the
reduced quantity $h$. From eqs. (\ref{ncvsiete}) and (\ref{nctuno}), one gets after
a simple calculation that $h$ is given by:
\beq
h\,=\,r^{-1}\,f_p^{-{1\over 2}}\,\Bigg[\,r^2\,\pi_r^2\,+\,\rho^{2(6-p)}\,+\,
(1-\rho^2)\,\pi_{\rho}^2\,+\,
{\Big(\pi_{\phi}-\Lambda\rho^{7-p}\Big)^2\over 1-\rho^2}
\,\,\Bigg]^{{1\over 2}}\,\,.
\label{nctcuatro}
\eeq

\medskip
\subsubsection{Fixed  size configurations}
\medskip

We would like to obtain solutions of the equations of motion derived from the
hamiltonian (\ref{nctcuatro}) which correspond to a brane of fixed size. It
follows from eq. (\ref{ncocho}) that the coordinate $\rho$ plays the role of the
size of the system on the $S^{6-p}$ sphere. For this reason it is interesting
to look at solutions of the equations of motion for which $\rho$ is constant.
This same problem was analyzed in ref. \cite{DTV} for the case of brane probes
moving in the near-horizon Dp-brane background (see also refs. \cite{GST, GMT,
HHI, DJM}).  In ref. \cite{DTV} it was found that the $\rho$-dependent terms in
the hamiltonian can be arranged to produce a sum of squares from which the
$\rho={\rm constant}$ solutions can be found by inspection. By comparing the
right-hand side of eq. (\ref{nctcuatro}) with the corresponding expression in ref.
\cite{DTV}, one immediately realizes that the same kind of arrangement can be
done in our case if the condition:
\beq
\Lambda\,=\,1
\label{nctcinco}\,\,,
\eeq
is satisfied. Indeed, if eq. (\ref{nctcinco}) holds, one can rewrite $h$ as:
\beq
h\,=\,r^{-1}\,f_p^{-{1\over 2}}\,\Bigg[\,\pi_{\phi}^2\,+\,r^2\,\pi_r^2\,+\,
(1-\rho^2)\,\pi_{\rho}^2\,+\,
{\Big(\pi_{\phi}\rho-\rho^{6-p}\Big)^2\over 1-\rho^2}
\,\,\Bigg]^{{1\over 2}}\,\,.
\label{nctseis}
\eeq
The analysis of the hamiltonian (\ref{nctseis}) can be performed along the same
lines as in ref. \cite{DTV} (see below). Before carrying out this study, let us
explore the content of the condition (\ref{nctcinco}). By recalling the definition
of
$\Lambda$ (eq. (\ref{nctdos})), we realize that eq.  (\ref{nctcinco}) is equivalent
to require that $\lambda_1=\lambda_2$. Moreover, by using eq. (\ref{ncvcinco}) in
the definition of $\lambda_1$ (eq. (\ref{ncvocho})), one obtains:
\beq
\lambda_1^2\,=\,\cos^2\varphi\,F^2\,+\,f_p^{-1}\,\Big(\,F\sin\varphi\,-\,
{1\over \cos\varphi}\Big)^{2}\,\,.
\label{nctsiete}
\eeq
By comparing the right-hand side of eq. (\ref{nctsiete}) with the definition of 
$\lambda_2$ (eq. (\ref{ncvocho})), one concludes immediately that the condition
$\Lambda=1$ is equivalent to have the following constant value of the
worldvolume gauge field $F$:
\beq
F\,=\,{1\over \sin\varphi\cos\varphi}\,=\,2\csc (2\varphi)\,\,.
\label{nctocho}
\eeq
By substituting this value of $F$ on the right-hand side of eq.
(\ref{nctsiete}) one gets that $\lambda_1$ is also constant and given by:
\beq
\lambda_1\,=\,{1\over \sin\varphi}\,\,.
\label{nctnueve}
\eeq
In the next section we will show that for the value of $F$ displayed in eq.
(\ref{nctocho}) the brane probe breaks the  supersymmetry of the background
exactly in the same way as a massless particle. In this section this value of
the worldvolume gauge field should be considered as an ansatz which allows us
to find a class of fixed size solutions of the equations of
motion of the brane which are particularly interesting. In the remaining
of this section we will assume that $F$ is given by (\ref{nctocho}).

It is now straightforward to find the configurations of the system with
constant $\rho$. From the second expression in eq. (\ref{nctuno}) one concludes
that one must have:
\beq
\pi_{\rho}\,=\,0\,\,.
\label{nccuarenta}
\eeq
Then, the hamiltonian equation of motion for $\pi_{\rho}$, \ie\ 
$\dot\pi_{\rho}\,=-\,\partial h / \partial \rho$, implies that the last term on
the right-hand side of eq. (\ref{nctseis}) must vanish. For $p<6$ this happens
either for:
\beq
\rho=0\,\,,
\label{nccuno}
\eeq
or else when $\pi_{\phi}$ is given by:
\beq
\pi_{\phi}\,=\,\rho^{5-p}\,\,.
\label{nccdos}
\eeq 
(when $p=6$ only eq. (\ref{nccdos}) gives rise to a constant $\rho$
configuration). Notice that, as $h$ does not depend on $\phi$, the
momentum $\pi_{\phi}$ is a constant of motion and, thus,  for $p\not= 5$, eq.
(\ref{nccdos}) only makes sense when $\rho$ is constant. Moreover, when 
$p\not= 5$, eq. (\ref{nccdos}) determines $\rho$ in terms of $\pi_{\phi}$. On the
contrary, when $p= 5$ eq. (\ref{nccdos})  fixes $\pi_{\phi}\,=\,1$, independently
of the value of $\rho$. 

It is not difficult to translate the constant $\rho$ condition into 
a relation involving the time derivatives of $\phi$ and $r$. Let us, first of
all, invert the relation (\ref{nctuno}) between $\pi_{\phi}$ and $\dot\phi$. By
taking $\Lambda=1$ as in eq. (\ref{nctcinco}) one easily finds after a simple
calculation that:
\beq
\dot\phi\,=\,
{\pi_{\phi}-\rho^{7-p}\over 1-\rho^2}\,\,
{\Bigg[\,r^{-2}\big(f_p^{-1}\,-\,\dot r^2\,\big)\,-\,
{\dot\rho^2\over 1-\rho^2}\,\Bigg]^{{1\over 2}}\over
\Bigg[\,\pi_{\phi}^2\,+\,
{\big(\,\pi_{\phi}\rho\,-\,\rho^{6-p}\,\big)^2
\over 1-\rho^2}\,\Bigg]^{{1\over 2}}}\,\,.
\label{nccdosextra}
\eeq
By taking $\dot\rho=0$ on  eq. (\ref{nccdosextra}), and imposing one of the two
conditions (\ref{nccuno}) or (\ref{nccdos}), one gets that in both cases $\dot\phi$
and $\dot r$ satisfy the following relation:
\beq
f_p\,\big[\,r^2\,\dot\phi^2\,+\,\dot r^2\,]\,=\,1\,\,.
\label{ncctres}
\eeq
For the configurations we are considering the last two terms of the reduced
hamiltonian $h$ of eq. (\ref{nctseis}) vanish and, then, our configurations
certainly minimize the energy. These configurations are characterized by eq.  
 (\ref{nctocho}), which fixes the value of
the worldvolume gauge field, and  (\ref{ncctres}), which is a consequence of
the vanishing of the last two terms of the hamiltonian. Remarkably, eq.
(\ref{ncctres}) is the condition satisfied by a particle moving in the $(r,\phi)$
plane at $\rho=0$ along a null trajectory, \ie\ with
$ds^2=0$, in the metric (\ref{strncbs1}). Thus, our brane probe configurations have
the  characteristics of a massless particle : the so-called giant graviton.
Notice that the point $\rho=0$ can be considered as the ``center of mass" of
the expanded brane. In
order to confirm this picture let  us introduce the two-component vector 
${\bf v}$, defined as:
\beq
{\bf v}\,=\,(v^{\underline{r}}, v^{\underline{\phi}})\,\equiv\,
f_p^{{1\over 2}}\,(\dot r\,,\,r\dot\phi)\,\,,
\label{ncccuatro}
\eeq
which is nothing but the velocity of the particle in the $(r,\phi)$ plane. From
eq. (\ref{ncctres}) one clearly has:
\beq
(v^{\underline{r}})^2\,+\,(v^{\underline{\phi}})^2\,=\,1\,\,,
\label{ncccinco}
\eeq
which simply states that the center of mass of the giant graviton moves
at the speed of light.  The corresponding value of the momentum density ${\cal
P}_{\phi}$ can be  straightforwardly obtained from eqs. (\ref{nccdos}) and
(\ref{nctreinta}). Indeed, by using eqs.  (\ref{ncdocho}), (\ref{ncvcuatro}),
(\ref{strncbs3}) and (\ref{nctnueve}), one gets:
\beq
{\cal P}_{\phi}\,=\,{T_f\over 2\pi}\,F\,N\,\rho^{5-p}\,\,,
\label{nccseis}
\eeq
where $F$ is given in eq. (\ref{nctocho}). 
The momenta $p_{\phi}$ and $p_r$ can be obtained from the densities
${\cal P}_{\phi}$ and ${\cal P}_{r}$ by integrating them in the $x^{p-1}x^p$
plane:
\beq
p_\phi\,=\,\int dx^{p-1}\,dx^p\,\,{\cal P}_{\phi}\,\,,
\,\,\,\,\,\,\,\,\,\,\,\,\,\,\,\,\,\,\,\,\,\,
p_r\,=\,\int dx^{p-1}\,dx^p\,\,{\cal P}_{r}\,\,.
\label{nccsiete}
\eeq
By using the value of the momentum density ${\cal P}_{\phi}$ given in eq. 
(\ref{nccseis}) and the flux condition (\ref{ncvnueve}), one gets the following
value  of $p_\phi$ for a giant graviton:
\beq
p_\phi\,=\,N\,N'\,\rho^{5-p}\,\,.
\label{nccocho}
\eeq
It follows from eq. (\ref{nccocho}) that, when $p<5$, the size $\rho$ of the
wrapped brane increases with the momentum $p_\phi$. Moreover,  it is interesting
to point out that, as $0\le \rho\le 1$, when
$p<5$ the momentum $p_\phi$ has a maximum value $p_\phi^{max}$ given by:
\beq
p_\phi^{max}\,=\,N\,N'\,\,,
\label{nccnueve}
\eeq
which is reached when $\rho=1$. The existence of such a maximum for $p_\phi$ is
a manifestation of the so-called stringy exclusion principle. Notice that when
$p=5$ the momentum $p_\phi$ is independent of $\rho$, whereas for $p=6$ the
value of $p_\phi$ written in eq. (\ref{nccnueve}) is in fact a minimum. 

Let us now study the energy of the giant graviton solution. First of all, we
define the metric elements ${\cal G}_{MN}$ as:
\beq
{\cal G}_{MN}\,=\,{G_{MN}}_{{\,\big |}_{\rho=0}}\,\,,
\label{nccincuenta}
\eeq
where the $G_{MN}$'s correspond to the metric displayed in eqs. (\ref{strncbs1}) and
(\ref{ncocho}). The hamiltonian for the giant graviton configurations, which we
will denote by $H_{GG}$, can be easily obtained from eq. (\ref{nctseis}). In
terms of the $\rho=0$ metric it can be written as:
\beq
 H_{GG}\,=\,\sqrt{-{\cal G}_{tt}}\,\Bigg[\,
 {p_{\phi}^2\over {\cal G}_{\phi\phi}}\,+\,
{ p_{r}^2\over {\cal G}_{rr}}\,\,\Bigg]^{{1\over 2}}\,\,,
\label{ncciuno}
\eeq
which, according to our expectations, is exactly the hamiltonian of a massless
particle which moves in the metric ${\cal G}_{MN}$ along a trajectory contained
in the $(r,\phi)$ plane. Interestingly, one can use the hamiltonian 
(\ref{ncciuno}) to find and solve the equations of motion for the giant graviton.
Let us, first of all, rewrite $H_{GG}$ as:
\beq
 H_{GG}\,=\,R^{{p-7\over 2}}\,\,
\Bigg[\,r^{7-p}\, p_{r}^2\,+\,r^{5-p}\, p_{\phi}^2
\,\Bigg]^{{1\over 2}}\,\,.
\label{nccidos}
\eeq
We will study the equations of motion of this system by using the conservation
of energy. In this method we first put $H_{GG}\,=\,E$, for constant $E$, and
then we use the relation between $p_{r}$ and $\dot r$, namely:
\beq
p_{r}\,=\,{R^{7-p}\over r^{7-p}}\,\,E\,\,\dot r\,\,.
\label{nccitres}
\eeq
By substituting (\ref{nccitres}) in the condition $H_{GG}\,=\,E$, we get:
\beq
\dot r^2\,+\,{r^{7-p}\over R^{7-p}}\,\Bigg[\,
{p_{\phi}^2\over E^2 R^{7-p}}\,r^{5-p}\,-\,1\,\Bigg]\,=\,0\,\,.
\label{nccicuatro}
\eeq
Eq. (\ref{nccicuatro}) determines the range of values that $r$ can take. Indeed,
by consistency of eq. (\ref{nccicuatro}), the second term in this equation must
be negative or null. The points for which this term vanishes are the turning
points of the system. For $p<5$ these points are $r=0$ and $r_{*}$, with 
$r_{*}$ being:
\beq
(r_{*})^{5-p}\,=\,{E^2\over p_{\phi}^2}\,R^{7-p}\,\,.
\label{nccicinco}
\eeq
In this $p<5$ case, $r$ can take values in the range $0\le r\le r_{*}$. If 
$p=5$ only the $r=0$ turning point exists and $r$ is unrestricted, \ie\ $r$
can take any non-negative value. Finally if $p=6$ the $r=0$ turning point is
missing and $r\ge r_{*}$, where $r_{*}$ is the value given in eq. 
(\ref{nccicinco}) for $p=6$. 

It is not difficult to find the explicit dependence of $r$ on $t$. Let us
consider first the $p\not= 5$ case. From eq. (\ref{nccicuatro}) it follows that
$t$ as a function of $r$ is given by the following indefinite integral:
\beq
t\,-\,t_{*}\,=\,R^{{7-p\over 2}}\,\int
{dr\over r^{{7-p\over 2}}\,
\sqrt{1\,-\,\Big(\,{r\over r_{*}}\,\Big)^{5-p}}}\,\,,
\,\,\,\,\,\,\,\,\,\,\,\,\,\,\,\,\,\,\,\,\,\,\,\,\,\,\,\,\,\,\,\,
(\,p\not= 5\,)\,\,,
\label{ncciseis}
\eeq
where $t_{*}$ is a constant of integration. The integral (\ref{ncciseis}) can be
easily performed by means of the following trigonometric change of variables:
\beq
\Big(\,{r\over r_{*}}\,\Big)^{5-p}\,=\,\cos^2\theta\,\,,
\,\,\,\,\,\,\,\,\,\,\,\,\,\,\,\,\,\,\,\,\,\,\,\,\,\,\,\,\,\,\,\,
(\,p\not= 5\,)\,\,.
\label{nccisiete}
\eeq
The result of the integration  is:
\beq
\Big(\,{r_{*}\over r}\,\Big)^{5-p}\,=\,1\,+\,
(5-p)^2\,(r_{*})^{5-p}\,R^{p-7}\,
\Big(\,{t\,-\,t_{*}\over 2}\,\Big)^{2}\,\,,
\,\,\,\,\,\,\,\,\,\,\,\,\,\,\,\,\,\,\,\,\,\,\,\,\,\,\,\,\,\,\,\,
(\,p\not= 5\,)\,\,.
\label{ncciocho}
\eeq 
It follows from eq. (\ref{ncciocho}) that $t_{*}$ is precisely the value of $t$ 
at which $r=r_{*}$. Moreover if $p<5$ the coordinate $r\rightarrow 0$ as
$t\,-\,t_{*}\rightarrow \pm \infty$, \ie\ the giant graviton always falls
asymptotically into the black hole. On the contrary, for $p=6$ the coordinate
$r$ diverges asymptotically and, thus, in this case the particle always escapes
away from the $r=0$ point.

The  $p=5$ case needs a special treatment since, in this case, eq.
(\ref{ncciseis}) is not valid any more. One can, however, easily integrate eq. 
(\ref{nccicuatro}) with the result:
\beq
r\,=\,r_0\,e^{\pm{t\over R}\,\,\sqrt{1\,-\,{p_{\phi}^2\over E^2 R^2}}\,\,},
\,\,\,\,\,\,\,\,\,\,\,\,\,\,\,\,\,\,\,\,\,\,\,\,\,\,\,\,\,\,\,\,
(\,p= 5\,)\,\,.
\label{nccinueve}
\eeq
It follows from (\ref{nccinueve}) that, in this case, the solution connects  
asymptotically the points $r=0$ and $r=\infty$. 

In order to complete the integration of the equations of motion one has to
determine $\phi$ as a function of $t$. This can be easily achieved by
substituting $r(t)$ from eq. (\ref{ncciocho}) or (\ref{nccinueve}) into the
condition (\ref{ncctres})  to get $\dot\phi$, followed by an integration over
$t$. The result for $p\not= 5$ is:
\beq
\tan\Bigg[\,{5-p\over 2}\,(\phi\,-\,\phi_{*})\,\Bigg]\,=\,
{5-p\over 2}\,\Bigg(\,{r_{*}\over R}\,\Bigg)^{{5-p\over 2}}\,\,\,
{t-t_{*}\over R}\,\,,
\,\,\,\,\,\,\,\,\,\,\,\,\,\,\,\,\,\,\,\,\,\,\,\,\,\,\,\,\,\,\,\,
(\,p\not= 5\,)\,\,,
\label{ncsesenta}
\eeq
whereas for $p= 5$ one gets:
\beq
\phi\,=\,\phi_0\,+\,{p_{\phi}\over E\,R^2}\,\,t\,\,,
\,\,\,\,\,\,\,\,\,\,\,\,\,\,\,\,\,\,\,\,\,\,\,\,\,\,\,\,\,\,\,\,
(\,p= 5\,)\,\,.
\label{ncsuno}
\eeq
It is also interesting to find the value of the velocity ${\bf v}$ along the
trajectory. Actually, a simple calculation shows that, for $p\not= 5$, it depends
on the coordinate $r$ as:
\beq
v^{\underline{r}}\,=\,-\Big[\,1\,-\,\Big(\,{r\over r_*}\,\Big)^{5-p}
\,\Big]^{{1\over 2}}\,\,,
\,\,\,\,\,\,\,\,\,\,\,\,\,\,\,\,\,\,\,\,\,\,\,\,\,\,\,
v^{\underline{\phi}}\,=\,\Big(\,{r\over r_*}\,\Big)^{{5-p\over 2}}\,\,,
\,\,\,\,\,\,\,\,\,\,\,(p\not= 5)\,\,,
\label{ncsunoextrauno}
\eeq
where, on the right-hand side, it should be understood that $r$ is the function
of $t$ given in eq. (\ref{ncciocho}) for $t\ge t_*$.  Curiously, when $p=5$ the
vector
${\bf v}$ is constant and has the following components:
\beq
v^{\underline{r}}\,=\,\pm\Big[\,1\,-\,{p_{\phi}^2\over E^2\,R^2}
\,\Big]^{{1\over 2}}\,\,,
\,\,\,\,\,\,\,\,\,\,\,\,\,\,\,\,\,\,\,\,\,\,\,\,\,\,\,
v^{\underline{\phi}}\,=\,{p_{\phi}\over E\,R}\,\,,
\,\,\,\,\,\,\,\,\,\,\,(p= 5)\,\,.
\label{ncsunoextrados}
\eeq
To finish this section, 
let us now discuss the extensivity of the brane probe on the noncommutative
plane $x^{p-1}\,x^p$. As argued above, for fixed worldvolume flux $N'$, this
extensivity depends on the value of the gauge field $F$. Actually, by using in
eq. (\ref{ncvnueve}) the  value of $F$ given in eq. (\ref{nctocho}), one gets
that the volume occupied by the brane probe along the $x^{p-1}\,x^p$ plane is:
\beq
\int\,\, dx^{p-1}\,dx^p\,=\,{\pi N'\over T_f}\,\,\sin(2\varphi)\,\,.
\label{ncsdos}
\eeq
This volume clearly goes to zero as $\varphi\rightarrow 0$ for fixed $N'$. This
means that if we switch off the $B$ field in the background, the D(8-p)-brane
probe is effectively converted into a D(6-p)-brane, in agreement with the
results of refs. \cite{GST}-\cite{DTV}. Actually, one can check that in this
limit our results agree with those corresponding to the Dp-brane background if we
replace everywhere $N$ by $NN'$.

\medskip
\subsubsection{Supersymmetry}
\medskip

The objective of this section is to analyze the  supersymmetry behaviour of the 
brane configurations studied in  the previous section. Actually, we will verify that 
these configurations break the supersymmetry of the background
just as  a massless particle which moves precisely along the trajectories 
we have found for the probe.

The number of supersymmetries preserved by a Dp-brane is the number of
independent solutions of the constraint equation imposed by kappa-symmetry 
\beq
\Gamma_{\kappa}\,\epsilon\,=\,\epsilon\,\,,
\label{ncstres}
\eeq
on the Killing spinors of the background $\epsilon$. For a Dp-brane 
$\Gamma_{\kappa}$ was given in eq. (\ref{kappamatrix}). Actually, for the sake of
simplicity, we will restrict ourselves to the analysis of the $(D1,D3)$
background, although one can find similar results for the general case by
making the appropriate modifications in our equations.

In order to solve eq. (\ref{ncstres}) it is clear that we must first determine
the Killing spinors of the background, which is equivalent to characterize the
supersymmetry preserved by our supergravity solution.  We
are considering bosonic  backgrounds, which are supersymmetric iff the
supersymmetry variation of the fermionic supergravity fields vanishes.
Actually, this only occurs for some class of transformation parameters, which
are precisely the Killing spinors we are interested in. 

The analysis of the supersymmetry preserved by the (D1,D3) background by using
the transformation rules of the type IIB supergravity is performed in the appendix A. 
Here we will characterize the Killing spinors of this background by means
of an alternative and more simplified method \cite{Kim}, which makes use of the
$\kappa$-symmetry matrix  (\ref{kappamatrix}). Let us place a test D3-brane parallel
to the background and choose its worldvolume coordinates as
$(\xi^{0}, \xi^{i})\,=\,(t,x^i)$ for $i\,=1,2,3$. The equations
which determine the embedding of the parallel D3-brane are :
\beq
X^{0}\,=\,t\,=\xi^0\,,
\,\,\,\,\,\,\,\,\,\,\,\,\,\,\,\,\,\,\,\,\,\,\,
X^{i}\,=\,x^i\,=\xi^i\,\,,\,\,\,\,\,\,\,\,\,\,\,\,(i\,=1,2,3)\,\,,
\label{ncsocho}
\eeq
with the other spacetime coordinates being independent of the 
$(\xi^{0}, \xi^{i})$. We will study the supersymmetry preserved by this
test brane when the worldvolume gauge field $F$ is zero, \ie\ when 
${\cal F}\,=\,-P[B]$. This means that the only non-zero component of ${\cal F}$
is:
\beq
{\cal F}_{x^2x^3}\,=\,-\tan\varphi\,f_3^{-1}\,h_3\,\,.
\label{ncsnueve}
\eeq
The test brane configuration we are considering is the same as the one that
creates the background. Therefore, it is natural to expect that such a parallel
test brane preserves the same supersymmetries as the supergravity background.
Then, let us consider the $\kappa$-symmetry condition 
$\Gamma_k\epsilon\,=\,\epsilon$ for this case. The induced gamma matrices for
the embedding (\ref{ncsocho}) are:
\bear
\gamma_{x^{\alpha}}\,&=&\,f_3^{-{1\over 4}}\,\Gamma_{\underline x^{\alpha}}\,\,,
\,\,\,\,\,\,\,\,\,\,\,\,\,\,\,\,\,\,\,\,\,\,\,\,\,\,\,\,
(\alpha\,=0,1)\,\,,\rc\rc
\gamma_{x^{i}}\,&=&\,f_3^{-{1\over 4}}\,h_3^{{1\over 2}}\,
\Gamma_{\underline x^{i}}\,\,,
\,\,\,\,\,\,\,\,\,\,\,\,\,\,\,\,\,\,\,\,\,\,\,(i\,=2,3)\,\,.
\label{ncsetenta}
\eear
By using eq. (\ref{ncsetenta}) in the expression of $\Gamma_k$ for $p=3$ (eqs. 
(\ref{kappamatrix})-(\ref{Gamma0})), one gets that the $\kappa$-symmetry matrix for
this case is:
\beq
\Gamma_k\,=\,h_3^{{1\over 2}}\,\Big[\,\cos\varphi\,(i\sigma_2)\,
\Gamma_{\underline{x^0\cdots x^3}}\,-\,
f_3^{-{1\over 2}}\,\sin\varphi\,\sigma_1\Gamma_{\underline{x^0x^1}}
\,\Big]\,\,.
\label{ncstuno}
\eeq
In eq. (\ref{ncstuno}), and in what follows, we have suppressed the tensor
product symbol. It is now straightforward to check that the condition
$\Gamma_k\epsilon\,=\,\epsilon$ can be written as:
\beq
(i\sigma_2)\,
\Gamma_{\underline{x^0\cdots x^3}}\,\epsilon\,=\,
e^{-\alpha\,\Gamma_{\underline{x^2x^3}}\sigma_3}\,\epsilon\,\,,
\label{ncstdos}
\eeq
where $\alpha$ is the function of $r$ given by:
\beq
\sin\alpha\,=\,f_3^{-{1\over 2}}\,h_3^{{1\over 2}}\sin\varphi\,\,,
\,\,\,\,\,\,\,\,\,\,\,\,\,\,\,\,\,\,\,\,\,\,
\cos\alpha\,=\,h_3^{{1\over 2}}\cos\varphi\,\,.
\label{ncsttres}
\eeq
In the appendix A we will verify that the supersymmetry invariance of the dilatino
of type IIB supergravity in the $(D1,D3)$ background requires that the
supersymmetry parameter $\epsilon$ satisfies eq. (\ref{ncstdos}) with $\alpha$
given by eq. (\ref{ncsttres}), which is is a confirmation of the correctness of our
$\kappa$-symmetry argument. Moreover, in view of eq. (\ref{ncstdos}), the Killing
spinor $\epsilon$ can be written as:
\beq
\epsilon\,=\,e^{{\alpha\over 2}\,\Gamma_{\underline{x^{2}x^{3}}}\,
\sigma_3}\,\,\tilde\epsilon\,\,,
\label{ncstcuatro}
\eeq
where $\tilde\epsilon$ is a spinor which satisfies:
\beq
(i\sigma_2)\,
\Gamma_{\underline{x^0\cdots x^3}}\,\,\tilde\epsilon\,\,=\,\,
\tilde\epsilon\,\,.
\label{ncstcinco}
\eeq

In our study of the supersymmetry preserved by the giant graviton solution we
will need to know the dependence of $\epsilon$ on the coordinate $\rho$. Again,
this dependence can be extracted  by solving the corresponding supergravity
equations. Here, however, we will present a simpler argument which, as we will
check in appendix A, gives the right answer. The starting point of our argument
is to consider the supergravity solution in the asymptotic region 
$r\rightarrow\infty$ without making the near-horizon approximation
(\ref{ncquince}). In this case the harmonic function $f_p$ is given by eq. 
(\ref{strncbs2}) and, when $r\rightarrow\infty$,  the metric is flat
and the different forms vanish. The supersymmetry preserved in this region is
just the one corresponding to covariantly constant spinors, \ie\ spinors
which satisfy  $D_M\epsilon\,=\,0$ in the flat asymptotic metric. 
Taking $M\,=\,\rho$, we get that:
\beq
\partial_{\rho}\epsilon\,=\,-{1\over 4}\,\omega_{\rho}^{\underline {MN}}\,
\Gamma_{\underline {MN}}\,\epsilon\,\,,
\label{ncstseis}
\eeq
where $\omega_{\rho}^{\underline {MN}}$ are the components of the spin
connection. As the only non-zero component of the
form $\omega_{\rho}^{\underline {MN}}$ of the spin connection in our coordinates
(\ref{strncbs1}) and (\ref{ncocho}) for the the asymptotic metric is:
\beq
\omega_{\rho}^{\underline {\rho r}}\,=\,{1\over \sqrt{1-\rho^2}}\,\,,
\label{ncstsiete}
\eeq
we obtain from eq. (\ref{ncstseis})
that the $\rho$ dependence of $\tilde\epsilon$ can be
parametrized as:
\beq
\tilde\epsilon\,=\,
e^{-{\beta\over 2}\,\,
\Gamma_{\underline{\rho r} }}\,\,\,\hat\epsilon\,\,,
\label{ncstocho}
\eeq
with $\beta$ being the following function of $\rho$:
\beq
\sin\beta\,=\,\rho\,\,,
\,\,\,\,\,\,\,\,\,\,\,\,\,\,\,\,\,\,\,\,\,\,
\cos\beta\,=\,\sqrt{1-\rho^2}\,\,,
\label{ncstnueve}
\eeq
and $\hat\epsilon$ satisfying the same equation as $\tilde\epsilon$, namely:
\beq
(i\sigma_2)\,
\Gamma_{\underline{x^0\cdots x^3}}\,\,\hat\epsilon\,\,=\,\,
\hat\epsilon\,\,.
\label{ncochenta}
\eeq
The supergravity analysis of the $(D1,D3)$ background provides the explicit
dependence of $\hat\epsilon$ on the other coordinates (see eq.
(\ref{ncapadoce})). However, in our analysis of the $\kappa$-symmetry preserved by
the giant graviton solution we will only need to know that $\hat\epsilon$
satisfies eq. (\ref{ncochenta}). Actually, we will rewrite this equation in a form
more convenient for our purposes. Let us, first of all, recall that all spinors
(and in particular 
$\hat\epsilon$) have fixed chirality and thus $\hat\epsilon$ satisfies:
\beq
\Gamma_{\underline{x^0\cdots x^3}}\,\Gamma_{\underline{\rho r}}\,
\Gamma_{\underline{\phi}}\,\Gamma_{*}\,\hat\epsilon\,=\,\hat\epsilon\,\,,
\label{ncouno}
\eeq
where we have denoted:
\beq
\Gamma_{*}\,=\,\Gamma_{\underline{\theta^{1}\cdots \theta^{3}}}\,\,.
\label{ncodos}
\eeq
Taking eq. (\ref{ncouno}) into account, eq. (\ref{ncochenta}) can be written as:
\beq
\Gamma_{\underline{\rho r}}\,\hat\epsilon\,=\,
\Upsilon\,\hat\epsilon\,\,,
\label{ncotres}
\eeq
where $\Upsilon$ is the following matrix:
\beq
\Upsilon\,=\,(i\sigma_2)\,\,\Gamma_{\underline{\phi}}\,\Gamma_{*}\,\,.
\label{ncocuatro}
\eeq
Using eq. (\ref{ncotres}) in eq. (\ref{ncstocho}), we can  reexpress
$\tilde\epsilon$ as:
\beq
\tilde\epsilon\,=\,
e^{-{\beta\over 2}\,\,\Upsilon}\,\,
\,\,\,\hat\epsilon\,\,,
\label{ncocinco}
\eeq
which is the parametrization of $\tilde\epsilon$ which we will use in our
analysis of the $\kappa$-symmetry for the giant graviton.

We are now ready to determine the supersymmetry preserved by the giant graviton
in the $(D1,D3)$ background\footnote{Similar methods have been applied in
refs. \cite{Imamura, kappa} to study the supersymmetry of the baryon vertex.}. 
According to the formalism developed in the preceding section,
we have to consider a D5-brane probe extended along the directions 
$(x^2, x^3, \theta^1, \theta^2, \theta^3)$. The $\kappa$-symmetry matrix 
$\Gamma_{\kappa}$ for such a probe can be obtained from eq. (\ref{kappamatrix})
by taking $p=5$. For a D5-brane embedding of the type (\ref{ncvdos}) with
$\dot\rho=0$ the induced Dirac matrices are:
\bear
\gamma_{x^0}&=&\sqrt{-G_{tt}}\,\Gamma_{\underline{x^0}}\,+\,
\dot\phi\sqrt{G_{\phi\phi}}\,\Gamma_{\underline{\phi}}\,+\,
\dot r\sqrt{G_{rr}}\,\Gamma_{\underline{r}}\,\,,\rc\rc
\gamma_{x^{2}x^{3}}&=&f_3^{-{1\over 2}}\,h_3\,
\Gamma_{\underline{x^{2}x^{3}}}\,\,,\rc\rc
\gamma_{\theta^i}&=&f_3^{{1\over 4}}\,r\,\rho \,e_{i}^{\,\underline i}
\,\Gamma_{\underline{\theta^{i}}}\,\,,
\label{ncoseis}
\eear
where $e_{j}^{\,\underline i}$ denotes the $S^{3}$ vielbein. We will assume, as
in section 2, that the brane probe has a worldvolume gauge field whose only
non-vanishing component is ${\cal F}_{x^2 x^3}$, which we simply will denote by
${\cal F}$. By substituting the values given in eq. (\ref{ncoseis}) in eq. 
(\ref{kappamatrix}) one readily verifies that the contribution of the gauge field
${\cal F}$ exponentiates and, as a consequence, the matrix 
$\Gamma_{\kappa}$ can be written as:
\bear
\Gamma_{\kappa}&=&{\,\,i\sigma_2\over
\sqrt{-G_{tt}\,-\,G_{\phi\phi}\dot\phi^2\,-\,G_{rr}\dot r^2}}\,\,\times\rc\rc
&&\times\Bigg[\,\sqrt{-G_{tt}}\,\Gamma_{\underline{x^0}}\,+\,
\dot\phi\sqrt{G_{\phi\phi}}\,\Gamma_{\underline{\phi}}\,+\,
\dot r\sqrt{G_{rr}}\,\Gamma_{\underline{r}}\,\Bigg]\,\Gamma_{*}\,\,
e^{-\eta\,\Gamma_{\underline{x^{2}x^{3}}}
\sigma_3}\,\,.
\label{ncosiete}
\eear
In eq. (\ref{ncosiete}) $G_{MN}$ denotes the elements of the metric 
(\ref{strncbs1}) for $p=3$, $\Gamma_{*}$ is the same as in eq. (\ref{ncodos}) and
$\eta$ is defined as:
\beq
\sin\eta\,=\,{f_3^{-{1\over 2}}\,h_3^{{1\over 2}}\over \lambda_1}\,\,,
\,\,\,\,\,\,\,\,\,\,\,\,\,\,\,\,\,\,\,\,\,\,
\cos\eta\,=\,{{\cal F}\,h_3^{-{1\over 2}}\over \lambda_1}\,\,,
\label{ncoocho}
\eeq
where $\lambda_1$ has been defined in eq. (\ref{ncvocho}).

Let us now compute the action of $\Gamma_{\kappa}$ on  spinor
$\epsilon$, which we will parametrize as the Killing spinors  of the $(D1,D3)$
background, namely (see eqs.  (\ref{ncstcuatro}) and  (\ref{ncocinco})):
\beq
\epsilon\,=\,e^{{\alpha\over 2}\,\Gamma_{\underline{x^{2}x^{3}}}\,
\sigma_3}\,\,\tilde\epsilon\,=\,
e^{{\alpha\over 2}\,\Gamma_{\underline{x^{2}x^{3}}}\,
\sigma_3}\,\,
e^{-{\beta\over 2}\,\,\Upsilon}\,\,
\,\,\,\hat\epsilon\,\,,
\label{ncoochoextra}
\eeq
where $\alpha$ and $\beta$ are given in eqs. (\ref{ncsttres}) and (\ref{ncstnueve}),
$\Upsilon$ is the matrix written in (\ref{ncocuatro}) and $\hat \epsilon$ is
independent of
$\rho$. By using this representation, one immediately gets:
\bear
\Gamma_{\kappa}\,\epsilon&=&e^{(\eta-{\alpha\over 2})\,\,
\Gamma_{\underline{x^{2}x^{3}}}\, \sigma_3}\,\,\,
{i\sigma_2\over
\sqrt{-G_{tt}\,-\,G_{\phi\phi}\dot\phi^2\,-\,G_{rr}\dot r^2}}\,\,\times\rc\rc
&&\times\Bigg[\,\sqrt{-G_{tt}}\,\Gamma_{\underline{x^0}}\,+\,
\dot\phi\sqrt{G_{\phi\phi}}\,\Gamma_{\underline{\phi}}\,+\,
\dot r\sqrt{G_{rr}}\,\Gamma_{\underline{r}}\,\Bigg]\,\Gamma_{*}\,\,
\tilde\epsilon\,\,.
\label{nconueve}
\eear
Then, making use again of eq. (\ref{ncoochoextra}), one concludes that 
the equation $\Gamma_{\kappa}\,\epsilon=\epsilon$ is equivalent to
the following condition for $\tilde\epsilon$:
\bear
{i\sigma_2\over
\sqrt{-G_{tt}\,-\,G_{\phi\phi}\dot\phi^2\,-\,G_{rr}\dot r^2}}
\Bigg[\,\sqrt{-G_{tt}}\,\Gamma_{\underline{x^0}}+
\dot\phi\sqrt{G_{\phi\phi}}\,\Gamma_{\underline{\phi}}+
\dot r\sqrt{G_{rr}}\,\Gamma_{\underline{r}}\,\Bigg]\,\Gamma_{*}\,\,
\tilde\epsilon\,=
e^{(\alpha-\eta)\,\,
\Gamma_{\underline{x^{2}x^{3}}}\, \sigma_3}\,\tilde\epsilon\,\,.\rc
\label{ncnoventa}
\eear
Notice that in eq. (\ref{ncnoventa}) the matrix $\Gamma_{\underline{x^{2}x^{3}}}$
only appears on the right-hand side. Actually, if $\alpha=\eta$ this dependence 
on $\Gamma_{\underline{x^{2}x^{3}}}$ disappears. Moreover, by comparing the
definitions of $\alpha$ and $\eta$ (eqs. (\ref{ncsttres}) and (\ref{ncoocho}), 
respectively) one immediately realizes that $\alpha=\eta$ if and only if
$\lambda_1\,=\,1/\sin\varphi$. On the other hand, eq. (\ref{nctsiete}) tells us
that this only happens when $F\,=\,2\csc(2\varphi)$, \ie\ precisely when the
worldvolume gauge field strength takes the same value as the one we have found in
our hamiltonian analysis (see eq. (\ref{nctocho})). Let us assume
that this is the case and let us try to find out what are the consequences of
this fact. Actually, we will also assume that our second condition 
(\ref{ncctres}) holds. It is a simple exercise to verify that, when (\ref{ncctres})
is satisfied, the denominator of the left-hand side of eq. (\ref{ncnoventa})
takes the value:
\beq
\sqrt{-G_{tt}\,-\,G_{\phi\phi}\dot\phi^2\,-\,G_{rr}\dot r^2}\,=\,
\rho\,r\,f_3^{{1\over 4}}\,\dot\phi\,\,.
\label{ncnuno}
\eeq
Then, for a configuration satisfying eq. (\ref{nctocho}) and (\ref{ncctres}), 
the equation $\Gamma_{\kappa}\,\epsilon=\epsilon$ 
becomes:
\beq
\Bigg[\,\sqrt{-G_{tt}}\,\Gamma_{\underline{x^0\phi}}-
\dot\phi\sqrt{G_{\phi\phi}}+
\dot r\sqrt{G_{rr}}\,\Gamma_{\underline{r\phi}}\,\Bigg]\,\,
\tilde\epsilon\,=\,\rho\, r\, f_3^{{1\over 4}}\,\dot\phi\,\Upsilon
\,\,\tilde\epsilon\,\,,
\label{ncndos}
\eeq
where $\Upsilon$ has been defined in eq. (\ref{ncocuatro}). Let us now rewrite
eq. (\ref{ncndos}) in terms of the $\rho=0$ spinor $\hat\epsilon$. By
substituting eq. (\ref{ncocinco}) on both sides of eq. (\ref{ncndos}), and using the
fact that $\Upsilon$ anticommutes with $\Gamma_{\underline{x^0\phi}}$ and 
$\Gamma_{\underline{r\phi}}$, one gets:
\beq
\Bigg[\,f_3^{-{1\over 4}}\,e^{\beta\,\Upsilon}\,
\Gamma_{\underline{x^0\phi}}-
\dot\phi\,r\,f_3^{{1\over 4}}\,\sqrt{1-\rho^2}+
\dot r f_3^{{1\over 4}}\,e^{\beta\,\Upsilon}\,
\Gamma_{\underline{r\phi}}\,\Bigg]\,\,
\hat\epsilon\,=\,\rho\, r\, f_3^{{1\over 4}}\,\dot\phi\,\Upsilon
\,\,\hat\epsilon\,\,.
\label{ncntres}
\eeq
Let us consider now eq. (\ref{ncntres}) for the particular case $\rho=0$. When
$\rho=0$ the right-hand side of eq. (\ref{ncntres}) vanishes and, as 
$\beta=0$ is also zero in this case (see eq. (\ref{ncstnueve})), one has:
\beq
\Bigg[\,f_3^{-{1\over 4}}\,\Gamma_{\underline{x^0\phi}}
-\dot\phi\,r\,f_3^{{1\over 4}}+
\dot r f_3^{{1\over 4}}\Gamma_{\underline{r\phi}}\,\Bigg]\,\,
\hat\epsilon\,=\,0\,\,.
\label{ncncuatro}
\eeq
For a general value of $\rho$ the $\kappa$-symmetry condition can be obtained
by substituting in (\ref{ncntres}) $e^{\beta\,\Upsilon}$ by:
\beq
e^{\beta\,\Upsilon}\,=\,\sqrt{1-\rho^2}\,+\,\rho\,\Upsilon\,\,.
\label{ncncinco}
\eeq
By so doing one obtains two types of terms, with and without $\Upsilon$, which,
amazingly, satisfy the equation independently for all $\rho$ as a consequence of
the $\rho=0$  condition (\ref{ncncuatro}). Thus, eq. (\ref{ncncuatro}) is
equivalent to the $\kappa$-symmetry condition 
$\Gamma_{\kappa}\epsilon=\epsilon$ and is the constraint we have to
impose to the Killing spinors of the background in order to define a
supersymmetry transformation preserved by our brane probe configurations. In
order to obtain a neat interpretation of (\ref{ncncuatro}), let us define the
matrix $\Gamma_v$ as:
\beq
\Gamma_v\,\equiv\,v^{\underline{r}}\,\Gamma_{\underline{r}}\,+\,
v^{\underline{\phi}}\,\Gamma_{\underline{\phi}}\,\,,
\label{ncnseis}
\eeq
where $v^{\underline{r}}$ and $v^{\underline{\phi}}$ are the components of the
velocity vector ${\bf v}$ defined in eq. (\ref{ncccuatro}). By using eq. 
(\ref{ncccinco}), which is a consequence of (\ref{ncctres}), one readily proves that
the matrix $\Gamma_v$ satisfies:
\beq
(\,\Gamma_v\,)^2\,=\,1\,\,.
\label{ncnsiete}
\eeq
Moreover, from the explicit expression of the components of ${\bf v}$ (see eq. 
(\ref{ncccuatro})) it is straightforward to demonstrate that the $\kappa$-symmetry
condition (\ref{ncncuatro}) can be written as:
\beq
\Big[\Gamma_{\underline{x^0}}\,+\,\Gamma_v\,\Big]\,\hat\epsilon\,=\,0\,\,.
\label{ncnocho}
\eeq
Taking into account eq. (\ref{ncnsiete}), 
one can  rewrite eq. (\ref{ncnocho}) in the form:
\beq
\Gamma_{\underline{x^0}}\,\Gamma_v\,\hat\epsilon\,\,=\,\,
\hat\epsilon\,\,.
\label{ncnochoextra}
\eeq
Moreover, recalling the relation (\ref{ncoochoextra}) between 
$\hat\epsilon$ and $\epsilon$, and taking into account that 
$\Gamma_{\underline{x^0}}\,\Gamma_v$ commutes with
$\Gamma_{\underline{x^{2}x^{3}}}$, eq. (\ref{ncnochoextra}) is equivalent to:
\beq
\Gamma_{\underline{x^0}}\,\Gamma_v\,\epsilon_{{\,\big |}_{\rho=0}}\,\,=\,\,
\epsilon_{{\,\big |}_{\rho=0}}\,\,.
\label{ncnnueve}
\eeq
Eq. (\ref{ncnnueve}) is the condition satisfied by the parameter of the
supersymmetry preserved by a massless particle
which moves  in the direction of the vector ${\bf v}$ at $\rho=0$, \ie\ by a
gravitational wave which propagates precisely along the trajectories found 
above. However, the background projector 
$(i\sigma_2)\, \Gamma_{\underline{x^0\cdots x^3}}$ and the one of the probe,
$\Gamma_{\underline{x^0}}\,\Gamma_v$ do not commute (actually, they
anticommute). This means that the conditions (\ref{ncochenta}) and
(\ref{ncnochoextra}) cannot be imposed at the same time and, therefore, the probe
breaks completely the supersymmetry of the background. The most relevant
aspect of this result is that the supersymmetry breaking  produced by the probe
is just identical to the one corresponding to a massless particle which moves
in the direction of ${\bf v}$. This fact is a confirmation of our
interpretation of the giant graviton configurations as blown up gravitons.

Notice that we have found the supersymmetry projection for the
giant graviton configurations from the conditions (\ref{nctocho}) and
(\ref{ncctres}). Clearly, we could have done our reasoning in reverse order and,
instead, we could have imposed first that our brane probe breaks
supersymmetry as a massless particle at $\rho=0$. In this case we would arrive
at the same conditions as those obtained by studying the hamiltonian and,
actually, this would be an alternative way to derive them.


\setcounter{equation}{0}
\section{M-Theory Giant Gravitons with C field} \label{MTGG}
\medskip

In this section we extend the analysis of previous section to M-theory
backgrounds generated by a stack of non-threshold bound states of the type
(M2,M5), which was given in section \ref{strsecmtbs}. The probe we will consider is, 
as mentioned in the introduction, an M5-brane wrapped on an $S^2$ transverse sphere 
and extended along three directions parallel to the background. It will be shown that, 
after switching on a particular value of the worldvolume gauge field, one can find 
giant graviton solutions of the corresponding worldvolume equations of motion. It 
will also be analyzed the supersymmetry of the problem, and it will be shown that 
our M5-brane configurations break supersymmetry exactly in the same way as a wave 
which 
propagates with the velocity of the center of mass of the M5-brane probe.

The metric in the one given in (\ref{strmtuno}) and the four-form field strength 
$F^{(4)}\,=\,dC^{(3)}$ was given in (\ref{strmttres}). In order to
obtain the explicit form of $C^{(3)}$, let us introduce again the particular set of
coordinates for the transverse $S^4$ given by (\ref{ads5}). Then, the line element 
$d\O_{4}^2$ can be written as:
\be
d\O_{4}^2\,=\,{1\over 1-\rho^2}\,d\rho^2\,+\,
(1\,-\,\rho^2\,)\,d\phi^2\,+\,\rho^2\,d\O_{2}^2\,\,,
\label{mtcuatro}
\ee
where $d\O_{2}^2$ is the metric of a unit $S^2$ (which we will parametrize
by means of two angles $\theta^1$ and $\theta^2$) and  $\rho$ and $\phi$ take
values in the range $0\le\rho\le 1$ and $0\le\phi\le 2\pi$ respectively. 
In these coordinates one can
take $C^{(3)}$ as:
\bear
C^{(3)}\,&=&\,-\sin\varphi\,f^{-1}\,dx^0\wedge dx^1\wedge
dx^2\,-\, R^3\cos\varphi \,\rho^3\,d\phi\wedge\epsilon_{(2)}\,+\cr\cr
&&+\tan\varphi \,hf^{-1}\,dx^3\wedge dx^4\wedge dx^5\,\,,
\label{mtcinco}
\eear
where $\epsilon_{(2)}$ is the volume form of the $S^2$. It is not difficult to
verify from eq. (\ref{strmttres}) that  ${}^*F^{(4)}$ satisfies:
\beq
d\,{}^*F^{(4)}\,=\,-{1\over 2}\,F^{(4)}\wedge F^{(4)}\,,
\label{mtseis}
\eeq
where the seven-form ${}^*F^{(4)}$ is the Hodge dual of $F^{(4)}$ with respect to
the metric  (\ref{strmtuno}).  Eq.
(\ref{mtseis}) implies that ${}^*F^{(4)}$ can be represented in terms of a
six-form potential 
$C^{(6)}$ as follows:
\beq
{}^*F^{(4)}\,=\,d\,C^{(6)}\,-\,{1\over 2}\,C^{(3)}\wedge dC^{(3)}\,\,.
\label{mtsiete}
\eeq
By taking the exterior derivative of both sides of (\ref{mtsiete}), one
immediately verifies eq. (\ref{mtseis}). Moreover, it is not difficult to find the
potential $C^{(6)}$ in our coordinate system. Actually, one can easily check 
that one can take $C^{(6)}$ as:
\bear
C^{(6)}\,&=&\,{1\over 2}\,\sin\varphi\cos\varphi\,f^{-1}\,R^3\,\rho^3\,
dx^0\wedge dx^1\wedge dx^2\wedge d\phi\wedge\epsilon_{(2)}\,-\,\rc\rc
&&-\,{1\over 2}\,{1+h\cos^2\varphi\over \cos\varphi}\,f^{-1}\,
dx^0\wedge dx^1\wedge dx^2\wedge dx^3\wedge dx^4\wedge dx^5\,-\,\rc\rc
&&-\,{1\over 2}\,\sin\varphi\,R^3\,\rho^3\,h\,f^{-1}\,
dx^3\wedge dx^4\wedge dx^5\wedge d\phi\wedge \epsilon_{(2)}\,\,.
\label{mtocho}
\eear

\medskip
\subsubsection{The M5-brane probe}
\medskip
We shall now consider the near-horizon region of the $(M2,M5)$ geometry. In
this region the radial coordinate $r$ is small and one can approximate the
function $f$ appearing in the supergravity solution as 
$f\,\approx\,R^{3}/ r^{3}$. Following the analysis of the preceding section, we
place an M5-brane probe in this geometry in such a way that it shares three
directions  $(x^3, x^4, x^5)$ with the branes of the background and wraps the
$S^2$ transverse sphere parametrized by the angles $\theta^1$ and $\theta^2$. The
dynamics of the M5-brane probe is determined by its worldvolume action, \ie\ by
the PST action, given in (\ref{actPST}). 

The worldvolume coordinates $\s^{i}$ ($i=0,\dots, 5$) will be taken as 
$\s^{i}=(\,x^0,x^3,x^4,x^5,\theta^1,\theta^2\,)$. In this system of
coordinates the configurations we are interested in are described  by functions
of the type $r=r(t)\,\,$, $\rho=\rho(t)\,\,$ and $\phi=\phi(t)\,\,$, where
$t\equiv x^0$. Moreover, we will assume that the only non-vanishing components
of the field $H$, defined in (\ref{actH}), are those of $P[C^{(3)}]$, \ie\ 
$H_{x^3x^4x^5}\equiv H_{345}$ and
$H_{x^0\theta^1 \theta^2}\equiv H_{0*}$. As discussed in section \ref{secPSTact}, the
scalar field $a$ is an auxiliary field which, by fixing its gauge symmetry, can
be eliminated from the action at the expense of loosing manifest covariance. In
this paper we will work in the gauge $a=x^0$. In this gauge the only non-zero
component of $\tilde H$ is:
\beq
\tilde H_{\theta^1\theta^2}\,=\,f^{7/6}\,h^{-4/3}\,r^2\rho^2\,
\sqrt{\hat g^{(2)}}\,\,\,H_{345}\,\,,
\label{mttrece}
\eeq
where $\hat g^{(2)}$ is the determinant of the metric of the two-sphere. 
By using (\ref{mttrece}) one can easily obtain ${\cal L}_{DBI}$ for our
configurations. Indeed, after a short calculation one gets:
\beq
{\cal L}_{DBI}\,=\,-R^3\rho^2\,\sqrt{\hat g^{(2)}}\,\lambda_1\,
\sqrt{r^{-2}\,f^{-1}\,\,-\,r^{-2}\dot r^2\,-\,
{\dot\rho^2\over 1-\rho^2}\,-\,
(1-\rho^2)\,\dot\phi^2}\,\,,
\label{mtcatorce}
\eeq
where the dot denotes time derivative and $\lambda_1$ is defined as:
\beq
\lambda_1\,\equiv\,\sqrt{h\,f^{-1}\,+\,\big(\,H_{345}\big)^2\,h^{-1}}\,\,.
\label{mtquince}
\eeq
It is also very easy to prove that the remaining terms of the action are:
\bear
{\cal L}_{H\tilde H}\,+\,{\cal L}_{WZ}&=&{1\over 2}\,F_{345}\,F_{0*}\,-\,
F_{345}\,P[C^{(3)}]_{0*}\,+\,\rc\rc
&&+\,P[C^{(6)}]_{0345*}\,+\,{1\over 2}\,P[C^{(3)}]_{345}\,P[C^{(3)}]_{0*}\,\,,
\label{mtdseis}
\eear
with $F_{0*}\,\equiv\,F_{x^0\theta^1\theta^2}$ and similarly for the pull-backs
of $C^{(6)}$ and $C^{(3)}$. From eqs. (\ref{mtcinco}) and (\ref{mtocho}) it
follows that:
\bear
P[C^{(6)}]_{0345*}&=& {1\over 2}R^3\rho^3\sin\varphi\,h\,f^{-1}\,
\sqrt{\hat g^{(2)}}\,\dot\phi\,\,,\rc\rc
P[C^{(3)}]_{345}&=&\tan\varphi\,h\,f^{-1}\,\,,\rc\rc
P[C^{(3)}]_{0*}&=&-R^3\rho^3\cos\varphi\,\sqrt{\hat g^{(2)}}\,\dot\phi\,\,.
\label{mtdsiete}
\eear
By using eq. (\ref{mtdsiete}) it is straightforward to demonstrate that the sum
of the  last two terms in ${\cal L}_{H\tilde H}\,+\,{\cal L}_{WZ}$ vanishes
and, thus,  we can write:
\beq
{\cal L}_{H\tilde H}\,+\,{\cal L}_{WZ}\,=\,R^3\rho^3\,F_{345}\cos\varphi\,
\sqrt{\hat g^{(2)}}\,\dot\phi\,+\,{1\over 2}\,F_{345}\,F_{0*}\,\,.
\label{mtdocho}
\eeq
Let us assume that $F_{0*}\,=\,\sqrt{\hat g^{(2)}}\,f_{0*}$ with $f_{0*}$
independent of the angles of the $S^2$. With this ansatz for the electric
component of $F$, the action can be written as:
\beq
S\,=\,\int dt\,dx^3\,dx^4\,dx^5\,\,{\cal L}\,\,,
\label{mtdnueve}
\eeq
with the lagrangian density ${\cal L}$ given by:
\bear
{\cal L}&=&4\pi R^3\,T_{M5}\,\Bigg[\,-\rho^2\,
\lambda_1\, \sqrt{r^{-2}\,f^{-1}\,\,-\,r^{-2}\dot r^2\,-\,
{\dot\rho^2\over 1-\rho^2}\,-\, (1-\rho^2)\,\dot\phi^2}\,+\,\rc\rc
&&+\,\lambda_2\,\rho^3\dot\phi\,+\,{1\over 2 R^3}\,
F_{345}\,f_{0*}\,\Bigg]\,\,.
\label{mtveinte}
\eear
In eq. (\ref{mtveinte}) we have defined $\lambda_2$ as:
\beq
\lambda_2\,\equiv\,F_{345}\,\cos\varphi\,\,.
\label{mtvuno}
\eeq
It is interesting to characterize the spreading of
the M5-brane in the $x^3x^4x^5$ directions by means of the flux of the
worldvolume gauge field $F$. We shall parametrize this flux as follows:
\beq
\int dx^3\,dx^4\,dx^5\,\,F\,=\,{2\pi\over T_{M2}}\,\,N'\,\,,
\label{mtvdos}
\eeq
where $T_{M2}\,=\,1/(2\pi)^2\,l_{p}^3$ in the tension of the M2-brane. Notice
that, when the coordinates $x^3x^4x^5$ are compact, the condition (\ref{mtvdos})
is just the M-theory flux quantization condition found in section \ref{flM2}, with
the flux number $N'$ being an integer for topological reasons. 

In order to perform a canonical hamiltonian analysis of this system, let us
introduce the density of momenta:
\bear
{\cal P}_r&=&{\partial {\cal L}\over \partial \dot r}\,\equiv\,
4\pi R^{3}\,T_{M5}\,\lambda_1\,\pi_r\,\,,\rc\rc
{\cal P}_{\rho}&=&{\partial {\cal L}\over \partial \dot \rho}\,\equiv\,
4\pi R^{3}\,T_{M5}\,\lambda_1\,\pi_{\rho}\,\,,\rc\rc
{\cal P}_{\phi}&=&{\partial {\cal L}\over \partial \dot \phi}\,\equiv\,
4\pi R^{3}\,T_{M5}\,\lambda_1\,\pi_{\phi}\,\,,
\label{mtvtres}
\eear
where we have defined the reduced momenta $\pi_r$, $\pi_{\rho}$ and
$\pi_{\phi}$.  From the explicit value of ${\cal L}$ (eq. (\ref{mtveinte})), we
get:
\bear
 \pi_r&=&{\rho^{2}\over r^2}\,\,{\dot r\over
\sqrt{r^{-2}\,f^{-1}\,\,-\,r^{-2}\dot r^2\,-\,
{\dot\rho^2\over 1-\rho^2}\,-\,(1-\rho^2)\,\dot\phi^2}}\,\,,\rc\rc
\pi_{\rho}&=&{\rho^{2}\over 1- \rho^2}\,\,{\dot \rho\over
\sqrt{r^{-2}\,f^{-1}\,\,-\,r^{-2}\dot r^2\,-\,
{\dot\rho^2\over 1-\rho^2}\,-\,(1-\rho^2)\,\dot\phi^2}}\,\,,\rc\rc
 \pi_{\phi}&=&(1-\rho^2)\rho^{2}\,\,{\dot \phi\over
\sqrt{r^{-2}\,f^{-1}\,\,-\,r^{-2}\dot r^2\,-\,
{\dot\rho^2\over 1-\rho^2}\,-\,(1-\rho^2)\,\dot\phi^2}}
\,\,+\,\,\Lambda\,\rho^{3}\,\,,
\label{mtvcuatro}
\eear
where we have introduced the quantity  
$\Lambda \equiv \lambda_2 /\lambda_1$. The hamiltonian density of the system is:
\beq
{\cal H}\,=\,\dot r\,{\cal P}_r\,+\,\dot\rho\,{\cal P}_{\rho}\,
+\,\dot\phi{\cal P}_{\phi}\,
+\,F_{0*}\,{\partial {\cal L}\over \partial F_{0*}} \,-\,{\cal L}\,\,.
\label{vcinco}
\eeq
After a short calculation one can prove that ${\cal H}$ is given by:
\beq
{\cal H}\,=\,4\pi R^3\,T_{M5}\,\lambda_1\,
r^{-1}\,f^{-{1\over 2}}\,\Bigg[\,r^2\,\pi_r^2\,+\,\rho^{4}\,+\,
(1-\rho^2)\,\pi_{\rho}^2\,+\,
{\Big(\pi_{\phi}-\Lambda\rho^{3}\Big)^2\over 1-\rho^2}
\,\,\Bigg]^{{1\over 2}}\,\,.
\label{mtvseis}
\eeq

\medskip
\subsubsection{Giant graviton configurations}
\medskip

We are interested in finding configurations of fixed size,
\ie\ those solutions of the equations of motion with constant $\rho$. By
comparing the hamiltonian density written in (\ref{mtvseis}) with the one studied
in ref. \ref{NCGG}, it is not difficult to realize that these fixed size
solutions exist if the quantity $\Lambda$ takes the value $\Lambda=1$. Indeed,
if this condition holds, the hamiltonian density ${\cal H}$ can be put as:
\beq
{\cal H}\,=\,4\pi R^3\,T_{M5}\,\lambda_1\,
r^{-1}\,f^{-{1\over 2}}\,\Bigg[\,\pi_{\phi}^2\,+\,r^2\,\pi_r^2\,+\,
(1-\rho^2)\,\pi_{\rho}^2\,+\,
{\Big(\pi_{\phi}\rho-\rho^{2}\Big)^2\over 1-\rho^2}
\,\,\Bigg]^{{1\over 2}}\,\,,
\label{mtvsiete}
\eeq
and, as we will verify soon, one can easily find constant $\rho$ solutions of the
equations of motion for the hamiltonian (\ref{mtvsiete}). Moreover, by using the
value of $P[C^{(3)}]_{345}$ given in eq. (\ref{mtdsiete}), one can write
$\lambda_1$ as:
\beq
\lambda_1^2\,=\,\cos^2\varphi\,F_{345}^2\,+\,f^{-1}\,
\Big(\,F_{345}\sin\varphi\,-\,{1\over \cos\varphi}\Big)^{2}\,\,.
\label{mtvocho}
\eeq
Taking into account the definition of $\lambda_2$ (eq. (\ref{mtvuno})), it
follows that the condition  $\Lambda\,=\,1$ (or $\lambda_1=\lambda_2$) is
equivalent to have the following constant value of the worldvolume gauge field:
\beq
F_{345}\,=\,{1\over \sin\varphi\cos\varphi}\,=\,2\csc (2\varphi)\,\,.
\label{mtvnueve}
\eeq
It follows  from eq. (\ref{mtvcuatro}) that for a configuration with $\dot\rho=0$
the momentum $\pi_{\rho}$ necessarily vanishes and, in particular, one must
require that $\dot\pi_{\rho}=0$. Then, the hamiltonian equations of motion
imply that $\partial {\cal H}/\partial \rho$ must be zero, which happens if the
last term inside the square root of the right-hand side of eq. (\ref{mtvsiete})
vanishes, \ie\ when $\pi_{\phi}\rho\,-\,\rho^{2}=0$. This occurs either when
$\rho=0$ or else when the angular momentum  $\pi_{\phi}$ is:
\beq
\pi_{\phi}\,=\,\rho\,\,.
\label{mttreinta}
\eeq
In order to clarify the nature of these solutions, let us invert the relation
between $\pi_{\phi}$ and $\dot\phi$ (eq. (\ref{mtvcuatro})). After a simple
calculation one gets:
\beq
\dot\phi\,=\,
{\pi_{\phi}-\rho^{3}\over 1-\rho^2}\,\,
{\Bigg[\,r^{-2}\big(f^{-1}\,-\,\dot r^2\,\big)\,-\,
{\dot\rho^2\over 1-\rho^2}\,\Bigg]^{{1\over 2}}\over
\Bigg[\,\pi_{\phi}^2\,+\,
{\big(\,\pi_{\phi}\rho\,-\,\rho^{2}\,\big)^2
\over 1-\rho^2}\,\Bigg]^{{1\over 2}}}\,\,.
\label{mttuno}
\eeq
By taking $\dot\rho=\pi_{\phi}\rho\,-\,\rho^{2}=0$ on the right-hand side of
eq. (\ref{mttuno}), one finds the following relation between $\dot\phi$ and
$\dot r\,$:
\beq
f\,\big[\,r^2\,\dot\phi^2\,+\,\dot r^2\,]\,=\,1\,\,.
\label{mttdos}
\eeq
Remarkably, eq. (\ref{mttdos}) is the condition satisfied by a particle which
moves in the $(r,\phi)$ plane at $\rho=0$ along a null trajectory (\ie\ with 
$ds^2=0$) in the metric (\ref{strmtuno}). Therefore, our brane probe configurations
behave as a massless particle: the so-called giant graviton. The point $\rho=0$
can be interpreted as the ``center of mass" of the expanded brane. Actually,
if one defines the velocity vector  ${\bf v}$ as 
${\bf v}\,=\,(v^{\underline{r}}, v^{\underline{\phi}})\,\equiv\,
f^{{1\over 2}}\,(\dot r\,,\,r\dot\phi) $, eq. (\ref{mttdos}) is equivalent to
the condition $(v^{\underline{r}})^2\,+\,(v^{\underline{\phi}})^2\,=\,1$ and,
thus, the center of mass of the giant graviton moves at the speed of light. On
the other hand,  the angular momentum density  ${\cal P}_{\phi}$ for the
$\pi_{\phi}=\rho$ solution can be obtained from eq. (\ref{mtvtres}), namely:
\beq
{\cal P}_{\phi}\,=\,{T_{M2}\over 2\pi}\,F_{345}\,N\,\rho\,\,.
\label{mtttres}
\eeq
Moreover, by integrating the densities ${\cal P}_{\phi}$ and 
${\cal P}_{r}$ along the $x^{3}x^4x^5$ directions, one gets the values of the
momenta $p_\phi$ and $p_r$:
\beq
p_\phi\,=\,\int dx^{3}\,dx^4\,\,dx^5\,{\cal P}_{\phi}\,\,,
\,\,\,\,\,\,\,\,\,\,\,\,\,\,\,\,\,\,\,\,\,\,
p_r\,=\,\int dx^{3}\,dx^4\,\,dx^5\,\,{\cal P}_{r}\,\,.
\label{mttcuatro}
\eeq
By using the value of the momentum density ${\cal P}_{\phi}$ displayed in eq. 
(\ref{mtttres}), together with the flux quantization condition (\ref{mtvdos}), one
gets the following value of $p_\phi$:
\beq
p_\phi\,=\,N\,N'\,\rho\,\,,
\label{mttcinco}
\eeq
which implies that the size $\rho$ of the wrapped brane increases with its
angular momentum $p_\phi$. As $0\le\rho\le 1$, the momentum $p_\phi$ has a
maximum given by  $p_\phi^{max}\,=\,N\,N'$. This maximum is reached when
$\rho=1$ and its existence is a manifestation of the stringy exclusion
principle. 

In order to analyze the energy of the giant graviton solution, let 
${\cal G}_{MN}$ be the metric elements of eqs. (\ref{strmtuno}) and (\ref{mtcuatro})
at the point $\rho=0$. Then, it is straightforward to verify that the
hamiltonian $H_{GG}$ of the giant graviton configurations is:
\beq
 H_{GG}\,=\,\sqrt{-{\cal G}_{tt}}\,\Bigg[\,
 {p_{\phi}^2\over {\cal G}_{\phi\phi}}\,+\,
{ p_{r}^2\over {\cal G}_{rr}}\,\,\Bigg]^{{1\over 2}}\,\,,
\label{mttseis}
\eeq
which is exactly the one corresponding to a massless particle which moves in
the $(r,\phi)$ plane under the action of the metric ${\cal G}_{MN}$. By
substituting in eq. (\ref{mttseis}) the explicit values of the ${\cal G}_{MN}$'s,
one can write 
$H_{GG}$ as:
\beq
 H_{GG}\,=\,R^{-{3\over 2}}\,\,
\Bigg[\,r^{3}\, p_{r}^2\,+\,r\, p_{\phi}^2
\,\Bigg]^{{1\over 2}}\,\,.
\label{mttsiete}
\eeq
By using the conservation of energy, one can integrate the equations of motion
and get the functions $r(t)$ and $\phi(t)$. It turns out that the corresponding
equations coincide with one of the cases studied in section \ref{NCGG}.
Therefore, we simply write the results of this integration and refer to
section \ref{NCGG} for the details of the calculation. One gets:
\bear
r\,&=&{r_{*}\over 1\,+\,{r_{*}\over 4R^3}\,(t\,-\,t_{*})^2}\,\,,\rc\rc\rc
\tan\Big[{\phi\,-\,\phi_*\over 2}\Big]&=&{1\over 2R}\,\,
\Big({r_{*}\over R}\Big)^{{1\over 2}}\,(t\,-\,t_{*})\,\,,
\label{mttocho}
\eear
where $r_{*}$, $\phi_*$ and $t_{*}$ are constants. Notice that $r\le r_{*}$ and
that $r\rightarrow 0$ as $t\rightarrow\infty$, which means that the giant
graviton always falls asymptotically to the center of the potential. 

It is also interesting to study the volume occupied by the M5-brane probe along
the $x^{3}x^4x^5$ directions. By plugging the value (\ref{mtvnueve}) of the
worldvolume gauge field into the flux quantization condition (\ref{mtvdos}), one
gets that this volume is:
\beq
\int dx^{3}\,dx^4\,\,dx^5\,=\,{\pi N'\over T_{M2}}\,\,\,\sin(2\varphi)\,\,.
\label{mttnueve}
\eeq
When $\varphi\rightarrow 0$, the M2-brane component of 
the background bound
state disappears and we are left with a M5-brane background. For fixed $N'$, it
follows from eq. (\ref{mttnueve}) that the three directions of the M5-brane probe
which are parallel to the background collapse and, therefore, the M5-brane
probe is effectively converted into a M2-brane, in agreement  with the results
of section \ref{ADSGG}. 

The gauge field $F$ of the PST action satisfies a generalized self-duality
condition which relates its electric and magnetic components. In order to get
this self-duality constraint  one must  use both the equations of motion and the
symmetries of the  PST action. In our case, this condition
reduces to:
\beq
{\partial {\cal L}\over \partial F_{345}}\,=\,0\,\,.
\label{mtcuarenta}
\eeq
Indeed, after using the explicit expression of ${\cal L}$ (eq. (\ref{mtveinte})),
and solving eq. (\ref{mtcuarenta}) for $f_{0*}$, one gets:
\beq
f_{0*}\,=\,2R^3\Bigg[\,\rho^2\,{H_{345}\over \lambda_1\,h}\,
\sqrt{r^{-2}\,f^{-1}\,\,-\,r^{-2}\dot r^2\,-\,
{\dot\rho^2\over 1-\rho^2}\,-\,(1-\rho^2)\,\dot\phi^2}\,-\,
\cos\varphi\,\,\rho^3\,\dot\phi\,\Bigg]\,\,.
\label{mtcuno}
\eeq
By substituting on the right-hand side of eq. (\ref{mtcuno}) the values
corresponding to our giant graviton configurations, one gets a vanishing
result, \ie:
\beq
{f_{0*}}_{\big|{GG}}\,=\,0\,\,.
\label{mtcdos}
\eeq
Thus, our expanded graviton solutions have zero electric  
field on the M5-brane worldvolume.

\medskip
\subsubsection{Supersymmetry}
\medskip
Let us now examine the supersymmetry of our configurations. First of all, we
consider the supersymmetry preserved by the background. As the solution of
D=11 supergravity we are dealing with is purely bosonic, it is only invariant
under those supersymmetry transformations which do no change the gravitino field 
$\psi_M$. This calculation is detailed in appendix B. 
Actually, if we define the matrix
$\Upsilon\,=\,\Gamma_{\underline{\phi}}\,\Gamma_{*}$ 
with $\Gamma_{*}\,\equiv\,\Gamma_{\underline{\theta^1\theta^2}}$, they
can be parametrized as follows: 
\beq
\epsilon\,=\, e^{{\alpha\over 2}\,\Gamma_{\underline{x^{3}x^{4}x^{5}}}}\,\,
e^{-{\beta\over 2}\,\,\Upsilon}\,\,
\,\,\,\hat\epsilon\,\,,
\label{mtccuatro}
\eeq
where $\alpha$ and $\beta$ are:
\bear
\sin\alpha&=&\,f^{-{1\over 2}}\,h^{{1\over 2}}\sin\varphi\,\,,
\,\,\,\,\,\,\,\,\,\,\,\,\,\,\,\,\,\,\,\,\,\,
\cos\alpha\,=\,h^{{1\over 2}}\cos\varphi\,\,,\rc\rc
\sin\beta&=&\rho\,\,,
\,\,\,\,\,\,\,\,\,\,\,\,\,\,\,\,\,\,\,\,\,\,
\,\,\,\,\,\,\,\,\,\,\,\,\,\,\,\,\,\,\,\,\,\,\,\,\,\,\,\,\,\,
\cos\beta\,=\,\sqrt{1-\rho^2}\,\,,
\label{mtccinco}
\eear
and $\hat\epsilon$ is independent of $\rho$ and satisfies:
\beq
\Gamma_{\underline{x^{0}\cdots x^{5}}}\,\,\hat\epsilon\,=\,\hat\epsilon\,\,.
\label{mtcseis}
\eeq
By working out the condition $\delta\psi_M\,=\,0$ one can determine
$\hat\epsilon$ completely. We will do this calculation in appendix B. 
Nevertheless representation (\ref{mtccuatro}) is enough for our purposes. Let us 
however
mention that that it follows from this analysis that the (M2,M5) background is
$1/2$ supersymmetric. 

The number of supersymmetries preserved by the M5-brane probe is the number
independent solutions of the equation $\Gamma_{\kappa}\epsilon=\epsilon$, where 
$\epsilon$ is one of the Killing spinors (\ref{mtccuatro}) and 
$\Gamma_{\kappa}$ is the $\kappa$-symmetry matrix of the PST
formalism (\ref{kappaM5}). In our case the vector $t^m$ defined in (\ref{kM5def}) 
is zero and the only non-zero component of $\n_m$ is: $\n_0\,=\,\sqrt{-G_{tt}}$. 
Using these facts, after some calculation, one can represent $\G_{\k}$ as:
\bear
\Gamma_{\kappa}&=&{\,\,1\over
\sqrt{-G_{tt}\,-\,G_{\phi\phi}\dot\phi^2\,-\,G_{rr}\dot r^2}}\,\,\times\rc\rc
&&\times\Bigg[\,\sqrt{-G_{tt}}\,\Gamma_{\underline{x^0}}\,+\,
\dot\phi\sqrt{G_{\phi\phi}}\,\Gamma_{\underline{\phi}}\,+\,
\dot r\sqrt{G_{rr}}\,\Gamma_{\underline{r}}\,\Bigg]\,\Gamma_{*}\,\,
e^{-\eta\,\Gamma_{\underline{x^{3}x^{4}x^{5}}}}\,\,,
\label{mtcnueve}
\eear
with $\eta$ given by:
\beq
\sin\eta\,=\,{f^{-{1\over 2}}\,h^{{1\over 2}}\over \lambda_1}\,\,,
\,\,\,\,\,\,\,\,\,\,\,\,\,\,\,\,\,\,\,\,\,\,
\cos\eta\,=\,{H_{345}\,h^{-{1\over 2}}\over \lambda_1}\,\,. 
\label{mtcincuenta}
\eeq
By using eqs. (\ref{mtcnueve}) and (\ref{mtccuatro}), the equation 
$\Gamma_{\kappa}\epsilon=\epsilon$ takes the form:
\bear
{1\over
\sqrt{-G_{tt}\,-\,G_{\phi\phi}\dot\phi^2\,-\,G_{rr}\dot r^2}}
&&\Bigg[\,\sqrt{-G_{tt}}\,\Gamma_{\underline{x^0}}+
\dot\phi\sqrt{G_{\phi\phi}}\,\Gamma_{\underline{\phi}}+
\dot r\sqrt{G_{rr}}\,\Gamma_{\underline{r}}\,\Bigg]\,\Gamma_{*}\,\,
e^{-{\beta\over 2}\,\Upsilon}\hat\epsilon\,=\rc\rc
&&=\,e^{(\alpha-\eta)\,\,
\Gamma_{\underline{x^{3}x^{4}x^{5}}}}\,
e^{-{\beta\over 2}\,\Upsilon}\hat\epsilon\,\,.
\label{mtciuno}
\eear
Let us now evaluate eq. (\ref{mtciuno}) for our solution. First of all, one can
verify that, when the worldvolume gauge field $F_{345}$ takes the value 
(\ref{mtvnueve}), the angles $\alpha$ and $\eta$ are equal and, thus, the
dependence on $\Gamma_{\underline{x^{3}x^{4}x^{5}}}$ of the right-hand side of 
(\ref{mtciuno}) disappears. Moreover, using the condition (\ref{mttdos}), and
performing some simple manipulations, one can convert eq.  (\ref{mtciuno}) into:
\beq
\Bigg[\,f^{-{1\over 4}}\,e^{\beta\,\Upsilon}\,
\Gamma_{\underline{x^0\phi}}-
\dot\phi\,r\,f^{{1\over 4}}\,\sqrt{1-\rho^2}+
\dot r f^{{1\over 4}}\,e^{\beta\,\Upsilon}\,
\Gamma_{\underline{r\phi}}\,\Bigg]\,\,
\hat\epsilon\,=\,\rho\, r\, f^{{1\over 4}}\,\dot\phi\,\Upsilon
\,\,\hat\epsilon\,\,.
\label{mtcidos}
\eeq
If, in particular, we take $\rho=0$ in eq. (\ref{mtcidos}), one arrives at:
\beq
\Bigg[\,f^{-{1\over 4}}\,\Gamma_{\underline{x^0\phi}}
-\dot\phi\,r\,f^{{1\over 4}}+
\dot r f^{{1\over 4}}\Gamma_{\underline{r\phi}}\,\Bigg]\,\,
\hat\epsilon\,=\,0\,\,.
\label{mtcitres}
\eeq
Remarkably, if eq. (\ref{mtcitres}) holds, then eq. (\ref{mtcidos}) is satisfied for
an arbitrary value of $\rho$. Thus, eq. (\ref{mtcitres}) is equivalent to the
$\kappa$-symmetry condition $\Gamma_{\kappa}\epsilon=\epsilon$.  In order to
interpret (\ref{mtcitres}), let us define the matrix 
$\Gamma_v\,\equiv\,v^{\underline{r}}\,\Gamma_{\underline{r}}\,+\,
v^{\underline{\phi}}\,\Gamma_{\underline{\phi}}$, where $v^{\underline{r}}$ and 
$v^{\underline{\phi}}$ are the components of the center of mass velocity vector
${\bf v}$ defined above. This matrix is such that 
$(\,\Gamma_v\,)^2\,=\,1$, and one can prove that eq. (\ref{mtcitres}) can
be written as:
\beq
\Gamma_{\underline{x^0}}\,\Gamma_v\,\hat\epsilon\,\,=\,\,
\hat\epsilon\,\,.
\label{mtcicuatro}
\eeq
Taking into account the relation (\ref{mtccuatro}) between $\hat\epsilon$ and
$\epsilon$, and using the fact that $\Gamma_{\underline{x^0}}\,\Gamma_v$
commutes with $\Gamma_{\underline{x^{3}x^{4}x^{5}}}$, one can recast eq. 
(\ref{mtcicuatro}) as:
\beq
\Gamma_{\underline{x^0}}\,\Gamma_v\,\epsilon_{{\,\big |}_{\rho=0}}\,\,=\,\,
\epsilon_{{\,\big |}_{\rho=0}}\,\,,
\label{mtcicinco}
\eeq
which is the supersymmetry projection induced by a massless particle 
moving in the direction of ${\bf v}$ at $\rho=0$. Notice that, however,  the
background projector 
$\Gamma_{\underline{x^{0}\cdots x^{5}}}$ does not commute with 
$\Gamma_{\underline{x^0}}\,\Gamma_v$ and, therefore, eq. (\ref{mtcseis}) and
(\ref{mtcicuatro})  cannot be imposed at the same time. Thus, the M5-brane probe
breaks completely the supersymmetry of the background. The interesting point in
this result is that this supersymmetry breaking is just identical to the one
corresponding to a massless particle, which constitutes a confirmation of our
interpretation of the giant graviton configurations.

\section{Discussion}
In this chapter we have found configurations of a brane probe on the 
backgrounds of (D(p-2), Dp) and (M2,M5) bound states which behave as 
massless particles. We have checked this fact by
studying the motion of the probe and the way in which breaks
supersymmetry. These giant graviton configurations admit the interpretation of a
set of massless quanta polarized by the gauge fields of the background. Actually,
by recalling the arguments of refs. \cite{GST}-\cite{DTV}, one can argue that
there are two possible descriptions of this system, as a point-like particle or
as an expanded brane, which are valid for different ranges of momenta and
cannot be simultaneously valid. 

In the cases studied in refs. \cite{GST}-\cite{DTV} the blow up of the gravitons
takes place on a (fuzzy) sphere. In our case the brane probe shares some
dimensions with the branes of the background and, thus, our gravitons are also
expanded along them. We have parametrized the volume occupied
by the probe in the common directions by means of the flux of the worldvolume
gauge field or the self-dual field strength, respectively. 
If this flux is fixed and finite, the angular momentum of the brane
is bounded for $p<5$ and one realizes the stringy exclusion principle.

\medskip
\section{Appendix 5A. Supergravity analysis of (D1,D3)}                                 
\medskip                                             
\renewcommand{\theequation}{\rm{A}.\arabic{equation}}  
\setcounter{equation}{0}  

In this appendix we shall study the supersymmetry preserved by the $(D1,D3)$
solution of the type IIB supergravity equations. In order to perform this
analysis it is more convenient to work in the Einstein frame, in which the metric
$ds^2_E\,=\,e^{-\phi_D/2}\,ds^2$ takes the form:
\bear
ds^2_E&=&f_3^{-1/2}\,h_3^{-1/4}\,\Big[\,-(\,dx^0\,)^2\,+\,(\,dx^1\,)^2\,+
\,h_3\,\Big((\,dx^{2}\,)^2\,+\,(\,dx^{3}\,)^2\Big)\,\Big]\,+\rc\rc
&&+\,f_3^{1/2}\,h_3^{-1/4}\,dr^2\,+\,
f_3^{1/2}\,h_3^{-1/4}\,r^2\,\Big[\,{1\over 1-\rho^2}\,d\rho^2\,+\,
(1-\rho^2)d\phi^2\,+\,\rho^2d\Omega_3^2\,\Big]\,\,,\rc
\label{ncapauno}
\eear
where $f_3$ and $h_3$ are given in eq. (\ref{strncbs2}) and, for simplicity, we have
taken the string coupling constant $g_s$ equal to one.

Using complex spinors and with our notations for the gauge forms, 
the supersymmetry transformations of the dilatino $\lambda$ and gravitino 
$\psi$ in type IIB supergravity are the ones given in (\ref{strsusyiib}). 
The solutions of the supergravity equations we are dealing with are purely
bosonic and, thus, they are only invariant under those supersymmetry
transformations which do not change the fermionic fields $\lambda$ and $\psi$.
Let us consider first the variation of the dilatino $\lambda$ for the (D1,D3)
background. From eqs. (\ref{strncbs1}), (\ref{strncbs4}) and (\ref{nctrece}) it follows
that the non-vanishing components of the complex three-form $F$ are:
\bear
F_{01r}&=&i\sin\varphi \,\,h_3^{{1\over 4}}\,\partial_r\,f_3^{-1}\,\,,\rc\rc
F_{23r}&=&\sin\varphi \cos\varphi\,\,h_3^{{7\over 4}}\,\partial_r\,f_3^{-1}\,\,.
\label{ncapaseis}
\eear
By using eq. (\ref{ncapaseis}), and by computing $P_M$ from eq. (\ref{ncapaccinco}), 
one
easily finds from the first equation in (\ref{strsusyiib}) that the
supersymmetry variation of $\lambda$ is:
\beq
\delta\lambda\,=\,{\sin\varphi\over 4}\,f_3^{{1\over 2}}\,h_3^{{5\over 8}}
\partial_r\,f_3^{-1}\,\Gamma_{\underline {r}}\,\Big[\,
-i\sin\varphi\,h_3^{{1\over 2}}\,f_3^{-{1\over 2}}\epsilon^*\,+\,
\Gamma_{\underline {x^0x^1}}\epsilon\,+\,i\cos\varphi\,h_3^{{1\over 2}}\,
\Gamma_{\underline {x^2x^3}}\epsilon\,\Big]\,\,.
\label{ncapasiete}
\eeq
Clearly $\delta\lambda\,=\,0$ if and only if the term inside the brackets on the
right-hand side of eq. (\ref{ncapasiete}) vanishes. It is an easy exercise to
verify, by using the correspondence (\ref{striibcn}),  that the condition so
obtained coincides with the $\kappa$-symmetry condition (\ref{ncstdos}). Thus, we
can represent the spinor $\epsilon$ as in eq. (\ref{ncstcuatro}) with $\alpha$
given in eq. (\ref{ncsttres}) and $\tilde \epsilon$ satisfying (\ref{ncstcinco}).

More information about the spinors which leave invariant the (D1,D3) solution
can be gathered by looking at the gravitino transformation rule
(\ref{strsusyiib}). Let us consider first the components of $\psi$ along the
directions $x^{\mu}$ ($\mu=0,1,2,3$) parallel to the D3-brane. Due to the
presence of a covariant derivative on the expression of $\delta\psi_M$, in order
to obtain the variation of the gravitino, we  need to know the value of the spin
connection. It is straightforward to check that the only non-vanishing
components of the latter of the type 
$\omega_{x^{\mu}}^{\underline{mn}}$ are:
\bear
\omega_{x^0}^{\underline{x^0 r}}&=&\omega_{x^1}^{\underline{x^1 r}}\,=\,
{1\over 4}\,f_3^{{1\over 2}}\partial_r
f_3^{-1}\,+\,{1\over 8}\,\sin^2\varphi\,h_3\,f_3^{-{1\over 2}}\partial_r
f_3^{-1}\,\,,\rc\rc
\omega_{x^2}^{\underline{x^2 r}}&=&\omega_{x^3}^{\underline{x^3 r}}\,=\,
{1\over 4}\,f_3^{{1\over 2}}\,h_3^{{1\over 2}}\partial_r
f_3^{-1}\,-\,{3\over 8}\,\sin^2\varphi\,h_3^{{3\over 2}}
\,f_3^{-{1\over 2}}\partial_r
f_3^{-1}\,\,.
\label{ncapaocho}
\eear
By using  eq. (\ref{ncapaocho}), the values of the forms given in eqs. 
(\ref{nctrece}) and (\ref{ncapaseis}) and the constraint imposed on $\epsilon$ by
the invariance of the dilatino (eq. (\ref{ncstdos})), one concludes after some
calculation that the gravitino components   $\psi_{x^{\mu}}$ ($\mu=0,1,2,3$)
are invariant under supersymmetry if and only if the Killing spinor is
independent of the $x^{\mu}$'s, namely:
\beq
\partial_{x^{\mu}}\,\epsilon\,=\,0\,\,,
\,\,\,\,\,\,\,\,\,\,\,\,\,
(\mu=0,1,2,3)\,\,.
\label{ncapanueve}
\eeq

The condition $\delta\psi_{\rho}\,=\,0$ determines the dependence of $\epsilon$
on $\rho$. The relevant component of the spin connection needed in the
calculation of $\delta\psi_{\rho}$ is:
\beq
\omega_{\rho}^{\underline{\rho r}}\,=\,
{1\over \sqrt{1\,-\,\rho^2}}\,\,\Bigg[\,
1\,-\,{1\over 4}\,r\,f_3\,\partial_r\,f_3^{-1}\,+\,
{1\over 8}\,r\,\sin^2\varphi\,h_3\,\partial_r\,f_3^{-1}\,\Bigg]\,\,.
\label{ncapadiez}
\eeq
(Notice that when $r\rightarrow\infty$ we recover eq. (\ref{ncstsiete})). Using
again the condition imposed by the dilatino invariance, one can easily prove
that the dependence on $\rho$  of $\epsilon$ is the same as in eq. 
(\ref{ncstocho}). Thus, we have verified the representation of $\epsilon$ in
terms of the spinor $\hat\epsilon$ written in eq. (\ref{ncstcuatro}) and 
(\ref{ncstocho}). An explicit representation of $\hat\epsilon$ can be obtained by
looking at the transformation of the other components of the gravitino. For
instance, one can consider the equation for the radial component of $\psi$.
Using the fact that $\omega_{r}^{\underline{mn}}\,=\,0$, one gets that the
dependence of $\hat\epsilon$ on $r$ is determined by the  equation:
\beq
\partial_r\hat\epsilon\,=\,-{1\over 8}\,\partial_r\,
\Big[\,\ln\big(f_3h_3^{{1\over 2}}\big)\,\Big]\,\,\,\hat\epsilon\,\,,
\label{ncapaonce}
\eeq
which can be immediately integrated. One can proceed similarly with the
remaining components of $\psi$. The final result of this analysis is the
complete determination of the form of $\hat\epsilon$ and, therefore, of the
complete Killing spinor $\epsilon$.  
One gets:
\bear
\epsilon\,&=&\,\Big[\,f_3\,h_3^{1/2}\,\Big]^{-{1\over 8}}\,\,
e^{{1\over 2}\,\alpha\,\Gamma_{\underline {x^2x^3}}\,\sigma_3}\,\,
e^{-{1\over 2}\,\beta\,\Gamma_{\underline {\rho r}}}\,\,\,
e^{-{1\over 2}\,\phi\,\Gamma_{\underline {\phi r}}}\,\,\,\times\rc\rc
&&\times\,e^{-{1\over 2}\,\theta_1\,\Gamma_{\underline {\theta_{1}\rho}}}\,\,
e^{-{1\over 2}\,\theta_2\,\Gamma_{\underline {\theta_{2}\theta_{1}}}}\,\,
e^{-{1\over 2}\,\theta_3\,\Gamma_{\underline {\theta_{3}\theta_{2}}}}
\,\,\epsilon_0\,\,,
\label{ncapadoce}
\eear
where $\epsilon_0$ is a constant spinor satisfying the condition
\beq
(i\sigma_2)\Gamma_{\underline{x^0\cdots x^3}}\,\epsilon_0\,=\,
\epsilon_0\,\,.
\label{ncapatrece}
\eeq
It follows from eqs. (\ref{ncapadoce}) and (\ref{ncapatrece}) that the 
(D1,D3) background is ${1\over 2}$ supersymmetric.

\section{Appendix 5B. Supergravity analysis of (M2,M5)}                                
\medskip                                             
\renewcommand{\theequation}{\rm{B}.\arabic{equation}}  
\setcounter{equation}{0}

In this appendix we shall study the supersymmetry preserved by the $(M2,M5)$
solution of the eleven dimensional supergravity equations. The metric was given 
in (\ref{strmtuno}). As the solution of
D=11 supergravity we are dealing with is purely bosonic, it is only invariant
under those supersymmetry transformations which do no change the gravitino field 
$\psi_M$. The transformation of this field was given in (\ref{mtapbuno}).

The form $F^{(4)}$ is the four-form given by eq. (\ref{strmttres}). 
Let us consider first the components of $\psi$ along the
directions $x^{\mu}$ ($\mu=0,1,\dots,5$) parallel to the M5-brane. The only 
non-vanishing
components of the spin connection of the type 
$\omega_{x^{\mu}}^{\underline{mn}}$ are:
\bear
\omega_{x^0}^{\underline{x^0 r}}&=&\omega_{x^1}^{\underline{x^1 r}}\,=
\,\omega_{x^2}^{\underline{x^2 r}}\,=\,
{1\over 6}\,f^{{1\over 2}}\partial_r
f^{-1}\,+\,{1\over 12}\,\sin^2\varphi\,h\,f^{-{1\over 2}}\partial_r
f^{-1}\,\,,\rc\rc
\omega_{x^3}^{\underline{x^3 r}}&=&\omega_{x^4}^{\underline{x^4 r}}\,=
\,\omega_{x^5}^{\underline{x^5 r}}\,=\,
{1\over 6}\,f^{{1\over 2}}\,h^{{1\over 2}}\partial_r
f^{-1}\,-\,{1\over 3}\,\sin^2\varphi\,h^{{3\over 2}}
\,f^{-{1\over 2}}\partial_r
f^{-1}\,\,.
\label{mtapbdos}
\eear
By using  eq. (\ref{mtapbdos}), the value of the 4-form given in eq. 
(\ref{strmttres}) and the fact that the eleven dimensional 
gamma matrices satisfy that $\G_{\underline{x^0\dots x^5r\r\,\p\,\th^1\,\th^2}}=1$ 
we see that the gravitino components   
$\psi_{x^{\mu}}$ ($\mu=0,1,2$)
are invariant under supersymmetry if and only if the Killing spinor is
independent of the $x^{\mu}$'s, namely:
\beq
\partial_{x^{\mu}}\,\e\,=\,0\,\,,
\,\,\,\,\,\,\,\,\,\,\,\,\,
(\mu=0,1,2)\,\,,
\label{mtapbtres}
\eeq
but for $\d\psi_{x^{\mu}}$ ($\mu=3,4,5$) 
to vanish we get 
the condition
\be
\G_{\underline{x^0\dots x^5}}\,\e\,=
\,e^{-\a\,\G_{\underline{x^3\,x^4\,x^5}}}\,\e\,,
\label{mtapbcuatro}
\ee
where $\a$ is a function of $r$ given by:
\be
\sin\a\,=\,f^{-{1\over 2}}\,h^{{1\over 2}}\sin\varphi\,\,,
\,\,\,\,\,\,\,\,\,\,\,\,\,\,\,\,\,\,\,\,\,\,
\cos\a\,=\,h^{{1\over 2}}\cos\varphi\,\,.
\ee
We see that eq. (\ref{mtapbcuatro}) is solved by a spinor 
of the form 
\be
\e\,=\,e^{{\a\over 2}\,\G_{\underline{x^{3}x^{4}x^{5}}}\,}
\,\,\tilde\e\,\,,
\label{mtapbcinco}
\ee
where $\tilde\e$ is a spinor which satisfies:
\be
\G_{\underline{x^0\dots x^5}}\,\tilde\e\,=\,\tilde\e\,.
\label{mtapbseis}
\ee
For the condition $\d\psi_{\r}\,=\,0$ the relevant 
component of the spin connection needed is
\beq
\omega_{\rho}^{\underline{\rho r}}\,=\,
{1\over \sqrt{1\,-\,\rho^2}}\,\,\Bigg[\,
1\,-\,{1\over 3}\,r\,f\,\partial_r\,f^{-1}\,+\,
{1\over 6}\,r\,\sin^2\varphi\,h\,\partial_r\,f_3^{-1}\,\Bigg]\,\,,
\label{mtapbsiete}
\eeq
and using again that $\G_{\underline{x^0\dots x^5\,r\,\r\,\p\,\th^1\,\th^2}}=1$ we 
arrive at
\be
\d\psi_{\r}\,=\,\partial_{\r}\e\,+\,\frac{1}{2}\,
\frac{1}{\sqrt{1\,-\,\r^2}}\,\G_{\underline{\r r}}\,\e\,.
\label{mtapbocho}
\ee
Using the form (\ref{mtapbcinco}) of the spinor we find 
a solution of (\ref{mtapbocho}) by taking
\be
\tilde\e\,=\,e^{-\frac{\b}{2}\,\,\G_{\underline{\r r}}}\,\hat\e\,,
\label{mtapbnueve}
\ee
where now $\hat\e$ is a spinor independent of $\r$ 
and satisfies
\be
\Gamma_{\underline{x^{0}\cdots x^{5}}}\,\,\hat\epsilon\,=\,\hat\epsilon\,\,.
\label{mtapbdiez}
\ee

For completeness we can obtain explicit form of $\hat\e$. The condition 
$\d\psi_r=0$ is equivalent to
\be
\partial_r\hat\e\,=\,-\,\partial_r\,
\Big[\,\ln\big(f\,h \big)^{1\over 12}\,\Big]\,\,\,\hat\e\,\,,
\label{mtapbonce}
\ee
and doing the same for the two angular components  of $\psi$ 
we find the complete Killing spinor which is:
\bear
\epsilon\,&=&\,\Big[\,f\,h\,\Big]^{-{1\over 12}}\,\,
e^{{1\over 2}\,\alpha\,\Gamma_{\underline {x^3x^4x^5}}}\,\,
e^{-{1\over 2}\,\beta\,\Gamma_{\underline {\rho r}}}\,\,\,
e^{-{1\over 2}\,\phi\,\Gamma_{\underline {\phi r}}}\,\,\,
\times \rc \rc
&&\times\,e^{-{1\over 2}\,\theta_1\,\Gamma_{\underline {\theta_{1}\rho}}}\,\,
e^{-{1\over 2}\,\theta_2\,\Gamma_{\underline {\theta_{2}\theta_{1}}}}\,\,
\,\,\epsilon_0\,\,,
\label{mtapbdoce}
\eear
where $\e_0$ is a constant spinor satisfying the condition
\be
\Gamma_{\underline{x^{0}\cdots x^{5}}}\,\,\e_0\,=\,\e_0\,\,,
\ee
and the (M2,M5) background is $1/2$ supersymmetric.


\chapter{Resume}
\medskip

\renewcommand{\theequation}{{\rm\thesection.\arabic{equation}}}

\setcounter{equation}{0}

A teor\'\i a de cordas \'e unha candidata para seres unha teor\'\i a 
fundamental que unifique todas as forzas presentes na Natureza. Tres 
delas, as interacci\'ons electromagn\'etica, feble e forte est\'an 
ben descritas por unha teor\'\i a cu\'antica consistente, o Modelo 
Standard, que describe axeitadamente os resultados experimentais. 
Nembargantes, 
a cuarta interacci\'on, a gravidade, non est\'a incluida no Modelo Standard. 
A gravidade \'e descrita pola teor\'\i a da Relatividade  Xeral, que \'e 
unha teor\'\i a cl\'asica, como a din\'amica do espacio-tempo. Esta 
interacci\'on pode ser despreciada \'a escala de enerx\'\i as das 
interacci\'ons do Modelo Standard pero a enerx\'\i as mais altas tanto os 
efectos cu\'anticos como os gravitacionais son relevantes, facendo 
necesaria unha teor\'\i a cu\'antica da gravidade. A enerx\'\i a na que 
os efectos cu\'anticos non son despreci\'abeis na descripci\'on do 
espacio-tempo define a escala de Planck, que \'e do orden de $10^{19} GeV$. 
Nembargantes, cando un tenta cuantizar a gravidade at\'opase cunha 
teor\'\i a de campos non-renormalizable, polo que \'e de agardar que 
unha nova f\'\i sica xurda a esta escala de enerx\'\i as.

Na teor\'\i a de cordas as part\'\i culas aparecen como modos de vibraci\'on 
da corda, e entre estes modos hai un non masivo con spin 2 que pode ser 
identificado co gravit\'on. As cordas prop\'aganse  no espacio-tempo 
describindo superficies, chamadas follas de universo, que xeralizan a 
li\~na de universo que describen as part\'\i culas puntuais. Nesta teor\'\i a 
a corda \'e cuantizada perturbativamente pero non existe unha segunda 
cuantizaci\'on de cordas.

A formulaci\'on dunha teor\'\i a cu\'antica de cordas relativistas xeraliza 
tam\'en a das part\'\i culas puntu\'ais. A acci\'on que describe a 
propagaci\'on 
dunha corda no espacio-tempo ten como campos din\'amicos un conxunto de 
funci\'ons reais que desciben o embebemento (``embedding") da corda no 
espacio-tempo. Esta teor\'\i a ten que ter as mesmas simetr\'\i as que a 
Relatividade Xeral e o Modelo Standard, \ie, invariancia baixo difeomorfismos 
e simetr\'\i a gauge. Non obstante, aparecen mais simetr\'\i as  
nesta formulaci\'on como a simetr\'\i a conforme e a supersimetr\'\i a, esta 
\'ultima introducida para incluir fermions na folla de universo. A teor\'\i a 
de supercordas e s\'o consistente nun espacio-tempo de dez dimensi\'ons e 
un at\'opase con cinco teor\'\i as de supercordas, todas diferentes 
no nivel perturbativo. Inda m\'ais, para obter unha f\'\i sica efectiva en 
catro dimensi\'ons tense que compactificar as outras seis dimensi\'ons, o 
que pode ser levado a cabo de moitos xeitos non equivalentes sen que haxa un 
preferido. Polo tanto, todo isto semella estar en contra de considerar a 
teor\'\i a de cordas como unha teor\'\i a fundamental da Natureza.

O problema do n\'umero de teor\'\i as de cordas \'e resolto pola dualidade. 
As dualidades en teor\'\i as de cordas relacionan diferentes teor\'\i as e 
amosan que estas poden unificarse nunha \'unica teor\'\i a. As dualidades 
tam\'en dan informaci\'on sobre os r\'eximes de acoplamento forte das 
diferentes teor\'\i as, xa que algunhas destas dualidades aplican o r\'exime 
de acoplamento feble dunha teor\'\i a de supercordas \'o r\'exime de 
acoplamento 
forte doutra. Isto leva a conxeturar que as diferentes teor\'\i as de cordas 
son diferentes l\'\i mites dunha \'unica teor\'\i a fundamental chamada 
teor\'\i a M.

Un pode acadar as acci\'ons efectivas que describen a din\'amica dos modos da 
corda. Se se calcula a acci\'on efectiva para os modos non masivos para 
enerx\'\i as  pequenas comparadas coa tensi\'on da corda obt\'en as diferentes 
supergravidades en dez dimensi\'ons. Estas acci\'ons te\~nen  soluci\'ons 
cl\'asicas  que poden ser identificadas como solit\'ons da  teor\'\i a de cordas 
completa. Alg\'uns destes solit\'ons de cordas  poden ser interpretados como 
obxetos extendidos p-dimensionais que son chamados p-branas. Estes obxetos son 
non perturbativos xa que a s\'ua masa vai como o inverso da constante de 
acoplamento das cordas. De feito as p-branas son \'utiles porque son obxetos 
supersim\'etricos e algunhas das s\'uas propiedades non var\'\i an cando 
cambia 
a constante de acoplamento. Un caso particular de p-branas son as Dp-branas, 
que son obxetos extendidos cargados baixo os campos de Ramond-Ramond das 
acci\'ons de supergravidade.

Por outra banda, as Dp-branas poden ser definidas como hiperplanos sobre os 
que as cordas abertas con certas condici\'ons de contorno poden rematar. De 
feito, cando se calcula as masas e cargas destes obxetos coinciden cos das 
soluci\'ons solit\'onicas correspondentes. As Dp-branas son polo 
tanto obxetos din\'amicos e as s\'uas fluctuaci\'ons est\'an determinadas 
pola din\'amica das cordas que rematan na brana. Un pode usar isto para 
calcular unha acci\'on que describa a fluctuaci\'ons das branas. Nesta Tese 
empr\'eganse as acci\'ons efectivas para as branas de teor\'\i a M e da 
teor\'\i a 
de cordas co prop\'osito de estudiar algunhas configuraci\'ons de branas de 
proba 
en diferentes fondos, e tam\'en ser\'an discutidos alg\'uns aspectos da 
din\'amica de branas.

Esta Tese est\'a baseada nos artigos \cite{Baryon, Flux1, Flux2, NCgrav} 
e \cite{MTgrav}. Est\'a estructurada como segue. No cap\'\i tulo 1 dase 
unha breve introducci\'on \'a teor\'\i a de cordas, discutindo brevemente as 
dualidades entre as distintas teor\'\i as e \'o xurdemento  da teor\'\i a M. 
Son presentadas as acci\'ons efectivas de baixa enerx\'\i a para as d\'uas 
teor\'\i as posibles de cordas pechadas e a acci\'on efectiva de baixa enerx\'\i a 
da teor\'\i a M, list\'andose algunhas das s\'uas soluci\'ons solit\'onicas. O 
maior interese est\'a nas soluci\'on extrem\'ais 
correspondentes a un conxunto de branas  coincidentes e alg\'uns estados 
ligados 
de branas 'non-threshold', xa que estas soluci\'ons ser\'an empregadas como 
fondos de supergravidade nos que se van propagar as branas de proba. Para 
rematar o cap\'\i tulo disc\'utense alg\'uns aspectos relacionados coas D-branas 
e tam\'en se presenta a conxetura de Maldacena.

No cap\'\i tulo 2 constr\'uese a acci\'on efectiva supersim\'etrica 
para Dp-branas bos\'onicas. Esta acci\'on pos\'ue unha simetr\'\i a local 
fermi\'onica, chamada simetr\'\i a kappa que garante que o n\'umero de 
grados de liberdade bos\'onicos e fermi\'onicos coinciden. Tam\'en se presenta 
unha proposta de xeralizaci\'on non-abeliana da devandita acci\'on que 
describe 
a din\'amica de m\'ultiples branas coincidentes. A din\'amica que resulta 
desta acci\'on implica que as Dp-branas poden ser polarizadas por campos 
externos. Este \'e o chamado efecto de polarizaci\'on de Myers. Para rematar 
o cap\'\i tulo dase unha acci\'on no vol\'umen de universo para a M5-brana 
bos\'onica da teor\'\i a M que \'e de tipo Born-Infled e pos\'ue invariancia 
baixo simetr\'\i a kappa.

No cap\'\i tulo 3, est\'udiase o embedding de D(8-p)-branas na xeometr\'\i a 
de fondo creada por Dp-branas paralelas con $p\le 6$. Os embeddings que se 
obte\~nen representan branas conectadas por tubos. Analizando a enerx\'\i a 
destes tubos concl\'uese que poden ser considerados como feixes de cordas 
fundamentais. A soluci\'on tam\'en da unha realizaci\'on do efecto de 
Hanany-Witten. 
 
No cap\'\i tulo 4, est\'udianse algunhas configuraci\'ons de branas 
arrolladas na xeometr\'\i a pr\'eto do horizonte creada por un conxunto 
de branas coincidentes. A caracter\'\i stica com\'un de todos os casos  
analizados \'e a regra de cuantizaci\'on e o xurdemento dun n\'umero finito 
de configuraci\'ons nas que as branas est\'an parcialmente arrolladas en
esferas. A enerx\'\i a destas configuraci\'ons pode ser dada de forma pechada
e a an\'alise das pequenas oscilaci\'ons amosa que son estables.
As configuraci\'ons de branas atopadas admiten unha interpretaci\'on 
como estados ligados de cordas (ou M2-branas na teor\'\i a M) que se 
extenden \'o longo de direcci\'ons non arrolladas. Isto compr\'obase 
directamente 
nun caso particular enpregando o mecanismo de polarizaci\'on de Myers. 

No cap\'itulo 5, est\'udianse configuraci\'ons de branas de proba 
correspondentes a gravit\'ons xigantes na xeometr\'\i a de fondo creada por 
un conxunto de estados ligados non-threshold.  Am\'osase que para un valor 
particular do campo gauge do volume de universo existen configuraci\'ons
da brana de proba que se comportan como part\'\i culas non masivas e que 
poden ser interpretadas como gravit\'ons inflados en esferas difusas e 
extendidos sobre un plano non conmutativo. Compr\'obase este comportamento 
estudiando o movemento e enerx\'\i a da brana de proba e determinando o xeito 
no que se rompe supersimetr\'\i a cando a proba se move na xeometr\'\i a 
de fondo.

\medskip
\section{Teor\'\i a de cordas e teor\'\i a M} 
\medskip                                             
\setcounter{equation}{0}

Un obxeto unidimensional varre unha superficie bidimensional \'o moverse, 
a chamada folla ou superficie de universo, que pode ser descrita en termos de 
dous par\'ametros 
$X^{\m}(\t,\s)$. A acci\'on mais simple que un pode construir que \'e 
invariante baixo repamarametrizaci\'ons, a acci\'on de Nambu-Goto, \'e 
proporcional \'a area da superficie de universo,
\be
S\,=\,{-1\over{2\pi \a^{\prime}}}\,\int_{\S}\,d\t d\s\,
\sqrt{-det(h_{ab})}\,,
\label{strngr}
\ee
onde $\S$ denota a superficie de universo e $h_{ab}=\partial_aX^{\m}\,
\partial_bX^{\n}\,\eta_{\m\n}$ \'e a m\'etrica inducida en $\S$. A 
tensi\'on da corda \'e $T_f=1/2\pi\a^{\prime}$. A acci\'on de Nambu-Goto 
pode simplificarse introducindo unha m\'etrica independente $\g_{ab}(\t,\s)$ 
na superficie de universo. De feito, a acci\'on
\be
S\,=\,{-1\over{4\pi \a^{\prime}}}\,\int_{\S}\,d\t d\s\,
\sqrt{-\g}\,\g^{ab} \,\partial_aX^{\m}\,\partial_bX^{\n}\,\eta_{\m\n}\,,
\label{strpolr}
\ee
\'e equivalente a (\ref{strpolr}) cando se sustit\'ue a ecuaci\'on de 
movemento de $\g_{ab}$ en (\ref{strpolr}). A acci\'on (\ref{strpolr}) 
\'e a acci\'on de Polyakov. Aparte da invariancia baixo reparametrizaci\'ons, 
esta acci\'on ten invariancia de Poincar\'e D-dimensional e invariancia 
Weyl en d\'uas dimensions
\br
{X^{\prime}}^{\m}(\t,\s)&=&X^{\m}(\t,\s)\,, \nonu\\
\g^{\prime}_{ab}(\t^{\prime},\s^{\prime})&=&
e^{2\omega(\t,\s)}\g_{ab}(\t,\s)\,, 
\er
para $\omega(\t,\s)$ arbitraria. A invariancia Weyl non ten an\'alogo en 
(\ref{strngr}) e temos que m\'etricas equivalentes Weyl corresponden \'o mesmo 
embedding 
no espacio-tempo. P\'odense usar d\'uas reparametrizaci\'ons e un 
rescalamento de Weyl para fixar os tres par\'ametros independentes de 
$\g_{\m\n}$ de tal xeito que $\g_{\m\n}=\eta_{\m\n}$, onde $\eta_{\m\n}$ \'e 
a m\'etrica de Minkowski en d\'uas dimensi\'ons. A acci\'on (\ref{strpol}) 
conv\'\i rtese en:
\be
S\,=\,-{T_f\over 2}\,\int_{\S}\,d\t d\s\,
\,\eta^{ab} \,\partial_a X^{\m}\,\partial_b X^{\n}\,\eta_{\m\n}\,.
\label{strgfpolr}
\ee
Variando $X^{\m}$ en (\ref{strgfpolr}) obtense
\be
\d S=\frac{1}{2\pi\a^{\prime}}\int d\t d\s\,
(-\g)^{1/2}\d X_{\m}\,{\dal}\,X^{\m}\,-\,
\frac{1}{2\pi\a^{\prime}} \int d\t \,
(-\g)^{1/2}\d X^{\m}\partial_{\s}X_{\m}\vert_{\s=0}^{\s=\pi}\,,
\ee
onde tomamos a rexi\'on de coordenadas pertencente \'a superficie de universo 
como $-\infty\le \t\le\infty, 0\le \s \le\pi$. O termo de volume da a ecuaci\'on 
de movemento
\be
{\dal}\,X^{\m}=0\,.
\ee
O segundo termo \'e un termo de fronteira. An\'ulase autom\'aticamente para 
unha corda pechada, xa que en este caso os campos son peri\'odicos:
\br
X^{\m}(\t,\pi)&=&X^{\m}(\t,0) \rc
\partial_{\s}X^{\m}(\t,\pi)&=&\partial_{\s}X^{\m}(\t,0) \rc
\g_{ab}(\t,\pi)&=&\g_{ab}(\t,0)
\er
Deste xeito os extremos est\'an unidos formando unha curva pechada e
non hai fronteira. Para a corda aberta un pode ter d\'uas posibles condici\'ons 
de fronteira:
\br
\partial_{\s}X^{\m}(\t,0)=\partial_{\s}X^{\m}(\t,\pi)=0\,,\ \ \ \ &&{\rm 
Neumann}\,,\nonu\\
\nonu\\
\d X^{\m}(\s=0)=\d X^{\m}(\s=\pi)=0\,,\ \ \ \ &&{\rm Dirichlet}\,. \nonu\\
\label{strbcr}
\er
Teres condici\'ons de fronteira tipo Neumann para unha corda aberta significa 
que non hai transmisi\'on de momento nos extremos da corda. As condici\'ons 
de fronteira tipo Dirichlet rompen a invariancia Poincar\'e e requiren 
que os extremos da corda estean fixos nun hiperplano. 

Diferentes condici\'ons de fronteira corresponden a diferentes espectros. 
O espectro non masivo da corda pechada cont\'en unha part\'\i cula de spin 2, 
que pode ser identificada co gravit\'on, un campo escalar $\p$ 
(o dilat\'on) e unha 2-forma antisim\'etrica $B_{\m\n}$, a 2-forma NSNS. 
O espectro da corda aberta mov\'endose nun espacio-tempo plano con condici\'ons 
de fronteira tipo Neumann 
en todalas direcci\'ons cont\'en un campo vectorial $A_{\m}$. Unha teor\'\i a 
de cordas abertas cont\'en necesariamente cordas pechadas xa que os dous 
extremos da corda aberta p\'odense unir para formar unha corda pechada.

P\'odense incluir fermi\'ons na folla de universo facendo a acci\'on 
supersim\'etrica. Os espinores da folla de universo $\psi^{\m}=\psi^{\m}(\s)$ 
levan un \'\i ndice vectorial do grupo de Lorentz espacio-temporal SO(1,9). 
En d\'uas dimensi\'ons o espinor $\psi$ pode ser Majorana-Weyl e ent\'on 
descomponse en d\'uas compo\~nentes quir\'ais, $\psi=\psi_L+\psi_R$ ou, 
en notaci\'on matricial
\be
\psi=\left(\begin{array}{c} \psi_R \\ 
                            \psi_L 
           \end{array}
     \right)\,.
\ee
O termo fermi\'onico da acci\'on de cordas podese po\~ner como
\be
-{i\over 2\pi \a'}\int d\t d\s\,\left\{
\psi^\m_R \partial_+ \psi^\n_R+ \psi^\m_L \partial_- \psi^\n_L \right\}
\,\eta_{\m\n}\,,
\ee
en termos das coordenadas do cono de luz $\s^{\pm}=\t\pm\s$. As ecuaci\'ons 
de movemento $\partial_+\psi_R^{\m}=0,\ \partial_-\psi_L^{\m}=0$ 
implican que os fermi\'ons con quiralidades positiva ou negativa 
m\'ovense respectivamente \'a dereita ou \'a esquerda. A condici\'on de 
fronteira para a supercorda aberta \'e
\be
0=(\psi_L^{\m}\d{\psi_L}_{\m}-\psi_R^{\m}\d{\psi_R}_{\m})|_{\s=0}^{\s=\pi}\,\,.
\ee 
P\'odense impo\~ner dous tipos de 
condici\'ons de fronteira 
para os campos fermi\'onicos: $\psi_R=\pm \psi_L$. Si nos dous extremos 
da corda temos a condici\'on co mesmo signo os campos fermi\'onicos $\psi^\mu$ 
ter\'an modos enteiros. Este \'e o chamado sector de Ramond (R). Si, polo 
contrario, te\~nen signo oposto, os $\psi^\mu$ ter\'an modos semienteiros. 
Este sector \'e o chamado sector de Neveu-Schwarz (NS).

Para unha corda pechada hai d\'uas posibles elecci\'ons das condici\'ons de 
periodicidade para os fermi\'ons: $\psi^\mu_R(\sigma=2\pi)=\pm \psi^\mu_R
(\sigma=0)$, e similarmente para $\psi^\mu_L$. Os signos $+$ e $-$ 
corresponden respectivamente \'os sectores R o NS respectivamente. Nembargantes, 
agora temos d\'uas posibilidades independentes para elixir. Haber\'a polo 
tanto catro sectores diferentes, NS-NS, R-R, R-NS e NS-R.

No sector R da corda aberta o estado fundamental ten que ser unha  
representaci\'on da \'alxebra de modos cero dos campos fermi\'onicos 
$\psi^\mu_0$, que \'e a \'alxebra de Clifford para o grupo de  Lorentz 
$SO(1,9)$. Se clasificamos os estados en representaci\'ons do grupo pequeno 
$SO(8)$, vemos que o estado fundamental do sector R pode escollerse que estea 
na representaci\'on ${\bf 8}_+$ ou ben na ${\bf 8}_-$ (que difiren na s\'ua 
quiralidade). Por outra parte, o estado fundamental do sector NS ten unha 
enerx\'\i a do vac\'\i o de $-{1\over 2}$, e \'e un singlete de $SO(8)$. 
Este estado \'e polo tanto un taqui\'on. Cando se act\'ua sobre el cun 
modo $\psi_{-{1\over 2}}$, crease un estado non masivo cos \'\i ndices na 
representaci\'on vectorial ${\bf 8}_v$ de $SO(8)$.

A supersimetr\'\i a no espacio-tempo require que haxa o mesmo n\'umero 
de grados de liberdade bos\'onicos e fermi\'onicos na capa de masas 
en cada nivel de masa. Esto se acada coa proxecci\'on GSO, que corresponde 
b\'asicamente a po\~ner a cero todolos estados no sector NS que son 
creados por un n\'umero par de operadores de creaci\'on fermi\'onicos no 
vac\'\i o , e fixa a quiralidade do estado fundamental do sector R. O taqui\'on 
\'e eliminado do espectro. O sector non masivo da supercorda aberta  
consiste nos estados ${\bf 8}_v+{\bf 8}_+$ e realiza a simetr\'\i a quiral 
${\cal N}=1$ $D=10$. Corresponden a un bos\'on vectorial non-masivo e o 
seu superasociado fermi\'onico. O superasociado fermi\'onico do bos\'on gauge 
\'e un espinor en dez dimensi\'ons con 16 compo\~nentes independentes que se 
reducen a 8 cando se usan as ecuaci\'ons de movemento, coincidindo coas 8 
componentes f\'\i sicas  do campo gauge.

Nas teor\'\i as de supercordas pechadas os sectores de movemento \'a 
dereita e \'a esquerda desac\'oplanse, e cada un deles \'e cuantizado 
exactamente no mesmo xeito que para as cordas abertas.
O sector non masivo dunha teor\'\i a de cordas pechadas \'e polo tanto 
o producto tensorial do sector non masivo da supercorda aberta. Nembargantes, 
temos dous productos tensoriais diferentes que se poden facer, dependendo 
de si un toma estados fundamentais de Ramond de quiralidade oposta 
nos dous lados ou da mesma quiralidade. Xa que a teor\'\i a efectiva 
da supercorda aberta ten supersimetr\'\i a ${\cal N}=1$ no espacio-tempo, 
teremos unha teor\'\i a con supersimetr\'\i a ${\cal N}=2$. Para a corda 
pechada teremos 32 componentes espinoriais independentes. A proxecci\'on 
GSO act\'ua independentemente en cada sector das d\'uas copias da supercorda 
aberta, e polo tanto un pode elixir os sectores R nos campos que se moven 
\'a dereita e \'a esquerda de tal xeito que te\~nan a misma quiralidade ou 
oposta.

A teor\'\i a tipo IIA corresponde a escoller os estados R con quiralidade 
oposta e polo tanto a teor\'\i a no espacio-tempo non \'e quiral. 
As part\'\i culas bos\'onicas do sector NSNS son respectivamente un campo 
escalar relacionado co dilat\'on $\p$, un campo tensorial antisim\'etrico de 
rango 2, $B^{(2)}_{\mu\nu}$, e un campo tensorial sim\'etrico e sen traza 
relacionado co graviton. No sector de RR atopamos un vector e un tensor 
antisim\'etrico de tres \'\i ndices, que corresponden \'os campos 
$C_\mu^{(1)}$ e $C_{\m_1\m_2\m_3}^{(3)}$. Os fermi\'ons est\'an nos sectores 
RNS e NSR, e son dous gravitinos. Polo tanto a teor\'\i a ten supersimetr\'\i a 
${\cal N}=2$ no nivel non masivo. Inda m\'ais, como os dous gravitinos te\~nen 
distinta quiralidade a teor\'\i a non \'e quiral. A acci\'on para unha 
teor\'\i a non quiral en dez dimensi\'ons con ${\cal N}=2$ e con gravidade 
est\'a completamente fixada e ch\'amase teor\'\i a IIA. Esta acci\'on 
reproduce as interacci\'ons entre os campos non masivos calculadas a partires 
de amplitudes de teor\'\i as de cordas. Esta acci\'on tam\'en pode ser obtida 
da acci\'on de cordas en fondos curvos requerindo que a invariancia conforme 
non estea rota cu\'anticamente,\ie , a funci\'on $\b$ ten que se anular. 
A acci\'on para o sector bos\'onico ven dada en (\ref{striiaact}).

Na teor\'\i a IIB os sectores dereita e esquerda te\~nen a mesma quiralidade.
As part\'\i culas non masivas clasif\'\i canse en representaci\'ons de $SO(8)$. 
O sector NSNS \'e exactamente o mesmo que o da supercorda IIA.
O sector RR \'e diferente. Cont\'en un potencial escalar (ou 0-forma) $\chi$, 
unha potencial 2-forma $C_{\m_1\m_2}^{(2)}$ e un potencial 4-forma 
$C_{\m_1\dots\m_4}^{(4)}$. A intensidade de campo do potencial 4-forma ten 
que ser autodual para que $C^{(4)}$ encaixe na representaci\'on ${\bf 35}_+$.
Os fermi\'ons dos sectores NS-R and R-NS son dous gravitinos, agora da mesma 
quiralidade. Temos polo tanto unha teor\'\i a dez dimensional quiral 
${\cal N}=2$ con gravidade. Esta \'e a supergravidade IIB dada en 
(\ref{striibact}). Nembargantes a condici\'on de autodualidade ten que 
ser imposta a man.

A teor\'\i a tipo I \'e unha teor\'\i a de cordas abertas e pechadas non 
orientadas. O espectro non masivo da teor\'\i a tipo I cont\'en o espectro 
de cordas abertas non orientadas e m\'ais o de cordas pechadas non orientadas. 
As cordas pechadas tipo II non se poden acoplar consistentemente a cordas abertas, 
xa que te\~nen supersimetr\'\i a ${\cal N}=2$, namentras que os modos non masivos 
da supercorda aberta realizan supersimetr\'\i a  ${\cal N}=1$. C\'ompre polo 
tanto dividir por dous a supersimetr\'\i a das cordas pechadas. Isto pode facerse 
se as cordas pechadas son non orientadas. A teor\'\i a cont\'en un gravit\'on, 
un escalar e un tensor antisim\'etrico no sector bos\'onico e un gravitino 
Majorana-Weyl con un fermi\'on Majorana-Weyl que completan o multiplete 
quiral ${\cal N}=1$ de supergravidade en dez dimensi\'ons. Un ten que engadir 
cordas abertas para acadar consistencia a nivel cu\'antico. Isto orixina un 
sector gauge na teor\'\i a xa que as cordas abertas poden estar cargadas nos 
seus extremos de xeito consistente coa invariancia de Poincar\'e e a invariancia 
conforme na folla de universo. Isto se acada engadindo $n$ grados de liberdade 
non din\'amicos nos extremos da corda etiquetados cun \'\i ndice $i=1\dots n$. 
Estes son os factores de Chan-Paton. Pode definirse unha acci\'on cunha 
simetr\'\i a $U(n)$ nestes factores. Si estes factores est\'an na representaci\'on 
fundamental do grupo $U(n)$, un estado de corda aberta conv\'ertese nunha 
matriz $n\times n$ que se transforma baixo $U(n)$ na representaci\'on adxunta. 
O vector non masivo do espectro conv\'ertese nun campo de Yang-Mills $U(n)$ 
na acci\'on de baixa enerx\'\i a. Se a corda non \'e orientada o grupo pode 
ser $SO(n)$ ou $Sp(n)$ pero a \'unica ceive de anomalias \'e a $SO(32)$. A 
acci\'on de baixa enerx\'\i a das teor\'\i as de supercordas tipo I \'e 
supergravidade ${\cal N}=1$ en dez dimensi\'ons acoplada a unha teor\'\i a de 
super Yang-Mills \cite{Sgrav+YM} con grupo gauge $SO(32)$.

As cordas heter\'oticas son teor\'\i as de cordas pechadas orientadas. 
A corda heter\'otica \'e un h\'\i brido de corda bos\'onica nun sector 
e de supercorda no outro. Para que a cuantizaci\'on sexa consistente 
hai que engadir 32 fermi\'ons quir\'ais no sector a esquerdas que estean 
na representaci\'on escalar de $SO(8)$. Neste novo sector fermi\'onico 
un pode impo\~ner condici\'ons de contorno peri\'odicas ou antiperi\'odicas. 
Se escollemos as mesmas para todos os fermi\'ons o sistema \'e invariante 
baixo rotaci\'ons $SO(32)$. Esta \'e a corda heter\'otica $SO(32)$. Se 
impux\'eramos condici\'ons de contorno peri\'odicas na metade dos fermi\'ons e 
antiperi\'odicas na outra ter\'\i amos a corda heter\'otica $E_8\times E_8$. 
A teor\'\i a cont\'en o multiplete de supergravidade ${\cal N}=1$  
m\'ais o multiplete vector ${\cal N}=1$ dunha teor\'\i a gauge $SO(32)$ ou 
$E_8\times E_8$. O contido en campos non masivos \'e o mesmo que o da 
teor\'\i a tipo I. A acci\'on efectiva \'e polo tanto supergravidade 
${\cal N}=1$ acoplada a super Yang-Mills cos campos de Yang-Mills na 
representaci\'on adxunta. Unha diferencia importante coas outras teor\'\i as 
de cordas \'e que non hai campos RR.

Isto foi unha curta presentaci\'on das cinco teor\'\i as de cordas consistentes, 
que semellan non ter relaci\'on nengunha. Nembargantes, a an\'alise anterior 
red\'ucese \'o r\'exime perturbativo e non temos unha descrici\'on a acoplamento 
forte. Asemade as teor\'\i as de cordas est\'an relacionadas entre elas por 
dualidades. D\'uas teor\'\i as est\'an relacionadas por unha dualidade se hai 
unha correspondencia un a un entre os seus espectros f\'\i sicos e se a s\'ua 
din\'amica \'e 
equivalente. As diferentes dualidades son presentadas brevemente na secci\'on 
\ref{secdual}. Algunhas destas dualidades relacionan a descrici\'on perturbativa 
dunha teor\'\i a co r\'exime non perturbativo doutra, dando informaci\'on deste 
r\'exime sen unha formulaci\'on expl\'\i cita. Os r\'eximes de acoplamento 
forte de calqueira das teor\'\i as de cordas poden ser aplicados por dualidade 
U co l\'\i mite de acoplamento feble doutra teor\'\i a de cordas. De feito, 
a dualidade S mapea os r\'eximes de acoplamento forte e feble das teor\'\i a I 
e heter\'otica $SO(32)$ e a autodualidade S da teor\'\i a IIB aplica cada un 
destes dous r\'eximes da mesma teor\'\i a co outro. Nembargantes, o acoplamento 
forte das teor\'\i as IIA e $E_8\times E_8$ \'e descrito por unha teor\'\i a 
en once dimensi\'ons que ten como acci\'on de baixa enerx\'\i a a supergravidade 
en once dimensions, dada na ecuaci\'on (\ref{str11act}). Esta teor\'\i a once 
dimensional \'e a chamada teor\'\i a M. A imaxe deixada polas dualidades 
entre as distintas teor\'\i as de cordas leva \'a conxetura de que as cinco 
teor\'\i as poden ser interpretadas como diferentes l\'\i mites dunha teor\'\i a 
fundamental, que \'e a teor\'\i a M.

A secci\'on \ref{secstrsol} amosa soluci\'on cl\'asicas solit\'onicas das 
supergravidades IIA, IIB e once dimensional, que poden ser interpretadas como 
obxetos extendidos. Estas soluci\'ons te\~nen  carga el\'ectrica ou magn\'etica 
respecto dos campos tensori\'ais que aparecen nas acci\'ons de baixa enerx\'\i a e son 
obxetos supersim\'etricos (estados BPS), que preservan unha certa fracci\'on 
das supersimetr\'\i as. Estas 
propiedades supersim\'etricas das p-branas son cruci\'ais para relacionar 
estas soluci\'on c\'asicas con obxetos cu\'anticos. Isto \'e, as branas 
BPS son solit\'ons da teor\'\i a de cordas completa. Os estados BPS (Bogomoln'yi
- Prasad-Sommerfeld) satisfacen unha cota nas s\'uas masas (a cota BPS), o que 
implica que a s\'ua masa est\'a relacionada coa s\'ua carga, de tal xeito que 
estes estados est\'an caracterizados por un s\'o par\'ametro. As branas BPS 
son obxetos fundament\'ais, xa que poden ser aplicados \'a corda fundamental por 
dualidade U. Tam\'en hai soluci\'ons cargadas baixo varios campos de RR. Estas 
soluci\'ons representan p-branas elementales que intersectan con certas 
regras. As intersecci\'ons poden ser ortogon\'ais, no senso de que cada 
brana se extende \'o longo dunha direcci\'on determinada, pero tam\'en poden 
formar un estado ligado cando as distintas branas non est\'an ben separadas. 
Alg\'uns destes estados ligados son configuraci\'ons de branas que sinten unha 
forza atractiva entre elas, e polo tanto te\~nen unha enerx\'\i a de 
ligadura non nula. Estes son os estados ligados non-marxinales e 
alg\'uns deles son listados en \ref{strsecbs}. Estes estados ligados 
empr\'eganse na tese como fondos de supergravidade nos que estudiar configuraci\'ons 
de branas de proba.

\medskip
\subsubsection{D-branas} 
\medskip

As D-branas def\'\i nense como os hiperplanos nos que rematan as cordas 
abertas con condici\'ons de contorno de Dirichlet (ver figura \ref{D-branas}) 
e xurden de xeito natural cando a dualidade T act\'ua sobre as cordas abertas. 
Por outro lado, unha teor\'\i a de cordas abertas  ten que conter tam\'en cordas 
pechadas, \'as que deber\'\i an acoplarse as D-branas. De feito, as D-branas 
levan carga de RR e son BPS. No artigo \cite{Polchi} am\'osase como d\'uas 
D-branas paralelas trocan bos\'ons de RR o que cancela o troco de part\'\i culas 
NSNS causando que a forza entre as d\'uas branas est\'aticas sexa nula.
\begin{figure}
\centerline{\hskip -.8in \epsffile{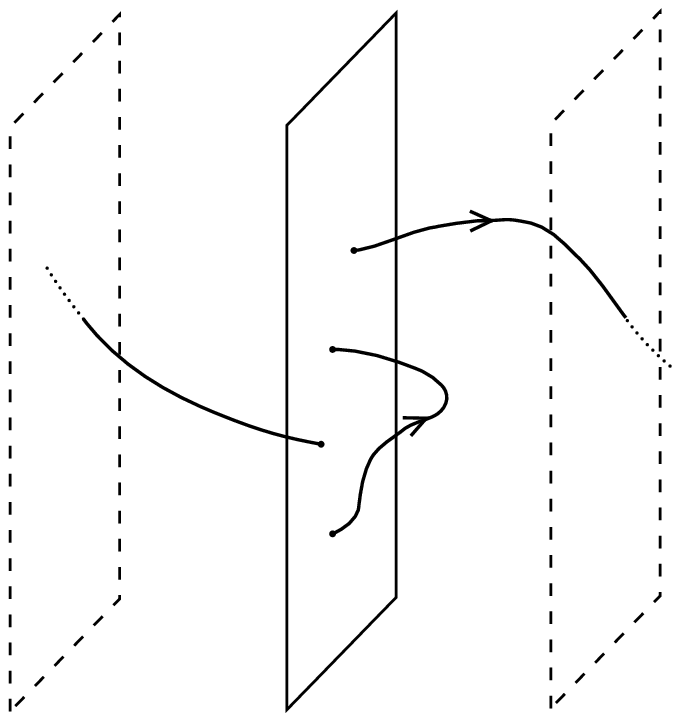}}
\caption{Cordas abertas con remates en hiperplanos. As li\~nas de puntos  
son identificadas peri\'odicamente. As cordas amosadas te\~nen n\'umeros 
de arrollamento cero e un.}
\label{D-branas}
\end{figure}
As D-branas son obxetos din\'amicos e a s\'ua din\'amica ven determinada 
pola din\'amica das cordas abertas que rematan nelas. A acci\'on efectiva 
de baixa enerx\'\i a dunha Dp-brana  \'e unha teor\'\i a supersim\'etrica 
en $(p+1)$ dimensi\'ons onde as fluctuaci\'ons da brana est\'an descritas 
por campos escalares $X^{\m}$ transversos \'as direcci\'ons do volume de 
universo da brana. Esta acci\'on pode ser calculada se se require que o 
modelo sigma non-lineal que describe a propagaci\'on dunha corda aberta 
con condici\'ons de contorno de Dirichlet nun fondo de supergravidade xeral 
sexa invariante conforme. Isto da un conxunto de ligaduras nos campos 
que son igu\'ais \'as ecuaci\'ons de movemento que seguen da seguinte 
acci\'on efectiva:
\be
S_{Dp}=-T_{p}\int d^{p+1}\s e^{-\p}\sqrt{-det(g+{\cal F})}\,,
\label{strDBIactr}
\ee
onde $g$ \'e a m\'etrica inducida no volume de universo e ${\cal F}=dA-B$, 
sendo $B$ o pull-back da d\'uas-forma de NSNS e $A$ o campo gauge $U(1)$.
Esta \'e a acci\'on de Dirac-Born-Infeld (DBI) que aparte de conter os modos 
non masivos da corda aberta tam\'en cont\'en o acoplo \'os modos non masivos 
da corda pechada. A acci\'on DBI \'e unha teor\'\i a gauge $U(1)$ que se 
reduce, a primeira orde en $\a'$, a super Yang-Mills en $p+1$ dimensi\'ons 
con $9-p$ campos escalares, cando o espacio-tempo \'e plano. Se se incl\'uen 
fermi\'ons, ent\'on a acci\'on efectiva de baixa enerx\'\i a para as D-branas 
conv\'ertese en Yang-Mills supersim\'etrica ${\cal N}=1$ en $p+1$ simensi\'ons. 
A acci\'on DBI para unha Dp-brana xeral pode seres obtida por reducci\'on 
dimensional a $p+1$ dimensi\'ons da teor\'\i a $U(1)$ SYM en dez dimensi\'ons. 
Neste procedemento as compo\~nentes do potencial gauge $A_{\m}$ nas direcci\'ons 
nas que se reduce v\'olvense escalares. Unha xeralizaci\'on de todo isto \'e 
considerar un conxunto de N D-branas do mesmo tipo. Isto introduce un novo 
n\'umero cu\'antico que distingue as distintas branas nas que poden rematar 
as cordas abertas. Estes n\'umeros son os factores de Chan-Paton. Si as $N$ D-branas 
son coincidentes, a acci\'on efectiva v\'olvese SYM $U(N)$ en $p+1$ dimensi\'ons.
  
As Dp-branas est\'an cargadas baixo os potenciais de RR e a s\'ua acci\'on 
deber\'\i a conter un termo que acople a brana a estes campos. Nembargantes, 
os campos de RR est\'an suxeitos a transformaci\'ons gauge e o acoplamento 
deber\'\i a ser invariante gauge. Se requerimos invariancia gauge respecto 
\'os campos do espacio de fondo e que as D-branas pos\'uan carga el\'ectrica 
determ\'\i nase un acoplamento do tipo 
\be
\int_{\S_{p+1}}\,C\wedge e^{\cal F}\,,
\label{WZactr}
\ee
onde $C$ denota o pull-back no volume de universo dos correspondentes potenciais 
do fondo. Este \'e un termo topol\'oxico de tipo Wess-Zumino que imos denotar 
por $S_{WZ}$. Finalmente acadamos a seguinte acci\'on para unha brana BPS
\be
S\,=\,-\,T_p\,\int_{\S_{p+1}}\,d^{p+1}\s\,e^{-\p}\,
\sqrt{-det(g+\cal F)}\,\pm\,T_p\,
\int_{\S_{p+1}}\,C\wedge e^{\cal F}\,,
\label{fulldpbactr}
\ee
onde o signo $+$ \'e para branas e o signo $-$ para anti-branas.

Para facer a extensi\'on supersim\'etrica da acci\'on (\ref{fulldpbactr}) un 
pode adoptar dous tipos de mecanismos: a formulaci\'on de Neveu-Schwarz-Ramond, 
con supersimetr\'\i a manifesta no volume de universo ou a formulaci\'on de 
Green-Schwarz (GS) con supersimetr\'\i a manifesta no espacio de fondo. 
\'Ambalas d\'uas 
levan \'a mesma teor\'\i a pero a ventaxa da segunda \'e que fai manifesta a 
supersimetr\'\i a do fondo. O ingrediente crucial na formulaci\'on GS \'e unha 
simetr\'\i a fermi\'onica local chamada ``simetr\'\i a kappa''. O papel 
desenvolvido pola simetr\'\i a $\k$ \'e eliminar os grados de liberdade 
fermi\'onicos que sobran para completar supermultipletes. Esta simetr\'\i a 
reflexa o feito de que a presencia da brana rompe a metade da supersimetr\'\i a 
en $D$ dimensi\'ons, xa que fixando o gauge da simetr\'\i a $\k$ red\'ucense 
\'a metade os grados de liberdade fermi\'onicos que son f\'\i sicos. No 
cap\'\i tulo 2 am\'osase que a acci\'on (\ref{fulldpbactr}) \'e invariante $\k$ 
nun superespacio plano primeiro e logo nun superespacio curvado xeral. Para 
configuraci\'ons bos\'onicas os grados de liberdade bos\'onicos sobrantes 
son eliminados escollendo o ``gauge est\'atico'', que emprega a invariancia 
baixo difeomorfismos en $(p+1)$ dimensi\'ons da teor\'\i a no volume de universo 
para identificar $p+1$ das coordenadas do espacio-tempo coas coordenadas do 
volume de universo. Neste gauge  as coordenadas espaciais restantes son $9-p$ 
campos escalares  que representan as excitaci\'ons transversas da brana. A 
condici\'on que fixa o gauge de simetr\'\i a $\k$ \'e a seguinte:
\be
\G_{\k} \epsilon = \epsilon\,,
\label{kappafix4r}
\ee
onde a matriz $\G_{\k}$ (a matriz de simetr\'\i a kappa) dada en 
(\ref{kappamatrix}) satisface que $(\G_{\k})^2=1$, de tal xeito que 
(\ref{kappafix4r}) \'e unha condici\'on de proxecci\'on sobre o par\'ametro 
de supersimetr\'\i a $\epsilon$. 

A ecuaci\'on (\ref{kappafix4r}) determina a fracci\'on de supersimetr\'\i as 
preservada por configuraci\'ons de branas bos\'onicas. Para configuraci\'ons 
de supergravidade con branas como fontes temos a acci\'on da D-brana acoplada 
\'a supergravidade e a condici\'on que fixa o gauge ten que ser complementada 
coas ecuaci\'ons dos espinores de Killing  da supergravidade correspondente.

\medskip
\subsubsection{A conxetura de Maldacena} 
\medskip

As D-branas son obxetos masivos e polo tanto crean unha xeometr\'\i a curvada. 
A din\'amica dos campos neste fondo \'e descrita polas ecuaci\'ons de 
movemento cl\'asicas de supergravidade. Por outra banda, a din\'amica de baixa 
enerx\'\i a das D-branas ben descrita pola acci\'on DBI, que se  reduce a $SYM$ 
cando a constante da corda $\a'$ se vai a cero. A conxetura de Maldacena 
identifica \'ambalas d\'uas descrici\'ons tendo en conta a s\'ua validez no 
espacio de par\'ametros da teor\'\i a de cordas. O exemplo m\'ais simple 
\'e a teor\'\i a IIB compactificada en $AdS_5\times S^5$. Isto corresponde 
\'a xeometr\'\i a pr\'eto do horizonte dun conxunto de D3-branas paralelas 
e coincidentes. A fronteira  de $AdS_5$ \'e o espacio-tempo de Minkowski en 
catro dimensi\'ons. Ent\'on, a din\'amica de baixa enerx\'\i a dos modos 
non masivos da corda pechada \'e descrita pola supergravidade dez-dimensional 
en $AdS_5\times S^5$ e a dos modos da corda aberta pola acci\'on de SYM 
${\cal N}=4$. Os modos da corda pechada prop\'aganse no fondo mentras que 
os modos da corda aberta est\'an confinados no volume de universo catro 
dimensional das D3-branas. A interacci\'on entre os modos da brana e do fondo 
obtense a partires da dependencia expl\'\i cita da acci\'on de DBI no gravit\'on 
a trav\'es do pull-back da m\'etrica do fondo. Esta v\'olvese a m\'etrica plana 
de Minkowski na fronteira mais fluctuaci\'ons asociadas \'o gravit\'on. 
Isto \'e, $G=\eta+\kappa_{10}\,h$, onde $h$ \'e o gravit\'on.

P\'odese ver que no l\'\i mite $\a^{\prime}=\ls^2\rightarrow 0$, cos par\'ametros 
$g_s$ e $N$ fixos, ambas teor\'\i as desac\'oplanse. Isto \'e porque neste 
l\'\i mite $\kappa_{10}\sim\a^{\prime}\rightarrow 0$. O l\'\i mite 
$\a^{\prime}\rightarrow 0$ corresponde, pola parte de supergravidade, a 
pequeno radio de curvatura medido en unidades da lonxitude da corda $\ls$. O 
radio para a xeometr\'\i a de D3-branas ven dado por $R=4\pi g_sN\ls^4$. Pola 
parte da teor\'\i a gauge a acci\'on de DBI red\'ucese a ${\cal N}=4$ U(N) SYM 
xa que todas as correcci\'ons con derivadas de orde superior te\~nen potencias 
positivas de $\a'$. Polo tanto, ficamos con d\'uas teor\'\i as desacopladas, 
supergravidade en $AdS_5\times S^5$ e SYM ${\cal N}=4$ na fronteira de $AdS$. No 
l\'\i mite $\a^{\prime}\rightarrow 0$ estamos a ampliar un pequeno anaco do 
espacio-tempo moi pr\'eto das branas qu\'e \'e expandido, xa que a m\'etrica 
ten que ser reescalada para obter un resultado finito.

A teor\'\i a de cordas tipo IIB en $AdS_5\times S^5$ \'e unha teor\'\i a cu\'antica 
cun grupo de isometr\'\i as $SO(2,4)\times SO(6)$. Por outra banda, ${\cal N}=4$ 
SYM en 3+1 dimensi\'ons \'e unha teor\'\i a de campos conformes  con grupo 
conforme $SO(2,4)$. O $SO(6)$ da esfera pode identificarse co grupo de simetr\'\i a 
R da teor\'\i a gauge. Polo tanto, ambalas d\'uas teor\'\i as te\~nen as mesmas 
supersimetr\'\i as espacio-temporais. A conxetura de Maldacena afirma que 
a teor\'\i a de supercordas tipo IIB en $AdS_5\times S^5$ e ${\cal N}=4$ U(N) 
super Yang-Mills (definida na fronteira de $AdS$) son equivalentes 
matem\'aticamente. Nembargantes as d\'uas descrici\'ons son v\'alidas en diferentes 
r\'eximes do espacio de par\'ametros ou espacio de m\'odulos (moduli space). 
En termos da constante de acoplamento de 't Hooft $\l\equiv g_{YM}^2N$, que \'e 
o par\'ametro relevante da teor\'\i a de Yang-Mills con $N$ grande, o radio 
$R$ de $AdS$ e da esfera ven dado por
\be
R^4\sim 4\pi g_sN\ls^4\sim g_{YM}^2N\ls^4=\ls^4\l\,.
\ee
No l\'\i mite $N\rightarrow\infty$ con $\l$ fixo, e ent\'on $g_s\sim g_{YM}^2
\rightarrow 0$, un pode calcular en teor\'\i a de cordas perturbativa. 
Obter\'\i ase a descrici\'on cu\'antica non-perturbativa de ${\cal N}=4$ SYM 
en termos do l\'\i mite cl\'asico de teor\'\i a de cordas. Se, asemade, 
consideramos o l\'\i mite de acoplamento forte de SYM ($\l\gg 1$) con $R$ fixo, 
ent\'on
\be
\frac{R^4}{\ls^4}\sim g_sN\sim g_{YM}^2N=\l\gg 1\,,
\ee  
e estamos a tratar ent\'on co l\'\i mite $\a^{\prime}=\ls^2\rightarrow 0$ de 
teor\'\i a de cordas, e neste l\'\i mite, a teor\'\i a cl\'asica de cordas 
\'e a supergravidade cl\'asica.

\medskip
\section{Bari\'ons a partires das acci\'ons de branas} 
\medskip                                             
\setcounter{equation}{0}

Como comentamos na secci\'on anterior a relaci\'on entre as teor\'\i as gauge 
e a supergravidade est\'a na base da conxetura de Maldacena. Maldacena argumentou 
que existe unha dualidade extre supergravidade cl\'asica na rexi\'on pr\'eto 
do horizonte e o l\'\i mite de 't Hooft de gran $N$ da teor\'\i a de Yang-Mills 
$SU(N)$. En particular, temos a chamada correspondencia $AdS/CFT$ entre 
supercordas tipo IIB en $AdS_5\times S^5$ e SYM non abeliana ${\cal N}=4$ en 
catro dimensi\'ons \cite{Gubser,
Witten, Gross}. Neste contexto un pode calcular valores agardados de bucles
de Wilson considerando unha corda aberta fundamental colocada no interior 
do espacio $AdS_5$ tendo os seus remates na fronteira. Este formalismo foi 
empregado para obter potenci\'ais de quarks tanto na teor\'\i a supersim\'etrica 
como na non supersim\'etrica. Inda m\'ais, Witten \cite{Wittenbaryon} propuxo 
un xeito de incorporares 
bari\'ons considerando unha D5-brana arrollada nunha $S^5$ da que xurden as 
cordas 
fundamentais que rematan na D3-brana \cite{Branbaryon, Imabaryon}. Este \'e 
o correspondente en teor\'\i a de cordas \'o v\'ertice bari\'onico da teor\'\i a 
gauge SU(N), que representa un estado ligados de N quarks externos.

O v\'ertice bari\'onico pode ser descrito dende un punto de vista do volume 
de universo, que permite unha descrici\'on unificada das D5-branas e das cordas. 
Na referencia \cite{CGS1}, foi considerado o problema da D5-brana mov\'endose 
baixo a influencia dun fondo de D3-branas. As ecuaci\'ons de movemento da 
D5-brana est\'atica res\'olvense se un asume que os campos satisfacen unha certa 
ecuaci\'on diferencial BPS de primeira orde que foi obtida previamente por 
Imamura \cite{Imamura}. Esta condici\'on BPS pode seres integrada exactamente 
na rexi\'on pr\'eto do horizonte da xeometr\'\i a do fondo de D3-branas e as 
soluci\'ons son picos do volume de universo da D5-brana que poden ser 
interpretados coma feixes de cordas fundament\'ais que rematan nas D3-branas. 
De feito, esta clase de picos son unha caracter\'\i stica xeral da teor\'\i a 
gauge non linear de Dirac-Born-Infeld e poden ser empregadas para describires  
cordas unidas a branas \cite{CM, Gibbons}.

A condici\'on BPS para a m\'etrica asint\'oticamente plana completa das 
D3-branas foi analizada num\'ericamente na referencia \cite{CGS1}. Coma 
resultado deste estudio obtense unha descrici\'on precisa do proceso de 
creaci\'on das cordas que ten lugar cando d\'uas branas se cruzan entre 
elas, \ie, do chamado efecto de Hanany-Witten \cite {HW, HWothers}. Estes 
resultados foron xeralizados en \cite{CGS2} para o caso de bari\'ons en 
teor\'\i as gauge confinantes. Asemade, na referencia \cite{Craps} a 
ecuaci\'on diferencial BPS foi obtida coma unha condici\'on que ten que 
ser satisfeita co elo de saturar unha cota inferior na enerx\'\i a. O  
cap\'\i tulo 3 est\'a baseado no artigo \cite{Baryon}. Nel est\'udiase a 
din\'amica do volume de universo dunha D(8-p)-brana mov\'endose baixo 
a acci\'on dos campos gravitacion\'ais e de Ramond-Ramond dun conxunto de 
Dp-branas de fondo con $p\le 6$. A D(8-p)-brana ext\'endese \'o longo das 
direcci\'ons ortogon\'ais \'o volume de universo das Dp-branas do fondo. Para 
$p=3$ o sistema \'e o analizado en \cite{CGS1}. Usando o m\'etodo de 
\cite{Craps}, atopamos unha ecuaci\'on diferencial BPS de primeira orde 
tal que as s\'uas soluci\'ons tam\'en satisfacen as ecuaci\'ons de 
movemento. Esta ecuaci\'on BPS foi integrada anal\'\i ticamente tanto 
na m\'etrica pr\'eto do horizonte coma na asint\'oticamente plana. Para 
$p\le 5$ as soluci\'ons que se atopan son similares \'as descritas en 
\cite{CGS1}, \ie , o volume de universo das D(8-p)-branas ten picos que 
poden ser interpretados coma tubos de fluxo feitos de cordas fundament\'ais. 
A soluci\'on anal\'\i tica permite dar unha descrici\'on da forma destes tubos, 
da s\'ua enerx\'\i a e do proceso de creaci\'on de cordas a partires de 
volumes de universo de D-branas (ver figura \ref{barfig5r}).
\begin{figure}
\centerline{\epsffile{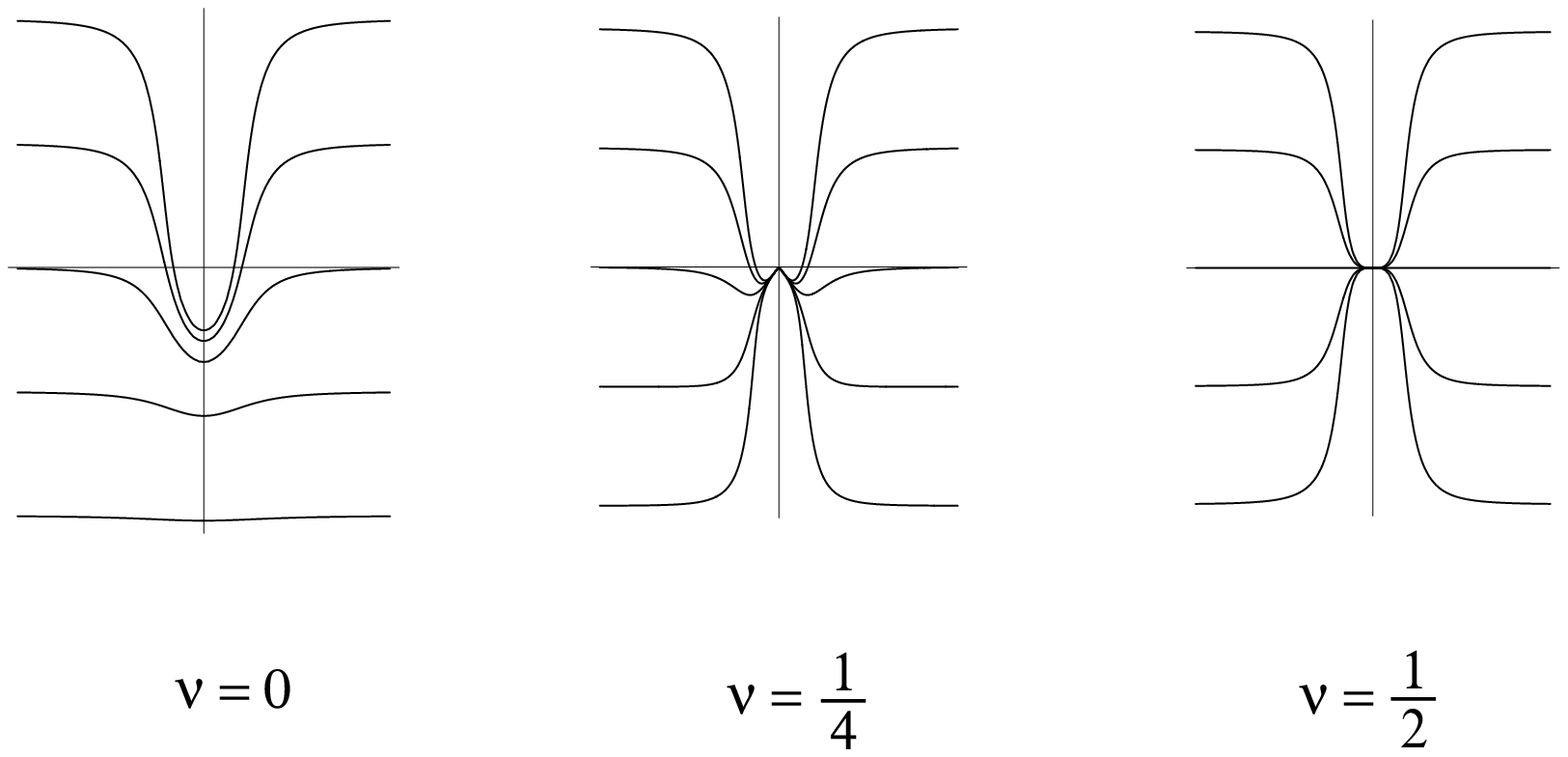}}
\caption{Soluci\'ons da ecuaci\'on diferencial BPS para a m\'etrica
asint\'oticamente plana (eq.~(\ref{baronueve})) para $p=3$ e distintos 
valores de $\nu$ e $z_{\infty}$. Os tubos son feixes de $(1-\n)N$ (tubo 
superior) e $\n N$ (tubo inferior) cordas fundament\'ais, con $0\le\n\le 1$.}
\label{barfig5r}
\end{figure}
Estas configuraci\'ons saturan unha cota BPS e son $1/4$ 
supersim\'etricas. Na secci\'on \ref{barsecsym} am\'osase que a ecuaci\'on 
diferencial BPS \'e precisamente o requerimento que un ten que impo\~ner 
\'o embebemento da brana no fondo de tal xeito que se preserve $1/4$ da 
supersimetr\'\i a do espacio-tempo.

\medskip
\section{Estabilizaci\'on por fluxo} 
\medskip                                             
\setcounter{equation}{0}

Nun artigo recente \cite{Bachas},  Bachas, Douglas e Schweigert amosaron 
como as D-branas en variedades de grupo son estabilizadas de tal xeito 
que non colapsan (ver tam\'en ref. \cite{Pavel}). O modelo concreto estudiado 
en \cite{Bachas} foi o movemento dunha D2-brana na xeometr\'\i a da variedade 
do grupo $SU(2)$. Topol\'oxicamente, $SU(2)$ \'e equivalente a unha tres-esfera 
$S^3$ e a D2-brana est\'a embebida nesta $S^3$ \'o longo dunha d\'uas-esfera 
$S^2$ que, nun sistema de coordanadas esf\'ericas, est\'a a \'angulo de latitude 
$\th$ constante. A din\'amica da D2-brana est\'a determinada pola acci\'on de 
Dirac-Born-Infeld, na que un campo gauge est\'a activado no volume de universo. 
Un ingrediente esencial nesta an\'alise \'e a condici\'on de cuantizaci\'on do 
fluxo no volume de universo. A condici\'on de cuantizaci\'on p\'odese acadar 
acoplando a D2-brana a unha corda F1. Para o caso dunha D2-brana a condici\'on 
de cuantizaci\'on ven dada por
\be
\int_{S^2}\,F\,=\,{2\pi n\over T_f}\,\,,
\,\,\,\,\,\,\,\,\,\,\,\,\,\,\,\,\,
n\in\ZZ\,\,.
\label{resqc1}
\ee

Empregando a condici\'on de cuantizaci\'on un pode atopar de xeito sinxelo a 
forma da intensidade de campo gauge no volume de universo en termos dun 
enteiro de cuantizaci\'on, e tam\'en a enerx\'\i a da D2-brana. O m\'\i nimo desta 
enerx\'\i a determina o embebemento da brana na variedade de grupo, e isto 
sucede para un conxunto finito de \'angulos $\th$. Resulta que as configuraci\'ons 
est\'aticas atopadas por este m\'etodo son estables baixo pequenas perturbaci\'ons 
e coinciden exactamente coas que se acadan considerando a CFT (teor\'\i a de campos 
conforme) correspondente a cordas abertas en variedades de grupo \cite{KS,KO}.

Seguindo a an\'alise de \cite{PR}, no cap\'\i tulo 4 est\'udiase o movemento 
dunha D(8-p)-brana no fondo dun conxunto de Dp-branas paralelas. A rexi\'on 
externa da m\'etrica das Dp-branas ten simetr\'\i a rotacional $SO(9-p)$, que 
\'e manifesta cando se escolle un sistema de coordenadas esf\'ericas. Neste 
sistema de coordenadas def\'\i nese naturalmente unha esfera $S^{8-p}$ e 
a condici\'on de latitude constante na $S^{8-p}$ determina unha $S^{7-p}$. 
Se unha D(8-p)-brana se embebe neste fondo de tal xeito que estea arrollada 
nesta esfera a latitude constante $\,S^{7-p}\subset S^{8-p}$ e extendida 
na direcci\'on radial at\'opanse configuraci\'ons con \'angulo $\th$ 
constante (ver figura \ref{flfig1r}), onde o \'angulo polar $\th$ s\'o pode 
tomar un conxunto discreto de valores que ven dado polas soluci\'ons da ecuaci\'on
\begin{figure}
\centerline{\hskip -.8in \epsffile{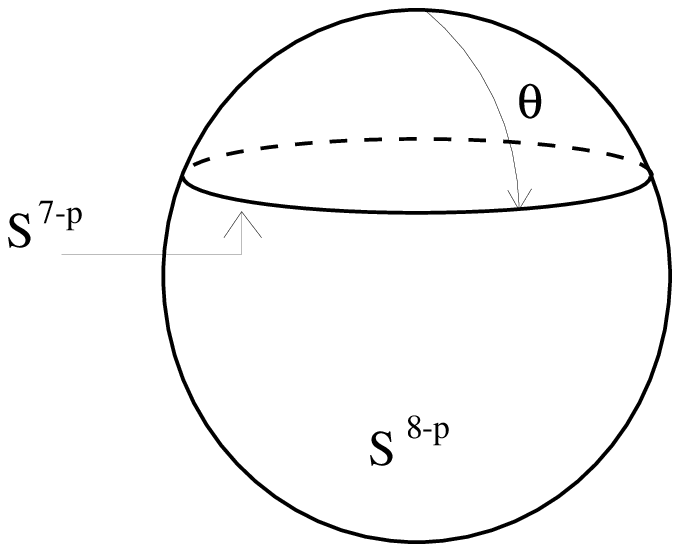}}
\caption{Os puntos da esfera $S^{8-p}$ co mesmo \'angulo polar 
$\th$ definen unha esfera $S^{7-p}$. O \'angulo $\th$ representa a 
latitude na $S^{8-p}$,  medida dende un dos polos.}
\label{flfig1r}
\end{figure}
\be
\L_{p,n}(\bar\th_{p,n})\,=\,0\,,\ \ \ n\le N\,.
\label{resangulos}
\ee
A condici\'on de cuantizaci\'on (\ref{resqc1}) involucra un campo magn\'etico 
no volume de universo. Neste caso o campo gauge no volume de universo \'e 
el\'ectrico e a condici\'on de cuantizaci\'on empregada foi 
\be
\int_{S^{7-p}}\,\,d^{7-p}\th\,\,\,
{\partial \,{\cal L}\over\partial F_{0,r}}\,=\, n\,T_{f}\,\,,
\label{fl1dsieter}
\ee
con $n\in\ZZ$.

A an\'alise para fondos de Dp-branas foi extendido a fondos de estados 
ligados non-threshold, empreg\'andose as condici\'ons de cuantizaci\'on 
correspondentes a campos el\'ectrico e magn\'etico no volume de universo de 
xeito combinado e como resultado acadouse de novo un  conxunto discreto de 
\'angulos nos que a D-brana de proba correspondente era estabilizada. Estes 
\'angulos son de novo as soluci\'ons da ecuaci\'on (\ref{resangulos}), no caso 
do estado ligado (NS5,Dp) con $p=5$. O mecanismo de estabilizaci\'on por fluxo 
na teor\'\i a  M foi estudiado na secci\'on \ref{secflM5}, onde se propuxo unha 
xeralizaci\'on da cuantizaci\'on dos campos gauge no volume de universo:
\be
\int_{S^3}\,F\,=\,{2\pi n\over T_{M2}}\,\,,
\,\,\,\,\,\,\,\,\,\,\,\,\,\,\,\,\,
n\in\ZZ\,\,.
\ee

\medskip
\section{Gravit\'ons xigantes} 
\medskip                                             
\setcounter{equation}{0}

Unha das cousas m\'ais interesantes que xurdiron recentemente na teor\'\i a 
de cordas \'e o feito de que un sistema pode incrementar o seu tama\~no cando 
o momento aumenta. Existen varias manifestaci\'ons deste fen\'omeno, que \'e 
oposto \'a intuici\'on de teor\'\i a de campos est\'andar, en varios contextos 
relacionados coa teor\'\i a de cordas, como a conexi\'on infravermella ultravioleta 
\cite{holography} ou a xeometr\'\i a non conmutativa \cite{NCG}.

Na referencia \cite{GST} McGreevy, Susskind e Toumbas atoparon outro exemplo de 
crecemento en tama\~no co incremento de enerx\'\i a. Estes autores consideraron 
unha part\'\i cula non masiva mov\'endose no espacio-tempo do tipo $AdS_m\times 
S^{p+2}$ e atoparon que existe unha configuraci\'on na que unha brana expandida 
(o gravit\'on xigante) ten exactamente os mesmos n\'umeros cu\'anticos que a 
part\'\i cula puntual. Esta brana expandida est\'a arrollada na parte esf\'erica 
do espacio-tempo sendo estabilizada contra o seu colapso polo fluxo do 
campo gauge de Ramond-Ramond. O tama\~no do gravit\'on xigante medra co 
momento angular e como o radio da brana non pode ser maior que o radio do 
espacio-tempo un at\'opase con que existe unha cota superior para o momento 
da brana. Isto \'e unha realizaci\'on do chamado principio de exclusi\'on da 
teor\'\i a de cordas. Asemade, nas referencias \cite{GMT, HHI} foi probado 
que os gravit\'ons xigantes de \cite{GST} son configuraci\'ons BPS que 
preservan a mesma supersimetr\'\i a que o gravit\'on puntual. Tam\'en foi 
amosado en \cite{GMT, HHI} que existen gravit\'ons expandidos na parte de 
$AdS$ do espacio-tempo que, nembargantes, non te\~nen unha cota superior no 
seu momento angular debido \'a natureza non compacta da variedade $AdS$.

A imaxe f\'\i sica xeral que un ten a partires destes resultados \'e que 
para momentos altos a aproximaci\'on linearizada de supergravidade non \'e 
v\'alida e un est\'a forzado a introducires interacci\'ons para describir a 
din\'amica dos modos non masivos da teor\'\i a. Un procedemento efectivo 
para representar estas interacci\'ons \'e supo\~ner que as part\'\i culas non 
masivas polar\'\i zanse e conv\'\i rtense nunha brana. O mecansimo responsable 
desta polarizaci\'on \'e o efecto diel\'ectrico de Myers \cite{Myers}.

O inflamento dos gravit\'ons en branas pode acontecer en fondos distintos de 
$AdS_m\times S^{p+2}$. De feito en \cite{DTV} foron atopadas configuraci\'ons 
de gravit\'ons xigantes de D(6-p)-branas mov\'endose  na xeometr\'\i a pr\'eto 
do horizonte dun fondo dilat\'onico creado por un conxunto de Dp-branas. No 
cap\'\i tulo 5 at\'opanse configuraci\'ons de gravit\'ons xigantes para branas 
de proba que se moven na xeometr\'\i a creada por un conxunto de estados 
ligados non-threshold do tipo (D(p-2), Dp) para $2\le p \le 6$, e ext\'endese 
esta an\'alise para fondos de teor\'\i a M xerados por un conxunto de estados 
ligados non-threshold do tipo (M2,M5). \'Ambolos dous fondos son $1/2$ 
supersim\'etricos. As soluci\'ons (D(p-2), Dp) est\'an caracterizadas por un 
campo $B$ de Kalb-Ramond distinto de cero dirixido \'o longo da Dp-brana, do 
sector de Neveu-Schwarz da supercorda, xunto cos correspondentes campos de RR. 
Neste fondo colocamos unha brana de proba de tal xeito que capture tanto o 
fluxo de RR como o fluxo de campo $B$. Este derradeiro requerimento implica 
que temos que extender a proba \'o longo de d\'uas direcci\'ons paralelas 
\'as branas de fondo, e polo tanto a proba ser\'a unha D(8-p)-brana arrollada 
nunha esfera $S^{6-p}$ transversa \'o fondo e extendida \'o longo do plano 
non conmutativo paralelo \'o fondo. No caso da teor\'\i a M o fondo est\'a 
caracterizado por unha tres-forma autodual $C$ e a brana de proba \'e unha M5 
arrollada nunha esfera $S^2$ e extendida \'o longo dun hiperplano paralelo \'o fondo 
de tal xeito que o fluxo do campo $C$ \'e capturado. Verif\'\i case que 
con certos valores do campo gauge no volume de universo p\'odense atopar 
configuraci\'ons da D(8-p) ou da M5-brana que se comportan como unha 
part\'\i cula non masiva. A enerx\'\i a destes gravit\'ons xigantes \'e 
exactamente a mesma que a de part\'\i culas non masivas mov\'endose na m\'etrica 
do fondo. Xen\'ericamente, a brana cae no centro do potencial gravitatorio 
\'o longo dunha traxectoria que \'e calculada expl\'\i citamente. A proxecci\'on 
de supersimetr\'\i a introducida pola brana rompe completamente a supersimetr\'\i a 
do fondo do mesmo xeito que unha onda que se propaga coa velocidade do centro de 
masas da brana. Polo tanto, tam\'en dende o punto de vista de supersimetr\'\i a, 
a brana expandida mimetiza \'a part\'\i cula non masiva.


\end{document}